\documentclass[11pt,twoside,dvips]{article}

\usepackage{graphicx,color}
\usepackage{times}
\usepackage{booktabs}
\usepackage{cite}
\usepackage{epsfig}
\usepackage{feynmp}
\usepackage{color}
\usepackage{mciteplus}

 \usepackage{a4wide}
 \newcommand{\figscale}{1.0}
 \newenvironment{eqn}{\begin{equation}}{\end{equation}}
 \newenvironment{eqnar}{\begin{eqnarray}}{\end{eqnarray}}
 \newcommand{\capt}[1]{\caption{#1}}

\newcommand{\mrm}[1]{\mathrm{#1}}
\newcommand{\mbf}[1]{\mathbf{#1}}
\newcommand{\tsc}[1]{\textsc{#1}}
\newcommand{\ttt}[1]{\texttt{#1}}

\newcommand{\pT}[1]{\ensuremath{p_{\perp #1}}}

\newcommand{\TeV}{\,\mbox{Te\kern-0.2exV}}
\newcommand{\GeV}{\,\mbox{Ge\kern-0.2exV}}
\newcommand{\MeV}{\,\mbox{Me\kern-0.2exV}}
\newcommand{\keV}{\,\mbox{ke\kern-0.2exV}}
\newcommand{\eV}{\,\mbox{e\kern-0.2exV}}
\newcommand{\qbar}{\ensuremath{\bar{q}}}
\renewcommand{\d}[1]{\ensuremath{\mrm{d}#1}\hspace*{0.2em} }
\newcommand{\obs}{\ensuremath{\mathcal{O}}}

\newcommand{\PS}{\ensuremath{\Phi}}
\newcommand{\dPS}[1]{\d{\PS_{#1}}}

\newcommand{\kqe}{\kappa_E}
\newcommand{\IntA}{{{\cal A}}}
\def\dash{\hbox{-\kern-.02em}}
\newcommand{\IntAtrialemit}{{{\cal A}}_{\mrm{trial\dash{}emit}}}
\newcommand{\IntAtrialsplit}{{{\cal A}}_{\mrm{trial\dash{}split}}}

\newcommand{\abar}{\ensuremath{\bar{a}}}
\newcommand{\atrial}{\ensuremath{a_{\mathrm{trial}}}}
\newcommand{\abartrial}{\ensuremath{\bar{a}_{\mathrm{trial}}}}
\newcommand{\atrialemit}{\ensuremath{a_{\mathrm{trial\dash{}emit}}}}
\newcommand{\abartrialemit}{\ensuremath{\bar{a}_{\mathrm{trial\dash{}emit}}}}
\newcommand{\atrialsplit}{\ensuremath{a_{\mathrm{trial\dash{}split}}}}
\newcommand{\abartrialsplit}{\ensuremath{\bar{a}_{\mathrm{trial\dash{}split}}}}
\newcommand{\qe}[1]{Q_{E#1}^2}
\newcommand{\bh}{b}
\newcommand{\sect}[1]{sec.~\ref{#1}}
\newcommand{\eq}[1]{eq.~(\ref{#1})}

\newcommand{\pll}{P^{\mrm{LL}}}
\newcommand{\pme}{P^{\mrm{ME}}}
\newcommand{\pari}{P_{\mrm{Ari}}}
\newcommand{\pps}{P^{\mrm{PS}}}
\newcommand{\pimp}{P_{\mrm{imp}}}
\newcommand{\Al}{\tsc{Alpgen}}
\newcommand{\Ar}{\tsc{Ariadne}}

\newcommand{\Fw}{\tsc{Mc@nlo}}
\newcommand{\Hw}{\tsc{Herwig}}

\newcommand{\Mg}{\tsc{MadGraph}}
\newcommand{\Py}{\tsc{Pythia}}
\newcommand{\Sh}{\tsc{Sherpa}}
\newcommand{\Vc}{\tsc{Vincia}}
\newcommand{\Pw}{\tsc{Powheg}}

\newcommand{\plotset}[2]{
\scalebox{\figscale}{
\mbox{\hspace*{-0.6cm}
\includegraphics*[scale=0.4]{#1}\hspace*{-12mm}
\raisebox{6mm}{
\includegraphics*[scale=0.4]{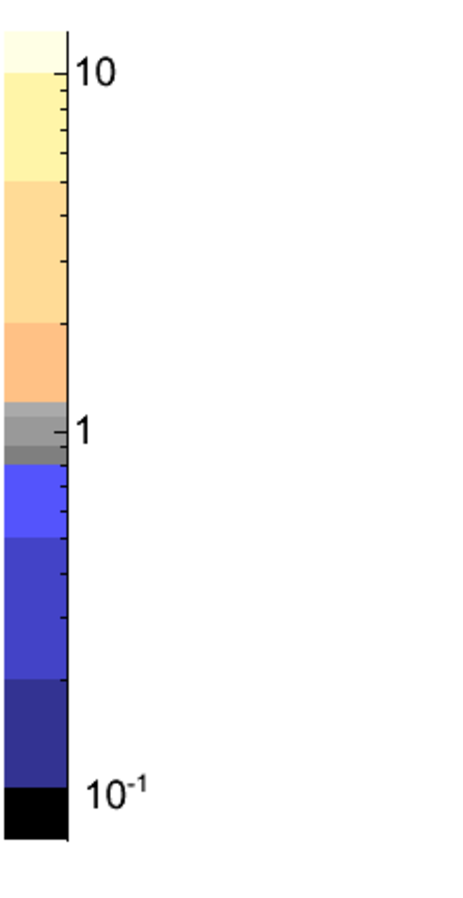}}
\hspace*{-2.3cm}
\includegraphics*[scale=0.4]{#2}\hspace*{-12mm}
\raisebox{6mm}{
\includegraphics*[scale=0.4]{scalez.eps}}
\hspace*{-2.3cm}}
}}

\begin{document}
{\vspace*{-1cm}\flushright\small
FERMILAB-PUB-10-515-T \\
CERN-PH-TH-2010-283\\
Saclay-IPhT--T10/196\\
MCNET-11-05\\
}

\begin{center}
\Large{\bf Higher-Order Corrections to Timelike Jets}\\[5mm]
{\normalsize W.~T.~Giele$\mbf{^1}$, D.~A.~Kosower$\mbf{^2}$,
  P.~Z.~Skands$\mbf{^{3}}$}\\
\end{center}
\begin{center}
{\small 
\noindent $\mbf{^1}$ Theoretical Physics, Fermilab, MS106, Batavia
60510-0500, USA\\
\noindent $\mbf{^2}$ Institut de Physique Th\'{e}orique, CEA--Saclay, F-91191 Gif-sur-Yvette cedex, France\\
\noindent $\mbf{^3}$ Theoretical Physics, CERN, CH-1211 Geneva 23, Switzerland\\}
\end{center}
\begin{abstract}
We present a simple formalism for the evolution of timelike jets in which
tree-level matrix element corrections can be systematically
incorporated, up to arbitrary parton multiplicities and over all of
phase space, in a way that exponentiates the matching corrections. 
The scheme is cast as a shower
Markov chain which generates one single unweighted event sample, 
that can be passed to standard hadronization models. 
Remaining perturbative uncertainties are estimated 
by providing several alternative weight sets for the same events, 
at a relatively modest additional overhead. 
As an explicit example, we consider $Z \to q\qbar$ evolution with 
unpolarized, massless quarks and include several formally subleading 
improvements as well as matching to tree-level matrix elements 
through $\alpha_s^4$. The resulting algorithm is 
implemented in the publicly available
\Vc\ plugin\footnote{Available from the
  web site: \ttt{http://projects.hepforge.org/vincia/}} to the \Py8
event generator.  
\end{abstract}

\section{Introduction}
\label{sec:intro}

The experimental program now underway at the Large Hadron Collider (LHC)
will make extensive demands on theorists' ability to predict background
processes, governed by Standard-Model physics.  These predictions
require perturbative computations in the electroweak sector, and
call on both perturbative and nonperturbative physics in quantum
chromodynamics.  The dependence on nonfactorized and nonperturbative
QCD is not yet amenable to a first-principles calculation, and must
therefore rely on experimental measurements of the parton distribution
functions, as well as models for hadronization and the underlying event.
An appropriate choice of observables --- infrared- and collinear-safe
ones --- can minimize (but not eliminate) the dependence on these models.
The factorizable perturbative component of the backgrounds can be
computed systematically. 

An essential class of observables for new-physics searches are multi-jet
differential cross sections, typically with events also required
to include decay products of one or more 
electroweak vector bosons. Jet shapes and jet substructure observables
are also important, both from a 
calibration point of view as well as to search for decays of boosted objects  
\cite{Butterworth:2007ke,Abdesselam:2010pt}.
The two basic approaches to computing perturbative contributions are via
a fixed-order expansion in powers of the strong coupling $\alpha_s$, 
and in a parton-shower approach which resums leading (LL) and possibly 
next-to-leading logarithms (NLL) of large ratios of scales multiplying the
strong coupling.  The former approach can be carried out
using widely-available tools
to leading order (LO) in $\alpha_s$ for essentially any jet multiplicity,
for a growing list of processes to next-to-leading order~(NLO), and for
a select list to next-to-next-to-leading order~(NNLO).  It
sacrifices a detailed picture of each event, and of jet substructure,
in favor of of a systematically improved description of the ``hard'' or
wide-angle radiation reflected in the several jets.  The parton-shower
approach  (see \cite{Buckley:2011ms} for a recent review) 
favors developing an accurate picture of the ``soft'' or
 ``collinear'' radiation that dresses the hard event, thereby providing
an event-by-event description of jet substructure as well as allowing
the incorporation of a nonperturbative model for the final-state
hadrons. 

From a theoretical point of view, the two approaches correspond to 
summing up different but overlapping sets of contributions to the
short-distance perturbative matrix elements.  Simply adding the two
would yield an overcounting of the common contributions, and the
separation of the two sets is somewhat delicate.
The recent decade has seen the appearance
of a number of strategies for combining or
``matching'' these two approaches for general final states. 
These can be broadly described as ``slicing'', ``subtraction'',
or ``unitarity'' approaches.  

In a slicing approach, the phase space for multiple emissions
is separated into
two regions.  In one, the calculation uses the leading-order
matrix element; in the other, the leading-log approximation as
given by a parton shower.  
The schemes introduced by Mangano (MLM) in the days
of the top quark discovery and formalized 
later~\cite{Mangano:2006rw,Hoche:2006ph}, and by Catani, Krauss, Kuhn, and Webber 
(CKKW)~\cite{Catani:2001cc} 
 are examples of a slicing approach. The CKKW approach has been 
implemented
within the \Sh{} framework~\cite{Gleisberg:2008ta}; the MLM one, using
\Al{}~\cite{Mangano:2002ea} interfaced~\cite{Boos:2001cv,Alwall:2006yp}
 to both \Py{}~\cite{Sjostrand:2006za} and \Hw{}~\cite{Corcella:2000bw}.  
Refined versions of the CKKW method have since been
introduced by L\"onnblad
\cite{Lonnblad:2001iq,Lavesson:2005xu} in the context of the color-dipole
model~\cite{Gustafson:1986db,Gustafson:1987rq} and
implemented in the \Ar\ generator \cite{Lonnblad:1992tz}, as well as
by Mrenna and Richardson~\cite{Mrenna:2003if} 
using \Mg~\cite{Maltoni:2002qb}, 
again interfaced to \Hw\ and \Py.
The original strategy~\cite{Seymour:1994we,Seymour:1994df}
for matching in \Hw{}, for one emission beyond a basic process, also follows
this approach and may be seen as a precursor to the CKKW
formalism.

In a subtractive approach, the shower approximation is subtracted
from the exact fixed-order matrix element, and two classes of events
are generated: standard events and counter-events.  The latter may
have negative weights.
The \Fw{} program built on \Hw{} is an example of subtractive
matching to one loop \cite{Frixione:2002ik}. 
Other subtraction implementations include
a \Sh{} implementation~\cite{Schumann:2007mg} and
one by Dinsdale, Ternick, and Weinzierl (DTW)~\cite{Dinsdale:2007mf}
based on the Catani-Seymour (CS) method~\cite{Catani:1996vz} of fixed-order
calculations. 
Finally, the \textsc{Menlops} scheme
\cite{Hamilton:2010wh} probably
represents the most advanced current matching method, combining \Pw{}
(a unitarity-based variant of \Fw, see below) with a slicing-based
matching for multijet emissions. 

Note, however, that in both slicing and subtraction, any subleading
divergencies in the matched matrix elements beyond one additional
parton are not regulated by the (LL) shower, and hence all the multileg
schemes --- MLM, CKKW, and \textsc{Menlops} --- are forced to
introduce a ``matching scale'' below which only the pure LL shower is
used. While such schemes therefore guarantee the rates of hard additional jets
to be correct to LO, the same is not guaranteed for jet
substructure, which can be explicitly sensitive to multiparton
correlations below the matching scale and beyond LL \cite{Almeida:2010pa}. 

In a unitarity approach, to maintain a sequence of unit-weight events,
the selection of branching events is modified by a veto depending on the ratio
of the exact matrix element to the shower approximation. Since the
correction is applied on the splitting probability itself, it is
automatically resummed to all orders by the shower Sudakovs, with real
and virtual corrections canceling order by order in perturbation
theory. 
Indeed, the original approach to matching, carried out for one additional 
emission beyond a basic process in \Py{}
\cite{Bengtsson:1986et,Bengtsson:1986hr}, follows this approach. It is
also used in \Pw{} \cite{Frixione:2007vw}, there combined with a
subtractive matching to NLO. 
The proposal of matching by Sudakov reweighting by Nagy and
Soper~\cite{Nagy:2006kb} is also within this class. However, 
the unitary approach to matching has so far only been worked out for a
single emission. We shall here generalize it to an arbitrary number of
emissions, arriving at the equivalent to \textsc{Menlops} but based on
unitarity instead of slicing for the additional emissions. This will
allow us to extend the matching over all of phase space, and should
therefore result in a more accurate modeling not only of jet rates but
also of jet substructure. 

Most showers, with the exception of \Ar{} and
the Winter--Krauss shower~\cite{Winter:2007ye},
 are based on collinear factorization,
which is to say $1\rightarrow2$ branching in shower evolution.
(\Py~8 combines a $1\rightarrow2$ splitting probability with
a $2\rightarrow3$ phase-space mapping.)
In the present paper, we continue the development
of a leading-log (LL) parton shower~\cite{Giele:2007di}
 based on dipole antenn\ae{}, that is $2\rightarrow3$
branching.  We choose a simpler context than hadron collisions,
that of electron--positron collisions.  This allows us to set
aside the questions of initial-state emission as well as those
of the underlying event. 

In \sect{sec:conventions}, we describe in greater detail the ingredients
needed for such a shower, as well as our normalization conventions,
and compare the origins of different singularities and corresponding
logarithms in different shower formalisms.  We also discuss the
different matching approaches in more detail. In \sect{sec:evolution},
we discuss the evolution integral, and show how to cast it in
a general form whose specializations correspond to a wide variety
of interesting evolution variables.  We then solve the resulting 
evolution equation.  In \sect{sec:showering}, we discuss the shower
algorithm, as well as improvements that can be made to its logarithmic
accuracy.
In \sect{sec:matching}, we discuss the
details of matching the dipole-antenna shower to tree-level matrix
elements, at both leading and subleading
color. The procedure we use to
evaluate the remaining perturbative uncertainties is described in
\sect{sec:uncertainties}, and 
in \sect{sec:hadronization}, we comment on hadronization;
in \sect{sec:LEP}, we compare the results of running the unitarity-based
approach implemented in \Vc{} to LEP data and to \Py~8.  We make some
concluding remarks in \sect{sec:conclusions}.

\section{Nomenclature and Conventions}
\label{sec:conventions}

In this section, we introduce the basic elements of our
perturbative formalism, which is largely based 
on ref.~\cite{Giele:2007di}. First, in \sect{sec:Markov}, we illustrate how 
the KLN theorem may be used to rewrite the coefficients of perturbation
theory as the expansion of an all-orders Markov chain, using NLO as an 
explicit example. Then, in \sect{sec:shower},
we briefly describe each of the ingredients that enter our 
dipole-antenna shower formalism. 

\subsection{Perturbation Theory with Markov Chains \label{sec:Markov}}
Consider the Born-level cross section for an arbitrary hard process,
$H$, differentially in an arbitrary infrared-safe observable $\cal O$, 
\begin{eqn}
\left.   \frac{\d{\sigma_H}}{\d{\cal O}}\right\vert_{\mbox{\textcolor{black}{Born}}}
 = \int \d{\Phi_H} \ |M_H^{(0)}|^2 \ \delta({\cal
   O}-{\cal O}(\{p\}_H))~,
\label{eq:starting}
\end{eqn}
where the integration runs over the full final-state on-shell phase space of
$H$ (this expression and those below would also apply to hadron collisions
were we to include integrations over the parton distribution functions
in the initial state), and the $\delta$ function projects out a
1-dimensional slice defined by $\cal O$ evaluated on the 
set of final-state momenta 
which we denote $\{p\}_H$ (without the $\delta$ function, the
integration over phase space would just give the total cross section, not the
differential one).
 
To make the connection to parton showers, and to discuss
all-orders resummations in that context, 
we may insert an operator, ${\cal S}$, that acts on the Born-level
final state \emph{before} the observable is evaluated, i.e., 
\begin{eqn}
\left.   \frac{\d{\sigma_H}}{\d{\cal
    O}}\right\vert_{\mbox{\textcolor{black}{${\cal S}$}}}
 = \int \d{\Phi_H} \ |M_H^{(0)}|^2 \ {\cal S}(\{p\}_H,{\cal O})~.
\end{eqn}
Formally, this operator --- the evolution operator --- will be
responsible for generating all (real and virtual) 
higher-order corrections to the Born-level expression. 
The measurement $\delta$ function appearing explicitly in
\eq{eq:starting} is now implicit in ${\cal S}$.
(Ultimately, non-perturbative corrections can also be included.)   

Algorithmically, we shall cast $\cal S$ as an iterative Markov chain, 
with an evolution parameter that formally 
represents the factorization scale of the event, below which all
structure is summed over inclusively. As the Markov chain develops,
the evolution parameter will go towards zero, and the event structure
will become more and more exclusively resolved. A transition from a
perturbative evolution to a non-perturbative one can also be
inserted, at an appropriate scale, typically 
around $1$~GeV. This scale thus 
represents the lowest perturbative scale that can appear in the
calculations, with all perturbative corrections below it
summed over inclusively.

It is instructive to begin by considering the first-order expansion 
the operator must have in order 
to agree with NLO perturbation theory, 
\begin{eqnar}
{\cal S}^{(1)}(\{p\}_H,{\cal O}) & = & \left( 1 + \frac{ 2
  \mrm{Re}[M_H^{(0)}M_H^{(1)*}]}{|M_H^{(0)}|^2}\right) \delta({\cal
   O}-{\cal O}(\{p\}_H))~\nonumber\\
& & + \int \frac{\dPS{H+1}}{\dPS{H}} \frac{|M_{H+1}^{(0)}|^2}{|M_H^{(0)}|^2}
\delta({\cal O}-{\cal O}(\{p\}_{H+1}))~,\label{eq:s1}
\end{eqnar}
with  $M_H^{(1)}$ the one-loop amplitude and the ratio
$\dPS{H+1}/{\dPS{H}}$ in the second line 
representing the phase space of one additional final-state
particle;
we shall return to the associated factorization below. 
The two correction terms  are separately
divergent and hence \eq{eq:s1} only has a symbolic
formal meaning. It requires a regulator for actual calculations. 
Introducing the factorization scale mentioned above, 
and introducing an $n+1\rightarrow n$ mapping of momenta by
summing inclusively over all emissions below it, we obtain, 
instead, the first-order expansion corresponding to an evolution from
the starting scale, $s$ (the c.m.\ energy
squared), down to the scale $Q_E^2$, 
\begin{eqnar}
{\cal S}^{(1)}(\{p\}_H,s,Q^2_E,{\cal O}) & = & \left( 1 + \frac{ 2
  \mrm{Re}[M_H^{(0)}M_H^{(1)*}]}{|M_H^{(0)}|^2} + \int_0^{Q^2_E}
\frac{\dPS{H+1}}{\dPS{H}} \frac{|M_{H+1}^{(0)}|^2}{|M_H^{(0)}|^2} \right) \delta({\cal
   O}-{\cal O}(\{p\}_H))~\nonumber\\
& & + \int_{Q^2_E}^{s} \frac{\dPS{H+1}}{\dPS{H}} \frac{|M_{H+1}^{(0)}|^2}{|M_H^{(0)}|^2}
\delta({\cal O}-{\cal O}(\{p\}_{H+1}))~, \label{eq:s1a}
\end{eqnar}
where the factorization scale, $Q_E$ (a.k.a.\ the ``evolution
scale''), separates resolved from unresolved regions. This expression
is well-defined if the functional form of $Q_E$ properly separates
singular from non-singular regions, i.e., is ``infrared
sensible'' \cite{Skands:2009tb}. 
(Corrections to this expression arising from scales below $Q_E$ 
will be taken into account by eventually letting $Q_E \to 0$.)
Due to the KLN theorem \cite{Kinoshita:1962ur,*Lee:1964is}, the real
and virtual singularities must 
be equal and of opposite sign, thus we can rewrite 
\begin{eqn}
\frac{ 2 \mrm{Re}[M_H^{(0)}M_H^{(1)*}]}{|M_H^{(0)}|^2} = K^{(1)}_H - \int_0^s
\frac{\dPS{H+1}}{\dPS{H}} \frac{|M_{H+1}^{(0)}|^2}{|M_H^{(0)}|^2} ~,
\end{eqn}
where $K^{(1)}_H$ is a non-singular function when the regulator
is removed, allowing us to express
eq.~(\ref{eq:s1a}) as
\begin{eqnar}
{\cal S}^{(1)}(\{p\}_H,s,Q^2_E,{\cal O}) & = & \left( 1 + K^{(1)}_H - 
\int_{Q^2_E}^s 
\frac{\dPS{H+1}}{\dPS{H}} \frac{|M_{H+1}^{(0)}|^2}{|M_H^{(0)}|^2} \right) \delta({\cal
   O}-{\cal O}(\{p\}_H))~\nonumber\\
& & + \int_{Q^2_E}^{s} \frac{\dPS{H+1}}{\dPS{H}} \frac{|M_{H+1}^{(0)}|^2}{|M_H^{(0)}|^2}
\delta({\cal O}-{\cal O}(\{p\}_{H+1}))~.\label{eq:s1b}
\end{eqnar}
In this form, the NLO correction to the total cross section is given
solely by the term $K^{(1)}_H$, with the remaining terms having been
written in an explicitly unitary construction. 

We can also rewrite the exact ratio $|M_{H+1}|^2/|M_{H}|^2$ as a
process-dependent term whose integral is non-singular,
plus a sum over universal singular
ones,
\begin{eqn}
\frac{\dPS{H+1}}{\dPS{H}} 
\frac{|M^{(0)}_{H+1}|^2}{|M^{(0)}_{H}|^2} = \frac{\dPS{H+1}}{\dPS{H}} K^{(0)}_{H+1} + 
\sum_{r} \frac{\d{\PS_{H+1}^{[r]}}}{\dPS{H}} S_r ~,
\end{eqn}
where $r$ runs over ``radiators'', whose precise definition, such as
partons or dipoles, depends on the chosen decomposition of the singular
structures in $|M_{H+1}|^2$, and the superscript $^{[r]}$ on the phase
space factors indicate that each radiator may in principle be
associated with a different phase space factorization.  

By the simple rewritings above, we 
have now obtained a form of the expansion in which the singularity
and unitarity structure of $\cal S$ are both explicitly manifest. 
Deviations from unitarity are associated solely with the non-singular term 
$K^{(1)}_H$, and deviations from the universal radiation functions are
associated solely with the non-singular term $K^{(0)}_{H+1}$. In both
cases, the generalization to higher orders is straightforward. 

In traditional parton showers, all the non-singular terms are dropped,
and hence \emph{only} the unitary singular structure remains, 
\begin{eqnar}
{\cal S}^{(1)}(\{p\}_H,s,Q^2_E,{\cal O}) & = & \left( 1 - 
\sum_r\int_{Q^2_E}^s 
\frac{\d{\PS_{H+1}^{[r]}}}{\dPS{H}} S_r \right) \delta({\cal
   O}-{\cal O}(\{p\}_H))~\nonumber\\
& & + \sum_r \int_{Q^2_E}^{s} \frac{\d{\PS_{H+1}^{[r]}}}{\dPS{H}} 
\ S_r \ \delta({\cal O}-{\cal O}(\{p\}_{H+1}))~.
\end{eqnar}
Exponentiating the leading singularities, we may replace them by the
Sudakov factor, 
\begin{eqn} 
\Delta(\{p\},s,Q^2_j) = \exp\left[-\sum_r\int_{Q^2_{j}}^{s} 
\frac{\d{\PS_{H+1}^{[r]}}}{\dPS{H}} S_r \right]~. \label{eq:sudakov}
\end{eqn}
We thereby obtain the all-orders pure-shower Markov chain, 
\begin{eqn}\begin{array}{rcl}
\displaystyle \hspace*{-1mm} {\cal S}(\{p\}_H,s,Q^2_E,\obs) \hspace*{-1mm}
&  \hspace*{-1mm}= \hspace*{-1mm} &\displaystyle \hspace*{-1mm}
  \underbrace{
\Delta(\{p\}_H,s,Q_E^2)\ 
\delta\left(\obs-\obs(\{p\}_{H})\right) 
}_{\mbox{$H+0$ exclusive above $Q_E$}}\\[9mm]
& & \hspace*{-0.7cm}+ \displaystyle 
\underbrace{
\sum_r \int_{Q_E^2}^{s}
\frac{\d{\PS^{[r]}_{H+1}}}{\d\PS_H} 
\ S_r \ \Delta(\{p\}_H,s,Q^2_{H+1})
 \ {\cal S}(\{p\}_{H+1},Q^2_{H+1},Q^2_{E},\obs)}_{\mbox{$H+1$ inclusive above $Q_E$}}
~.\label{eq:markov} \hspace*{-1mm}
\end{array}
\end{eqn}
The shower
may exponentiate the entire set of universal singular terms, or only
a subset of them (for example, the terms leading in the number of colors
$N_c$).
More on the Markov formalism can be found in ref.~\cite{Giele:2007di}. 
We hope this brief introduction serves 
to put the developments below in context, and note that we will return to the
restoration of the finite terms in the section on matching. 

\subsection{Dipole-Antenna Showers \label{sec:shower}}

In leading-log dipole-antenna showers, the fundamental step
is a Lorentz invariant $2\to 3$ branching process by which two
on-shell ``parent'' partons are replaced by three on-shell ``daughter''
partons.
This $2\to 3$ process makes use of three ingredients:
\begin{enumerate}
\item An \emph{antenna function} that captures the leading tree-level
  singularities of QCD matrix elements. This is the equivalent of the
  splitting functions used in traditional parton showers, with some
  important differences, as we discuss below.
\item A \emph{kinematics map}, specifying how the
post-branching momenta are related to the pre-branching ones. This is
the equivalent of the ``recoil strategy'' of traditional parton showers.  
\item An antenna \emph{phase space} --- an exact, momentum-conserving and 
  Lorentz-invariant factorization of the pre- and post-branching phase
  spaces. Traditional parton showers, on the other hand, are based on a 
 phase-space factorization which is only exact in the
  collinear limit, and momentum conservation may only be imposed 
  {\it a posteriori\/}.
\end{enumerate}
In the following paragraphs, we present the notation and normalization
conventions that we shall use in the rest of the article for 
each of these pieces.

\paragraph{Factorization:} Labeling the three daughter partons $i$, $j$,
and $k$, we write the integral over a three-body matrix element
corresponding to that final state in factorized form as follows,   
\begin{eqn}
 |M_3(p_i,p_j,p_k)|^2 \dPS{3} =|M_2(s)|^2  \dPS{2} \ 
\frac{|M_3(p_i,p_j,p_k)|^2}{|M_2(s)|^2}\frac{\dPS{3}}{\dPS{2}}~,\label{eq:psfac}
\end{eqn}
where the two-parton matrix element we have introduced corresponds to
the ``parent'' configuration, in which we label the partons $I$ and
$K$. The branching process represented by this factorization is thus
$IK\to ijk$, with total Lorentz invariant $s_{ijk}=s_{IK}=s$.  The
$|M_3|^2/|M_2|^2$ factor in \eq{eq:psfac} represents the evolution
kernel, whose (negative) exponential is the Sudakov form factor,
cf.~eq.~(\ref{eq:sudakov}). 

\paragraph{Phase Space:} The dipole-antenna \emph{phase-space measure} is thus
\cite{Giele:2007di} 
\begin{eqn} 
\frac{\dPS{3}}{\dPS{2}} =
\d{s_{ij}} \d{s_{jk}} \frac{\d{\phi}}{2\pi} \frac{1}{16\pi^2\sqrt{\lambda\left(
s,m_{{I}}^2,m_{{K}}^2\right)}}~, \label{eq:phasespace}
\end{eqn}
where the K\"all\'en function, 
\begin{eqn}
\lambda(s,m^2_{{I}},m_{{K}}^2) = 
  s^2+m^4_{{I}}+m^4_{{K} }
  - 2sm^2_{{I}} -2 sm^2_{{K}} - 2 m^2_{{I}}m^2_{{K}}~,
\end{eqn}
 in the denominator reduces to $\sqrt{\lambda} = s$ for massless
 particles.

\paragraph{Antenna Function:}
The ratio of matrix elements appearing in the integrand of
\eq{eq:psfac} is then
decomposed into a symmetry factor, a coupling factor, a color factor,
and an 
\emph{antenna function}, 
\begin{eqn}
\frac{|M_3(p_i,p_j,p_k)|^2}{|M_2(s)|^2} =  S_{IK\to ijk}~ \ g^2 \ \mathcal{C}_{ijk}
\ \bar{a}^0_{ijk}(s,s_{ij},s_{jk}) ~, 
\end{eqn}
where $S$ takes into account potential identical-particle factors as
well as the possible presence of more than one antenna in the parent
($IK$) configuration, $g^2$ is the relevant coupling factor,
$\mathcal{C}_{ijk}$ is a color factor, and $\bar{a}_{ijk}^0$ is a
generic color- and coupling-stripped dipole-antenna function, with
superscript $0$ to denote a tree-level quantity.  The three-particle
matrix element is averaged azimuthally (over $\phi$).
Note that our use of
lower-case letters for the antenna function is intended to signify
that it corresponds to what is called a \emph{sub-antenna} in
ref.~\cite{GehrmannDeRidder:2005cm} for which lower-case letters are
likewise used\footnote{Thus, in the notation of
  ref.~\cite{GehrmannDeRidder:2005cm}, our dipole-antenna functions would
  be $\bar{a}_3^0 = A_3^0$, $\bar{d}_3^0=d_3^0$, $\bar{e}_3^0=\frac12
  E_3^0$, $\bar{f}_3^0=f_3^0$, and $\bar{g}_3^0=\frac12 G_3^0$.}.
 
\begin{figure}
\begin{center}
\scalebox{\figscale}{\includegraphics*[scale=0.56]{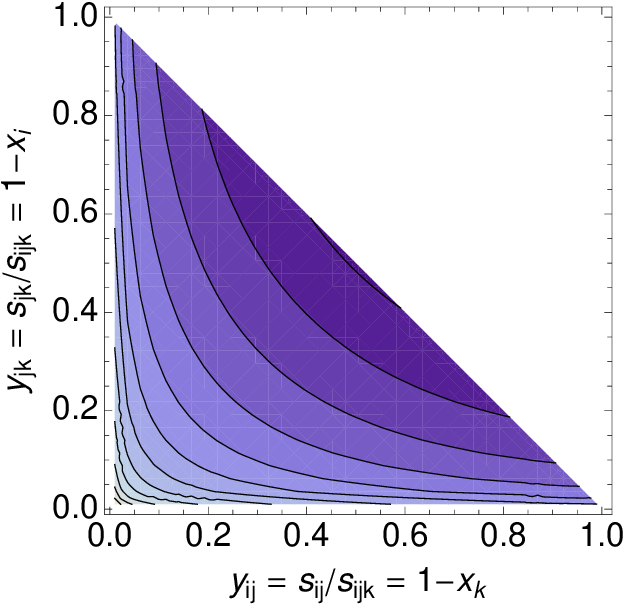}}
\capt{Contours of constant value of 
the antenna function, $\bar{a}^0_{ijk}$ for $q\qbar\to qg\qbar$
  derived from $Z$ decay as function of the two phase-space
  invariants, with an arbitrary normalization and a logarithmic color scale. 
  Larger values are shown in lighter shades.
  The (single) collinear divergences sit on the axes,
  while the (double) soft divergence sits at the origin.
\label{fig:zant}}
\end{center}
\end{figure}
For illustration, contours of constant value of
$\bar{a}^0_{qg\qbar}(s,s_{qg},s_{g\qbar})$  
as derived 
from $Z$ decay are shown in fig.~\ref{fig:zant}, over the
$2\to 3$ phase space, with an arbitrary
normalization and a logarithmic color scale. This function is
 called $A_3^0$ in ref.~\cite{GehrmannDeRidder:2005cm} and is 
identical to the radiation function 
used for $q\qbar\to qg\qbar$ splittings in \Ar. One clearly sees the
large enhancements towards the edges of phase space, with 
a double pole (the overlap of two singularities, usually called
soft \emph{and} collinear) sitting at
the origin, and single singularities (soft \emph{or} collinear) 
localized on the axes.

Writing the coupling factor as $g^2=4\pi \alpha_s$ and combining it with 
the phase space factor, eq.~(\ref{eq:phasespace}), we have the
following antenna function normalization
\begin{eqn}
a^0_{IK\to ijk}(s,s_{ij},s_{jk}) \equiv \frac{1}{\sqrt{\lambda\left(
s,m_{I}^2,m_{K}^2
\right)}}\frac{\alpha_s}{4\pi} \ 
\mathcal{C}_{ijk} \ \bar{a}^0_{ijk}(s,s_{ij},s_{jk}) ~.
\end{eqn}
That is, we use the notation $\bar{a}$ for the coupling- and
color-stripped antenna function, and the notation $a$ for the
``dressed'' antenna function, i.e., including its coupling, color, and
phase-space prefactors. 

Note
that $g^2 \times$(phase-space~normalization) leads to a factor
 $\alpha_s/(4\pi)$ independently of the type of branching.
As we believe that the formalism becomes
more transparent if the origin of each factor is kept clear
throughout, we shall therefore use this factor for \emph{all}
branchings, instead of the more traditional convention of using
$\alpha_s/(2\pi)$ for some branchings and $\alpha_s/(4\pi)$ for
others. Obviously, this convention choice will be compensated by 
our conventions for the color factors and antenna-function
normalizations, such that the final result remains independent of this
choice.

\paragraph{Color Factor:} 
For color factors, we shall systematically adopt a convention
optimized for  shower applications, in which
the color factor tends to $N_C$ in the large-$N_C$ limit whenever a new color
line is created (rather than to $N_C/2$ as is the case for $C_F$ in
the standard normalization) and to unity when no new color lines are
created (rather than $1/2$ which is the case for $T_R$ in the standard
normalization). That is, we shall use
\begin{eqnar}
C_A & = & N_C = 3 ~,\\
\hat{C}_F & = & \frac{N_C^2-1}{N_C} = 8/3 = 2 C_F~,\\
\hat{T}_R & = & 1 = 2 T_R~.
\end{eqnar}
With all other normalizations fixed,  the normalization
of the antenna functions is now 
also unique. Indeed, with
this choice, the radiation functions will turn out to be normalized in
a way that makes their similarities more readily apparent. Another useful
thing about this normalization is that the color factors 
now have a very simple interpretation. They 
provide a one-to-one count of the number of new color degrees of freedom that
have been summed over in any given process. This gives a simpler counting and
interpretation than in the standard normalization. 
Of course, the fact that there are only $N_C^2-1=8$ gluons, 
leads to corrections of order $1/N_C^2$, and here again 
it is trivial to let the 
``na\"ive'' color-line creation factor $C_A$ be replaced by 
$\hat{C}_F$ for, e.g., $q\qbar\to qg\qbar$, without artificially having
to compensate by a factor of 2 in the radiation function ---  the
eikonal part of the radiation function now remains invariant 
(as noted above in the discussion of the
normalization of the radiation functions), and the difference in color
factor is explicitly subleading, as it should be. 
A final argument is the question of which color factor to use,
e.g., for a $qg \to qgg$ emission. In the standard normalization, this could
lead to confusion, since one parent would seem to ``want'' $C_F$ and the
other $C_A$, which differ by a more than a factor of 2. 
In the normalization used here, the difference between
$\hat{C}_F$ and $C_A$ is explicitly subleading in color, and hence it
is clear that 
either of them could be used without any possibility for ambiguity at the
leading-color (LC) level, again placing the proper difference in the proper
place. 

To preempt confusion and 
illustrate how simple the translation between these convention
choices is, consider the
dipole-antenna function for gluon emission off a $q\qbar$ pair
used in the \Ar\ generator \cite{Lonnblad:1992tz}. As the
symmetry factor is unity, this is just the
matrix element squared for $Z\to 3$ divided by the one for $Z\to 2$
multiplied by the aforementioned phase-space factor, 
\begin{eqnar}
\frac{|M(Z\to q g
  \qbar)|^2}{16\pi^2 s|M(Z\to q\qbar)|^2} ~ = ~\frac{1}{s}\frac{2\alpha_s}{3\pi} \frac{(1-y_{ij})^2 +
  (1-y_{jk})^2}{y_{ij}y_{jk}}~, \label{eq:dipole}
\end{eqnar}
where $y_{ij} = s_{ij}/s = 1-x_k$.
The factor $2\alpha_s/(3\pi)$ in the first equation can be rewritten
in two  ways
\begin{eqn}
\frac{2\alpha_s}{3\pi} = \frac{4}{3} \frac{\alpha_s}{2\pi} = C_F
\frac{\alpha_s}{2\pi} ~~~~~~\mbox{or}~~~~~~
\frac{2\alpha_s}{3\pi} = \frac{8}{3} \frac{\alpha_s}{4\pi} = \hat{C}_F
\frac{\alpha_s}{4\pi} ~;
\end{eqn}
it is purely for our own convenience that we
choose the latter normalization. 

In a similar vein one could rewrite the DGLAP splitting kernels
\cite{Altarelli:1977zs}, which are used
in traditional parton showers
\cite{Sjostrand:2006za,Sjostrand:2007gs,Corcella:2000bw,Bahr:2008pv},
as  
\begin{eqnar}
\frac{P_{q_I\to q_ig_j}^0(z)}{s_{ij}} & = &\frac{1}{s_{ij}}\frac{\alpha_s}{2\pi} C_F 
\frac{1+z^2}{1-z} ~=~ \frac{1}{s}\frac{\alpha_s}{4\pi} \hat{C}_F
\frac{1+z^2}{y_{ij}(1-z)},\label{eq:Pz1}\\
\frac{P_{g_I\to g_ig_j}^0(z)}{s_{ij}} &= &\frac{1}{s_{ij}}\frac{\alpha_s}{2\pi} N_C
\frac{(1-z(1-z))^2}{z(1-z)} ~=~ \frac{1}{s}\frac{\alpha_s}{4\pi} 2 N_C \frac{(1-z(1-z))^2}{y_{ij}z(1-z)}~\label{eq:Pz2},
\end{eqnar}
where the gluon radiation function has absorbed a factor of 2 on the r.h.s.~of
the last line, due to the normalization choice.  
We note that, although these expressions
look quite different from the dipole
formula, eq.~(\ref{eq:dipole}), they lead to identical
singularities. This was shown in ref.~\cite{Bengtsson:1986hr} 
by identifying $z$ as the Lorentz invariant energy fraction taken by the
quark, $z=x_i/(x_i+x_k)$, and
adding the radiation from the antiquark, $\qbar_K\to g_j \qbar_k$. 

\paragraph{Shared Singularities: } 
This examination of the different presentations of singularities
brings us to the issue of ``shared singularities''. In
traditional parton showers, as we have just seen, 
the full leading-log radiation pattern 
can only be obtained after summing over pairs of 
partons (which each radiate as independent monopoles), 
and care must be taken in the construction
of the shower to make this sum  approximately
coherent to reproduce the correct singular behavior for soft
wide-angle radiation. 
This \emph{dipole} singularity is the
simplest case of what we 
shall generally refer to as a shared --- or multipole ---
singularity below; radiation whose full
singularity structure (in a particular phase-space limit) 
can only be recovered after summing over two or more
radiators. 
 
\begin{figure}[t]
\begin{center}
\scalebox{\figscale}{
\begin{fmffile}{fmfdual}
\begin{fmfgraph*}(450,45)

\fmfforce{0.08w,0.3h}{l1}
\fmfforce{0.12w,0.45h}{l2}
\fmfforce{0.12w,0.55h}{l3}
\fmfforce{0.08w,0.65h}{l4}

\fmfforce{w-0.08w,0.3h}{r1}
\fmfforce{w-0.12w,0.45h}{r2}
\fmfforce{w-0.12w,0.55h}{r3}
\fmfforce{w-0.08w,0.65h}{r4}

\fmfforce{0.16w,0.5h}{H}
\fmfforce{0.38w,0.5h}{I}
\fmfforce{0.62w,0.5h}{K}
\fmfforce{0.84w,0.5h}{L}

\fmf{plain,right=0.1,foreground=(0.8,,0.8,,0.8)}{l1,H}
\fmf{plain,right=0.05,foreground=(0.8,,0.8,,0.8)}{l2,H}
\fmf{plain,left=0.05,foreground=(0.8,,0.8,,0.8)}{l3,H}
\fmf{plain,left=0.1,foreground=(0.8,,0.8,,0.8)}{l4,H}

\fmf{plain,left=0.1,foreground=(0.8,,0.8,,0.8)}{r1,L}
\fmf{plain,left=0.05,foreground=(0.8,,0.8,,0.8)}{r2,L}
\fmf{plain,right=0.05,foreground=(0.8,,0.8,,0.8)}{r3,L}
\fmf{plain,right=0.1,foreground=(0.8,,0.8,,0.8)}{r4,L}

\fmf{plain,right=0.2,foreground=(0.6,,0.6,,0.6)}{H,I}
\fmf{plain,right=0.08,foreground=(0.6,,0.6,,0.6)}{H,I}
\fmf{plain,left=0.08,foreground=(0.6,,0.6,,0.6)}{H,I}
\fmf{plain,left=0.2,foreground=(0.6,,0.6,,0.6),label=${HI}$}{H,I}

\fmf{plain,right=0.2}{I,K}
\fmf{plain,right=0.08}{I,K}
\fmf{plain,left=0.08}{I,K}
\fmf{plain,left=0.2,label=$IK$}{I,K}

\fmf{plain,right=0.2,foreground=(0.6,,0.6,,0.6)}{K,L}
\fmf{plain,right=0.08,foreground=(0.6,,0.6,,0.6)}{K,L}
\fmf{plain,left=0.08,foreground=(0.6,,0.6,,0.6)}{K,L}
\fmf{plain,left=0.2,foreground=(0.6,,0.6,,0.6),label=${KL}$}{K,L}

\fmfv{d.sh=circ,d.siz=11,foreground=(0.6,,0.6,,0.6),d.filled=60,l.dist=0,label=\textcolor{white}{$H$}}{H}
\fmfv{d.sh=circ,d.siz=11,l.dist=0,label=\textcolor{white}{$I$}}{I}
\fmfv{d.sh=circ,d.siz=11,l.dist=0,label=\textcolor{white}{$K$}}{K}
\fmfv{d.sh=circ,d.siz=11,foreground=(0.6,,0.6,,0.6),d.filled=60,l.dist=0,label=\textcolor{white}{$L$}}{L}

\end{fmfgraph*}
\end{fmffile}}
\scalebox{\figscale}{\sl 
\begin{tabular}{p{9.2cm}ll}
& $\mrm{Coll}(I)$ & $\mrm{Soft}(IK)$ \\[1mm]
\small Parton Shower (DGLAP) &$a_I$& $a_{I} + a_{K}$ \\[1mm]
\small Coherent Parton Shower (\Hw~\cite{Marchesini:1983bm,Corcella:2000bw}, \Py{\sc6}~\cite{Sjostrand:2006za})&$\Theta_{I}a_I$& $\Theta_{I}a_{I} + \Theta_K a_K$ \\[1mm]
\small Global Dipole-Antenna (\Ar\ \cite{Lonnblad:1992tz}, GGG \cite{GehrmannDeRidder:2005cm}, WK~\cite{Winter:2007ye}, \Vc)&  $a_{IK} + a_{HI}$ & $a_{IK}$ \\[1mm]
\small Sector Dipole-Antenna (LP~\cite{Larkoski:2009ah}, \Vc)& $\Theta_{IK}a_{IK} +\Theta_{HI}
a_{HI}$ & $a_{IK}$ \\[1mm]
\small Partitioned-Dipole Shower (SK~\cite{Schumann:2007mg}, 
  NS \cite{Nagy:2007ty}, DTW~\cite{Dinsdale:2007mf}, \Py{\sc8}~\cite{Sjostrand:2007gs}, \Sh) &  $a_{I,K} +
a_{I,H}$ & $a_{I,K} + a_{K,I}$
\end{tabular}}
\capt{Schematic overview of how the full collinear 
  singularity of parton $I$ and the soft singularity of
  the $IK$ pair,
  respectively, originate in different shower
  types. ($\Theta_I$ and $\Theta_K$ represent angular vetos with
  respect to partons $I$ and $K$, respectively, and $\Theta_{IK}$ represents a
  sector phase-space veto, see text.)
\label{fig:gluonchain}}
\end{center}
\end{figure}

A chain of such uniquely labeled and color ordered gluons, which
could, e.g., represent a shower ``event 
record'' at a given point during its evolution, is 
illustrated in fig.~\ref{fig:gluonchain}. 
Below the schematic drawing we give an
overview of how 
the full collinear singularity of parton $I$, and the full soft
singularity of the $IK$ pair, would be obtained for five different kinds of parton
shower models, as follows.

In a traditional parton shower, the full collinear singularity of each
parton is contained in the DGLAP splitting kernel, $P(z)$, that
generates radiation off that parton. Since no other radiators share
that collinear direction, there is no double counting at the LL
level. (The kernel $P(z)$ constitutes a complete subtraction term for the
collinear singularities in real-emission contributions to an NLO calculation.) 
However, in this approach, the soft (eikonal) singularity between the
$IK$ pair must be obtained by summing the radiation
functions of partons $I$ and $K$ together, and therefore it is 
essential in this type of approach
that both the radiation functions and the shower 
phase-space factorization represent a correct partitioning of the soft
region, with no so-called dead or double-counted zones.

In the early eighties it was shown \cite{Marchesini:1983bm} that 
additional coherence effects can also be taken into account in this language,
albeit approximately, by imposing angular
ordering during shower evolution. 
This effectively represents a first step towards
treating color-connected partons as multipoles in the shower
context; partons $I$ and $K$ 
now effectively acquire some knowledge of each other, via their
relative opening angle, and hence they no longer act as completely
independent emitters. This improvement is denoted
``Coherent Parton Showers'' in the table in 
fig.~\ref{fig:gluonchain}. As indicated by the appearance of the
$\Theta$ function in the collinear term
in such approaches, it is important to construct the angular ordering 
in such a way that the effect of the veto disappears in the collinear
limit. 

In this paper, we follow the dipole-antenna approach to color coherence.
This is motivated by the observation that, whereas
 the collinear
singularities are associated with single logarithms, 
the parametrically larger double logarithms arise from soft
(eikonal) factors.
 It therefore makes sense to change the underlying
picture to a dual one in which parton \emph{pairs} are the fundamental
entities. Such pairs appear in the fixed-order literature under the
name of \emph{antenn\ae{}} and in the shower one, under the name of
\emph{dipoles}.  The latter term, however,
 usually means something else again in the
fixed-order literature. To avoid confusion, we therefore call these
pairs \emph{dipole-antenn\ae{}}. 
In this picture, the roles of soft and collinear singularities are
interchanged, with respect to the parton picture.  The soft double
logarithms between neigboring partons now come from a single term, which 
is thus guaranteed to be neither over- nor under-counted as no
other pairs become doubly singular in the same phase-space region.
The single logarithmic \emph{collinear} radiation off a given gluon 
must now be partitioned among the two neighboring
antenn\ae{} that share it. (Note that 
quarks are still unambiguous in this picture.)  
The gluon case is represented by the line labeled ``Global
Dipole-Antenna'' in the table in fig.~\ref{fig:gluonchain}.

There is considerable freedom in how to partition
the collinear radiation, because
terms can be shuffled back
and forth ``across the gluon'' while maintaining their sum
constant. Two convenient examples are furnished by
\Ar\  \cite{Lonnblad:1992tz} and by Gehrmann et al.\ (GGG)
\cite{GehrmannDeRidder:2005cm},  
which use different decompositions
(see ref.~\cite{Larkoski:2009ah} for
some additional discussion of this point).
 
The first important point concerns what to compare; obviously, the
individual shower functions differ by collinear
singular terms. Thus, if we naively subtract the \Ar\ functions for $qg\to qgg$
and $gg\to ggg$ \cite{Lonnblad:1992tz} from the corresponding GGG ones\footnote{
Note that, in the shower context, different 
color orderings of the final state 
are represented as separate events, wherefore the
shower function should only contain \emph{one} of the permutations, 
corresponding to what GGG label a \emph{sub-antenna}. } \cite{GehrmannDeRidder:2005cm}, 
we obtain  
\begin{eqnar}
\delta^{qg}_{\mrm{GGG-AR}}& = & \frac{1}{s} \left(\frac{2y_{ik}}{y_{ij} y_{jk}} +  \frac{y_{ij} y_{ik}}{y_{jk}} + 
  \frac{y_{ik} y_{jk} + y_{jk}^2}{y_{ij}} + \frac52 + \frac12
  y_{jk}\right) - \frac{1}{s} \left(\frac{(1-y_{ij})^3+(1-y_{jk})^2}{y_{ij} y_{jk}}\right)
\nonumber\\[1.5mm]
& = &
 \frac{1}{s}\left(\frac{1-2y_{ij}}{y_{jk}} -y_{ij}
+\frac52 - \frac12 y_{jk}\right)~, \label{eq:deltaqg} \\[3mm]
\delta^{gg}_{\mrm{GGG-AR}}& = & \frac{1}{s}
\left(\frac{2 y_{ik}}{y_{ij} y_{jk}} + \frac{y_{ij} y_{ik}}{y_{jk}}+\frac{y_{jk}y_{ik} }{y_{ij}}+\frac83\right)
 -  \frac{1}{s}\left(\frac{(1-y_{ij})^3+(1-y_{jk})^3}{y_{ij} y_{jk}}\right)
\nonumber\\
& = &  \frac{1}{s}\left(\frac{1-2 y_{jk}}{y_{ij}}-
y_{ij}+\frac{1 - 2 y_{ij}}{y_{jk}}-y_{jk}
+ \frac{8}{3}\right)~,\label{eq:deltagg}
\end{eqnar}
which differ by gluon-collinear singular terms. 
However, 
when we sum over the two possible orderings of the gluons in
eq.~(\ref{eq:deltaqg}) and the three orderings in 
eq.~(\ref{eq:deltagg}), the discrepancies become collinear-finite,
\begin{eqnar}
\Delta^{qg}_{\mrm{GGG-AR}}& = & \delta^{qg}_{\mrm{GGG-AR}} + (j
\leftrightarrow k) = \frac{6}{s}~, \\[3mm]
\Delta^{gg}_{\mrm{GGG-AR}}& = & \delta^{gg}_{\mrm{GGG-AR}} + (j
\leftrightarrow k) + (i
\leftrightarrow j) = \frac{12}{s}~.
\end{eqnar}
We see that, once summed over permutations, 
the \Ar\ functions have substantially
smaller finite terms than the GGG ones. The \Ar\ shower is
accordingly
somewhat softer than the default \Vc\ one, which uses the GGG functions, 
but the singular terms are
the same, in spite of the apparently singular differences
between the individual shower functions.

We shall usually choose the partitioning of GGG, which makes the
collinear terms of the gluon
antenn\ae{} appear almost identical to those involving quarks in our
parametrization,  
thus emphasizing their similarities. 
Note that for shower applications,
the partitioning must
be done in such a way that the resulting shower 
functions are positive definite.  This is indeed the case for all the
functions we consider in this paper --- a counter example is given
in ref.~\cite{Bern:2008ef}, where the positive-definite \textsc{Ariadne}
antenna functions are repartitioned \`a la GGG in a way that
introduces negative regions in the individual $gg\to ggg$ shower
functions, while maintaining their sum constant. 

\begin{figure}[t]
\begin{center}
\scalebox{\figscale}{\includegraphics*[scale=0.48]{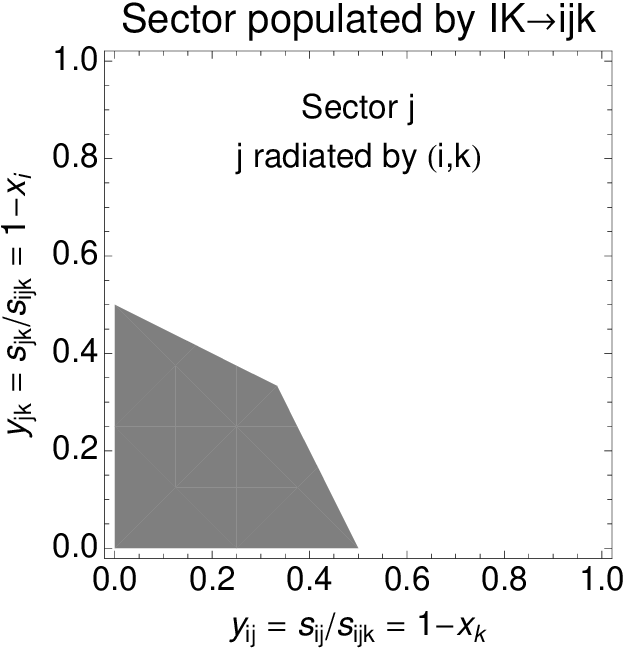}
\includegraphics*[scale=0.48]{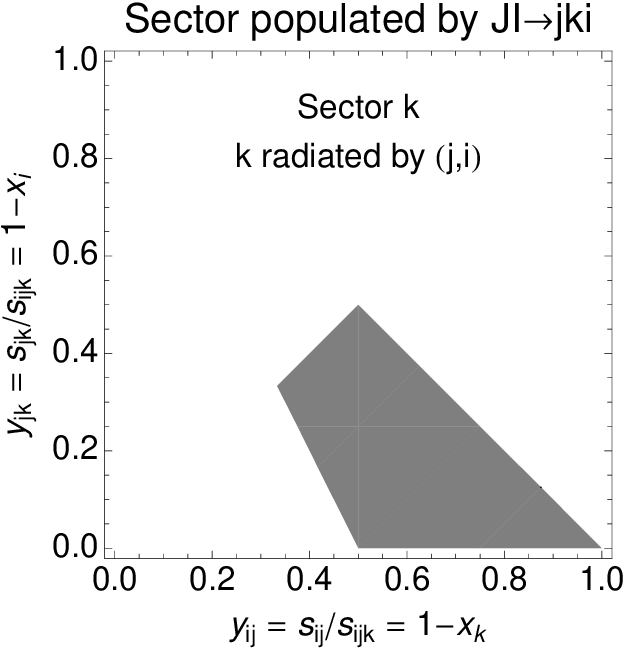}
\includegraphics*[scale=0.48]{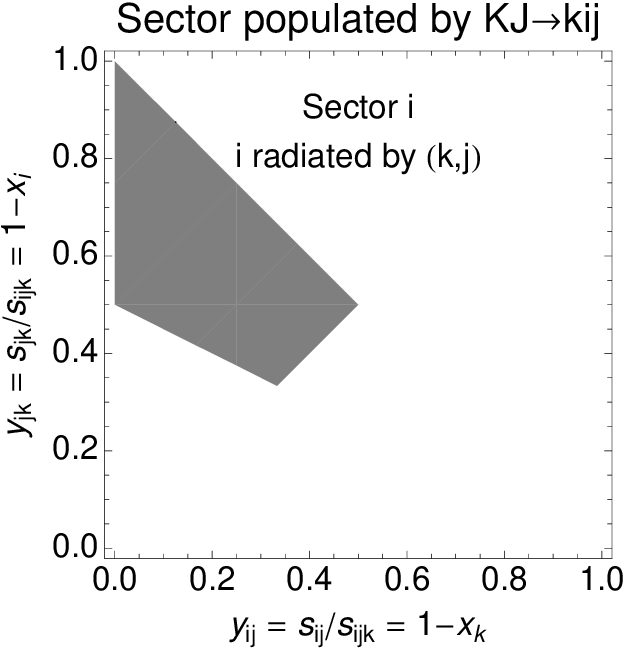}}
\capt{The three phase-space sectors in a color-singlet $g_ig_jg_k$
  configuration, using \pT{} as the disciriminator for which sector a
  given emission/clustering/history belongs to.\label{fig:sectors}}
\end{center}
\end{figure}

A different approach to the issue of how to partition the collinear
singularities is to retain the full collinear
singularity of the gluon in \emph{both} of the neighbouring antenn\ae{},
and combine this with phase-space vetos that
allow only one or the other antenna to contribute to each given phase-space
point, a possibility we have labeled ``Sector Dipole-Antenna'' in the
table in fig.~\ref{fig:gluonchain}. The distinction between global and
sector antenn\ae{} is thus that in the former, several antenn\ae{} (two
that are singular, and possibly more that are non-singular) are summed
over all of phase space, in a way such that their \emph{sum}
reproduces the full collinear singularity, and in the latter case
every single term contains the full singularities, but only one term
(the most singular) is allowed to contribute to any given phase-space
point. An illustration of the sectors appropriate to one color
ordering in the decay of
a scalar, $H^0\to g_ag_rg_b$,
is given in fig.~\ref{fig:sectors}. 
This approach has been suggested for use in NLO fixed-order calculations,
and can be used with \textsc{Vincia} as well.
Larkoski and Peskin (LP)~\cite{Larkoski:2009ah} have also considered
these kinds of antenn\ae{}, including polarization effects.

Finally, a different approach, which also treats dipole coherence exactly,
consists of systematically partitioning 
both the soft and collinear singularities of $I$ and $K$ into four terms, 
two of which treat parton $I$ as the emitter, with $K$ and $H$ acting
as spectator/recoiler, respectively, and the other two terms treating
parton $K$ as the emitter, now with partons $I$ and $L$ acting as
spectator/recoiler, respectively. Catani and Seymour
\cite{Catani:1996jh,Catani:1996vz}
labeled this a \emph{dipole} model, but as this usage differs from 
an older use of the term dipole in parton-shower calculations
to describe the sum of two such
terms (in the context of the Lund dipole
\cite{Gustafson:1986db,Gustafson:1987rq}), 
we avoid confusion by referring to the CS type as a 
\emph{partitioned-dipole}
shower and to the Lund-dipole/antenna type as a \emph{dipole-antenna}
shower.
In the case of partitioned dipoles, 
the radiation off parton $I$ is split into two terms
(``sides''), each of which 
contains ``half'' of the collinear singularity of parton $I$ and ``half'' of
the soft singularity with either of the neighboring partons, e.g., 
\begin{eqn}
a_{I,K} = \frac12 \mrm{Coll}(I) + \frac12\mrm{Soft}(IK)~, 
\end{eqn}
where the subscripts are intended to denote that this is the term for
$I$ emitting and $K$ recoiling. 
There are thus a total of 4 radiation functions involving partons $I$ and
$K$, but these terms can now all be constructed explicitly
so as to have the correct limiting behaviors. Obviously, there is
some ambiguity concerning which functional form to choose for how to
divide up the radiation among the various terms, which is why
``half'' is in quotation marks. 
 Recently, Nagy and Soper (NS) presented a proposal~\cite{Nagy:2005aa} 
for turning the CS subtraction scheme into a parton
shower; several
groups have since
developed CS-style showers, most notably Dinsdale, Ternick, and 
Weinzierl~\cite{Dinsdale:2007mf} and the \Sh\ group \cite{Schumann:2007mg}. 
Although not based on the CS formalism, we note that the $\pT{}$-ordered showers in
\textsc{Pythia 8} are also closely related to this approach. 

\paragraph{Kinematics Map:} 
The \emph{kinematics map} specifies the details of how
to reconstruct the parent momenta $IK$ from the daughter momenta
$ijk$ and is equivalent to what is referred to as the the ``recoil
strategy'' in parton shower language. In an old-fashioned parton
shower \cite{Sjostrand:2004ef}, or a Catani-Seymour one
\cite{Catani:1996vz,Nagy:2005aa,Nagy:2006kb,Dinsdale:2007mf,Schumann:2007mg}, 
for instance, the recoil strategy
usually implies classifying either $I$ or $K$ as the \emph{emitter}
and the other as the 
\emph{recoiler}, with the recoiler being constrained to experience a
momentum change only along its direction of motion in some frame
(e.g., the rest frame of the emitter + recoiler), say, $p_i ||
p_I$. In dipole-antenna approaches, $I$ and $K$ can be allowed to 
share the emitter/recoiler roles more smoothly over the resolved parts of
phase space, with a clear distinction only being made in the strictly singular
limits. 
 
In principle, allowing recoils
also outside the $2\to 3$ process itself, i.e., involving other partons
than $I$ and $K$, could be imagined, 
as long as the leading singular limits are
respected. This would change the subleading properties of the
resulting shower approach, which might be deemed desirable in some
contexts, although it of course would not alter the formal level of
precision. 
However, it does make the formalism more cumbersome, and hence
we shall here restrict our attention to ``local'' recoil strategies,
i.e., involving only the partons $I$ and $K$. 
With this restriction in mind, the constraints that must be fulfilled
to obey the singular limits and viable functional forms the kinematics
map 
were discussed in
refs.~\cite{Lonnblad:1992tz,Kosower:2003bh,GehrmannDeRidder:2003bm,GehrmannDeRidder:2005cm,Giele:2007di}
(including $2\to n$ generalizations in
refs.~\cite{Kosower:2003bh,GehrmannDeRidder:2003bm}).

For illustration, fig.~\ref{fig:phasespace} shows the branching phase
space together with examples of the orientation of the post-branching
partons in the CM of the branching antenna for various phase-space
points, using an antenna-like kinematics map, the ``\textsc{Ariadne}
  angle'', according to which the two parents
  share the transverse component of recoil. We shall return to kinematics
  maps later, but for the present merely note that the
  difference between an emitter-recoiler picture 
  and the map used in fig.~\ref{fig:phasespace} is just an
  overall rotation (about an axis perpendicular to the paper), 
  which vanishes in the soft and collinear limits.  
\begin{figure}[t]
\begin{center}
\vspace*{5mm}
\scalebox{\figscale}{\scalebox{0.62}{
\begin{fmffile}{fmfphase}
\begin{fmfgraph*}(320,320)
\fmfforce{0.0w,0.0h}{o}
\fmfforce{1.0w,0.0h}{r}
\fmfforce{0.0w,1.0h}{t}
\fmf{double,fore=blue}{o,b1,b2,b3,b4,b5,r}
\fmf{plain}{r,d1,d2,d3,d4,d5,t}
\fmf{double,fore=blue}{t,l5,l4,l3,l2,l1,o}
\fmfv{d.sh=circ,l.dist=12,lab={\Large Soft},fore=red}{o}
\fmfv{label=\rotatebox{90}{\Large Collinear with \qbar},l.ang=145}{l3}
\fmfv{label={\Large Collinear with $q$},l.ang=-65}{b3}
\fmfv{d.sh=tri,d.ang=30,lab={\Large $s_{ij}$}}{r}
\fmfv{d.sh=tri,d.ang=0,lab={\Large $s_{jk}$}}{t}
\fmffreeze
\fmf{dashes,foreground=(0.75,,0.75,,0.75)}{b1,l11,l21,l31,l41,d5}
\fmf{dashes,foreground=(0.75,,0.75,,0.75)}{b2,d4}
\fmf{dashes,foreground=(0.75,,0.75,,0.75)}{b3,d3}
\fmf{dashes,foreground=(0.75,,0.75,,0.75)}{b4,d2}
\fmf{dashes,foreground=(0.75,,0.75,,0.75)}{b5,d1}
\fmf{dashes,foreground=(0.75,,0.75,,0.75)}{l1,l11,l12,l13,l14,d1}
\fmf{dashes,foreground=(0.75,,0.75,,0.75)}{l2,l21,l22,l23,d2}
\fmf{dashes,foreground=(0.75,,0.75,,0.75)}{l3,l31,l32,d3}
\fmf{dashes,foreground=(0.75,,0.75,,0.75)}{l4,l41,d4}
\fmf{dashes,foreground=(0.75,,0.75,,0.75)}{l5,d5}
\fmffreeze
\fmfforce{0.167w+0.102w,0.167h-0.021h}{l11a}
\fmfforce{0.167w,0.167h+0.042h}{l11r}
\fmfforce{0.167w-0.102w,0.167h-0.021h}{l11b}
\fmfforce{0.333w+1.4*0.072w,0.333h-1.4*0.042h}{l22a}
\fmfforce{0.333w+0.000w,0.333h+1.4*0.083h}{l22r}
\fmfforce{0.333w-1.4*0.072w,0.333h-1.4*0.042h}{l22b}
\fmf{gluon}{l22,l22r}
\fmf{plain}{l22,l22a}
\fmf{plain}{l22,l22b}
\fmfforce{0.167w+0.077w,0.333h-0.032h}{l21a}
\fmfforce{0.167w+0.024w,0.333h+0.058h}{l21r}
\fmfforce{0.167w-0.101w,0.333h-0.026h}{l21b}
\fmfforce{0.333w+0.101w,0.167h-0.026h}{l12a}
\fmfforce{0.333w-0.024w,0.167h+0.058h}{l12r}
\fmfforce{0.333w-0.078w,0.167h-0.032h}{l12b}
\fmfforce{0.167w+0.049w,0.500h-0.039h}{l31a}
\fmfforce{0.167w+0.053w,0.500h+0.065h}{l31r}
\fmfforce{0.167w-0.101w,0.500h-0.025h}{l31b}
\fmfforce{0.500w+0.101w,0.167h-0.025h}{l13a}
\fmfforce{0.500w-0.053w,0.167h+0.065h}{l13r}
\fmfforce{0.500w-0.049w,0.167h-0.039h}{l13b}
\fmfforce{0.167w+0.016w,0.667h-0.039h}{l41a}
\fmfforce{0.167w+0.086w,0.667h+0.058h}{l41r}
\fmfforce{0.167w-0.102w,0.667h-0.020h}{l41b}
\fmfforce{0.667w+0.102w,0.167h-0.020h}{l14a}
\fmfforce{0.667w-0.086w,0.167h+0.058h}{l14r}
\fmfforce{0.667w-0.016w,0.167h-0.039h}{l14b}
\fmfforce{0.333w+0.033w,0.500h-0.053h}{l32a}
\fmfforce{0.333w+0.037w,0.500h+0.097h}{l32r}
\fmfforce{0.333w-0.070w,0.500h-0.045h}{l32b}
\fmfforce{0.500w+0.070w,0.333h-0.045h}{l23a}
\fmfforce{0.500w-0.037w,0.333h+0.097h}{l23r}
\fmfforce{0.500w-0.033w,0.333h-0.053h}{l23b}
\fmfforce{0.083w,0.833h}{lcol}
\fmfforce{0.083w+1.4*0.003w,0.833h-1.4*0.021h}{lcola}
\fmfforce{0.083w+1.4*0.111w,0.833h+1.4*0.026h}{lcolr}
\fmfforce{0.083w-1.4*0.115w,0.833h-1.4*0.005h}{lcolb}
\fmf{gluon}{lcol,lcolr}
\fmf{plain}{lcol,lcola}
\fmf{plain}{lcol,lcolb}
\fmfforce{0.833w,0.083h}{rcol}
\fmfforce{0.833w+1.4*0.115w,0.083h-1.4*0.005h}{rcola}
\fmfforce{0.833w-1.4*0.111w,0.083h+1.4*0.026h}{rcolr}
\fmfforce{0.833w-1.4*0.003w,0.083h-1.4*0.021h}{rcolb}
\fmf{gluon}{rcol,rcolr}
\fmf{plain}{rcol,rcola}
\fmf{plain}{rcol,rcolb}
\fmfforce{0.083w,0.083h}{soft}
\fmfforce{0.083w+1.4*0.114w,0.083h-1.4*0.010h}{softa}
\fmfforce{0.083w-1.4*0.000w,0.083h+1.4*0.021h}{softr}
\fmfforce{0.083w-1.4*0.114w,0.083h-1.4*0.010h}{softb}
\fmf{gluon}{soft,softr}
\fmf{plain}{soft,softa}
\fmf{plain}{soft,softb}
\fmfforce{0.083w,0.417h}{lc2}
\fmfforce{0.083w+1.4*0.073w,0.417h-1.4*0.026h}{lc2a}
\fmfforce{0.083w+1.4*0.045w,0.417h+1.4*0.043h}{lc2r}
\fmfforce{0.083w-1.4*0.113w,0.417h-1.4*0.017h}{lc2b}
\fmf{gluon}{lc2,lc2r}
\fmf{plain}{lc2,lc2a}
\fmf{plain}{lc2,lc2b}
\fmfforce{0.417w,0.083h}{rc2}
\fmfforce{0.417w+1.4*0.113w,0.083h-1.4*0.017h}{rc2a}
\fmfforce{0.417w-1.4*0.045w,0.083h+1.4*0.043h}{rc2r}
\fmfforce{0.417w-1.4*0.073w,0.083h-1.4*0.026h}{rc2b}
\fmf{gluon}{rc2,rc2r}
\fmf{plain}{rc2,rc2a}
\fmf{plain}{rc2,rc2b}
\fmfforce{0.250w,0.583h}{dl}
\fmfforce{0.250w+1.4*0.021w,0.583h-1.4*0.048h}{dla}
\fmfforce{0.250w+1.4*0.067w,0.583h+1.4*0.080h}{dlr}
\fmfforce{0.250w-1.4*0.088w,0.583h-1.4*0.033h}{dlb}
\fmf{gluon}{dl,dlr}
\fmf{plain}{dl,dla}
\fmf{plain}{dl,dlb}
\fmfforce{0.583w,0.250h}{dr}
\fmfforce{0.583w+1.4*0.088w,0.250h-1.4*0.033h}{dra}
\fmfforce{0.583w-1.4*0.067w,0.250h+1.4*0.080h}{drr}
\fmfforce{0.583w-1.4*0.021w,0.250h-1.4*0.048h}{drb}
\fmf{gluon}{dr,drr}
\fmf{plain}{dr,dra}
\fmf{plain}{dr,drb}
\fmfforce{0.8w,0.80h}{oc}
\fmfforce{0.8w-1.4*0.125w,0.80h}{ol}
\fmfforce{0.8w+1.4*0.125w,0.80h}{or}
\fmf{plain,fore=(0.5,,0.5,,0.5)}{ol,oc}
\fmf{plain,fore=(0.5,,0.5,,0.5)}{oc,or}
\fmfv{d.sh=circ,d.siz=0.01h,l.dist=12,l.ang=90,lab={\Large Original Dipole-Antenna:}}{oc}
\fmfv{l.dist=10,l.ang=180,lab={\Large $q$}}{ol}
\fmfv{l.dist=10,l.ang=0,lab={\Large $\qbar$}}{or}
\fmfforce{0.8w,1.0h}{tit}
\fmfv{label={\huge\textsc{Phase Space for $2\to3$}},l.ang=90,l.dist=0}{tit}
\fmfforce{0.8w,0.93h}{subtit}
\fmfv{label={\Large \textsc{kinematics including (E,,p) cons}},l.ang=90,l.dist=0}{subtit}
\fmfforce{0.4w,1.04h}{boxtl}
\fmfforce{0.4w,0.76h}{boxbl}
\fmfforce{1.2w,1.04h}{boxtr}
\fmfforce{1.2w,0.76h}{boxbr}
\fmf{plain}{boxtl,boxbl,boxbr,boxtr,boxtl}
\end{fmfgraph*}
\end{fmffile}}}\\[5mm]
\capt{Illustration of the branching phase space,
  eq.~(\ref{eq:phasespace}), for $q\qbar\to
  qg\qbar$, with the original dipole-antenna oriented
  horizontally, an antenna-like kinematics map (the ``\textsc{Ariadne}
  angle'') in which the two parents
  share the transverse component of recoil, and $\phi$ chosen such
  that the gluon is radiated upwards. \label{fig:phasespace} }
\end{center}
\end{figure}

\section{A Shower Generator Based on Dipole-Antenn\ae{}}
\label{sec:evolution}

A parton shower algorithm can be constructed from knowledge of 
the Sudakov form factor $\Delta(Q_{E1}^2,Q_{E2}^2)$ representing 
the probability that no branching occurs between the scales 
$Q_{E1}^2$ and $Q_{E2}^2$.  The Sudakov form factor is in turn
given by the exponential of a branching integral,
\begin{eqn}
\Delta = \exp\bigl(-\IntA\bigr)\,,
\end{eqn}%
where, following the conventions laid down in the previous section, 
the integral $\IntA$ of an antenna function over the $2\to 3$ antenna phase
space is 
\begin{eqn}
\IntA = \int \d{s_{ij}} \d{s_{jk}}\frac{\alpha_s}{4\pi}\frac{ {\cal
    C}_{ijk} \ \bar{a}(s,s_{ij},s_{jk},m_{i}^2,m_j^2,m_k^2)}{\sqrt{\lambda\left(
s,m_{I}^2,m_{K}^2
\right)}} \stackrel{m\to 0}{=} \int \d{s_{ij}}
\d{s_{jk}} a(s,s_{ij},s_{jk}) \,. \label{eq:subterm}
\end{eqn}
(We have suppressed the trivial integration over $\phi$.) 
Here, we have set all masses to zero, which approximation we
adopt throughout this paper.
Performing such integrals over all of phase space yields exactly 
the subtraction terms used in antenna-based fixed-order
calculations, see, e.g., ref.~\cite{GehrmannDeRidder:2005cm}. 
(If the approach relies on phase-space vetos, such as in
the case of sector antenn\ae{}, we can treat these as step functions
that are part of the antenna function $a$, so that
eq.~(\ref{eq:subterm}) remains valid).  
Within the shower algorithm, we need to evaluate the integral
for a range of scales, and then invert the function to write the
lower scale as a function of the Sudakov form factor.  The two-dimensional
nature of the integral means that we have to define coordinates, of
which one will be the evolution scale of the shower.  We can use this
to our advantage, as we will see below, to allow for a number of
different evolution variables within the same formalism. A functional
inversion that is both flexible and efficient is accomplished by 
first using a simple overestimate of the antenna function, and
then vetoing to obtain the exact result.

\subsection{The Evolution Integral}

In eq.~(\ref{eq:subterm}) we left the integration boundaries
unspecified. For showering purposes, what is needed is not the integral
over the entire phase space, but over a region  bounded by two values
of the \emph{evolution variable}, 
\begin{eqn}
\IntA(s,Q^2_{E1},Q^2_{E2})  = \int_{Q^2_{E2}}^{Q^2_{E1}} 
\d{s_{ij}} \d{s_{jk}} \ a(s,s_{ij},s_{jk})~~~;~~~Q^2_{E2}<Q^2_{E1}~, \label{eq:ev1}
\end{eqn}
which represents the integrated tree-level splitting probability when going
\emph{from} the scale $Q_{E1}$ \emph{to} the (lower) scale $Q_{E2}$. 
This is the fundamental building block which we shall later exponentiate and
invert in order to find the evolution equation, but first we need to
recast it so that the evolution variable appears explicitly as an
integration variable.

We do this by a change of variables from the original
invariants to a new set, one of which is the evolution
variable, $Q_E$, and the other we may label $\zeta$,
since it will play a role similar to, but not identical with, 
that of the $z$ variable of traditional 
collinear-based shower algorithms. That is, our generic evolution integral
will have the form
\begin{eqn}
\IntA(s,Q^2_{E1},Q^2_{E2}) = \int_{Q^2_{E2}}^{Q^2_{E1}} 
\! \! \d{Q^2_E} \d{\zeta} |J|
 \ a(s,s_{ij},s_{jk})~, \label{eq:ev2b}
\end{eqn}
where $|J|$ is the Jacobian associated with the transformation from
$(s_{ij},s_{jk})$ to $(Q^2_E,\zeta)$. 

One immediate difference between our approach and traditional
collinear-based formalisms is that 
 there is here no explicit dependence on the precise definition of $\zeta$;
it merely
serves to (re)parametrize phase space, and from a set of generated $(Q^2_E,
\zeta)$, the set of invariants $(s_{ij},s_{jk})$ may be obtained without
ambiguity. By contrast, 
in classic parton-shower approaches one would usually start from 
the collinear-limit splitting functions $P(z)$ or similar objects, 
and since these are only accompanied by an
unambiguous definition of $z$ in the collinear limit,  the precise
definition of $z$ away from this limit (energy
or light-cone momentum fraction? in which frame? with
finite-momentum recoils?) results
in an ambiguity which is not present in our treatment. 

There \emph{is} of course a dependence on the functional form of $Q_E$,
which formally enters starting from second order in the
expansion for an IR safe observable (for IR sensitive ones, $Q_E$ is
needed as a regulator already at first order). 
Much effort has gone into debating and examining
the vices and virtues of individual choices. Our stance 
is to order preferably in the inverse of the radiation function, since
it is the singularities of this function which drive the logarithms; 
i.e., in $\pT{}$ for gluon emission and in the invariant 
$m_{q\qbar}$ for gluon splitting to $q\bar{q}$. Here, however, 
since we start directly from eq.~(\ref{eq:ev1}), 
the phase-space factorization expressed in
eq.~(\ref{eq:phasespace}) is preserved for \emph{any} choice of $Q_E$ and
$\zeta$, and so rather than restrict ourselves to one specific form, 
we may instead seek a general solution to the evolution integral which will 
apply to a continuous class of Lorentz invariant evolution
variables. By varying this unphysical parameter (which effectively amounts to
changing the factorization scheme since the evolution variable is what
separates ``resolved'' from ``unresolved'' parts of the calculation at
any given stage of the evolution) we obtain an estimate of the 
amount of scheme dependence which is generated by
this variable --- a dependence that higher-order matching will
explicitly reduce, as we shall see below. The remaining variation can
then be interpreted as an uncertainty estimate.

We stress that we  here only give the
\emph{possibility} to vary this choice --- further studies would 
be required to determine what a sensible \emph{range} of variation
would be, in the context of uncertainty estimates, in
the same way that one discusses variations of other unphysical
parameters for uncertainty estimates. The latter is obviously an art, not an
exact science, but an art for which we can at least furnish the tools.

\subsection{Trial Functions}
\label{sec:TrialFunctions}
To simplify the equations, we shall make use of the veto algorithm to
replace the integrand, $a$, by a simpler function, $\atrial$, which we
shall call the ``trial function''. Provided
our trial function is larger than the actual integrand, the veto
algorithm will allow us to recover the exact integral post facto,
and if the overestimates are not extremely wild, only a small loss of
efficiency should result. 
We will choose the trial function to have the same leading (double)
singularities as the full antenna function.  It will thus
provide us with a simple Lorentz-invariant phase-space generator which is
pre-weighted to take into account these leading double-logarithmic
singularities of QCD; we may then reweight back
down to more exact expressions efficiently.  These more exact
expressions may be
complete matrix elements or dipole-antenna splitting functions, depending on
what is available at a given order in perturbation theory. 
The overestimating function we shall use for
all branchings is the following, with a normalization depending on
whether the trial process is gluon emission or gluon splitting
to $q\qbar$,
\begin{eqnar}
\atrialemit & = & 
\frac{\hat{\alpha}_s}{4 \pi s_{ijk}} C_A \ \abartrialemit = \frac{\hat{\alpha}_s}{4 \pi s_{ijk}} C_A \ \frac{2
  s_{ijk}}{s_{ij}s_{jk}}~~~\left\{\begin{array}{lll}q\qbar\to
qg\qbar \\ qg\to qgg~\mbox{\&~c.c.}\\ gg\to
ggg\end{array}\right.\label{eq:Ahatg} \\[2mm]
\atrialsplit & = & \frac{\hat{\alpha}_s}{4 \pi s_{ijk}}
n_f \hat{T}_R \ \abartrialsplit = \frac{\hat{\alpha}_s}{4 \pi s_{ijk}}
n_f \hat{T}_R  \frac{
  s_{ijk}}{s_{ij}s_{jk}}~~~\left\{\begin{array}{lll}qg\to
q\qbar'q' ~\mbox{\&~c.c.}\\ gg\to g\qbar q + \qbar qg\end{array}\right.~\label{eq:Ahatq}.
\end{eqnar}
where $n_f$ is the number of kinematically accessible flavours and  
$\hat\alpha_s$ is a parameter representing a ``trial''
$\alpha_s$; it should be greater than or equal to the latter.  We
will explain its use further in \sect{sec:showering}.
For comparison, the Eikonal (soft) approximation for gluon emission is
\begin{eqn}
\bar{S}_{\mrm{emit}}(s,s_{ij},s_{jk}) =
\frac{2  
  (s-s_{ij}-s_{jk})}{s_{ij}s_{jk}} = \frac{s-s_{ij}-s_{jk}}{s}
\  \abartrialemit(s,s_{ij},s_{jk})~,
\end{eqn}
such that our approximation coincides with the Eikonal in the soft
limit, $s_{ij}\to 0$ and $s_{jk}\to 0$, 
and is larger than the Eikonal everywhere else.
Obviously, the normalization factor can be adjusted should extreme
variations exceeding these upper bounds be deemed interesting to
study; however we have not found that to be necessary in connection with any
of the studies in this paper. 
Note also that using the Eikonal
for gluon splitting is likely to give a very large overestimate over
most of phase space, as compared to the physical antenna functions
(which only generate single logs, while the Eikonal generates double logs), 
thus leading to low efficiency in the subsequent veto step. 
This may therefore be a technical point
to improve on in future work, but for the time being we prefer the
simplicity of having the same functional form to work with for all
trials. Let us emphasize that the veto algorithm ensures that there is
no trace of the overestimator present in the final results, either in
tree-level expansions or in the Sudakov exponentials. The only
sensitivity to the overestimator is in the \emph{speed} of the
calculation, which we have tested to be comparable to that of
standalone \textsc{Pythia 8}. 

Inserting the trial function $\atrialemit$ from eq.~(\ref{eq:Ahatg}) 
and the Jacobian for the transformation from $(s_{ij},s_{jk})$ to
arbitrary $(Q^2_E,\zeta)$,  eq.~(\ref{eq:ev1}) then becomes 
\begin{eqn}
\IntAtrialemit(s,Q^2_{E1},Q^2_{E2})  =
 \frac{C_A}{4\pi} \int_{Q^2_{E2}}^{Q^2_{E1}} 
\! \! \d{Q^2_E} \d{\zeta} |J| \frac{ 2 \hat{\alpha}_s}{s_{ij}s_{jk}}~, 
\label{eq:ev2}
\end{eqn}
where we have kept $\hat{\alpha}_s$ inside the integral since the
renormalization scale may vary over phase space. The last remaining
step before we can solve this equation 
is now to rewrite the term $|J|/(s_{ij}s_{jk})$ in terms of $Q_E$
and $\zeta$ for a sufficiently general class of functional forms of $Q_E$. 

We shall start by specifying $\zeta$. In principle, one could define a
different $\zeta$ for each possible choice of $Q_E$, but this would be
cumbersome and would not lead to any significant gains. We have
therefore settled on one particular form for $\zeta$ to use for all
evolution variables in this paper. There are two requirements for such
a $\zeta$ definition:
\begin{itemize} 
\item It should be
linearly independent of any $Q_E$ that we would conceivably
consider. I.e., $\zeta$ should not itself be a candidate for an evolution
variable. 
\item Curves of constant $\zeta$ should
intersect curves of constant $Q_E$ once and only once for all $Q_E>0$, such that the
Jacobian is well-defined. 
\end{itemize}

\begin{figure}[t]
\begin{center}
\begin{tabular}{cc}
$\zeta$ definition & $\zeta$ definition (Log) \\
\scalebox{\figscale}{\includegraphics*[scale=0.56]{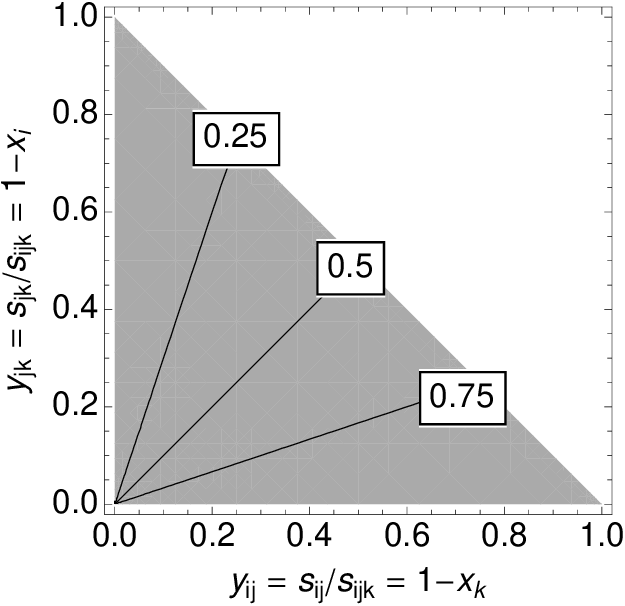}} & 
\scalebox{\figscale}{\includegraphics*[scale=0.56]{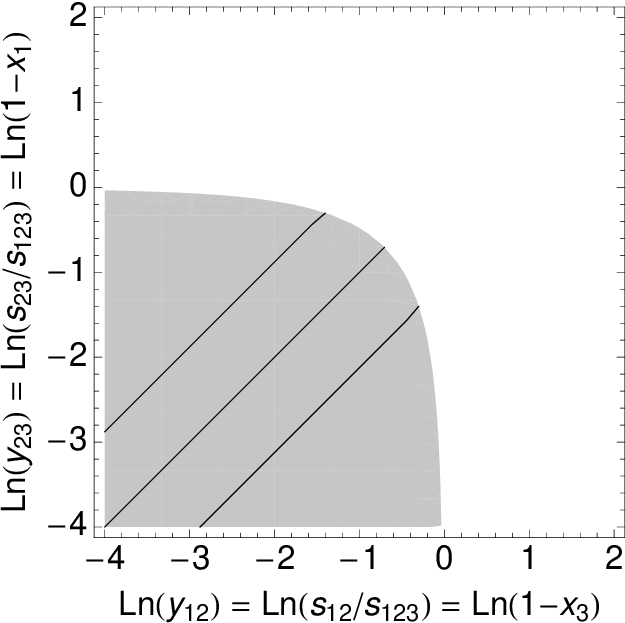}}
\end{tabular}
\capt{Illustration of the $\zeta$ definition, 
  eq.~(\ref{eq:z}). The physical phase space, shown in grey, is the
  same on both panes, and the $\zeta$ definition is also the same, 
but on the left the phase space is shown  on a linear scale in
  the branching invariants and on the right on a logarithmic one.  \label{fig:z}} 
\end{center}
\end{figure}

A $\zeta$
definition that fulfills these 
conditions for the entire class of $Q_E$ 
variables we shall consider is the following simple
ratio of invariants, illustrated in fig.~\ref{fig:z}:
\begin{eqn}
\zeta = \frac{s_{ij}}{s_{ij}+s_{jk}}~~~\Rightarrow~~~1-\zeta =
\frac{s_{jk}}{s_{ij}+s_{jk}} \label{eq:z}~.
\end{eqn}
We emphasize again that all we have done so far 
is recast the Lorentz invariant phase space in
terms of two new variables which are, themselves, arbitrary. There is
no explicit dependence on the particular form of $\zeta$. (There is a
dependence on $Q_E$ of course, but only through the 
boundaries of the integral.)

To compute the Jacobian, we will need the derivatives,
\begin{eqnar}
\frac{\partial{\zeta}}{\partial{s_{ij}}} = \frac{s_{jk}}{(s_{ij}+s_{jk})^2} =
\frac{\zeta(1-\zeta)}{s_{ij}} & ~,~ & 
\frac{\partial{\zeta}}{\partial{s_{jk}}} = \frac{-s_{ij}}{(s_{ij}+s_{jk})^2} =
\frac{-\zeta(1-\zeta)}{s_{jk}}~.
\end{eqnar}

The Jacobian is,
\begin{eqn}
\det \biggl(\frac{\partial\{s_{ij},s_{jk}\}}{\partial\{Q^2_E,\zeta\}}\biggr)
= \det{}^{-1} \biggl(\frac{\partial\{Q^2_E,\zeta\}}{\partial\{s_{ij},s_{jk}\}}\biggr)
= \frac{s_{ij} s_{jk}}{\zeta(1-\zeta)} 
\left[  s_{ij} \frac{\partial Q^2_E}{\partial s_{ij}} + s_{jk}
\frac{\partial Q^2_E}{\partial s_{jk}}\right]^{-1}\,.
\end{eqn}

Inserting this in eq.~(\ref{eq:ev2}) we get a master equation
for evolution in an arbitrary variable $Q_E$ (for trials distributed according to
the function $\abartrialemit$): 
\begin{eqn}
\IntAtrialemit(s,Q^2_{E1},Q^2_{E2})  =
\frac{C_A}{4\pi} \int_{Q^2_{E1}}^{Q^2_{E2}} 
\! \! \d{Q^2_E} \int_{\zeta_{\mrm{min}}(Q^2_E)}^{\zeta_{\mrm{max}}(Q^2_E)}\hspace*{-2mm}\d{\zeta} \frac{ 2 \hat{\alpha}_s}{\zeta(1-\zeta)} 
\left[  s_{ij} \frac{\partial Q^2_E}{\partial s_{ij}} + s_{jk}
\frac{\partial Q^2_E}{\partial s_{jk}}\right]^{-1}~, \label{eq:ev3}
\end{eqn}
where the function in square brackets represents the leftovers from the
Jacobian, and the functions $\zeta_{\mrm{min}}(Q^2_E)$ and
$\zeta_{\mrm{max}}(Q^2_E)$ re-express the phase-space triangle in terms of
$Q_E$ and $\zeta$. Given any specific form of $Q_E$ these three functions
can be derived, and hence a particular evolution equation
obtained. 

\subsection{Evolution Variables}
Given the structure of eq.~(\ref{eq:ev3}), one sees that 
the evolution integral will become particularly simple for any 
evolution variable which satisfies the following simple differential equation,
\begin{eqn}
Q_E^2(s,s_{ij},s_{jk}) = \kqe ~\left( s_{ij} \frac{\partial Q_E^2}{\partial s_{ij}} + s_{jk}
\frac{\partial Q_E^2}{\partial s_{jk}}\right)~. 
\label{eq:dif}
\end{eqn}
Rather than base our formalism on one particular choice of 
evolution variable, as is usually done, we shall therefore instead derive our
formalism so that it applies to \emph{any} evolution variable which
satisfies eq.~(\ref{eq:dif}). Making only this requirement, the
evolution integral simplifies to
\begin{eqn}
\IntAtrialemit(s,Q^2_{E1},Q^2_{E2})  =
\kqe\frac{C_A}{4\pi} \int_{Q^2_{E2}}^{Q^2_{E1}} 
\! \! \frac{\d{Q^2_E}}{Q^2_E} \int_{\zeta_{\mrm{min}}(Q^2_E)}^{\zeta_{\mrm{max}}(Q^2_E)}\hspace*{-2mm}\d{\zeta} \frac{ 2 \hat{\alpha}_s}{\zeta(1-\zeta)}~, 
\label{eq:ev4}
\end{eqn}
which may be solved once and for all, independently of the
choice of $Q_E$. (The explicit dependence on the form of 
$Q_E$ will reemerge, as it should, when 
translating from a generated set of $(Q_E,\zeta)$ back to the 
branching invariants, $(s_{ij},s_{jk})$, but this is a separate step,
to be treated below.)  As explained later, we will take $\hat{\alpha}_s$ to
depend solely on $Q_E^2$, and to be independent of $\zeta$.

Though any symmetric power series in the branching invariants is a
solution to eq.~(\ref{eq:dif}), for the purpose of this paper we shall
impose two additional 
``reasonable'' boundary conditions. Firstly, that an infinitely soft
branching should always be classified as ``unresolved'' for any finite value
of the evolution variable, i.e., the evolution variable must go to zero
when both invariants vanish, $Q_E(0,0) = 0$. We note that 
this apparently mild restriction will nonetheless prevent us from
considering a specific class of variables, which in fact include angular
ordering, since in an angular-ordered cascade, a large-angle emission
may be resolved, even when soft. We shall also require that the evolution
variable be symmetric in the two invariants,
$Q_E(y_{ij},y_{jk})=Q_E(y_{jk},y_{ij})$. 

We can put the differential equation~(\ref{eq:dif}) in dimensionless form
by dividing by $s$,
\begin{eqn}
Y_E(y_{ij},y_{jk}) = \kqe \biggl(
{\partial Y_E\over \partial \ln y_{ij}} 
+{\partial Y_E\over \partial \ln y_{jk}} \biggr)\,,
\label{eq:difdimensionless}
\end{eqn}
where $Y_E = Q_E^2/s$, $y_{ij} = s_{ij}/s$, and $y_{jk} = s_{jk}/s$.

If we define
\begin{eqn}
L_+ = (\ln y_1+\ln y_2)/2\,,\qquad
L_- = (\ln y_1-\ln y_2)/2\,,
\end{eqn}
then eq.~(\ref{eq:difdimensionless}) takes the form,
\begin{eqn}
Y_E(L_+,L_-) = \kqe {\partial Y_E\over \partial L_+} \,,
\end{eqn}
which has the general solution,
\begin{eqnar}
Y_E(s_{ij},s_{jk}) &=& \hat f_0 (L_-)\exp\left(L_+/\kqe\right) +c_0\nonumber\\
&=& f_0 (s_{ij}/s_{jk}) (s_{ij} s_{jk}/s^2)^{1/(2 \kqe)}+c_0\,.
\label{eq:generalsoln}
\end{eqnar}
where $f_0$ is a dimensionless function satisfying 
$f_0(x) = f_0(1/x)$ and the requirement that
$Y_E(0,0)=0$, and $c_0$ is a constant (which we
uniformly set to zero).

For illustrative purposes, it will be convenient to 
introduce additional parameters --- 
a main shape parameter, $a$, and two auxiliary parameters
$b$ and $p$ --- and take
\begin{eqn}
f_0(x) = \left(
\left[\left(\sqrt{x}+\sqrt{\frac1x}\right)^2\right]^a - b
\left[\left(\sqrt{x}-\sqrt{\frac1x}\right)^2\right]^a  \right)^p~.
\end{eqn}
This yields a 
continuously deformable class of evolution variables that fulfill
eq.~(\ref{eq:dif}) subject to our two additional conditions; the
corresponding expression for the evolution variable is,
\begin{eqn}
Q^2_E(s,s_{ij},s_{jk}) = s \left(
\left[\left(\frac{s_{ij}+s_{jk}}{s}\right)^2\right]^a - b
\left[\left(\frac{s_{ij}-s_{jk}}{s}\right)^2\right]^a  \right)^p~,
\label{eq:evvar}
\end{eqn}
subject to the constraints,
\begin{eqn}
a>0 ~,~~ b \le 1~,~~ p>0~,~~
2 a p  = \frac{1}{\kqe}~.~
\end{eqn}
The overall normalization is fixed so that 
the maximum of the evolution variable is the invariant mass of the
dipole: $a$ sets the relative soft/collinear resolution power (to be
explained further below); $b\ne1$ allows for
variables which do not go to zero on the axes (i.e., for which purely
collinear branchings may appear resolved, as is the case, e.g., for
energy ordering), and thus requires an additional infrared regulator
independent of the evolution variable; 
and $p$ allows
modifying the overall speed of the evolution
over phase space. 

\begin{figure}[t]
\begin{center}
{\footnotesize
\scalebox{\figscale}{\begin{tabular}{cccc}
\multicolumn{4}{l}{\normalsize \textsc{Vincia}}  \\
\toprule
\includegraphics*[scale=0.32]{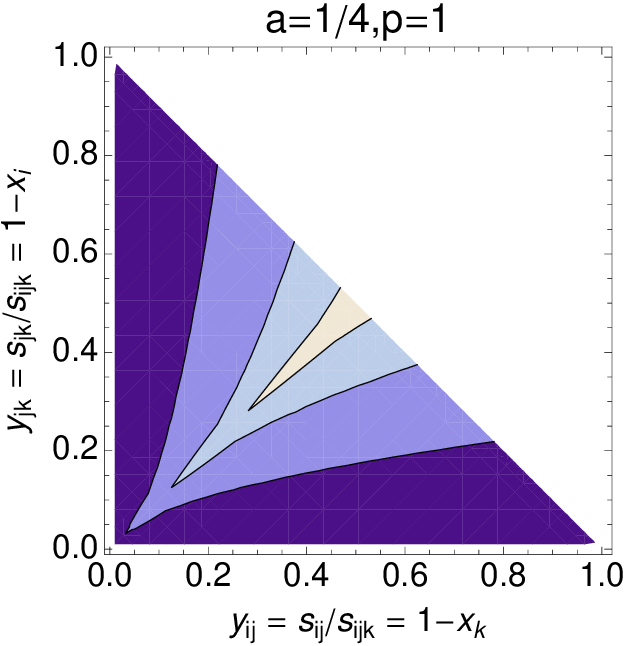} &
\includegraphics*[scale=0.32]{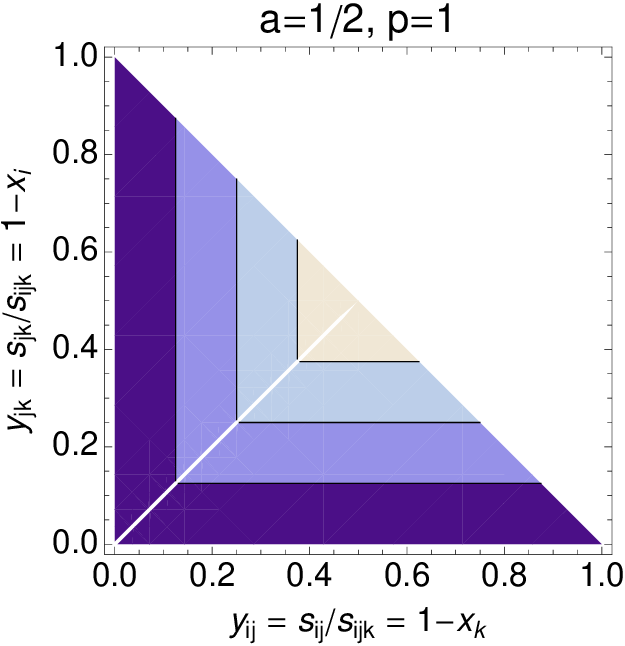} &
\includegraphics*[scale=0.32]{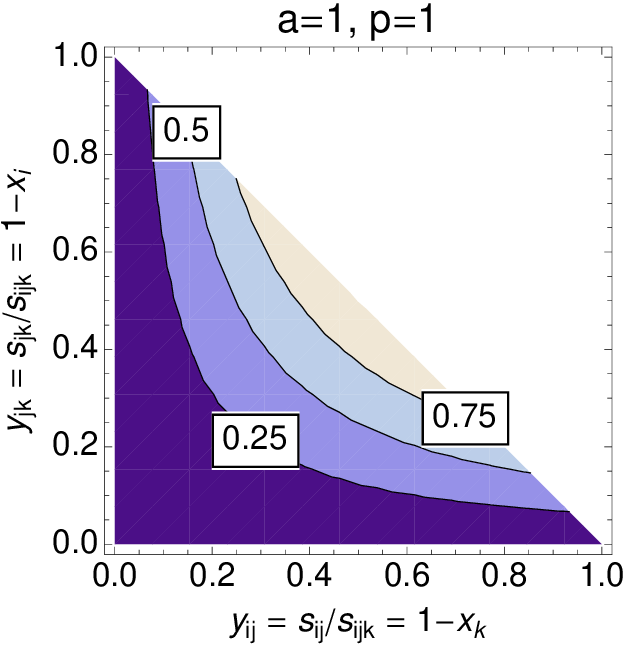} &
\includegraphics*[scale=0.32]{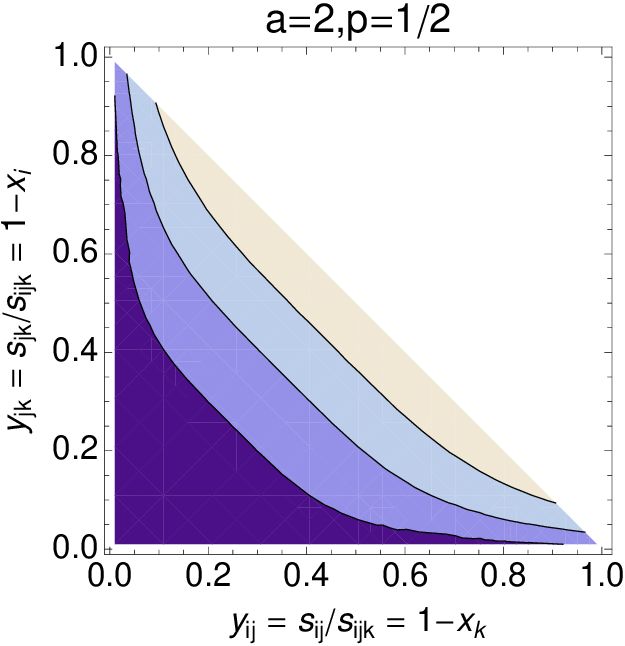} \\
{\sl a)} $V$-ordering &
{\sl b)} $m_{D}$-ordering &
{\sl c)} $\pT{}$-ordering &
{\sl d)} $E^*_{T}$-ordering\\\bottomrule
\end{tabular}}}
\capt{Illustration of the $a$ parameter in
  eq.~(\ref{eq:evvar}). Note that the second plot from the left
  corresponds to choosing the daughter antenna mass as the evolution
  variable, and the third pane corresponds to the \textsc{Ariadne}
  definition of $\pT{}$. Contour labels indicate 
 values of $y_E = Q_E^2/s$. \label{fig:orda}}
\end{center}
\end{figure}

The effect of varying $a$ is illustrated
in fig.~\ref{fig:orda}, with $b=1$.  
We increase $a$ from left to right in the figure.
Small values, toward the left,
yield evolution variables that are ``better'' at resolving 
phase-space points towards the origin (corresponding to soft branchings)
whereas large values of $a$ yield variables that are ``better'' at resolving
points near the axes (corresponding to purely collinear
ones). A shower 
based on the evolution variable to the far left in the figure would
generate soft branchings earlier in the shower than collinear ones,
while a complementary ordering would result in 
a shower based on the variable illustrated on
the far right. We stress that the leading limiting behavior in the soft
and collinear regions are in all cases the same, but these differences
will lead to {\it subleading\/} differences between the showers.
The size of these differences
may be estimated by varying the evolution variable used to
generate the showers. 

\begin{figure}[t]
\begin{center}
{\footnotesize
\scalebox{\figscale}{\begin{tabular}{ccc}
\multicolumn{3}{l}{\normalsize \textsc{Vincia}}  \\
\toprule
\includegraphics*[scale=0.4]{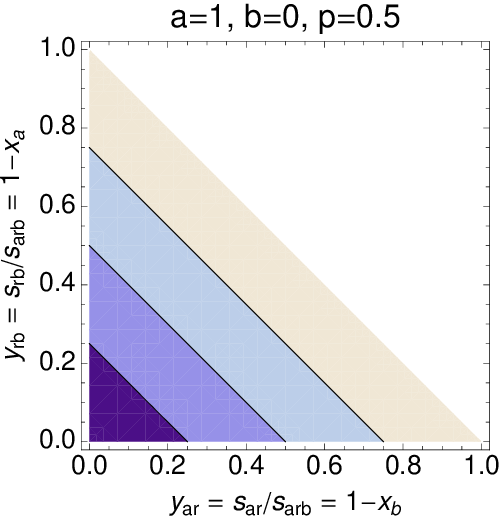} &
\includegraphics*[scale=0.4]{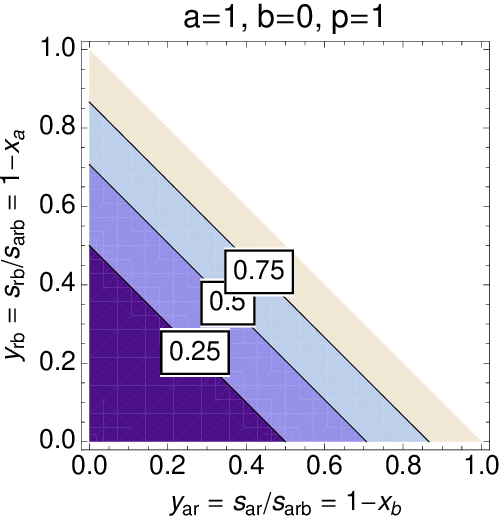} &
\includegraphics*[scale=0.4]{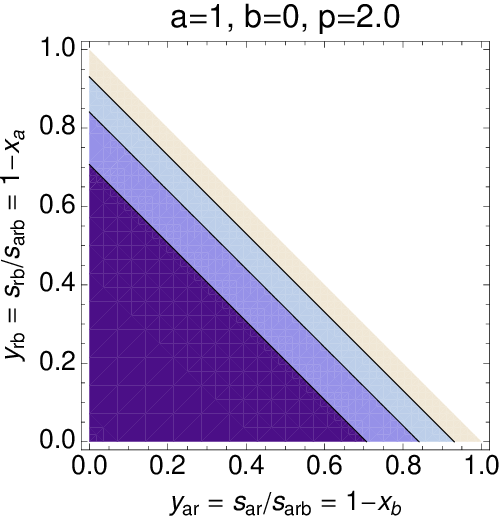}\\
{\sl e)} ${E^*}^\frac12$-ordering & 
{\sl e)} $E^*$-ordering & 
{\sl e)} ${E^*}^2$-ordering \\\bottomrule
\end{tabular}}}
\capt{Illustration of the $b$ and $p$ parameters in
  eq.~(\ref{eq:evvar}). Note that all of these variables are
  proportional to the energy of the emitted gluon (in the CM of the
  dipole-antenna) to some power. Contours indicate 
 values of $y_E = Q_E^2/s$. \label{fig:ordp}}
\end{center}
\end{figure}

An illustration of the $b$ parameter is obtained by comparing the plots
in fig.~\ref{fig:orda} to those in fig.~\ref{fig:ordp}, where the former
all have $b=1$ and the latter all $b=0$. 
As mentioned above,
choosing $b\ne 1$ gives evolution variables that do not go to zero on
the axes.  Accordingly, the contours on the plots in fig.~\ref{fig:ordp} 
intersect the axes, while the ones in fig.~\ref{fig:orda} do
not. The corresponding showers would generally have greater sensitivity
to the infrared region and thus to hadronization effects.  These
variables correspond to those that are used in some analytic resummation
calculations~\cite{Dokshitzer:2008ia}.
Finally, going from left
to right in fig.~\ref{fig:ordp}, we see 
that small values of the speed parameter, $p$, correspond to a faster
progress of the evolution variable over phase space, whereas large $p$
values give the opposite. This speed has no effect on the generation
of the first branching, but it does affect the value of the restart
scale for subsequent branchings, which we will return to below. 

Some specific examples of evolution variables of this form that could
be useful for Monte Carlo purposes are given in table
\ref{tab:evvar}, where we also give the corresponding $\zeta$ limit 
functions and show how the generic functional form,
eq.~(\ref{eq:evvar}), simplifies considerably in several cases. For
instance, setting $a=1,~b=1,~p=1$ gives the $\pT{}$-ordering variable
used both in \textsc{Ariadne} as well as in more recent work
\cite{Giele:2007di,Winter:2007ye}. Contours of constant value of 
this variable are shown on the third
pane of fig.~\ref{fig:orda}. Contours illustrating  the other
variables in tab.~\ref{tab:evvar} can also be found in
figs.~\ref{fig:orda} and \ref{fig:ordp}. ($E^*_{Tn}$-ordering is
shown for $n=1$ only, for which case we leave out the
explicit subscript $n$.) 
\begin{table}[tbp]
\begin{center}
\scalebox{\figscale}{\begin{tabular}{rl|rrr|cl|cc}
 & Name & $a$ & $b$ & $p$ & \multicolumn{2}{l}{Resulting form for $Q^2_E$} & $\zeta_{\mrm{min}}(Q^2_E)$ &
  $\zeta_{\mrm{max}}(Q^2_E)$\\\toprule
1 & $\pT{}$-ordering &$1$ & $1$ & $1$ & $\frac{4s_{ij}s_{jk}}{s}$ &
$=4\pT{\mrm{Ariadne}}^2$ &\multicolumn{2}{c}{
  $\frac{1\mp\sqrt{1-Q^2_E/s}}{2}$ } \\[2mm]
2 & $m_D$-ordering & $\frac12$ & $1$ & $1$ & 
  $2\mrm{min}(s_{ij},s_{jk})$& $=2m^2_D$ &  $\frac{Q^2_E}{2s}$ &
$1 - \frac{Q^2_E}{2s}$ 
\\[2mm]
3 & $E^*$-ordering & $1$ & $0$ & $1$ & $\frac{(s_{ij}+s_{jk})^2}{s}$ &
$=4E^{*2}_j$ & $0$ & $1$\\[2mm]
4 & $V$-ordering & $\frac14$ & $1$ & $1$ & \multicolumn{2}{c|}
{ $\sqrt{s (s_{ij}+s_{jk})}-\sqrt{s |s_{ij}-s_{jk}|}$ } & 
\multicolumn{2}{c}{
$\frac{1\mp(1-Q^2_E/s)^2}{2}$}\\[2mm]
5 & $E^*_{Tn}$-ordering $(n\ge1)$& $2n$ & $1$ & $\frac{1}{2n}$ 
& \multicolumn{2}{c|}{$n=1$ : $\frac{\sqrt{8s_{ij}s_{jk}(s_{ij}^2+s_{jk}^2)}}{s}$}
 & \multicolumn{2}{c}{$\frac{1\mp\left(1-(Q^2_E/s)^{2n}\right)^{\frac{1}{4n}}}{2}$}\\
\bottomrule
\end{tabular}}
\capt{\label{tab:evvar}
Examples of evolution variables in the form of eq.~(\ref{eq:evvar})
and corresponding to the illustrations in figs.~\ref{fig:orda} and
\ref{fig:ordp}. The nominal $\zeta$ boundaries for
$E^*$ ordering would lead to infinities, so for practical applications
the bounds implied by the hadronization cutoff should be used
instead.}
\end{center}
\end{table}
Note that
the name $E^*_{Tn}$ does not imply that
this variable represents a physical transverse energy, but rather that
it represents an interpolation between $\pT{}$ and $E^*$, with lower
values of the $n$ parameter making it closer to $\pT{}$ and higher
values making it closer to
$E^{*}$. 
The $V$  variable is named simply for the
shape it has 
over phase space, like a $V$ pointing towards the soft region, cf.\
the leftmost pane in fig.~\ref{fig:orda}. 

\begin{figure}[tp]
\begin{center}
\scalebox{\figscale}{
\begin{tabular}{lclcl}
\textsc{Jetset \& Fortran Pythia} & &\textsc{Pythia 6.3+
  \& Pythia 8} & & \textsc{Herwig++} \\
\toprule
\includegraphics*[scale=0.4]{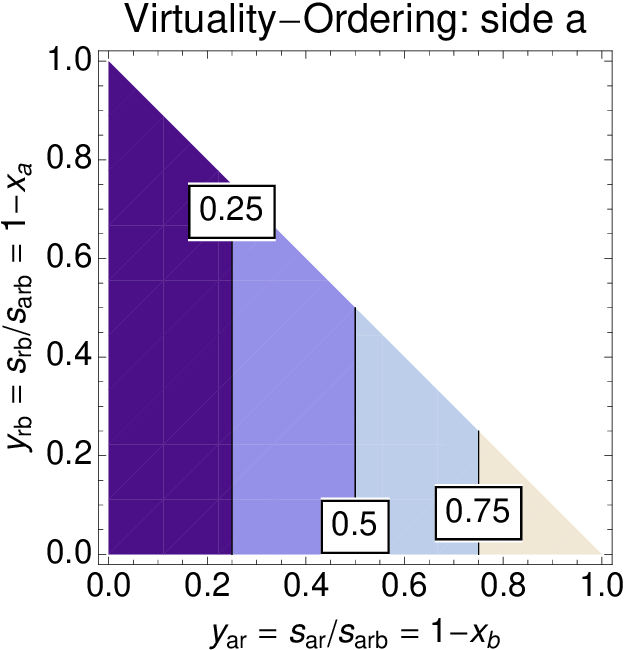} & & 
\includegraphics*[scale=0.4]{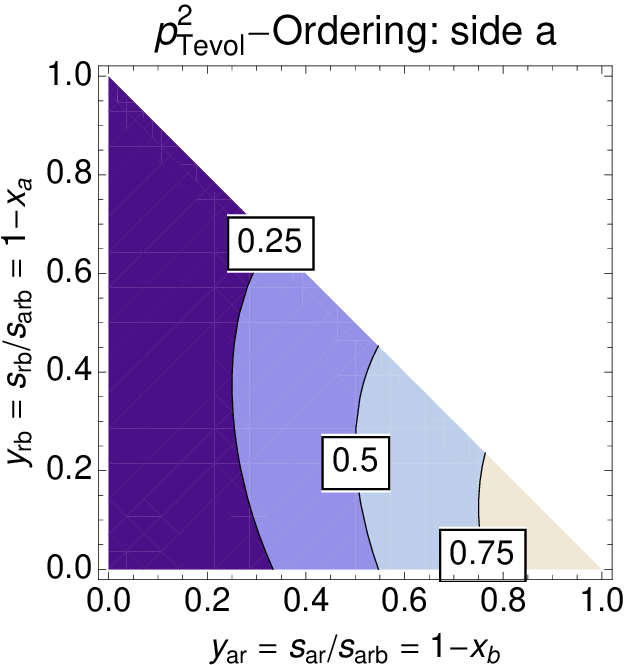} & & 
\includegraphics*[scale=0.4]{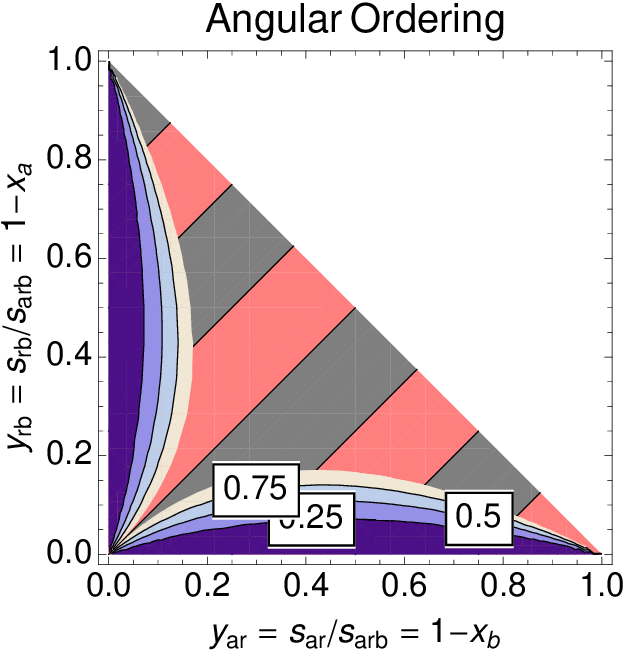}\\
\\[-4.6cm]
\hspace*{2.3cm}\includegraphics*[scale=0.16]{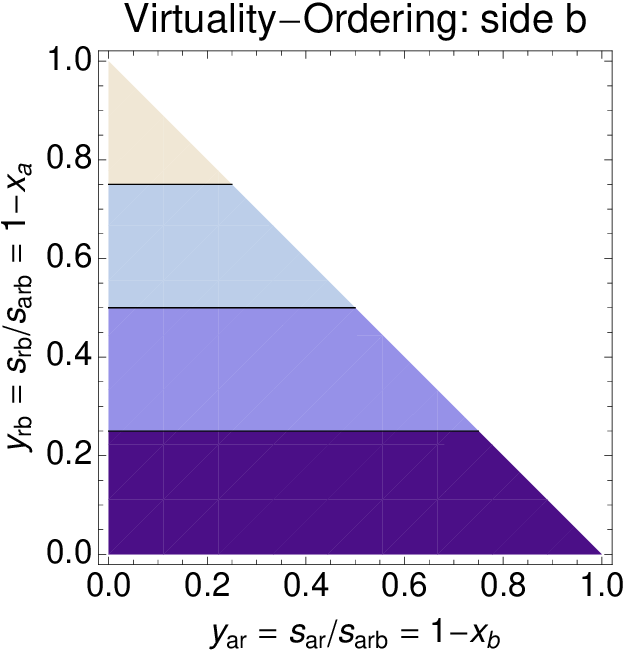} & &
\hspace*{2.3cm}\includegraphics*[scale=0.16]{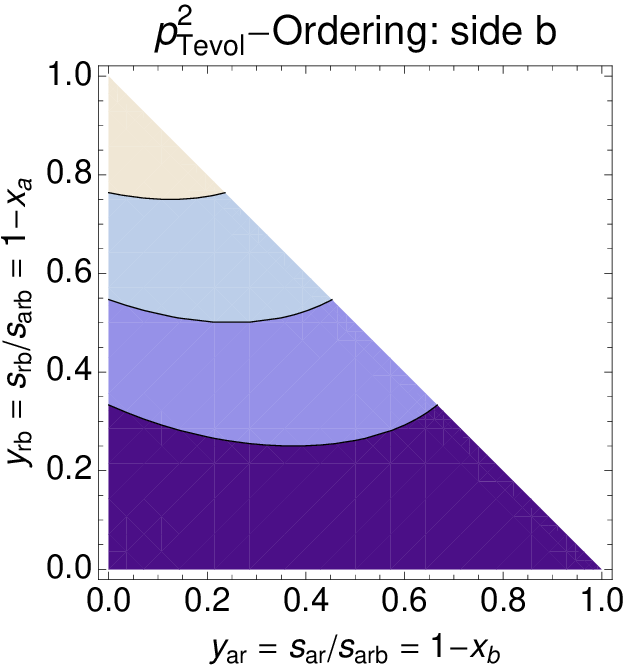} \\[2.1cm]
\bottomrule
\end{tabular}}
\capt{In an old-fashioned parton shower or partioned-dipole shower, 
  the process $\hat{a}\hat{b}\to arb$ is divided onto two terms, one representing
  emission off parton $\hat{a}$ and the other emission off parton
  $\hat{b}$.  {\sl Left Pane:} 
  contours of constant $s_{ar}$, i.e., the virtuality that corresponds
  to $a^*\to ar$ in a virtuality-ordered parton shower. The inset
  shows the equivalent contours for emission off side $b$. {\sl Middle
    Pane:} contours of constant $p_{T\mrm{evol}}$, the variable used
  in the $\pT{}$-ordered \textsc{Pythia} shower. Note that for the
  virtuality-ordered shower, additional vetos on the emission angle,
  not shown here, must be
  imposed to  enforce coherence, while in the $\pT{}$-ordered case,
  this is less crucial due to the use of dipole kinematics. 
  In an \emph{angular}-ordered parton shower (right pane), each parton is
  still evolved separately, but the potential for double counting 
  has been removed by effectively restricting the emission
  from each parton to non-overlapping regions, here angular-ordered
  cones, and hence we can represent the two terms on
    one and the same plot. (Note: while the original \textsc{Herwig}
    implementation of angular ordering did imply some overlap in the soft region, 
    this has been removed in \textsc{Herwig++}.) The price to pay
    is that this introduces an artificially unpopulated dead zone 
    in the phase space, illustrated by the striped area.   The contour
    labels denote values of $y_E = Q_E^2/s$. 
\label{fig:oldmc}}
\end{center}
\end{figure}

For completeness, a few important 
examples of evolution variables that are \emph{not} covered using
our formalism are the traditional $1\to 2$ parton-shower ones in
\textsc{Pythia} and \textsc{Herwig}, illustrated in
fig.~\ref{fig:oldmc}. 
For the forms of the \Hw\ and \Py\ evolution
variables, translated to our phase-space notation, we used the
following (for evolution with parton $I$ as the emitter):
\begin{eqn} \ 
\begin{array}{p{3.0cm}crcl}
{\Py\newline virtuality-ordering} & : & 
  \displaystyle m_{I^*}^2 & = & 
    \displaystyle s_{ij}  \\[2mm]
{\Py\newline \pT{\mrm{evol}}-ordering} & : & 
  \displaystyle 4\pT{\mrm{evol},I}^2 & = &  
    \displaystyle 4\frac{s_{ij}(s - s_{jk})(s_{ij} + s_{jk})}{(s +
      s_{ij})^2} 
\\[2mm]
{\Hw++\newline angular-ordering} & : & 
  \displaystyle q_{\theta,I}^2 & = &  
    \displaystyle 
      4s\left(\frac{s s_{ij}}{(s - s_{jk})(s_{ij} + s_{jk})}\right)^2 
      =  4s \left(\frac{1-x_k}{x_ix_j}\right)^2~, \end{array}
\end{eqn}
where we used $x_i = (1-s_{jk}/s)$ in the last expression.  
For the \Py\ \pT{\mrm{evol}} variable, $j$ is the emitted parton,
and $k$ is the recoiling one.
Analogous expressions hold when parton $K$ is the emitter, with the
substitutions $s_{ij} \leftrightarrow s_{jk}$. 

\begin{figure}[tp]
\begin{center}
\scalebox{\figscale}{
\includegraphics*[scale=0.4]{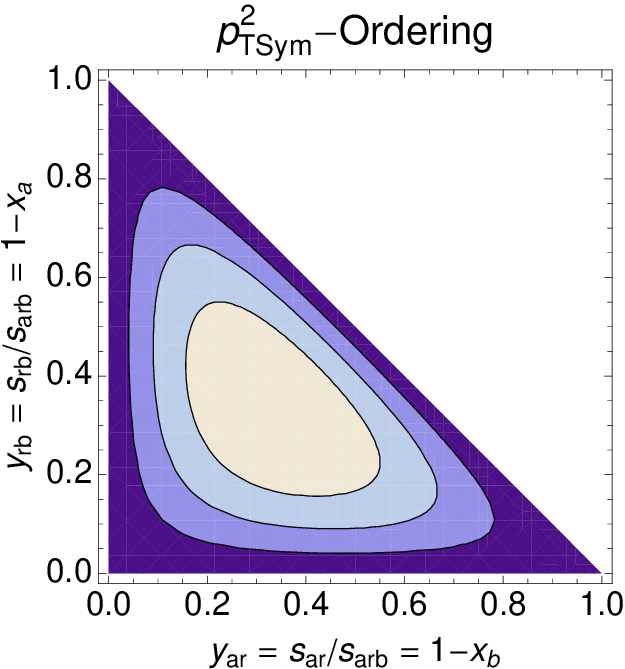}\hspace*{0.4cm}
\includegraphics*[scale=0.4]{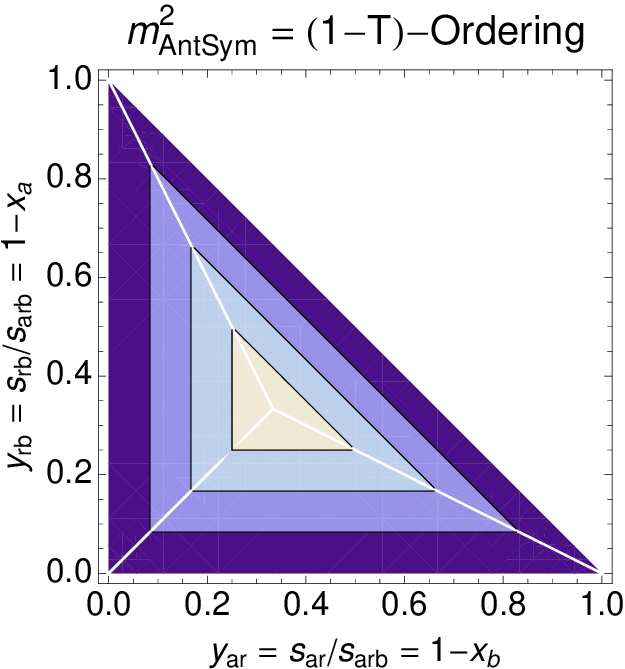}} \\
\capt{Examples of evolution variables for which our $\zeta$ definition
  would be multi-valued for each value of $Q_E$. These particular
  ones were obtained by merely symmetrizing $\pT{}$ (left) and
  $m_D$ (right) in all three branching
  invariants.  \label{fig:sym}}
\end{center}
\end{figure}

Likewise, variables that do not have well-defined Jacobians with our
$\zeta$ definition, such as those in fig.~\ref{fig:sym}, cannot be used
without additional work.
Examples include,
\begin{eqnar}
\pT{\mrm{Sym}}^2 & = & 27 \frac{s_{ij}s_{jk}s_{ik}}{s^2}\\
m_{D\mrm{Sym}}^2 & = & 3\mrm{min}(s_{ij},s_{jk},s_{ik}) = 3s(1-T_3)~,
\end{eqnar}
where $T_3$ is the Thrust event shape variable for a three-parton
configuration. The $\pT{\mrm{Sym}}$ variable is similar to the
$C$-parameter for a 3-parton configuration,  
\begin{eqn}
C_3=6\frac{s_{ij} s_{jk} s_{ki}}{(s-s_{ij})(s-s_{jk})(s-s_{ki})}~\le \frac34,
\end{eqn}
which one could therefore also imagine using as the basis for an 
evolution variable, with $Q_E^2 = \frac43 s C_3$. 

\subsection{The $\mathbf{\zeta}$ integral}
In general, one might expect the running coupling $\alpha_s$ to 
depend on $\zeta$ as well as $Q_E$ and $s$.
If we simplify the dependence by taking it to depend
only on $Q_E$ and/or $s$, 
then the integral over $\zeta$ is straightforward:
\begin{eqn}
I_\zeta(\zeta_{\mrm{min}}(Q^2_E),\zeta_{\mrm{max}}(Q_E^2))
  =\int_{\zeta_{\mrm{min}(Q^2_E)}}^{\zeta_{\mrm{max}}(Q^2_E)} \!\! \d{\zeta}
\frac{1}{\zeta(1-\zeta)} =
\ln\left[\frac{\zeta_{\mrm{max}}(Q^2_E)(1-\zeta_{\mrm{min}}(Q^2_E))}{\zeta_{\mrm{min}}(Q^2_E)(1-\zeta_{\mrm{max}}(Q^2_E))}\right]~.  \label{eq:zintegral}
\end{eqn}
One would expect any $\zeta$ dependence in $\alpha_s$ to be reduced by computing
to higher orders in perturbation theory.  At a fixed order, 
a $\zeta$-dependent $\alpha_s$ could still be accommodated,
for example, by choosing a $\zeta$-independent (large) 
$\hat{\alpha}_s$ for trials and then applying the veto
algorithm. The case of a $Q_E$-dependent $\alpha_s$ will be treated
explicitly below. 

%

\subsection{Evolution Windows}
\begin{table}[t]
\begin{center}
\scalebox{\figscale}{
\begin{tabular}{rlcrc}
\multicolumn{5}{c}{\textsc{Vincia Evolution Windows}}\\\toprule
$i$ & $[Q_{E\mrm{min}}$&,&$Q_{E\mrm{max}}]$ & $n_f$\\\midrule
0   & $[0$&,&$m_c]$ & 3 \\
1   & $[m_c$&,&$m_b]$ & 4 \\
2   & $[m_b$&,&$\sqrt{m_bm_t}]$ & 5 \\
3   & $[\sqrt{m_bm_t}$&,&$m_t]$ & 5 \\
4   & $[m_t$&,&$\infty]$ & 6  \\\bottomrule
\end{tabular}}
\capt{The evolution windows used in \textsc{Vincia}, with the $Q_{E}$ boundaries and active number of flavors corresponding to each. The number of active 
flavors is the same for windows 2 and 3, but the $\zeta$ boundaries for trials
 are different, due to the different $Q_{E\mrm{min}}$
 values. This improves the efficiency of the generator.
The first window will not actually extend down to 
zero in practice, but will instead be cut off by the hadronization scale. 
\label{tab:windows}}  
\end{center}
\end{table}
To simplify the trial generation further we shall generate trial branchings
in a larger phase-space region than the physically allowed
one, again using the veto algorithm to avoid generating any actual
branchings in the unphysical region. 
Specifically, we divide the generation of trial branchings into
discrete windows in $Q_E$ --- given in table \ref{tab:windows} --- 
and, in each such window, replace the $\zeta$ limits in the previous  
equation by constant ones, 
\begin{eqn}
\zeta_{\mrm{min}}(Q^2_E) = \zeta_{\mrm{min}}(Q^2_{E\mrm{min}})~~~,~~~
\zeta_{\mrm{max}}(Q^2_E) = \zeta_{\mrm{max}}(Q^2_{E\mrm{min}})~
\label{eq:zetaminmax}
\end{eqn}
where $Q_{E\mrm{min}}$ is the value of $Q_E$ at the end of the current
window (e.g., the next flavor threshold or, 
ultimately, the hadronization scale). If none of the generated trials
fall within the current evolution window, the evolution should be
restarted at $Q_E=Q_{E\mrm{min}}$, upon which the $Q_{E\mrm{min}}$ and $\zeta$
boundaries should be updated to correspond to those of the next
evolution window. If the crossing corresponds to a flavor threshold, the
normalization of the trial function for gluon splitting should be
updated with the new number of active flavors, and if $\alpha_s$
depends explicitly on $Q_E$, then $\Lambda_{\mrm{QCD}}$ should,
likewise, be updated.

\subsection{$\mathbf{Q_E}$ integral for $\mathbf{Q_E}$-independent $\mathbf{\alpha_s}$}
With the $\zeta$ integral for trial branchings having now effectively
become a constant
depending only on the current ``evolution window'' (i.e., the current
$Q_{\mrm{min}}$), we may perform the $Q_E$ integration independently
of the $\zeta$ one. We do this first for the case where the
renormalization scale in $\alpha_s$ is constant over the branching
phase space.  

In this case, the evolution integral, eq.~(\ref{eq:ev4}), becomes
\begin{eqn}
\IntAtrialemit(Q^2_{E1},Q^2_{E2}) 
=  2 \hat{\alpha}_s \kqe\frac{C_A}{4\pi}
I_\zeta(\zeta_{\mrm{min}},\zeta_{\mrm{max}})
\ln\left(\frac{Q^2_{E1}}{Q^2_{E2}}\right) \label{eq:ev5}~,
\label{eq:IAconst}
\end{eqn}
where 
$\zeta_{\mrm{min},\mrm{max}}$ 
are the ones appropriate to the current evolution
window, as given in tabs.~\ref{tab:evvar} \& \ref{tab:windows}. 

The equivalent expression for the trial function for gluon
splitting, $\IntAtrialsplit$, only differs by an overall normalization
factor, cf.~eqs.~(\ref{eq:Ahatg}) \& (\ref{eq:Ahatq}), and hence we do
not reproduce it explicitly here. 

In addition to a zeroth order (fixed) $\alpha_s$ or a running 
$\alpha_s$ that depends only on the parent dipole-antenna mass, $\alpha_s(s)$, 
these expression will be needed, for instance, to use
$\alpha_s(\pT{})$ together with any non-$\pT{}$ evolution
variable. Technically, we accomplish this by
setting the trial $\hat{\alpha}_s$ equal to unity (or some other relatively
large number) in the above equation 
and then accepting the generated $(Q_E,\zeta)$ pair with
the probability $\alpha_s(\pT{})/\hat{\alpha}_s$.  

\subsection{$\mathbf{Q_E}$ integral for first-order $\mathbf{Q_E}$-dependent $\mathbf{\alpha_s}$}
We shall also allow for the possibility to use a first-order running
$\alpha_s$,
 with a renormalization scale
that depends explicitly on $Q_E$,
\begin{eqn}
\alpha_s(k_\mu^2 Q^2_E) = \frac{1}{\bh_0\ln(k_\mu^2Q_E^2/\Lambda^2)}~,
\end{eqn}
where 
\begin{eqnar}
b_0 &=& \frac{11 C_A - 2n_f\hat{T}_R}{12\pi}\,,
\label{eq:b0}
\end{eqnar}
and where $k_\mu$ is an overall scaling factor, 
$n_f$ is the active number of flavors, and $\Lambda$ is the
appropriate ($n_f$-dependent) value of $\Lambda_{\mrm{QCD}}$. 

The evolution integral then becomes
\begin{eqnar}
\IntAtrialemit(s,Q^2_{E1},Q^2_{E2}) 
& = & 2 \kqe\frac{C_A}{4\pi}
I_\zeta(\zeta_{\mrm{min}},\zeta_{\mrm{max}}) 
 \frac1{\bh_0}\int_{Q^2_{E2}}^{Q^2_{E1}}
\frac{\d{Q^2_E}}{Q^2_E} \frac{1}{\ln(k_\mu^2Q^2_E/\Lambda^2)} \nonumber\\
& = & 2 \kqe\frac{C_A}{4\pi}
I_\zeta(\zeta_{\mrm{min}},\zeta_{\mrm{max}}) 
 \frac1{\bh_0}\int_{x_{E2}^2}^{x_{E1}^2}
\frac{\d{x_E^2}}{x_E^2} \frac{1}{\ln(x_E^2)} \nonumber \\ 
& = & 2 \kqe\frac{C_A}{4\pi}
I_\zeta(\zeta_{\mrm{min}},\zeta_{\mrm{max}}) 
 \frac1{\bh_0}\ln\left(\frac{\ln(x_{E1}^2)}{\ln(x_{E2}^2)}\right)~, 
\label{eq:IArun}
\end{eqnar}
where 
\begin{eqn}
x^2_E = \frac{k^2_\mu Q^2_E}{\Lambda^2} ~.
\end{eqn}

\subsection{$\mathbf{Q_E}$ integral for $\mathbf{Q_E}$-dependent $\mathbf{\alpha_s}$}

More generally, we can incorporate a $Q_E$-dependent $\alpha_s$ by changing
variables using the $\beta$ function,
\begin{eqn}
{d\alpha_s(Q_E^2)\over d\ln Q_E^2} = \beta(\alpha_s)\,.
\label{eq:beta}
\end{eqn}
This allows us to rewrite eq.~(\ref{eq:ev4}) as follows,
\begin{eqn}
\IntAtrialemit(s,\qe1, \qe2) = \kqe {C_A\over 4\pi}
\int_{\alpha_s(\qe1)}^{\alpha_s(\qe2)} {d\alpha_s\over \beta(\alpha_s)}
\,\int_{\zeta_{\min}(Q_E^2(\alpha_s))}^{\zeta_{\max}(Q_E^2(\alpha_s))} \!\d{\zeta}
{2 \alpha_s\over \zeta (1-\zeta)}
\label{eq:evrungenl}
\end{eqn}
When we make the substitutions of eq.~(\ref{eq:zetaminmax}) here, 
the inner integral will
be independent of $\alpha_s$.  
With the two-loop beta function, 
\begin{eqn}
\beta(\alpha_s) = -\bh_0 \alpha_s^2 -\bh_1 \alpha_s^3\,,
\label{eq:betafunction}
\end{eqn}
where $b_0$ is given by \eq{eq:b0} and
\begin{eqnar}
\bh_1 
 & = & \frac{17C_A^2-n_f\hat{T}_R(5C_A+\frac32\hat{C}_F)}{24 \pi^2} \,,
\label{eq:b1}
\end{eqnar}
the $\alpha_s$ integral is simple,
\begin{eqn}
\int_{\alpha_s(\qe1)}^{\alpha_s(\qe2)} {d\alpha_s\over \beta(\alpha_s)}=
{1\over\bh_0} \ln \biggl({\alpha_s(\qe1)\over\alpha_s(\qe2)}\biggr)
+{1\over\bh_0} \ln \biggl({1+\bh_1/\bh_0 \alpha_s(\qe2)
                               \over 1+\bh_1/\bh_0 \alpha_s(\qe1)}\biggr)\,.
\end{eqn}
This function can be inverted readily using a Newton-Raphson solver,
which can likewise be used to obtain $Q_E^2(\alpha_s)$.  It
can be extended readily to higher loops because additional orders only
introduce new denominator factors of the form $(1+c \alpha_s)$.

\subsection{The Evolution Equation}

We now have all the pieces in hand to construct the evolution
equation for a generic shower subject only to the conditions outlined
in the preceding paragraphs; that the individual trial branchings be $2\to
3$ mappings from on-shell momenta to on-shell momenta, respecting the
Lorentz-invariant phase-space decomposition, eq.~(\ref{eq:phasespace}), for any
evolution variable that satisfies the differential equation, eq.~(\ref{eq:dif}). 

The generating function for such a shower is the Sudakov form factor,
\begin{eqn}
\Delta(Q^2_{E1},Q^2_{E2}) = \exp\left(
- \IntA(Q^2_{E1},Q^2_{E2})
\right)~,
\end{eqn}
where we may substitute for $\IntA$ either of the expressions
eqs.~(\ref{eq:IAconst}) or (\ref{eq:IArun}). 

In order to generate trial branchings according to this Sudakov, we
must solve the equation
\begin{eqn}
R = \Delta(Q^2_{E1},Q^2_{E2}) \label{eq:QofR}
\end{eqn}
for $Q_{E2}$, where $R$ is a random number distributed uniformly
between zero and one and $Q_{E1}$ is the ``(re)starting scale''. The
latter represents the scale the shower is being restarted
at prior to the generation of the next trial branching. 
To give an idea, this can either be the full dipole 
center-of-mass energy, $\sqrt{s}$,
which will usually be the case for the very first branching following
a resonance decay, or, later on in the shower evolution,
the scale of the preceding trial branching.

The solution of this equation is the paramount reason we chose to use
a simplified antenna function for trial generations. It would have
been possible to solve the evolution integral itself for more
complicated trial functions, but the inversion of eq.~(\ref{eq:QofR}) to
solve for $Q_{E2}$ as a function of $R$ and $Q_{E1}$ 
would then have been much
more cumbersome.

Solving the evolution equation for a $Q_E$-independent
$\hat{\alpha}_s$, using eq.~(\ref{eq:IAconst}), yields
\begin{eqn}
Q^2_{E2} = Q_{E1}^2 R^{b} ~,
\end{eqn}
with
\begin{eqn}
b = \frac{4\pi}{2\hat{\alpha}_s\kqe C_AI_\zeta(\zeta_{\mrm{min}}(Q^2_{E\mrm{min}}),\zeta_{\mrm{max}}(Q^2_{E\mrm{min}}))} ~.
\end{eqn}

Solving the evolution for a first-order running $Q_E$-dependent
$\hat{\alpha}_s$, using eq.~(\ref{eq:IArun}), yields
\begin{eqn}
Q^2_{E2} = 
\frac{\Lambda^2}{k_\mu^2}\left(\frac{k_\mu^2Q^2_{E1}}{\Lambda^2}\right)^{R^{b'}} ~,
\end{eqn}
with
\begin{eqn}
b' = \frac{4\pi}{2\kqe
  C_AI_\zeta(\zeta_{\mrm{min}}(Q^2_{E\mrm{min}}),\zeta_{\mrm{max}}(Q^2_{E\mrm{min}}))}
\frac{11C_A - 2n_f \hat{T}_R}{12\pi}~.
\end{eqn}

As mentioned earlier, given any set of branching variables
$(Q^2_E,\zeta)$ we may obtain the invariants $(s_{ij},s_{jk})$ without
ambiguity. Thus, the next step is to generate 
a $\zeta$ value, given $Q_{E2}$. Since we defined $\zeta$
independently of $Q_E$, we may do this once and for all, with the
solution applicable to all evolution variables. To generate a random
$\zeta$ value distributed according to the integrand of the $I_\zeta$
integral, eq.~(\ref{eq:zintegral}), we must solve the equation 
\begin{eqn}
R_\zeta =
\frac{I_\zeta(\zeta_{\mrm{min}},\zeta)}{I_\zeta(\zeta_{\mrm{min}},\zeta_{\mrm{max}})} 
\label{eq:zGeneration}
\end{eqn}
for $\zeta$, where $R_\zeta$ is another random number uniformly
distributed between zero and one, the $I_\zeta$ integral given by
eq.~(\ref{eq:zintegral}), and $\zeta_{\mrm{min}}(Q_{E\mrm{min}})$ is given by the
evolution windows (Table~\ref{tab:windows}) and by the
evolution-variable-dependent $\zeta$ limits (Table~\ref{tab:evvar}). 

We solve eq.~(\ref{eq:zGeneration})  by first translating
to the variable $r$, 
\begin{eqn}
  r_{\mrm{max}} = \frac{\zeta_{\mrm{max}}}{1-\zeta_{\mrm{max}}}~~~,~~~
  r_{\mrm{min}} = \frac{\zeta_{\mrm{min}}}{1-\zeta_{\mrm{min}}}~~~,
\end{eqn}
generating a random value for $r$
\begin{eqn}
  r = r_{\mrm{min}} \left(\frac{r_{\mrm{max}}}{r_{\mrm{min}}}\right)^{R_\zeta}~,
\end{eqn}
and finally solving for $\zeta$,
\begin{eqn}
\zeta = \frac{r}{1+r} ~.
\end{eqn}

If the $\zeta$ generated in this way falls outside the physical phase
space, 
\begin{eqn}
\zeta < \zeta_{\mrm{min}}(Q^2_E) ~~~\vee~~~
\zeta > \zeta_{\mrm{max}}(Q^2_E)
\end{eqn}
the branching is vetoed. This occurs some fraction of the time for the
simple reason that we generated the trial
branchings in a hull encompassing the physical phase space. That is, the
trials are generated on a phase-space region bounded by
$\zeta_{\mrm{min,max}}(Q^2_{E\mrm{min}})$, whereas the physical phase
space at $Q_E$ is bounded by $\zeta_{\mrm{min,max}}(Q^2_{E})$. Since
the physical branching probability outside the physical phase space is
obviously zero, the probability to accept unphysical trial branchings should
be zero as well. This is accomplished by the veto. After a failed
trial branching, the evolution is restarted from the scale of the
failed trial, as mandated by the veto algorithm.

The last step is to obtain values for the pair of phase-space
invariants $(s_{ij},s_{jk})$ in terms of which we cast the original
evolution equation, eq.~(\ref{eq:ev1}). Since different forms of
$Q_E$ depend in a different way on these invariants, this step
is obviously $Q_E$-dependent. Here, we give the inversions relevant to
the four evolution variables so far implemented in \Vc. 
\begin{eqn}
s_{ij} \ = \ \zeta g_{\mrm{evolution}}(s,Q_E^2,\zeta)~~~~~~;~~~~~~s_{jk}\ = \ (1-\zeta) g_{\mrm{evolution}}(s,Q_E^2,\zeta)\ 
\end{eqn}
with
\begin{eqnar}
\mbox{\bf Type 1:} & g_{\pT{}} = &
\frac{Q_E\sqrt{s}}{2\sqrt{\zeta(1-\zeta)}}~,\\[3mm]\mbox{\bf Type 2:}
& g_{m_D}   = & \frac{Q_E^2}{\min(\zeta,1-\zeta)}~,\\[3mm]
\mbox{\bf Type 3:} & g_{E^*}   = & Q_E\sqrt{s}~,\\[3mm]
\mbox{\bf Type 4:} & g_{V}     = & 
  \frac{Q_E^4}{2s(\min(\zeta,1-\zeta)+\sqrt{\vert1-2\zeta\vert})}~.
\end{eqnar}
More generally, in terms of the function $f_0$ that parametrizes the
general solution~(\ref{eq:generalsoln}), the inversion takes the following form,
\begin{eqn}
g_{f_0}= \frac{s}{\sqrt{\zeta(1-\zeta)}} \left(\frac{Q_E^2}{s f_0}\right)^{\kappa_E}~.
\label{eq:genlinversion}
\end{eqn}

\subsection{The LL Shower}
In the previous subsections, the ingredients for
generating a single trial branching with a trial branching function
$\atrial$ were described.  To obtain an LL shower, it suffices to
accept each trial branching with a probability 
\begin{eqn}
\pll = \frac{\alpha_s}{\hat{\alpha}_s}  
                \frac{{\cal C}_{ijk}}{\hat{\cal C}_{ijk}}
\frac{\abar_\mrm{LL}(s,s_{ij},s_{jk})}{\abartrial(s,s_{ij},s_{jk})}~,
\label{eq:PLL}
\end{eqn}
where the $\alpha_s/\hat{\alpha}_s$ ratio takes into account the
possibility that the trial generator could be using a nominally larger
$\alpha_s$ than the physically desired one, 
the ${\cal C}/\hat{\cal C}$ factor represents the same for color factors, 
and the antenna function ratio matches the trial function onto
the desired physical splitting antenna for the relevant $2\to 3$
branching.  Because we chose $\abartrial$ to be larger than (or equal to)
the true 
antenna function $\abar_\mrm{LL}$ everywhere in the dipole-antenna phase
space, this probability is always less than (or equal to) 1.
We must also require $\abar_\mrm{LL}$ to be non-negative in order that the ratio
here be interpretable as probability.

We initiate the shower in electron--positron collisions with 
quark-pair creation from the intermediate vector boson.  At each stage
in the shower, a gluon will be emitted, or a gluon will split into 
a quark--antiquark pair.  The shower itself evolves in the leading-color
approximation, so after each emission, the number of different
antenn\ae{} grows by one, whereas each splitting leaves the number
of antenn\ae{} unchanged.  We must allow all the different
antenn\ae{} to branch, of course; we do this by computing trial branchings
for all of them, and picking the antenna with the highest trial branching scale.
For that antenna, we then apply a veto with the probability given in
eq.~(\ref{eq:PLL}).

When a branching is accepted, the physical replacement of partons $I$ and $K$
by $i$, $j$, and $k$ in the event record next has to be performed. It
is here that the dependence on the \emph{kinematics map} enters. Our
treatment of this point is identical to that described in detail in
ref.~\cite{Giele:2007di}, and the implementation in \Vc\ retains the 
possibility to choose between the three different maps defined
there. These maps all have identical limits in the LL singular regions, but differ
from each other elsewhere.

For a pure strongly-ordered LL shower (i.e., a level
of approximation comparable to all other
currently existing shower Monte Carlo implementations), the
evolution should then be restarted from the scale of the current
trial (regardless of whether that trial was accepted or not, as per
the veto algorithm; the phase space above that scale has already been
probed, and hence --- according to the strong-ordering condition --- 
should not be probed again.)

To examine the quality of this type of approximation independently of
the shower generator, we use \textsc{Rambo} \cite{Kleiss:1985gy} (an
implementation of which has been included in \textsc{Vincia}) to
generate a large number of evenly distributed 4-, 5-, and 6-parton
phase-space points.  For each phase-space point, we evaluate the $Z\to
n$ matrix element using \textsc{MadGraph} (suitably modified to be
able to switch subleading color terms on and off).  We then compute
the tree-level LL antenna-shower approximation corresponding to the
same phase-space point, based on nested products of $2\to 3$
branchings subjected to the condition of ordering in the chosen
evolution variable. Finally, we form the ratio between this
approximation and the full matrix element. This is similar to what was
done in ref.~\cite{Skands:2009tb}; where that study was limited to
the emission of two partons, the addition of a new
automated interface to \textsc{MadGraph} allows us here to extend the
corresponding comparisons through four emissions, thus making it
possible to illustrate in detail how the agreement or disagreement changes
with increasing number of emissions.

Using a clustering algorithm that contains the exact inverses of the
\textsc{Vincia} $2\to3$ kinematics maps \cite{Skands:2009tb}, we may
perform $m$ clusterings of the type $(i,j,k)\to(I,K)$ in a way
that exactly reconstructs the intermediate $(n-m)$-parton
configurations that would have been part of the shower history for
each $n$-parton test configuration, for any of the three kinematics
maps so far implemented in \textsc{Vincia}. This gives us an exact
 reconstruction of how the antenna shower would have
populated each path. The strong ordering condition corresponds to step
functions in the shower approximation. E.g., for $Z\to q_1 g_2 g_3
\qbar_4$, we have \cite{Skands:2009tb},
\begin{eqn}
R_{4} = \frac{
 \left(\Theta(Q_{3A}-Q_{4A}) a_{qg}(1,2,3) a_{q\qbar}(\widehat{12},\widehat{23},4)
 + \Theta(Q_{3B}-Q_{4B}) 
a_{g\qbar}(2,3,4) a_{q\qbar}(1,\widehat{23},\widehat{34})\right)|M_2(s)|^2}{|M_4(1,2,3,4)|^2}~,
\label{eq:R4E}
\end{eqn}
where hatted variables $\widehat{\i{}\j}$ denote clustered momenta, $a$
denote $2\to 3$ antenna functions, $|M_n|^2$ denote the 
color-ordered $n$-parton
matrix elements, $s$ is the total invariant mass squared of the
$n$-parton system, and the ordering conditions depend on 
\begin{eqn}
\begin{array}{rclcrcl}
Q_{4A} & = & Q_E(1,2,3) & ; & Q_{3A} & = & Q_E(\widehat{12},\widehat{23},4)\\
Q_{4B} & = & Q_E(2,3,4) & ; & Q_{3B} & = & Q_E(1,\widehat{23},\widehat{34})
\end{array}~.\label{eq:4pAB}
\end{eqn}
The numerator of eq.~(\ref{eq:R4E}) thus reproduces the shower approximation
expanded to tree level, phase-space point by phase-space point, 
for an arbitrary choice of kinematics map,
$(i,j,k)\to (\widehat{\i{}\j},\widehat{\j{}k})$, and evolution variable,
$Q_E$. $R$ thus gives a direct tree-level 
measure of the amount of over- or under-counting by
the shower approximation, with values greater than unity corresponding
to over-counting and vice versa.

\begin{figure}
\scalebox{\figscale}{
\includegraphics*[scale=0.79]{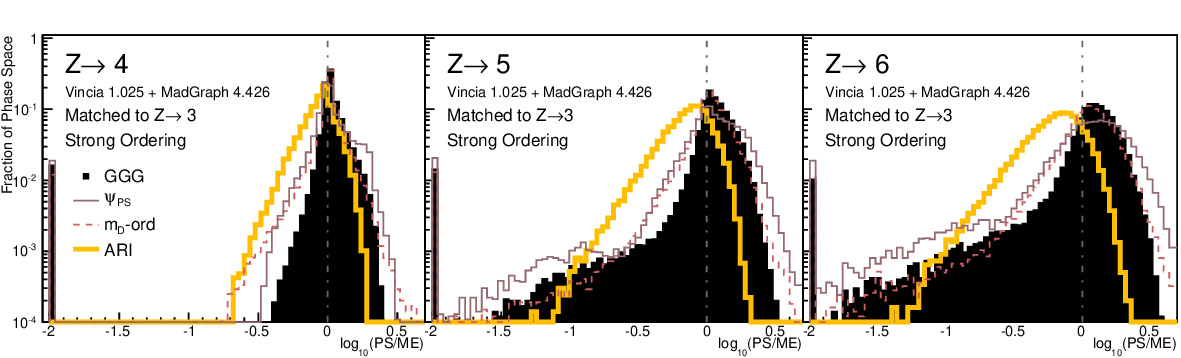}}
\capt{Strongly ordered parton showers compared to matrix elements. 
  Distribution of $\log_{10}(\mrm{PS}/\mrm{ME})$ in a flat phase-space
  scan. Contents normalized by the number of generated points. 
  Spikes on the far left represent the underflow bin ---
  dead zones in the shower approximations. 
  Gluon emission only. Matrix-element weights from
  \Mg~\cite{Alwall:2007st,Murayama:1992gi}, 
leading color (no sum over color permutations).  \label{fig:strongord}}
\end{figure}
We use the kinematics maps defined in ref.~\cite{Giele:2007di},
\begin{eqnar}
\psi_{\mrm{AR}} & = & \frac{E^2_k}{E_i^2+E_k^2}(\pi-\theta_{ik})\,, \\
\psi_{\mrm{PS}} & = & \left\{\begin{array}{cl}
0 & ; s_{ij} > s_{jk}\,, \\
\pi-\theta_{ik} & ; s_{ij} < s_{jk}\,. 
\end{array}\right\}
\end{eqnar} 
where $\theta_{ik}$ is the angle between the after-branching parents
in the CM frame of the branching.
We show the results of these comparisons in fig.~\ref{fig:strongord},
for four different shower approximations:
\begin{itemize}
 \item {\sl GGG:} \pT{}-ordering using default \Vc\ settings, i.e., 
the GGG antenna functions and the $\psi_{\mrm{AR}}$ kinematics map for all
branchings. I.e., the parents share the recoil in proportion to their
energies in the CM of the dipole-antenna. 
\item {\sl $\psi_{\mrm{PS}}$} \pT{}-ordering using
  the GGG antenna functions and the parton-shower-like (PS) longitudinal
  kinematics map. I.e., the parent with the largest invariant mass
  with respect to the emitted parton recoils only longitudinally.
\item {\sl $m_D$-ord:} $m_D$-ordering using the GGG antenna functions
  and the $\psi_{\mrm{AR}}$ kinematics map. 
\item {\sl ARI:}
  \pT{}-ordering using our best imitation of the
what the real \Ar\ program does. It uses $p_\perp$-ordering, but with 
the \Ar\ radiation functions instead of the GGG ones, and it 
also uses a special recoil strategy, as follows; for $qg$ dipoles,
the quark always takes the entire recoil (in the CM of the dipole), whereas
for $gg$ dipoles, the $\psi_{\mrm{AR}}$ angle is used to distribute the
recoil. 
\end{itemize}
In all cases, we compare to one leading-color (LC) matrix element, i.e.,
before summing over colors, and with all color factors having been divided out. 
We present an extensive set of comparisons for different ordering variables
in appendix~\ref{app:comparison}.

The two bins around zero correspond to the parton-shower approximation
having less than a 10\% deviation from the
full matrix element. At four partons, on the left-hand pane, these two
bins contain over 35-60\% of the sampled phase-space points, 
depending on the approximation, with tails extending out towards
larger deviations. 
The spikes at the extreme left edge of the plots 
represent the underflow bin, including $-\infty$,
which corresponds to zones in which all of the possible shower
histories have been removed by the strong-ordering condition.
Such dead zones are characteristic of (ordered) LL parton showers,
when the ordering variable is more restrictive than pure phase
space. We shall later discuss how to remove them while simultaneously
improving the approximation in the ordered region as well. 

For all multiplicities, the
default $p_\perp$-ordering with the antenna-like \Ar\ recoil map
appears to generate the best overall agreement (narrowest
distribution). The parton-shower-like 
longitudinal recoil map (thin solid line labeled $\psi_{\mrm{PS}}$),
following the spirit of \Py~6 and showers based on CS partitioned dipoles
and the dipole-mass ordering (dashed line labeled $m_D$-ord) 
give slightly worse agreement (wider distributions). 

\begin{figure}
\begin{center}
\scalebox{\figscale}{
\includegraphics*[scale=0.55]{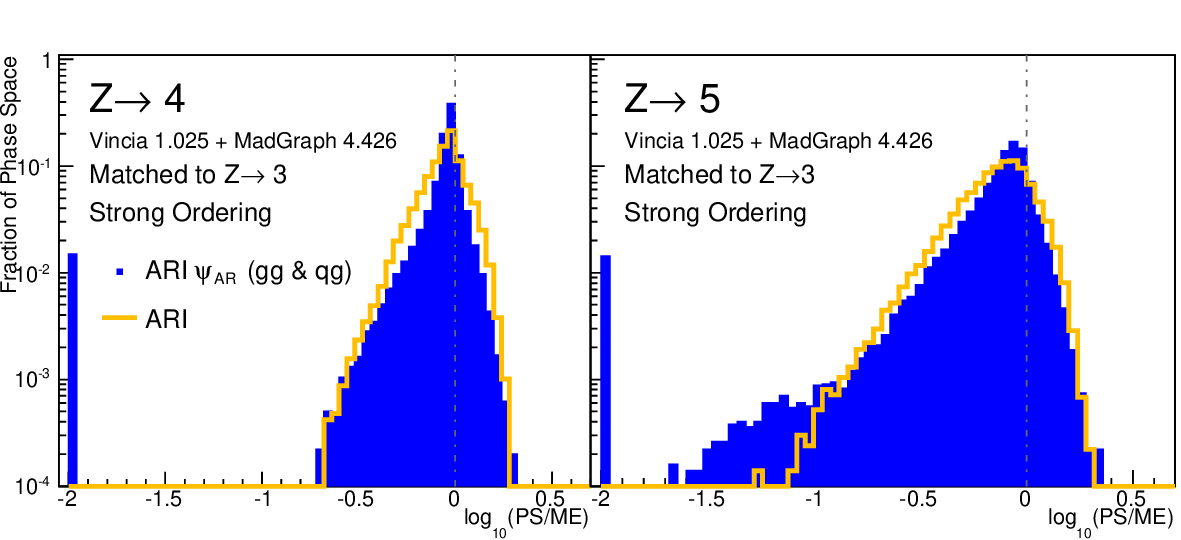}}
\capt{Strongly ordered parton showers, using the \Ar\ radiation
  functions and two different recoil strategies, compared to matrix elements. 
  Distribution of $\log_{10}(\mrm{PS}/\mrm{ME})$ in a flat phase-space
  scan.  Contents normalized by the number of generated points. Gluon
  emission only. Matrix-element weights from 
  \Mg~\cite{Alwall:2007st,Murayama:1992gi}, leading color (no sum over color permutations). 
 \label{fig:ariadne}}
\end{center}
\end{figure}

The ``ARI'' case (thick solid line) has no dead zone for this process
(due to the special kinematics map), 
but it also appears to generate a somewhat wider, and
systematically softer (shifted to the left) distribution, than the
GGG ones. To examine further whether this is an effect of the
intrinsically softer radiation functions used in \Ar\ (as shown in
ref.~\cite{Bern:2008ef}), or of the special recoil strategy employed by it, 
we plot in fig.~\ref{fig:ariadne} the result for 4 and 5 partons using the 
\Ar\ radiation functions, either with the default \Ar\ recoil strategy
(using $\psi_{\mrm{AR}}$ for $gg$ dipoles only, with the quark recoiling for
$qg$ ones --- thick solid line, as above) or just using $\psi_{\mrm{AR}}$ for all
dipoles (solid histogram). 
From this comparison, we conclude that there is no significant 
advantage in assigning all the recoil to the quark in $qg$ dipoles. 
Although it removes the dead zone, together with a tail of largely
undercounted events, the peak is actually degraded slightly,
which we interpret as indicating a worse agreement with the matrix 
elements close to the singular regions. 
Physically, this could be consistent with the
quark direction having to remain unchanged when the gluon branches
collinearly. The position of the peak, slightly
to the left of zero, however, is unchanged. This is consistent with 
the overall ``softness'' of the approximation being 
driven mainly by the finite terms in the radiation functions, 
and not by the choice of kinematics map. It is therefore quite
possible that one could find another process in which the finite terms
in the GGG functions would be too large, and the ones in the \Ar\ ones
just right. 

Returning to fig.~\ref{fig:strongord}, 
as the number of emissions grows, there remains a peak near 
ME/PS = 1 (note in particular the logarithmic $y$ axis), 
but the width of the distribution grows progressively
larger, indicating that there is a larger number of individual 
phase-space points in which the pure shower is not in agreement with
the matrix element. These could be due either to subleading logs, or to
finite terms in the higher-body matrix elements, not captured by the
pure shower. Recall, however, that, since we are looking at a flat
phase-space scan, we are biased towards the 
regions of phase space where matrix element corrections are
important, with the strongly ordered regions occupying a progressively
smaller volume of the total sampled space. Thus, we interpret this
broadening of the weight distribution 
not so much as a sign of any breakdown in the shower approximation itself,
but rather as illustrating why it is desirable to match to fixed-order
matrix elements, a point to which we return below. 
 
\section{Improved Showering}
\label{sec:showering}

\subsection{Improving the Logarithmic Accuracy: $2\to 3$ \label{sec:one-loop}}

A complete set of second-order $2 \to 3$ (one-loop) and $2\to 4$
(tree-level) dipole-antenna splitting functions is given in
ref.~\cite{GehrmannDeRidder:2005cm}. Ultimately, a matching to
these functions to all orders in the shower would be required to reach
formal precision beyond LL, but we note that a simpler, partial
matching can already be carried out at the $2\to 3$ level, to the
terms generated purely by the running of the coupling.

These terms can be
uniquely identified both in the shower expansion and in the
fixed-order antenna functions by the fact that they are 
proportional to the QCD
one-loop-running coefficient $b_0$~(\ref{eq:b0}), 
which appears in the expansion
\begin{eqn}
\alpha_s(\mu_1^2) = \alpha_s(\mu_2^2)\left[1 - \alpha_s b_0
  \ln\left(\frac{\mu_1^2}{\mu_2^2}\right) + \mathcal{O}(\alpha_s^2)+...\right]~.
\end{eqn}

By matching the terms arising from the expansion of $\alpha_s$ in 
the shower to the actual $b_0$-dependent pieces of the one-loop antenna
functions, we obtain a set of universal corrections at the one-loop
level which stabilize the scale dependence of the resulting calculation
to next-to-leading logarithmic (NLL) accuracy.

In particular, we may extract the relevant terms of the one-loop antenna
functions by isolating their $n_f$-dependent pieces, which are
generated purely by quark loops. 
For gluon emission, the one-loop 
antenn\ae{} in ref.~\cite{GehrmannDeRidder:2005cm} 
all contain the following $n_f$-dependent logarithms,
\begin{eqn}
n_f \frac{1}{6}\left(\ln y_{ij} + \ln y_{jk}\right) a_{ijk}^0 = n_f
\frac{1}{6}\ln\left(\frac{\pT{}^2}{s_{IK}}\right) a_{ijk}^0 ~, \label{eq:Nf}
\end{eqn}
where $a_{ijk}^0$ denotes the corresponding tree-level antenna
function, and $\pT{}$ is defined exactly as in \textsc{Ariadne} and
\textsc{Vincia}, i.e.,
\begin{eqn}
\pT{}^2 = \frac{s_{ij}s_{jk}}{s_{IK}}~. \label{eq:pt}
\end{eqn}
Because the default renormalization scale used in
ref.~\cite{GehrmannDeRidder:2005cm} is 
\begin{eqn}
\mu_{\mrm{GGG}}^2 = s_{IK}~,
\end{eqn}
a redefinition of the renormalization scale from
$s_{IK}$ to $\pT{}^2$ would absorb the entire term, eq.~(\ref{eq:Nf}),
into the definition of the coupling at tree level. 
Simultaneously, the $N_C$-dependent logarithms
generated by the same choice can easily be verified to 
cancel equivalent pieces in the one-loop function, however
the latter cancellation is not exact due to
the presence of additional terms in the one-loop function 
which do not originate from renormalization. (To absorb also these
terms would require full one-loop matching.)
We note that this is nothing but a ``renormalization-group-improved'' effective
redefinition of the tree-level coupling which has been known
for a long time~\cite{Catani:1990rr}
and is the reason why the default renormalization scale
both  in \textsc{Vincia} and in virtually all other Monte 
Carlos\footnote{Note, however, that eq.~(\ref{eq:pt}) is the
  only definition of \pT{} that \emph{exactly} matches the actual
  $b_0$-dependent one-loop terms. Parton shower models not based on
  the dipole-antenna picture, which make approximations to this
\pT{} definition, will therefore necessarily have small
$b_0$-dependent remainders left uncanceled.} 
is  $\pT{}^2$. 

However, rather than just hardcoding one particular choice, 
we shall here instead interpret it in the
context of a second-order matching condition, which will allow us the
flexibility to estimate the remaining uncertainties by varying the
scale freely and partly canceling the dependence on it via 
the matching condition. This effectively pushes the effects of
the scale variation one order higher in QCD. 

The relevant (partial) matching equation for gluon emission is 
\begin{eqn}
\alpha_s(\pT{}^2) a^0_{ijk} = \alpha_s(\mu_{\mrm{PS}}^2) P_g^{\mrm{NLL}_\mu}a^0_{ijk}~,
\end{eqn}
which, expanded to first order, easily yields the following form for the 
scale-stabilizing multiplicative factor:
\begin{eqn}
\mbox{gluon emission : }~~~P_g^{\mrm{NLL}_\mu} = \left(1+\alpha_s b_0 \ln \left( \frac{\mu_{\mrm{PS}}^2}{\pT{}^2}\right) \right)~,\label{eq:scalefixg}
\end{eqn}
where $\mu_{\mrm{PS}}$ is the renormalization scale used by
the shower evolution and the renormalization scale of $\alpha_s$
in the correction term constitutes an ambiguity of yet higher order. 
In order to be conservative, we wish to make the effects
of the scale cancellation produced by this term as small as
possible. We therefore evaluate the $\alpha_s$ in
eq.~(\ref{eq:scalefixg}) at the largest scale in the $2\to 3$
splitting, $s_{IK}$. Any further optimization would amount to a
beyond-NLL effect. 

Qualitatively, the scale stabilization works as follows. If
$\mu_{\mrm{PS}}$ is chosen large, then the correction factor,
eq.~(\ref{eq:scalefixg}), becomes greater than one, hence partly
compensating for the lower $\alpha_s$ value. Conversely, if a very
large $\alpha_s$ is used at the LL level, the logarithm in the 
correction term becomes negative, and again acts to stabilize the
result. We note that, in extreme cases, the correction term could 
in fact become larger than unity. As this 
would imply a divergent perturbative
expansion anyway,  \textsc{Vincia} therefore restricts the range of allowed 
values to $0<P^{\mrm{NLL}_\mu}<2$.

For gluon splitting to quarks, the one-loop antenna functions do not
contain universal logarithms in $\pT{}$. Instead, the universal
$n_f$-dependent terms are \cite{GehrmannDeRidder:2005cm}
\begin{eqn}
n_f\frac{2}{3}\ln\left(y_{q\bar{q}}\right)
= n_f\frac{2}{3}\ln\left(\frac{m_{q\bar{q}}^2}{s_{IK}}\right)~,
\end{eqn}
where $m_{q\bar{q}}$ is the invariant mass of the quark-antiquark pair
produced in the splitting. 
In this case, we see that the ``optimal choice'' of
renormalization scale is not $\pT{}^2$ but $m_{q\bar{q}}^2$. The
corresponding scale-stabilizing term for gluon splitting is therefore
\begin{eqn}
\mbox{gluon splitting : }~~~P_q^{\mrm{NLL}_\mu} = 
\left(1+\alpha_s b_0 \ln \left(
\frac{\mu_{\mrm{PS}}^2}{m_{q\qbar}^2}\right) \right)~. \label{eq:scalefixq}
\end{eqn}
Again, one can easily verify that the $N_C$-dependent logarithms
generated by this choice do partly cancel similar pieces in the same
one-loop antenna functions, and again, the latter cancellation is not
exact due to additional pieces unrelated to renormalization. Finally,
we note that since 
\begin{eqn}
m_{q\bar{q}}^2 > \pT{}^2~
\end{eqn}
over all of phase space, the effect of this stabilization is to reduce
the total number of gluon splittings slightly, as compared to what
would be obtained without the stabilization terms 
and using $\alpha_s(\pT{}^2)$ for
both gluon emissions and gluon splittings, as is traditionally 
done in shower Monte Carlos. 

These scale stabilizing terms have been implemented in the \textsc{Vincia}
code for all $2\to 3$ splittings 
since version 1.020, with an option to switch them on and off to
investigate their effects. The default in the code is to leave them
on.  

Let us emphasize again that this is not a complete one-loop
matching. With the scale variation, we seek only to
evaluate --- the scale variation. We do not make any assumption that this
variation is representative of the entire remaining uncertainty, on
which we have several other, independent, handles, to which we shall
return below. The procedure of employing scale 
variation alone as a (poor man's) estimate of the full 
uncertainty is obsolete in this framework.

\subsection{Improving the Logarithmic Accuracy: $2\to 4$}

While parton emission using trial branchings can easily be
made to cover the full phase space for a single emission, the
same is not true for multiple emissions. Due to the requirement of
strong ordering, some regions of phase
space may be inaccessible, leading to so-called dead zones.
This also happens in strongly ordered dipole-antenna showers, 
for example in regions where several emissions happen at 
closely similar emission scales, as shown in
ref.~\cite{Andersson:1991he,Skands:2009tb}. Since gluon emission and gluon
splitting processes have different singularity structures and 
are treated slightly differently, we first consider the dominant case,
that of gluon emission. We then give a few brief remarks about gluon
splitting, although we defer most of the details of that discussion to another
publication \cite{GehrmannDeRidder:2011dm}.

\subsubsection{Gluon Emission}
\begin{figure}
\begin{center}
\scalebox{\figscale}{
\mbox{\hspace*{-0.6cm}
\includegraphics*[scale=0.4]{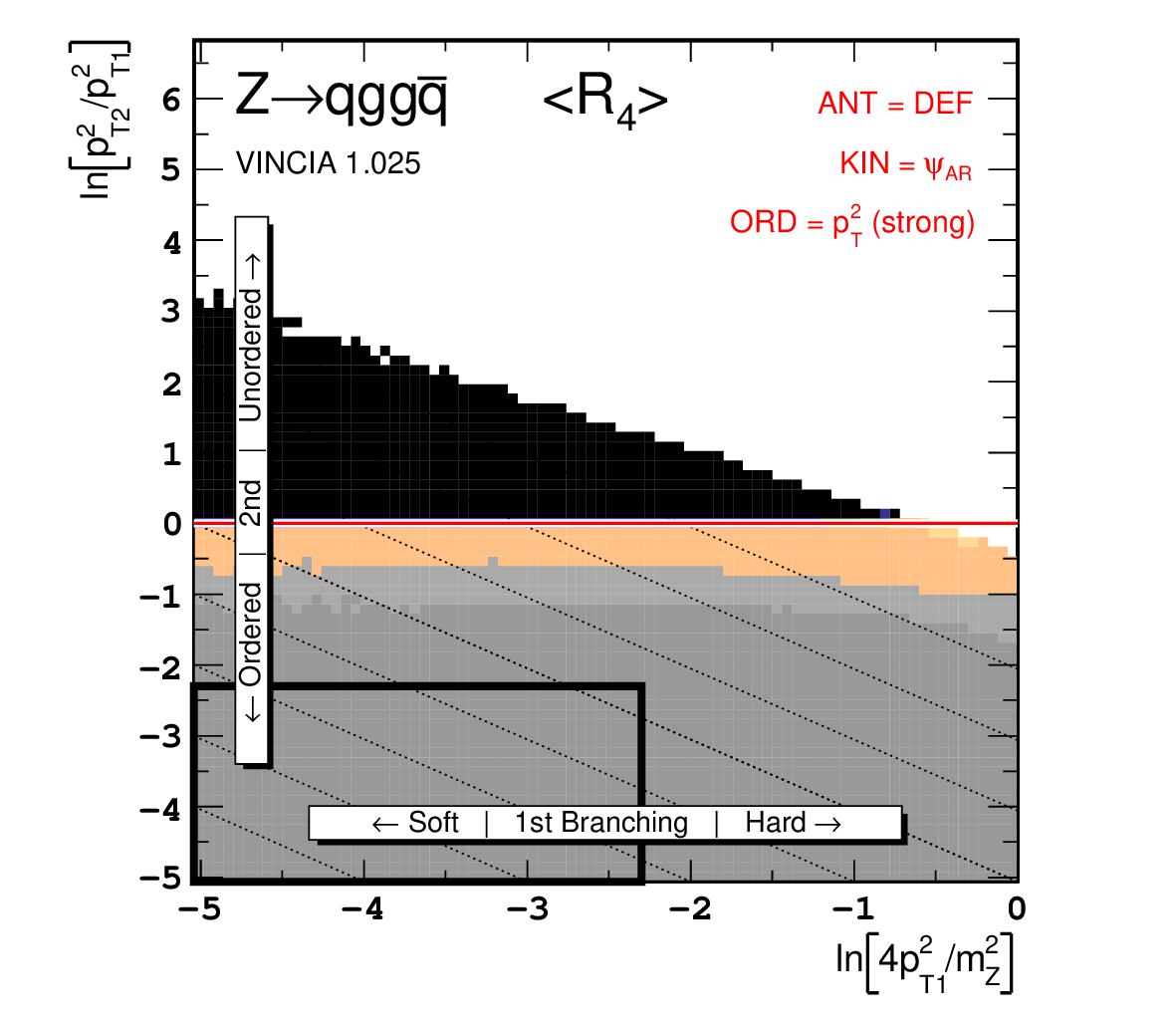}\hspace*{-12mm}
\raisebox{6mm}{
\includegraphics*[scale=0.4]{scalez.eps}}
\hspace*{-2.3cm}
\includegraphics*[scale=0.4]{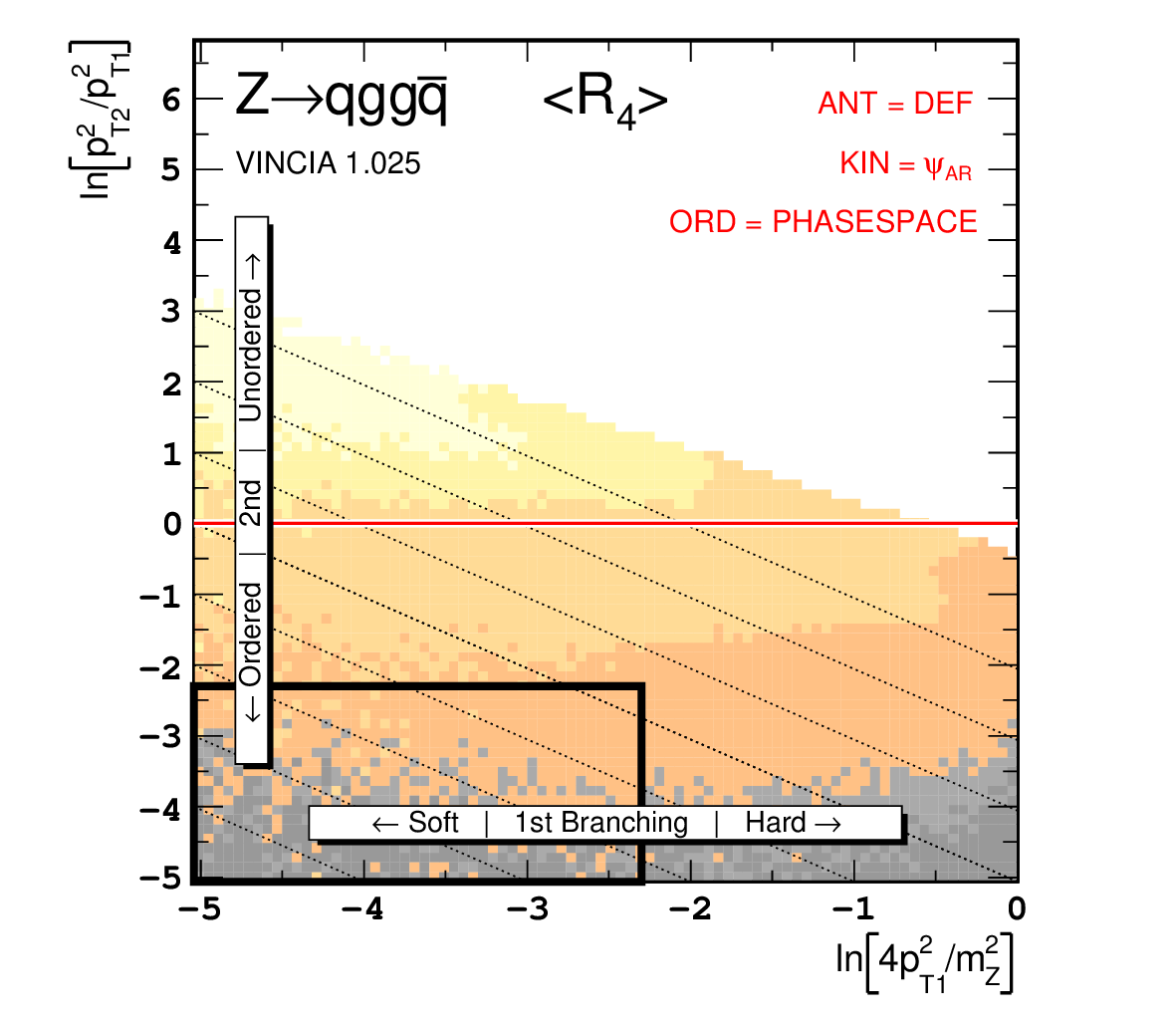}\hspace*{-12mm}
\raisebox{6mm}{
\includegraphics*[scale=0.4]{scalez.eps}}
\hspace*{-2.3cm}}}
\capt{\label{fig:deadzone}
The value of $\left<R_4\right>$ differentially over 4-parton phase
  space, with $p_\perp$ ratios characterizing the first
  and second emissions on the $x$ and $y$ axes, respectively.
Strong ordering in $p_\perp$ (left) compared to no ordering (right). 
Gluon emission only. Matrix-element weights from
  \Mg~\cite{Alwall:2007st,Murayama:1992gi}, leading color (no sum over color permutations). 
}
\end{center}
\end{figure}

A plot from ref.~\cite{Skands:2009tb}, showing the dead zone for $Z\to qgg\qbar$ 
in a $p_\perp$-ordered dipole-antenna shower, is reproduced in the
left-hand pane of Fig.~\ref{fig:deadzone}. Each bin of this 2D
histogram shows the value of $R_4$, eq.~(\ref{eq:R4E}), averaged over all
4-parton phase space points that populate that bin. 
The black zone above the strong-ordering line corresponds exactly to
the spike on the left-hand edge of the plots in
fig.~\ref{fig:strongord} (the underflow bin). 

If one simply removes the strong-ordering condition, equivalent to
ordering the emissions only by the nesting of the factorized phase
spaces, the dead zone obviously disappears. However, this comes at the price of
introducing a substantial double counting, which also extends deep
into the ordered region. We illustrate this on the plot in the
right-hand pane of fig.~\ref{fig:deadzone}, in which  $\left<R_4\right>
\gg 1$ both in the unordered region as well as in parts of the ordered region
where the agreement was previously good. Clearly, therefore, the 
ordered description, in the left-hand pane, is a better overall
approximation to QCD, even if it does include a dead zone. 

In order to go further, we must understand the physics behind the
ordered and unordered approximations to the matrix elements. Why is
ordering so important? The first exploration of this goes back to the early
eighties. Then, it was realized that parton showers ordered only in 
parton virtuality (which is equivalent to a pure phase-space ordering
in that language) represent an essentially incoherent addition of
independently radiating monopoles. In phase-space regions where the
contributions from each such monopole term are comparable, 
interference effects can become large. Without them, the
pure phase-space ordered shower gives a substantial overcounting of
those regions, as compared to matrix elements
\cite{Bengtsson:1986et,Bengtsson:1986hr}. As Marchesini and Webber 
showed \cite{Marchesini:1983bm}, 
this double counting can be approximately identified with
terms corresponding to non-angular-ordered emissions, and hence the
procedure to impose coherence on traditional parton showers has since
been to impose such an ordering, either implicitly by the choice of
evolution variable, as in \textsc{Herwig}, or explicitly as a veto on
the generated trial emissions, as in \textsc{Pythia}. 
 
In  dipole-based shower models, soft coherence inside each
dipole is guaranteed, regardless of the ordering variable, 
by using dipole-based radiation functions instead of the DGLAP ones,
but the problem still exists; it has  
just been pushed one order higher in the number of interfering
partons (see, e.g., refs.~\cite{Dokshitzer:2008ia,Skands:2009tb}).  
With pure phase-space ordering, dipole-antenna showers essentially
represent an incoherent addition of independently radiating
\emph{dipoles}. An independent addition of two such dipoles would
result in a substantial overcounting in all regions where several such 
 dipole terms contribute simultaneously at similar levels, 
i.e., in regions where dipole-dipole interference effects (or,
equivalently, multipole effects) would be important. 
Again, it would be interesting to investigate whether 
some variant of angular ordering could be used to restore a more
coherent behavior, but in the context of the dipole-antenna 
formalism we develop here, we have not been able to find such a
solution. In part, this owes to a strict ordering in angle having some
disadvantages in our language; being frame-dependent, it would not respect the
Lorentz-invariant dipole phase-space factorization we rely on, 
and it also classifies a subset of infinitely soft and/or collinear 
emissions as happening at finite values of the
evolution variable, which would lead to ill-defined evolution
integrals, see, e.g., ref.~\cite{Skands:2009tb}. 

Instead, let us recall the basic motivation for angular ordering: to
approximately remove the double counting caused by incoherent addition of
interfering diagrams of similar magnitudes. In the parton shower
language, each such term is associated with a divergence in the energy
times angle of the emission. 
In the region where several terms would nominally be large, the
angular ordering requirement forces at most one of them to contribute
--- approximating the destructive-interference effects by killing the
non-ordered contributions. In dipole-based approaches, however, the
leading divergence of the gluon radiation functions occurs unambiguously
in the $\pT{}$ of the emitted gluon. It therefore seems sensible to
use $\pT{}$ as the measure for the ordering, and thereby implicitly
for the size of each of the contributing terms. 
As can be seen from the left-hand pane of
fig.~\ref{fig:deadzone}, an ordering in $p_\perp$, 
yields a quite good average approximation to the full
$2\to 4$ matrix elements over most of phase space.

We included the above discussion to motivate that, while there is an
important physics issue behind strong ordering and also behind 
the choice of functional form of the ordering variable, there is
nothing particularly important about imposing it as a step
function in that variable. On the contrary, 
the actual destructive-interference terms in
the matrix elements exhibit a smooth behavior. 
To remove dead zones to all orders while
simultaneously improving the shower approximation also in the ordered
region, we therefore propose to change the condition of strong
ordering to a smoother condition with the same limiting behaviors.

Specifically, while we retain strong ordering as an option 
in \Vc, by default we replace the strong-ordering
condition of conventional parton showers in gluon emission
by the smooth suppression factor
\begin{eqn}
\mbox{Gluon Emission ~ : ~ }
\Theta_{\mrm{ord}} \ \pll
\to \pimp \pll = \frac{\hat{p}_{\perp}^2}{\hat{p}_{\perp}^2 + \pT{}^2} \ \pll~,
\label{eq:unord}
\end{eqn}
where $\hat{p}_{\perp}$ is the smallest $\pT{}$ scale among all the
color-connected parton triplets in the parent configuration (i.e., a
global measure of the ``current'' $\pT{}$ scale of that topology);
$\pT{}^2$ is the scale of the trial $2\to 3$ emission under
consideration; and $\pll$ is defined in \eq{eq:PLL}. Note that, if the
trial emission is not accepted,  the veto algorithm still mandates
that the evolution be continued starting from the scale of the failed
trial. Only after an accepted branching, when we change from $n$- to
$(n+1)$-parton phase space, do we allow the evolution to restart from
the full phase space.

Since the antenna function for the previous branching is proportional
to $1/\hat{p}_\perp^2$, the net effect of this term, in the 
unordered region, is to replace that divergence by a damped 
factor, $1/(\hat{p}_\perp^2+p_\perp^2)$. 
 The correction is thus constructed such that $\pll$ remains unmodified in
the strongly ordered limit $\pT{}\ll \hat{p}_\perp$,
and therefore will not affect the leading-logarithmic behavior of the
parton shower. It then drops off 
to $\frac12 \pll$ for $\pT{} = \hat{p}_\perp$, and finally tends smoothly to
zero in the limit of extreme unordering, $\pT{}\gg
\hat{p}_\perp$.  

\begin{figure}
\begin{center}
\scalebox{\figscale}{
\mbox{\hspace*{-0.6cm}
\includegraphics*[scale=0.4]{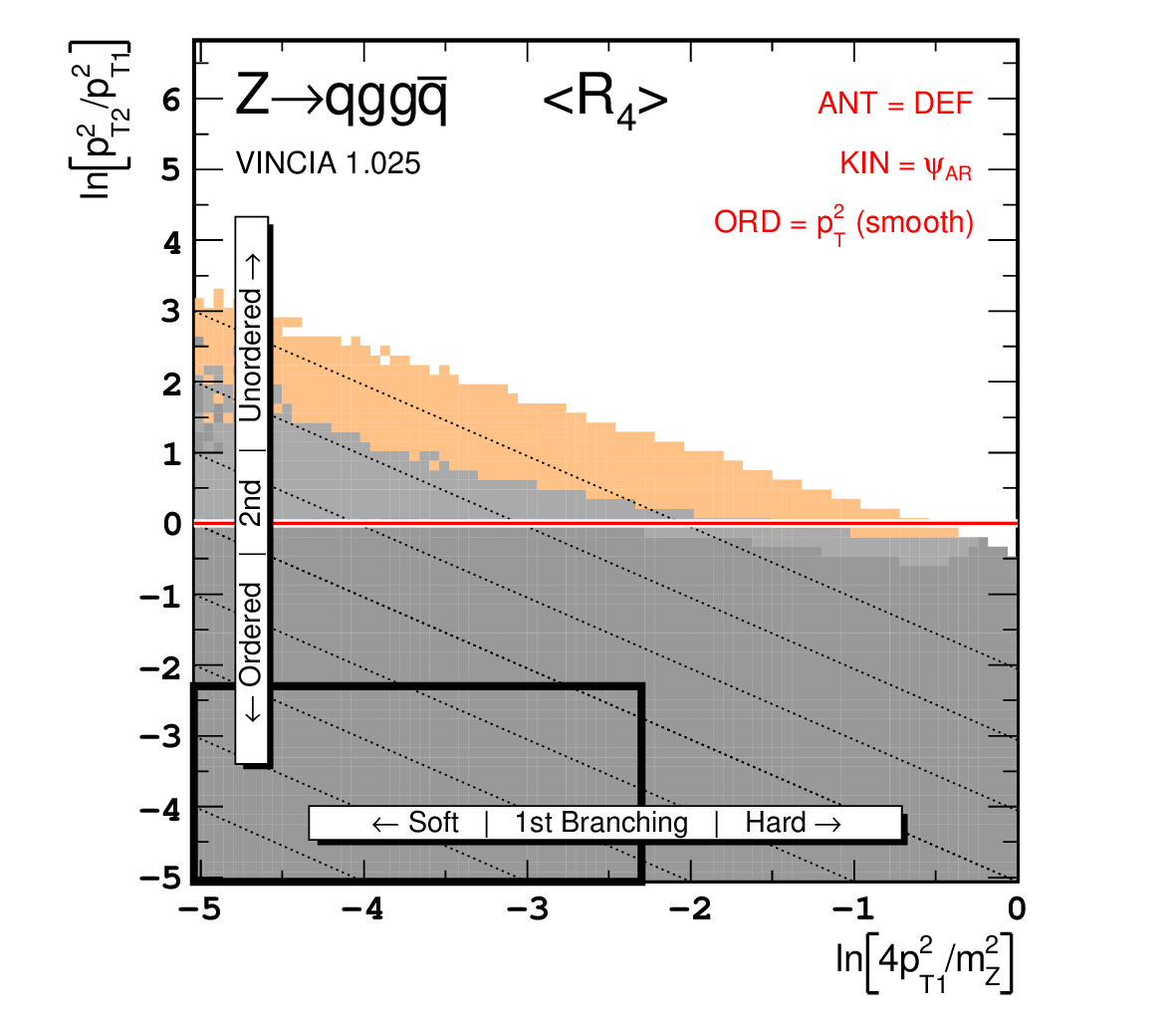}\hspace*{-12mm}
\raisebox{6mm}{
\includegraphics*[scale=0.4]{scalez.eps}}
\hspace*{-2.3cm}
\includegraphics*[scale=0.4]{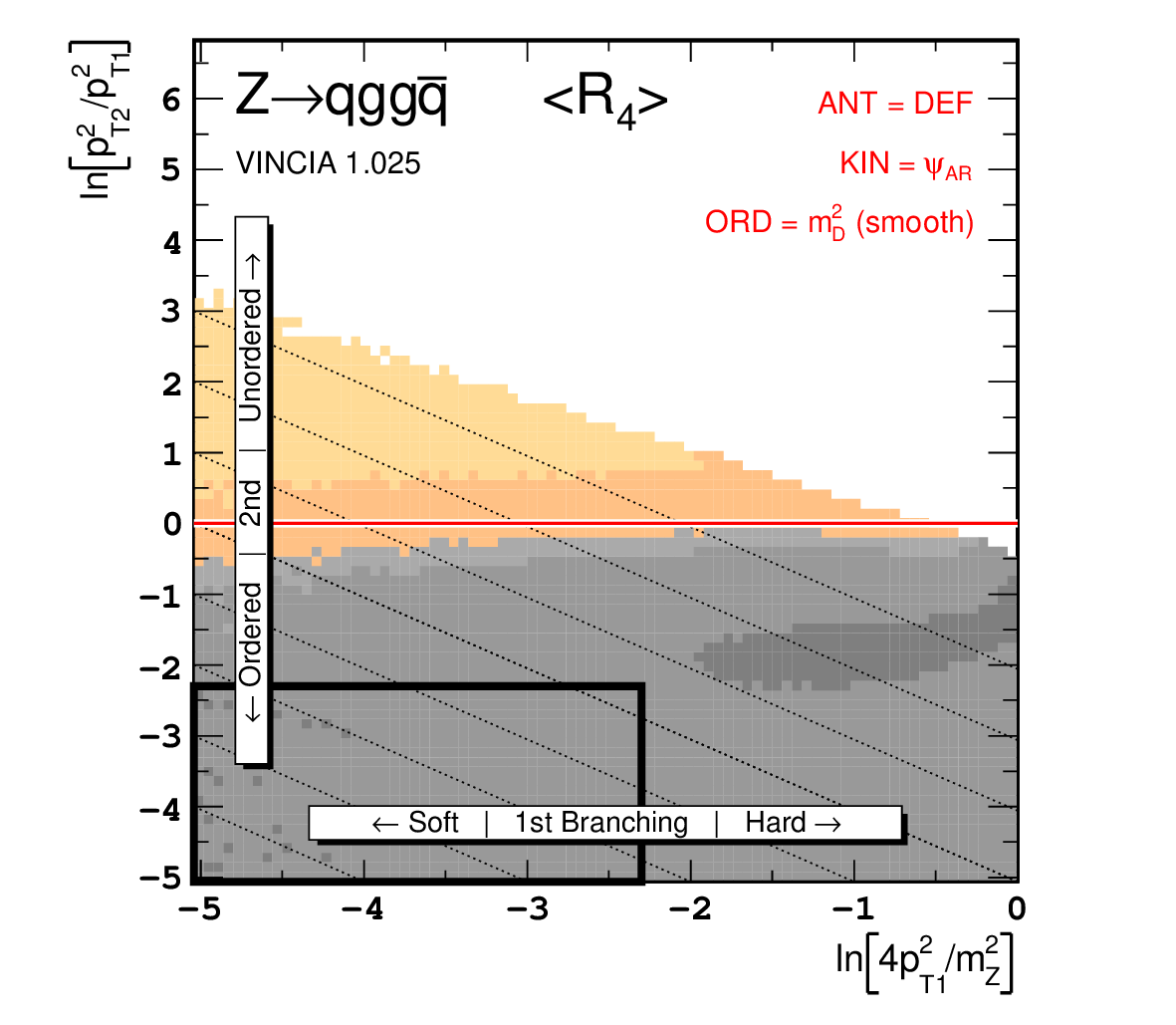}\hspace*{-12mm}
\raisebox{6mm}{
\includegraphics*[scale=0.4]{scalez.eps}}
\hspace*{-2.3cm}}}
\capt{The value of $\left<R_4\right>$ differentially over 4-parton phase
  space, with $p_\perp$ ratios characterizing the first
  and second emissions on the $x$ and $y$ axes, respectively.
Smooth ordering in $p_\perp$ (left) compared to smooth ordering in
$m_D$ (right). 
Gluon emission only. Matrix-element weights from
  \Mg~\cite{Alwall:2007st,Murayama:1992gi}, leading color (no sum over color permutations). 
\label{fig:smoothord}}
\end{center}
\end{figure}

The ratio of the resulting shower to matrix elements
is shown in the left-hand pane of
fig.~\ref{fig:smoothord}. Comparing this distribution with those in 
fig.~\ref{fig:deadzone},
we indeed see that not only has the dead zone been removed, without
introducing any serious overcounting of it, but the quality of the
approximation has also been improved inside the ordered region. 

For completeness, in the right-hand pane of
fig.~(\ref{fig:smoothord}), we also show how the approximation would
have looked if the alternative measure $m_D^2 = 2\,\mrm{min}(m_{ij}^2,m_{jk}^2)$
had been used instead of $\pT{}$ in the suppression factor
eq.~(\ref{eq:unord}). Although there is still clearly an improvement
over the pure phase-space-ordered case --- the dead zone has been
eliminated --- it is much less convincing
than for $\pT{}$, as the weights are larger in the region above the thin
horizontal red line.

\begin{figure}
\scalebox{\figscale}{
\includegraphics*[scale=0.79]{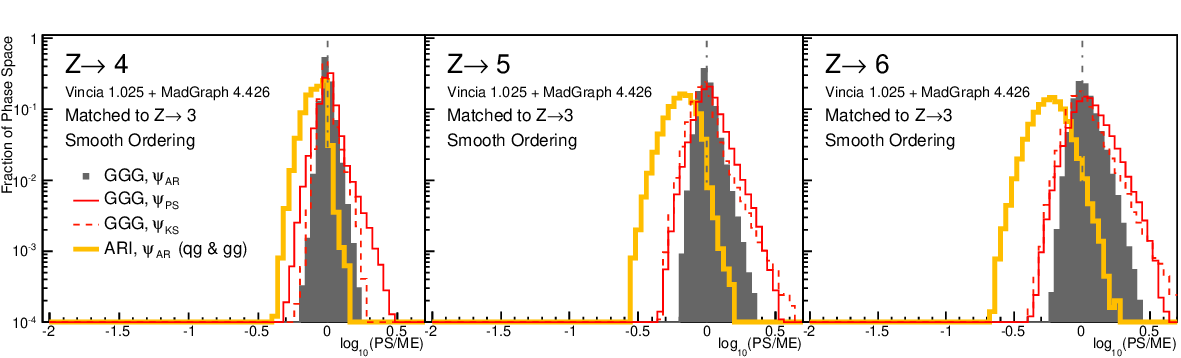}}
\capt{
Smoothly ordered parton showers compared to matrix elements. 
  Distribution of $\log_{10}(\mrm{PS}/\mrm{ME})$ in a flat phase-space
  scan.   Contents normalized by the number of generated points.
 Gluon emission only. Matrix-element weights from
  \Mg~\cite{Alwall:2007st,Murayama:1992gi}, leading color (no sum over color permutations). 
Compare to fig.~\ref{fig:strongord} for
  strong ordering. 
  \label{fig:multipole}}
\end{figure}

To illustrate how this approximation evolves with parton multiplicity,
we show the distribution of the log of the PS/ME ratio with this
modification, in fig.~\ref{fig:multipole}, for $Z\to 4$, $5$, and $6$
partons, including only leading-color 
gluon emission. One observes a marked
improvement with respect to the strongly ordered approximations, 
fig.~\ref{fig:strongord}, for all multiplicities. In particular, not
only the dead zones but also the
large tails towards low PS/ME ratios visible in the
higher-multiplicity plots in fig.~\ref{fig:strongord} have
disappeared, which we interpret as a confirmation that the logarithmic
accuracy of the approximation has indeed been improved. Notice,
however, that the \Ar\ functions (where we have here used the
$\psi_{\mrm{AR}}$ kinematics map for both $qg$ and $gg$ antenn\ae{}, hence
the explicit label on the plot) still tend to shift the shower
approximations systematically towards softer values, whereas the GGG
ones remain closer to the matrix elements. 

\subsubsection{Gluon Splitting}
\label{GluonSplittingSubsection}
For gluon splitting, there is no soft singularity, only a collinear
one. This means there is now only a single log-enhancement (instead of a
double log) driving the approximation and 
competing with the (uncontrolled) finite terms. It is therefore to be
expected that the LL approximation to gluon splitting is significantly
worse, over more of phase space, than is the case for gluon emission. 

Furthermore, if the two neighboring 
dipole-antenn\ae{} that share the splitting gluon 
are very \emph{unequal} in size, e.g., as a result of
a preceding close-to-collinear branching, then higher-order matrix elements and
splitting functions unambiguously indicate that the total gluon
splitting probability is significantly suppressed. 
This is not taken into account when treating the two antenn\ae{} as independent
radiators. This effect was
already noted by the authors of \textsc{Ariadne}, and a first attempt
at including it approximately was made by applying
the following additional factor to gluon splittings in
\textsc{Ariadne}, in addition to the strong-ordering condition, 
\begin{eqn}
\mbox{Gluon Splitting (\Ar) ~ : ~} 
\Theta_{\mrm{ord}} \ \pll~ \to \Theta_{\mrm{ord}}
\pari \pll = \Theta_{\mrm{ord}} \frac{2s_{N}}{s_{IK} +
  s_{N}} \ \pll~~, 
\label{eq:Pari}
\end{eqn}
where $s_N$ is the invariant mass squared of the neighboring dipole-antenna
that shares the gluon splitting, and $s_{IK}$ is the invariant mass
squared of the dipole-antenna in which the splitting occurs. The
additional factor reduces to unity when the two neighboring invariants
are similar; it suppresses splittings in an antenna whose
neighbor has a very small invariant mass; and it slightly enhances
splittings in dipole-antenn\ae{} whose neighbors have very large
invariant masses.   This modification is {\it ad hoc\/}, but formally
is beyond LL, and thus does not spoil the shower's properties at
this order.  In practice, it greatly improves the shower's approximation
to matrix elements.

For our purposes here, we first replace the $\Theta$ function in
\textsc{Ariadne} by the same smooth 
unordering suppression factor we used above, 
\begin{eqn}
\mbox{Gluon Splitting ~ : ~} 
\Theta_{\mrm{ord}} P_{LL} \to \pimp \pari P_{LL}~,
\end{eqn}
which gives an overall approximation at least as good as that of \Ar,
without any dead zones. The overall agreement
between even this improved gluon splitting and explicit matrix elements
is still far from perfect, however, due to the intrinsically smaller
relative size of the logs 
driving the approximation. We are in the process of 
preparing a more dedicated study of this issue, including quark mass
effects \cite{GehrmannDeRidder:2011dm}, and hence defer further
discussion of this 
topic for the time being.
For later convenience, we define this adjustment factor to be unity
for gluon emission,
\begin{eqnar}
\mbox{Gluon Emission~:}&& \pari = 1\,,\\
\mbox{Gluon Splitting~:}&& \pari = \frac{2 s_N}{s_{IK} + s_N}\,.
\end{eqnar}

\subsection{Subleading Color \label{sec:nlc}}
In the dipole-antenna formalism, a general result
\cite{Berends:1988zn} is that the
subleading-colour effects in a single $qgg...g\bar{q}$ chain can be
taken into account by including a subleading antenna spanned directly
between the $q$ and $\bar{q}$ associated with a color factor
$-1/N_{C}^2$ relative to the leading-color antennae (which are
proportional to $C_A$). 

However, in the context of a probabilistic framework, such as shower
Monte Carlo algorithms, the negative sign of this antenna means it
cannot be treated on the same footing as the (positive-definite) LC
ones\footnote{One could in principle imagine flipping the sign of the
  event weight when generating emissions with it, but such a procedure 
would drive the convergence rate of the resulting algorithm to become
infinitely slow at asymptotically large energies.}.  
Moreover, it is not  possible to define a unique LC color
assignment to the emissions generated by it, and hence the
subsequent shower evolution (and the infinite-order approximations
generated by it) would be ill-defined. 

Instead, in traditional parton-shower applications, this
correction is partly treated by associating quark emission terms with
$C_F$ instead  of $C_A$, thereby correctly absorbing the collinear
singularities of the 
correction term into the LC ones, at the price of introducing a
subleading-color ambiguity in the soft singularity
structure. 
To improve on this, one could imagine, e.g., trying to be
more clever about in which phase-space regions to use $C_F$ and in which
$C_A$ even for emissions off gluons \cite{Eden:1998ig}. But in both
cases the simplicity of the correction term would then be less explicit.  

Instead, we here attempt to reabsorb the subleading-color correction
systematically into the leading-color antenn\ae{}, using a smooth 
partitioning, which integrates to reproduce the double 
poles of the corresponding subleading-color one-loop antenna functions. 

Consider the evolution integral off an $n$-parton configuration in the massless
approximation:
\begin{eqn}
\sum_{IK\in LC} \int \d{s_{ij}} \d{s_{jk}}
\frac{\tilde{a}_{IK}(y_{ij},y_{jk})}{s_{IK}^2}~,  
\end{eqn} 
where $\tilde{a}$ is a dimensionless antenna function, and the scaled 
invariants are defined by $y_{ij} = s_{ij}/s_{ijk} = s_{ij}/s_{IK}$,
and where the sum is over all color-connected pairs, that is all LC
antenn\ae{}.
Changing integration variables to
dimensionless quantities, the integration measure becomes independent
of the size of the antenna phase spaces and we may replace the sum of
integrals by the integral of a sum, to which we can add a
subleading-color piece, 
\begin{eqn}
\int \d{y_{ij}} \d{y_{jk}} \left[ \sum_{IK\in LC}
  \tilde{a}_{IK}(y_{ij},y_{jk}) - \frac{1}{N_C^2}
  \tilde{a}_{NLC}(y_{ij},y_{jk})\right]~. 
\end{eqn}
Finally, we may partition the correction term among the LC pieces by
introducing a partitioning function, $f$, 
\begin{eqn}
= \int \d{y_{ij}} \d{y_{jk}} \left[ \sum_{IK\in LC}
  \tilde{a}_{IK}(y_{ij},y_{jk})\underbrace{\left(1 - \frac{f_{IK}}{N_C^2} 
  \frac{\tilde{a}_{NLC}(y_{ij},y_{jk})}{ \tilde{a}_{IK}(y_{ij},y_{jk})}\right)}_{P_{NLC}}\right]~,\label{eq:slccode}
\end{eqn}
where the underbraced term can now be implemented straightforwardly as
a veto probability, $P_{NLC}$, in the shower evolution. 

For $f$ to be a consistent partitioning, it must give unity when
summed over all the LC antennae. The prescription we use is 
to absorb corrections into each term in proportion to the relative
size of that term, i.e., 
\begin{eqn}
 f_{IK} = \frac{\tilde{a}_{IK}(y_{ij},y_{jk})}{\sum_{AB\in LC}
   \tilde{a}_{AB}(y_{ar},y_{rb})} ~. \label{eq:fnlc}
\end{eqn}
For the functional form of $\tilde{a}$, we give two options, 
\begin{enumerate}
\item Only the Eikonal part of the $q\bar{q}$ antenna function is
  included in $\tilde{a}_{NLC}$, corresponding to 
\begin{eqn}
\tilde{a}_{NLC}(y_{ij},y_{jk}) = 
\frac{2(1-y_{ij}-y_{jk})}{y_{ij}y_{jk}}
\end{eqn}
This integrates to give the correct $1/\varepsilon^2$ poles of the
$1/N_C^2$ piece of the one-loop antenna function (called $\tilde{A}_3^1$
in ref.~\cite{GehrmannDeRidder:2005cm}).
\item The full $q\qbar$ antenna function (called $A_3^0$ in
  \cite{GehrmannDeRidder:2005cm}) is included in $\tilde{a}_{NLC}$, 
corresponding to 
\begin{eqn}
\tilde{a}_{NLC}(y_{ij},y_{jk}) = 
\frac{2(1-y_{ij}-y_{jk})}{y_{ij}y_{jk}} + \frac{y_{ij}}{y_{jk}} +
\frac{y_{jk}}{y_{ij}} \label{eq:anlc2}
\end{eqn}
This integrates to give the correct $1/\varepsilon^2$ and $1/\varepsilon$
poles of the $1/N_C^2$ piece of the one-loop antenna function (called $\tilde{A}_3^1$
in ref.~\cite{GehrmannDeRidder:2005cm}). 
\end{enumerate}

\begin{figure}[t]
\begin{center}
\scalebox{\figscale}{
\includegraphics*[scale=0.65]{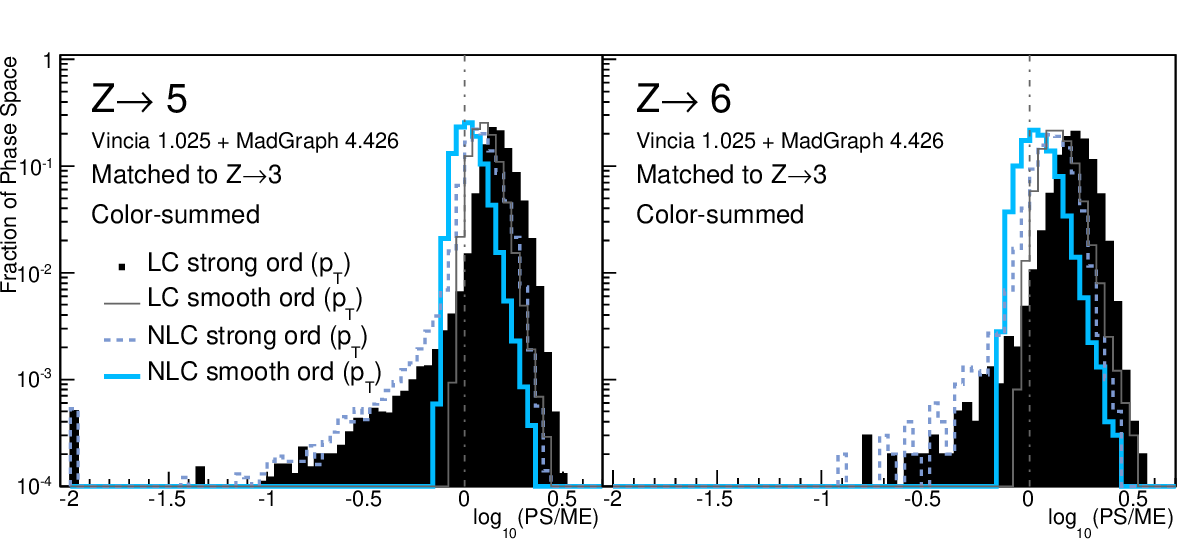}}
\capt{
LC and NLC parton showers compared to matrix elements. 
  Distribution of $\log_{10}(\mrm{PS}/\mrm{ME})$ in a flat phase-space
  scan.  Contents normalized by the number of generated points. 
  Gluon emission only. Matrix-element weights from
  \Mg~\cite{Alwall:2007st,Murayama:1992gi}, full color (summed over color permutations). 
Compare to figs.~\ref{fig:strongord} and
  \ref{fig:multipole}. Note: the black histogram here does not look
  identical to the one in fig.~\ref{fig:strongord}; this is because
  the distribution shown here is after summation over all 
  color-permutations. Also, the shower expansion here uses 
  $C_A$ for both $qg$ and $gg$ emission antenn\ae{}. The same is true 
  for the thin solid line, as compared to the solid histogram in
  fig.~\ref{fig:multipole}. 
  \label{fig:nlc}}
\end{center}
\end{figure}
For the corrected emission probability to be positive definite, the
condition  
\begin{eqn}
\tilde{a}_{NLC} < \frac{N_C^2}{f_{IK}} \tilde{a}_{IK} 
\end{eqn}
must be fulfilled. The Eikonal terms are guaranteed to
respect this by a wide margin, but there can in principle be 
subleading differences between $\tilde{a}_{IK}$ and $\tilde{a}_{NLC}$ 
at the level of single poles and/or finite terms, both of which can be
influenced to some extent by user-controlled settings in \Vc. 
Although we have not encountered any problems with such
corrections becoming large in our own studies so far, 
we have inserted a numerical safeguard in the code, limiting the
size of each NLC correction to be at most half of the corresponding LC
term. 

We compare the shower expansion corrected as in 
eq.~(\ref{eq:slccode}), using the definition of $\tilde{a}_{NLC}$ given by
eq.~(\ref{eq:anlc2}),
to color-summed matrix elements in
Fig.~\ref{fig:nlc}, for $Z\to 5$ and $Z\to 6$ partons (gluon emission
only), for both the strong- and smooth-ordering options. 
For reference, we show the pure LC shower expansions as well,
as in figs.~\ref{fig:strongord} and \ref{fig:multipole}, although we
here set both the  $qg$ and $gg$ color factors equal to $C_A$, to
better illustrate what happens when only the NLC correction is
switched on and off. Note also that
the previous comparisons in this paper were made to leading-color
matrix elements for each individual color structure separately. Such a
comparison is not meaningful once color interference effects are taken
into account, and hence we are here using the full color-summed matrix
elements, and are likewise summing over permutations of the gluon
momenta in the shower expansion.

Firstly, we notice that the LC shower expansions do indeed overcount the
matrix elements if we just use $C_A$ for both $qg$ and $gg$
antenn\ae{}, as expected (the solid filled histogram and thin solid
lines are shifted to the right of the vertical dot-dashed line
that represents perfect agreement). 
When we switch on the NLC correction in the manner described
above, both the strongly and smoothly ordered approximations are
noticeably improved, cf.~the blue dashed and thick solid curve,
respectively. The NLC correction is therefore switched on by default
in the \Vc\ code.

The procedure described above 
is correct for the case of a single $qgg...g\bar{q}$ chain and then
correctly takes into account the infrared singularities arising
from the first subleading-color term, which is 
proportional to $1/N_C^2$. We refer to this modification as
``next-to-leading color'' or NLC. (Strictly speaking, there are also
corrections proportional to $1/N_C$, but these are already taken into
account in leading-color showers by including $g\to q\bar{q}$
splittings.) However, due to the inherent ambiguity in
assigning a color flow to these corrections even in the soft limit, we
cannot be certain that further subleading terms are also correctly
described. I.e., we do not expect to reproduce the correct $1/N_C^4$
terms in all singular limits.

When there are several colour chains, such as after a $g\to q\bar{q}$
splitting (and hence already suppressed by at least $1/N_C$), we
generalize the treatment to each chain separately. That is, we do not
at this point attempt to include interference effects \emph{between}
different color-singlet systems. 

\section{Matching}
\label{sec:matching}

In this section, we describe our strategy for incorporating a detailed
matching to tree-level matrix elements. The philosophy is similar to
that pioneered by Sj\"ostrand in
refs.~\cite{Bengtsson:1986et,Bengtsson:1986hr}, 
and hence also to the \textsc{Powheg}
formalism, but we here generalize the method to include
arbitrary-multiplicity tree-level matrix elements. The inclusion of
the NLO virtual corrections to the lowest multiplicity was treated in 
ref.~\cite{Giele:2007di} for an arbitrary tree-level matching strategy. 

\subsection{Matching Strategies}

Given a parton shower and a matrix-element generator, there are
fundamentally three different ways in which we can consider matching the two:
\begin{enumerate}
\item {\sl Unitarity:} 
The oldest approach \cite{Bengtsson:1986et,Bengtsson:1986hr}
  consists of working out the 
 shower approximation to a given fixed order, and correcting
 the shower splitting functions at that order by a multiplicative
 factor given by the ratio of the matrix element
 to the shower approximation, phase-space point by phase-space
 point. We may sketch this as 
\begin{eqn}
\mbox{Matched} = \mbox{Approximate}
\frac{\mbox{Exact}}{\mbox{Approximate}}~. \label{eq:multiplicative}
\end{eqn}
That is, the shower approximation is essentially used as a 
pre-weighted (stratified) all-orders phase-space generator, on which a
more exact answer can subsequently be
imprinted order by order in perturbation theory.
In our notation \cite{Giele:2007di}, this translates to applying the following
correction factor to each antenna function $a_i$ (or any other kind of shower
splitting kernel) 
\begin{eqn}
a_i \to a_i P_n^{\mrm{ME}}~~~,~~~P_n^{\mrm{ME}} =
\frac{|M_n|^2}{\sum_j a_j |M_{n-1}|^2}~, 
\end{eqn}
where the sum over $j$ runs over all possible ways the
shower could have generated the $n$-parton state from $n-1$
partons\footnote{Note, however, that this gets substantially more
  complicated if the shower process is not completely Markovian, a
  point we shall return to.}. 
So long as the adjustment factors $\pme$ are less than or equal to
one, they can be interpreted as probabilities, and the adjustment
can be accomplished by  means of the veto algorithm 
Monte-Carlo technique.  This constraint can essentially always be
satisfied through appropriate choice of the finite terms in the
antenna functions $a_i$.\break
\indent When these correction factors are inserted back into the
shower evolution, they guarantee that the shower evolution off $n-1$
partons correctly reproduces the $n$-parton matrix elements, 
without the need to generate any separate $n$-parton samples. 
Moreover, since the corrections 
modify the actual shower evolution kernels, the corrections are
\emph{resummed} in the Sudakov exponential, 
and finally, since the shower is \emph{unitary},  an initially unweighted sample
of $(n-1)$-parton configurations remains unweighted, with 
no need for a separate event-unweighting or event-rejection
step. (Technically, the exponentiation allows beyond-LL
corrections to be resummed, thus improving the logarithmic accuracy of
the result, while the explicit constraint of 
unitarity ensures that the additional non-logarithmic terms that
are also exponentiated by this procedure do not lead
to disasters.) There are thus 
several quite desirable features to this kind of matching strategy,
which is currently employed by \textsc{Pythia}, \textsc{Powheg}, and
\textsc{Vincia}. However, since traditional shower expansions quickly get
more complicated as a function of the number of emissions, 
this strategy had only been worked out for a single additional
emission prior to this paper (although the \textsc{Menlops} strategy
\cite{Hamilton:2010wh} does allow to combine a unitary matching of the
first emission with traditional non-unitary methods for multi-jet
matching). Below, we shall generalize the unitarity 
method to arbitrary multiplicities and, as a proof of
concept, present a concrete implementation 
spanning four successive emissions, including all subleading color terms. 
\item {\sl Subtraction:} 
Another way of matching two calculations is by subtracting one
  from the other and correcting by the difference, schematically
\begin{eqn}
\mbox{Matched} = \mbox{Approximate} +
(\mbox{Exact}-\mbox{Approximate})~. \label{eq:additive}
\end{eqn}
This looks very much like the structure of an NLO
fixed-order calculation, in which the shower approximation plays the
role of subtraction terms, and indeed this is what is used  in
strategies like \Fw{} \cite{Frixione:2002ik,Frixione:2003ei,Frixione:2008ym}.
 In particular since
eq.~(\ref{eq:additive}) appears much simpler to the fixed-order
community than eq.~(\ref{eq:multiplicative}), this type of approach
has received much more attention than  the unitarity-based one
above (though, to be fair, the \textsc{Powheg} \cite{Frixione:2007vw} approach 
represents a kind of hybrid between the two).
In this approach, the corrections are \emph{not resummed}; the events are
\emph{not unweighted} --- we can even have negative weights, at 
phase-space points where the approximation is larger than the exact answer;
and we need a \emph{separate phase-space generator} for the $n$-parton correction
events. And finally, like for the unitarity-based case above, a systematic way of
extending this strategy beyond the first additional emission was not previously
available. All these issues are, however, less severe than in ordinary NLO
approaches, and hence they are not viewed as disadvantages if the
point of reference is an NLO computation. Since the
correction terms are applied by adding (or subtracting, depending on
the sign of the weight) events, we refer to this type of matching
strategy as \emph{subtraction}. 
\item {\sl Slicing:} The last matching type is 
  based on separating
  phase space into two regions, one of which is supposed to be
  mainly described by hard matrix elements and the other of which is
  supposed to be described by the shower. Basically, this amounts to a
  subtractive approach in which the shower approximation is set to zero
  above some scale (effectively a dead zone is forced on the shower by
  vetoing any emissions above a certain \emph{matching scale}),
  causing the matched result to be the unsubtracted matrix element in
  that region, modulo higher-order corrections,
\begin{eqn}
\mbox{Matched (above matching scale)} \sim \mbox{Exact}~ (1 +
\mathcal{O}(\alpha_s)) \label{eq:scalebased1}~,
\end{eqn}
and since the leading behavior of the matrix elements and the shower
approximation are assumed to be the same below the matching scale, the
small difference between them can be dropped, yielding the pure shower
answer in that region,
\begin{eqn}
\mbox{Matched (below matching scale)} = \mbox{Approximate} + (\mbox{Exact} -
\mbox{Approximate}) \sim \mbox{Approximate}~.
\end{eqn}
Since this strategy is discontinuous across phase space, a main point
here is to ensure that the behavior across the matching scale be as
smooth as possible. CKKW showed \cite{Catani:2001cc} that it is
possible to remove any  dependence on the matching scale through  NLL
precision by careful choices of all ingredients in the matching;
technical details of the implementation 
(affecting the 
$\mathcal{O}(\alpha_s)$ terms in eq.~(\ref{eq:scalebased1}))
are important~\cite{Lonnblad:2001iq}. 
The MLM \cite{Mangano:2006rw,Mrenna:2003if} 
approach is also an example of this type of
matching. However, we note 
that the slicing strategy inherits almost all of the problems that pure
subtraction has; 
the corrections are \emph{not resummed}, the events are 
\emph{not unweighted} (but at least we avoid the negative-weight issue),
and we need a \emph{separate phase-space generator} for the $n$-parton correction
events. In addition, the  dependence on the unphysical matching scale
has appeared, which in general is non-vanishing and may be larger than
NLL unless the implementation matches the theoretical algorithm 
precisely~\cite{Lonnblad:2001iq}.
However, due to the work of CKKW and others 
\cite{Catani:2001cc,Lonnblad:2001iq,Lavesson:2005xu,Mrenna:2003if}, 
a systematic way of
extending this strategy beyond the first additional emission \emph{is}
available, and a strategy has even been proposed whereby this matching
type could be extended to NLO precision \cite{Lavesson:2008ah}. The
\textsc{Menlops} approach \cite{Hamilton:2010wh} 
is also available to combine it with \Pw{}. Since this type
is already well developed, therefore, we shall not consider it 
further in this paper, but note that one could still obtain it from
our formulae for subtractive matching (eq.~(22) of ref.~\cite{Giele:2007di}), 
by inserting the appropriate
phase-space cutoffs at the matching scale. For this reason, we refer
to this strategy as \emph{scale-based} or \emph{slicing}.
\end{enumerate}
To summarize, in this paper, we focus on the extension of the 
\emph{unitary}  matching strategy to arbitrary numbers of emissions at
tree level. We shall also include an  
NLO matching to the Born multiplicity using the prescription from
ref.~\cite{Giele:2007di}. 

\subsection{Matching to Tree-Level Matrix Elements: Leading Color}

The formalism we shall describe here represents a generalization of
the one presented in ref.~\cite{Giele:2007di}. It is based on using 
the trial generator described in the previous section as a phase-space 
generator, whose phase-space weights can be expanded to the required
order and compared to the matrix-element answer at the same order. A 
correction can then be imposed before generating the next trial
emission. For comparison, most other 
approaches are currently based on 
generating separate event samples for different jet
multiplicities and then post facto attempting to remove the overlaps (``double
counting'') between them. Here, we generate one sample ab initio, 
where every event starts at the lowest multiplicity
 and is then successively
matched up to the desired orders. 

Compared to shower evolution, matrix-element (ME) evaluations are 
computationally intensive. It is therefore desirable that the matching
algorithm involve as few ME evaluations as possible. In an ordinary
shower approach, the effective weight of each phase-space point
depends on all the possible shower histories that could contribute to
it, resulting in a factorially increasing dependence on the
multiplicity (for each color configuration). 
In CKKW-type approaches, this is partly circumvented by always
selecting only one history, the ``most singular one'' according to
the $k_t$ algorithm. Since this does not exactly correspond to an
inversion of the parton shower, the matching 
only really addresses the LL overlaps, and the higher-order discrepancies are
``removed'' by introducing an explicit cut on the phase-space region
in which matching is applied, by the so-called ``matching scale'',
which limits the numerical size that any subleading divergence could attain. 

In contrast, since we match at each successive order, each emission
(up to the matched orders) will necessitate at least one
matrix-element evaluation as well as the corresponding shower weight,
which in turn will involve matrix-element evaluations at the preceding
orders. At first sight, this may sound extremely expensive. Note,
however, that we are doing the matching of ``all the samples'' once
and for all, so that only one ``run'' will be necessary, rather than a
separate one for each multiplicity.  (Separate ``runs'' for different
multiplicities would then spread a comparable number of matrix-element
evaluations across the different ``runs''.)
In that context, the scheme should
not be more expensive than current ones, provided that a
formalism can be found that minimizes the number of matrix-element
evaluations that are still necessary to determine the trial weight. 

One sufficient condition for minimizing the number of required
matrix-element evaluations is the Markov condition:
each shower step should depend only on the current $n$-parton configuration
and not on its previous history. This in turn implies that, in order
to compute the trial weight, only the histories one step back have to
be considered, rather than all possible clusterings all the way to the
Born. As mentioned above, this would not be
true of ordinary strongly-ordered parton showers, where the
\emph{restart} scale for each configuration would depend on
\emph{which} parton was the last emitted one.

A simple
prescription that \emph{does} obey the Markov condition is to generate
trial emissions for every antenna in the $n$-parton configuration over
their full phase space, irrespective of the current ordering
scale. Without matching, this would lead to a large overcounting, as
was illustrated in fig.~\ref{fig:deadzone}, but
with matching, the total shower weight can be calculated and the
corresponding matrix-element correction made, with two added 
benefits: 1) the removal of the strong-ordering condition explicitly
prevents any dead zones from appearing in the trial space, and 2)
since the trial weight generated this way will represent an
overestimate, 
it will be possible to impose the matching by
multiplying by a factor smaller than unity, which can be translated
into a probabilistic veto of the trial branching. 

Expanded to tree level (all Sudakovs set equal to unity, fixed
$\alpha_s$), the trial generator will produce the following
total weight for a specific color-ordered point in 
$(n+1)$-parton phase space when summed
over possible contributing $n$-parton ones,
\begin{eqn}
w^{\mrm{trial}}_{n+1}(\{p\}_{n+1}) = \sum_j 
a_{\mrm{trial}-j}(\{\hat{p}_n\}^{[j]} \leftarrow \{p\}_{n+1}) \
|M^{(0)}_{n\mrm{LC}}(\{\hat{p}\}^{[j]}_n)|^2~, \label{eq:wtrial}
\end{eqn}
where $\{p\}_{n+1}$ represents the color-ordered momenta of the $(n+1)$-parton
state, $j$ runs over the possible $n+1\to n$ clusterings, 
and $\{\hat{p}\}^{[j]}_{n}$ represents the
color-ordered set of $n$ momenta  obtained by the $j$'th
$3\to2$ clustering of the $(n+1)$-parton state 
according to the selected kinematics map. It is important to note 
that $|M^{(0)}_{n\mrm{LC}}|^2$ here represents the tree-level squared
amplitude for the particular color configuration under consideration,
i.e., without any color averaging performed. 

The improved matching to smoothly ordered LL 
antenna functions described in 
\sect{GluonSplittingSubsection} merely consisted
of multiplying $a_{\mrm{trial}-j}$ in eq.~(\ref{eq:wtrial}) 
by the LL smooth-ordering accept probability,
thus replacing the trial factors by their LL counterparts, 
\begin{eqn}
\mbox{LL Matching ~ : ~ } a_{\mrm{trial}-j} \ \to \ a_j^{\mrm{PS}} \ = \ 
\pps_{\mrm{accept\dash{}j}} \ a_{\mrm{trial}-j} ~,
\label{eq:ajPS}
\end{eqn}%
where $\pps_{\mrm{accept}-j}$
expresses the unmatched shower accept probability,
including the LL acceptance probability, $\pll$,
of \eq{eq:PLL} and the
improvement factors $\pimp$ of \eq{eq:unord}
and $\pari$ of \eq{eq:Pari},
\begin{eqn}
\pps_{\mrm{accept\dash{}j}} = \pimp \pari \pll\,.
\end{eqn}

We shall now apply a final multiplicative accept probability,
$P^{\mrm{ME}}_{n+1}$, defined such that it 
takes us from the approximation that would have
been generated by $\pps_{\mrm{accept\dash{}j}}$ alone to the full matrix
element. It has the simple definition
\begin{eqn}
P_{n+1}^{\mrm{ME}} =
\frac{|M_{n+1}(\{p\}_{n+1})|^2}
     {\sum_{k} a_k^{\mrm{PS}}(\{\hat{p}\}^{[k]}_{n}\leftarrow \{p\}_{n+1})
       \ |M_{n}(\{\hat{p}\}^{[k]}_{n})|^2}~, \label{eq:PME}
\end{eqn}
where $a_k^{\mrm{PS}}$ was defined in eq.~(\ref{eq:ajPS}). 
By summing over all shower histories, this 
can easily be verified to generate the correct total weight. 
Note also that, the ME accept probability 
does not have any dependence on $j$ and is  thus the same for
all contributing $n\to (n+1)$ branchings. 

Note that $P_{n+1}^{\mrm{ME}}$  involves
one evaluation of $|M_{n+1}(\{p\}_{n+1})|^2$ and one evaluation of
each of the reclustered configurations,
$|M_{n}(\{\hat{p}\}^{[k]}_{n})|^2$ (one for each possibility for
$[k]$, of which one has already been evaluated as part of the
current matched shower history), and hence the total number of
matrix-element evaluations required at each order 
grows linearly with the multiplicity, rather than factorially as would
have been the case for a non-Markov evolution.

Note also that, since we make no distinction between ``shower events'' and
``correction events'', we may use the shower $\alpha_s$ as a common
prefactor on all the accept probabilities. There is thus 
only one $\alpha_s$ for each history. Due to the scale-canceling 
partial one-loop matching discussed in \sect{GluonSplittingSubsection}, 
the default, to one-loop
accuracy, is thus to use $\alpha_s(\pT{})$ for all the terms, regardless of
other scale choices made in the generator. 

Another convenient way of writing $P_{n+1}^{\mrm{ME}}$ is the following,
\begin{eqn}
P_{n+1}^{\mrm{ME}} = 1 + 
\frac{|M_{n+1}(\{p\}_{n+1})|^2 - {\sum_{k} a_k^{\mrm{PS}}(\{\hat{p}\}^{[k]}_{n}\to \{p\}_{n+1})
       |M_{n}(\{\hat{p}\}^{[k]}_{n})|^2}}
     {\sum_{k} a_k^{\mrm{PS}}(\{\hat{p}\}^{[k]}_{n}\to \{p\}_{n+1})
       |M_{n}(\{\hat{p}\}^{[k]}_{n})|^2}~, \label{eq:PMEvar}
\label{eq:SubtractionReweighting}
\end{eqn}
where the connection to the subtraction approach described in
ref.~\cite{Giele:2007di} (as well as to other subtraction schemes)
becomes readily apparent, since the numerator in
eq.~(\ref{eq:PMEvar}) is nothing but a shower-subtracted matrix element.

In order to transform the unitary strategy described here to a
non-unitary subtractive one, it would therefore suffice to apply the
factor, \eq{eq:SubtractionReweighting},
as an event reweighting, rather than as a branch-accept
probability. The events then do not have unit weights any more, and a
subsequent unweighting step would be necessary, as in other
subtractive approaches. The fact that we are here doing the matching
phase-space point by phase-space point, however,  
  means that we here have the `$1+$' in front, which should mean
  that even the subtraction-based version could never generate negative
  weights. In current subtraction approaches such as \Fw, the `$1+$'
  is generated as a separate sample, and 
  in that case, the correction term by itself can of course yield negative
  correction events. The cancellation here is more elegant and not
  only yields positive-weight events but is also 
  better protected from fluctuations; in \Fw{} and \Pw{}, each event
  sample is uncorrelated and therefore the phase-space point of an
  event will never be hit exactly by the counter-events. 
  In limited-statistics samples, it is therefore
  events with slightly different momenta that have to compensate each
  other, whereas the proposal here achieves an exact cancellation 
  in one and the same phase-space point, event
  by event. Thus, instead of having one event with weight +3 and one
  with weight -1, we would here simply get one event with a total
  weight of +2.

A final technical note is that the \Mg\ matrix elements must be
evaluated on-shell, and hence one must first set the value of the
\Mg\ $Z$ mass equal to $\sqrt{s}$, even if this is not equal to the
physical $Z$ mass; the 
important thing is that the incoming momentum be on-shell. We also
set the value of the strong coupling $g_s=1$ in \Mg, equivalent to
factoring it out of the problem.  

\subsection{Matching to Tree-Level Matrix Elements: Subleading Color}

When summing over all events, the full answer contains an averaging
over all permutations of color orderings in every phase-space
point. In event generators, two different color
structures in one and the same phase-space point are  viewed as
two different events with the \emph{same} color structure, but in two
different phase-space points. Via the matching above, this sum now
reproduces the leading-color tree-level color-averaged matrix
element squared. We wish to extend this to include the
subleading-color contributions as well. Obviously, these cannot be
associated with any particular color structure, and we must therefore
here match across events in different phase-space points (or,
equivalently, the same momentum configuration, but different color
orderings).

When matching to subleading color, we shall use the specific structure
of matrix elements generated with \Mg\ and imported into the
\Vc\ code. The color structure of these matrix elements is cast as a
color matrix whose diagonal entries form the leading-color
contributions. Each diagonal entry corresponds to one particular color
ordering, hence summing over all the diagonal terms and dividing by
the number of rows (= number of permutations) is equivalent to
averaging over colors, in the leading-color limit. The leading-color
matching described above thus corresponds to matching to a \Mg\ matrix
element with only one diagonal term being non-zero in the entire color
matrix, representing the particular color structure used in the
matching. 

In order to include subleading color, a simple and
sufficient prescription is to compute also the full color-summed matrix
element and include a fraction of it in the matching to each color
structure, by modifying the LC matrix elements in
eqs.~(\ref{eq:PME}) and (\ref{eq:PMEvar}) as follows,
\begin{eqnar}
|M_i|^2 & \to & |M_i|^2 +  \frac{|M_{i}|^2}{\sum_{j} |M_{j}|^2} 
\sum_{j\ne k} M_j M_k^* \label{eq:slc}\\[3mm]
& = & |M_i|^2\left( 1 + \frac{\sum_{j\ne k} M_j M_k^* }{\sum_{j}
  |M_{j}|^2} \right)\\[3mm]
& = & |M_i|^2\left(\frac{\sum_{j,k} M_j M_k^* }{\sum_{j}
  |M_{j}|^2} \right)~,
\end{eqnar}
where $M_i$ is the amplitude for one specific color ordering. The
numerator in the last parenthesis is just the full color-summed matrix
element squared, and the denominator is the corresponding leading-color one. Since
both of these are well-defined leading-order physical matrix elements,
the term in parenthesis is positive definite and hence cannot generate
negative weights. 

Note that, in \Vc's interface to \Mg, we have so far been using the form
in eq.~(\ref{eq:slc}), since this is the fastest
in the context of that particular implementation.  In order to compare
between the leading- and full-color cases, we have implemented an
option to switch the subleading corrections  off, although they are on
by default in the program.  

\begin{figure}
\begin{center}
\scalebox{\figscale}{
\includegraphics*[scale=0.65]{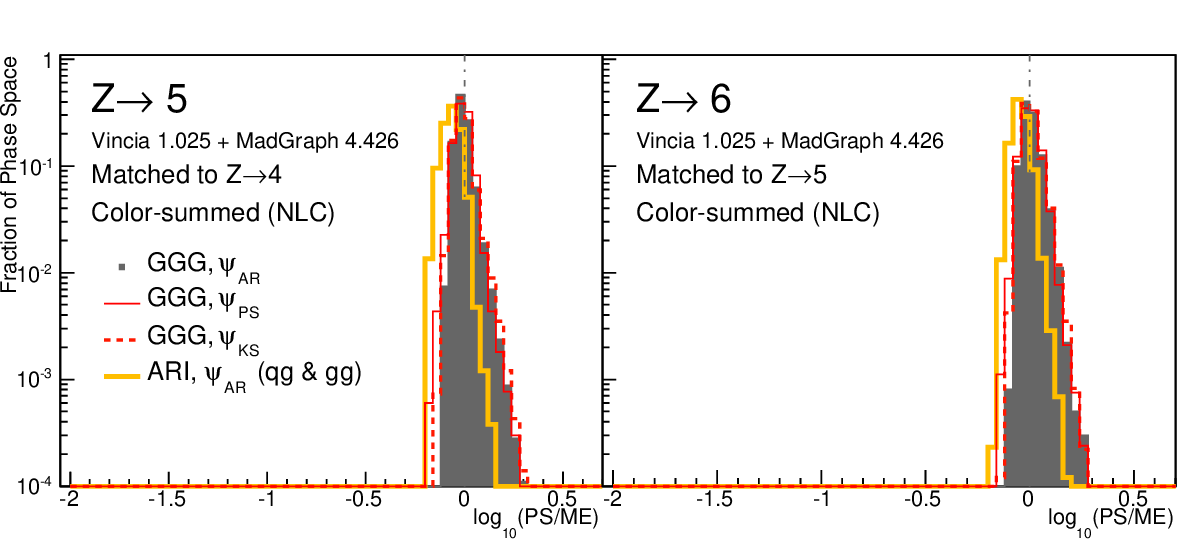}}
\capt{
Smoothly ordered matched parton showers (unordered showers using the
improvement factors $\pimp$ and $\pari$) compared to matrix elements. 
  Distribution of $\log_{10}(\mrm{PS}/\mrm{ME})$ in a flat phase-space
  scan.  Contents normalized by the number of generated points. 
Gluon emission only. Matrix-element weights from
  \Mg~\cite{Alwall:2007st,Murayama:1992gi}, full color (summed over color permutations). 
Compare to the unmatched shower distributions
  in figs.~\ref{fig:strongord}, \ref{fig:multipole}, and
  \ref{fig:nlc}. \label{fig:multi-matched}}
\end{center}
\end{figure}

In fig.~\ref{fig:multi-matched}, we show the weight ratios discussed
earlier (which are essentially just the inverses of $P^{\mrm{ME}}_n$), 
for $Z\to 5$ and $Z\to 6$ partons, now including 
matching at each preceding order. For the shower approximations, 
we use the default smoothly ordered NLC-improved GGG antenn\ae{}, with
  three  different kinematics maps (solid histogram, thin solid line,
  and dashed lines, respectively). We also compare to the same
  settings as the solid histogram but using 
  the \Ar\ radiation functions instead of the GGG ones (thick solid
  lines). 
  Comparing these distributions to those in fig.~\ref{fig:multipole}, we
  see that all the shower models reproduce the matrix elements very
  well, and hence the differences between the shower models are largely
  canceled by the matching to the preceding orders, as expected. At each order, now 
  only a relatively well-controlled and stable matching correction
  remains, which does not appear to exhibit any significant
  deterioration order by order. Note that we have not applied
  \emph{any} phase-space cuts here, and hence we find no evidence for
  any remaining subleading divergences in the matrix elements. This is
  in sharp contrast to 
  slicing- or subtraction-based approaches, where a non-zero 
matching scale is obligatory beyond the first matched order. 

\subsection{A note on \Mg\ and GGG color factor normalizations} 
As a final remark, we note that the subleading-color terms are not uniquely defined. 
Obviously, if the leading-color
pieces are not normalized the same way in two different approaches,
the subleading terms must likewise appear different. This, e.g., leads
to some apparent differences between \Mg\ and the GGG antenn\ae{}. With
color and coupling factors, the \Mg-GGG correspondence for the $Z\to qgg\qbar$
antenna is:
\begin{eqn}
g_s^4 A^{\mrm{GGG}}_4(0,1,2,3) =
\frac{2|M_{4\mrm{LC}}(0,1,2,3)|^2}{\hat{C}_F^2|M_2(s)|^2}~, \label{eq:lcA4}
\end{eqn}
where the factor $2$ on the \Mg\ matrix element cancels the color
averaging factor which is already present in $|M_{4\mrm{LC}}|^2$,
which represents a \Mg\ matrix element with only
 one element non-zero in the color matrix, the one corresponding to
the $(0,1,2,3)$ color flow squared. In particular, note that the LC
coefficient in \Mg\ comes with $\hat{C}_F^2$, whereas, 
in order to construct the full answer, i.e., including subleading
color, the GGG antenn\ae{} should be combined in the following way,
\begin{eqn}
\frac{|M_{4}|^2}{|M_2|^2} =g_s^4\hat{C}_F C_A \left(\frac12 a_4(0,1,2,3) + \frac12 a_4(0,2,1,3) 
- \frac{1}{N_C^2}\tilde{A}_4(0,1,2,3)\right)~.\label{eq:fullA4}
\end{eqn}
A direct comparison between what would be called subleading color
by GGG and by \Mg, respectively, 
would thus yield a different answer, simply because the piece called 
LC is normalized to $C_A\hat{C}_F$ in the former and to $\hat{C}_F^2$
in the latter. To verify these normalizations, 
the validity of  eqs.~(\ref{eq:lcA4}) and (\ref{eq:fullA4}) 
was tested numerically on a large number of phase-space points.

\section{Uncertainty Bands \label{sec:uncertainties}}

A calculation is only as good as the trustworthiness
of its uncertainty bands. Traditional methods for evaluating shower
uncertainties range from simple comparisons between different
models to more elaborate variations of salient model parameters within
some theoretically or phenomenologically justified ranges.  

The former kind is, at best, indicative, but can also be grossly  
misleading. As a classic example, consider two different parton 
showers with a cutoff at some factorization scale. They would both
agree there are no jets above that scale, even though a
matrix-element-based calculation would certainly produce jets in that
phase-space region. Comparisons of the \Hw\ $-$ \Py\ kind are therefore of
little value when pursuing rigorous uncertainty estimates.

Systematic variation of salient model parameters obviously gives a
more trustworthy idea of the overall uncertainty, and can also give
information about which particular sources dominate. 
However, it requires more careful preparation and more
expert input to set up: which parameters to vary, within what ranges, 
and how to make sure the variations are done consistently when
combining many tools in a long chain of event generation.
It also requires substantially more time and resources: 
for each variation, a new set of events must be generated, 
matched, unweighted, and possibly passed through detector
simulation. Finally, the ability of a single model to
span all possible variations is often limited ---
similarly to above, you still cannot use a strongly ordered shower to
estimate what the uncertainty associated with the strong-ordering
condition itself might be. There is also no way that, for instance,  
\Py's shower model could be varied to obtain an estimate of what an
angular-ordered shower would give.  

Here, we propose to combine the flexibility of the \Vc\ formalism to
take into account different ordering variables, radiation functions,
etc., with a treatment of uncertainties that   
only involves the generation of a \emph{single} event sample, with a time
requirement that is not greatly increased compared to the case without
uncertainty variations. We shall also automate the expert input to
some extent, reducing the number of choices the user must make.
 
The key question to ask is: if we use (matched) parton shower model A to
generate a set of unweighted events, what would the weight of each of
those events have been if we had instead used parton shower model B to
generate them? By answering this question, we can essentially 
use any parton shower model as a ``phase space primer'', 
provided it is still reasonably physical and that it 
does not have any dead zones, and then compute alternative
weights \emph{for the same  events} for any other set of
assumptions. 

The most trivial part is to note that, if a particular shower model
uses $\alpha_{s1} a_1$ as its 
radiation function for a particular branching, the same branching
would have happened with the relative probability
\begin{eqn}
P_2 = \frac{\alpha_{s2} a_2}{\alpha_{s1} a_1} \ P_1~, \label{eq:uAccept}
\end{eqn}
in a different model that uses $\alpha_{s2}$ as its coupling (e.g., with a
different renormalization scale or scheme) and $a_2$ as its radiation
function (e.g., with different finite terms, different 
partitioning of shared poles, different subleading or higher-order
corrections, or even a different ordering criterion). 

This, however, is not quite sufficient. Effectively, only the
tree-level expansion of the shower would be affected by keeping track of
such relative probabilities down along the shower chain; the Sudakov factors
would remain unmodified. Such a procedure would therefore explicitly break
the unitarity that is essential to resummation applications,
leading to possibly exponentially different weights between the sets,
which would be hard to interpret\footnote{For example, two models
that differ systematically by only a small amount on each branching, 
say 25\%, would, after 20 such branchings, differ by a factor $1.25^{20} =
100$. If they differ by a factor of 2
instead, the result would be a million, clearly not a reasonable
correction to the total event rate.}. More intuitively, a
big uncertainty on a very soft branching 
happening late in the shower should not be able to significantly 
change the entire event weight, jets and all. In the 
normal shower approach, it is the property of unitarity which
keeps such things from happening; as soon as any correction grows large,
its associated Sudakov factor must necessarily become small soon
thereafter, keeping the total size of any correction inside a unit-probability 
integral. 

The main part of our proposal therefore concerns a simple way to restore
unitarity explicitly also for the uncertainty variations, as follows. 
For each accepted branching, a number of trial branchings have usually
first been generated and discarded, to eliminate the overcounting done
by the trial function. In \Vc, we have so far not been particularly 
careful to optimize the choice of trial function (see 
\sect{sec:TrialFunctions}),
and hence we have
quite many failed trials. These are relatively cheap to generate,
however, so the code is not significantly slowed by
this inefficiency. Moreover, these failed trials actually turn out to 
be useful, even essential, in the present context. 

Just as eq.~(\ref{eq:uAccept}) expresses the relative probability
for a branching to be accepted under two different sets of model
parameters, 1 and 2, with 1 playing the role of phase-space generator
and 2 the role of uncertainty variation, 
it is also possible to ask what the probability of a 
\emph{failed} trial to have failed under different circumstances would
have been. Thus each failed trial can actually be used to compute
variations on the no-emission probability, i.e., the Sudakov
factors. 

Specifically, for each trial, the no-emission probability for the
model we use as our phase-space generator (which corresponds to the
settings chosen by the user in \Vc, including matching, subleading
corrections, etc.) is 
\begin{eqn}
P_{1;\mrm{no}} = 1- P_1~,
\end{eqn}
whereas the one for the alternative model should be 
\begin{eqn}
P_{2;\mrm{no}} = 1 - P_2 = 1 -  \frac{\alpha_{s2} a_2}{\alpha_{s1} a_1} \ P_1~.
\end{eqn}
Thus, by multiplying the relative event weight $w_2/w_1$ by $P_2/P_1$ for each
accepted branching and by $P_{2;\mrm{no}}/P_{1;\mrm{no}}$ for each
failed one, we explicitly restore the unitarity of the set of weights
$\{w_2\}$. In order to prevent extreme outliers from substantially
degrading the statistical precision of the variation samples, however, 
we limit the resulting weight adjustments to at most a factor of 2 \emph{per
  branching} in the code (in either direction).   The adjustment
of the weights for the failed branchings takes the place of `unfailing'
those which should have succeeded with model~2.

The accuracy of the approach obviously depends on the abundance
of failed branchings. If the trial function is completely exact, and 
no branching ever fails, then the tree-level problem
above will still occur. However, since \Vc\ typically generates
significantly higher numbers of failed branchings than accepted ones,
its effective numerical mapping of the changes in the Sudakov factors
during the no-branching evolution periods should be reasonably accurate. 

\begin{figure}[t]
\center
\vskip-3mm
\scalebox{\figscale}{
\includegraphics[scale=0.28]{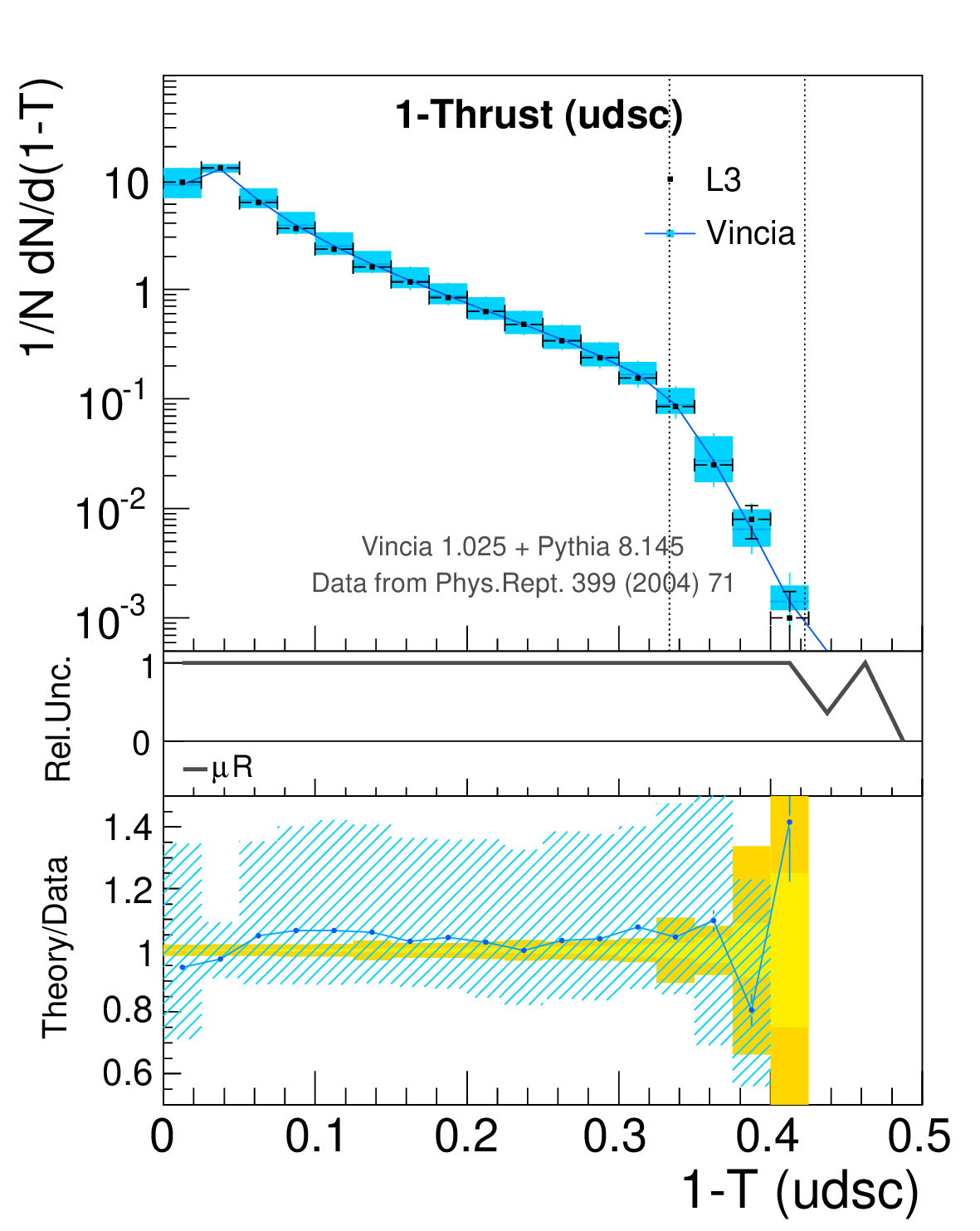}\hspace*{3mm}
\includegraphics[scale=0.28]{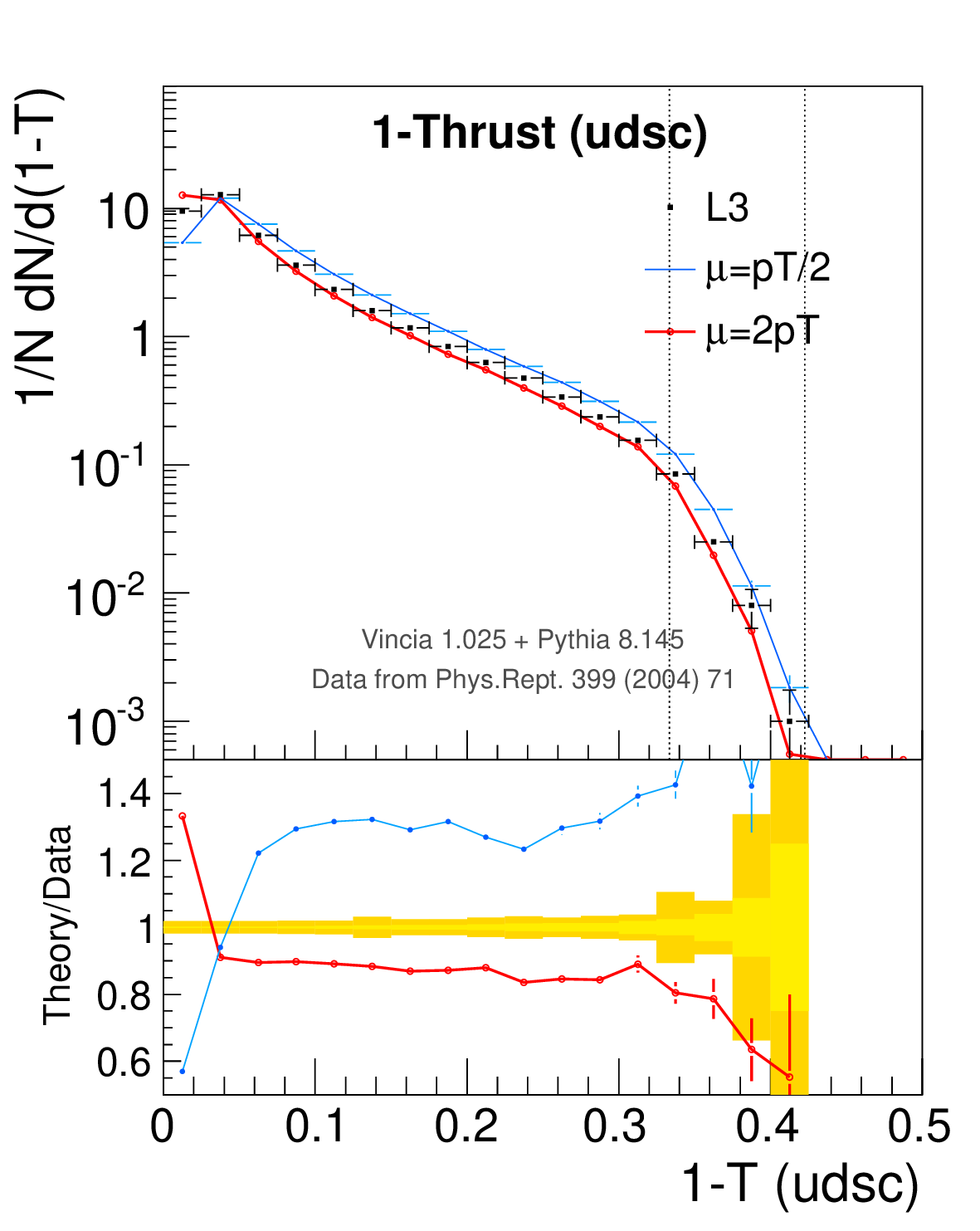}}
\vskip-1mm
\capt{
Comparison of {\it unmatched}
results from VINCIA's automatic uncertainty variations in the Thrust
observable around the default parameter set (left) with those from running the
generator for each variation separately (right), for variation of the
renormalization scale. The L3 data taken from ref.~\cite{Achard:2004sv} 
  is shown for
comparison. The yellow band in the lower plots represents the
experimental uncertainties on the thrust measurement.
\label{fig:uBands1}}
\end{figure}

\begin{figure}[t]
\center
\vskip-3mm
\scalebox{\figscale}{
\includegraphics[scale=0.28]{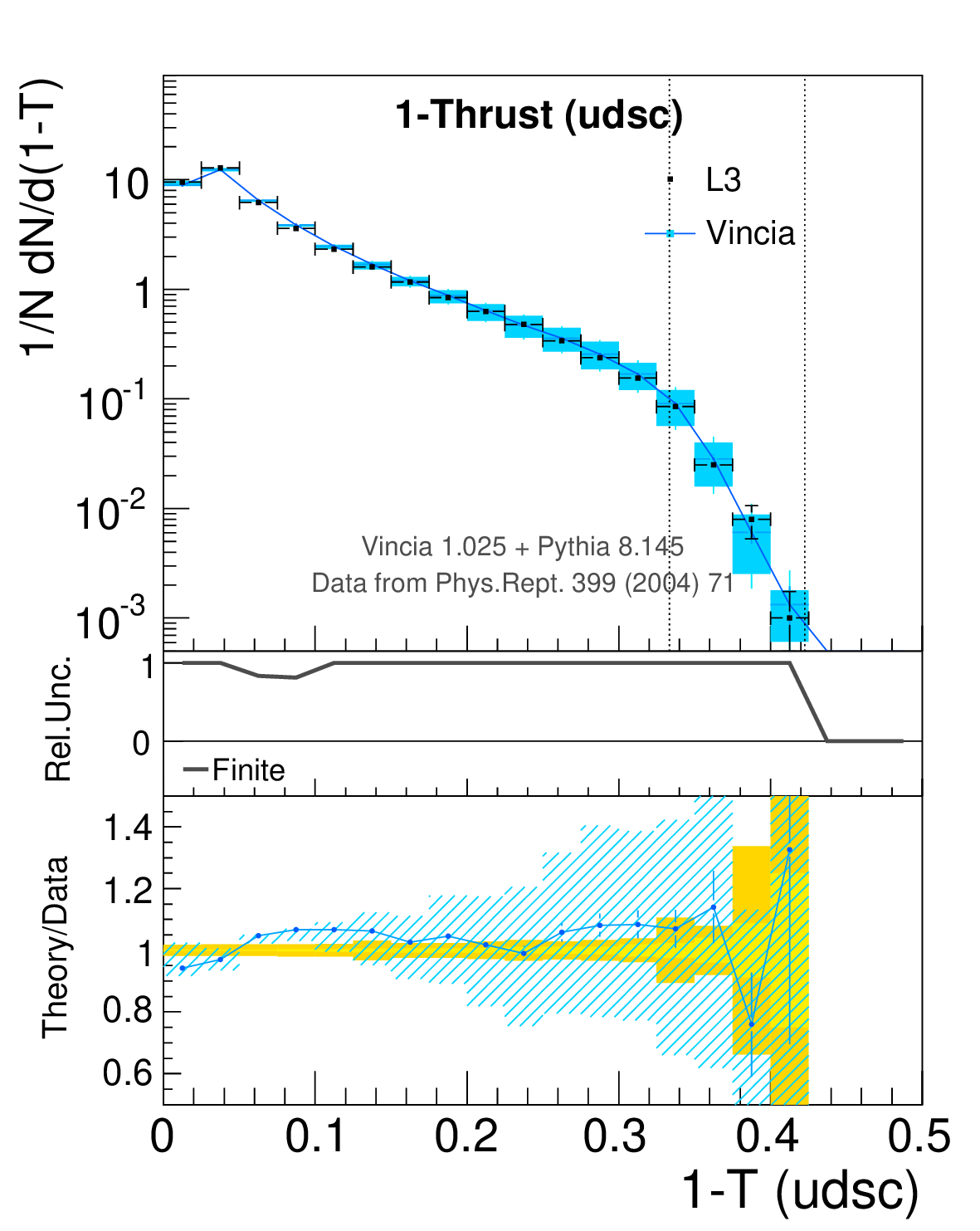}\hspace*{3mm}
\includegraphics[scale=0.28]{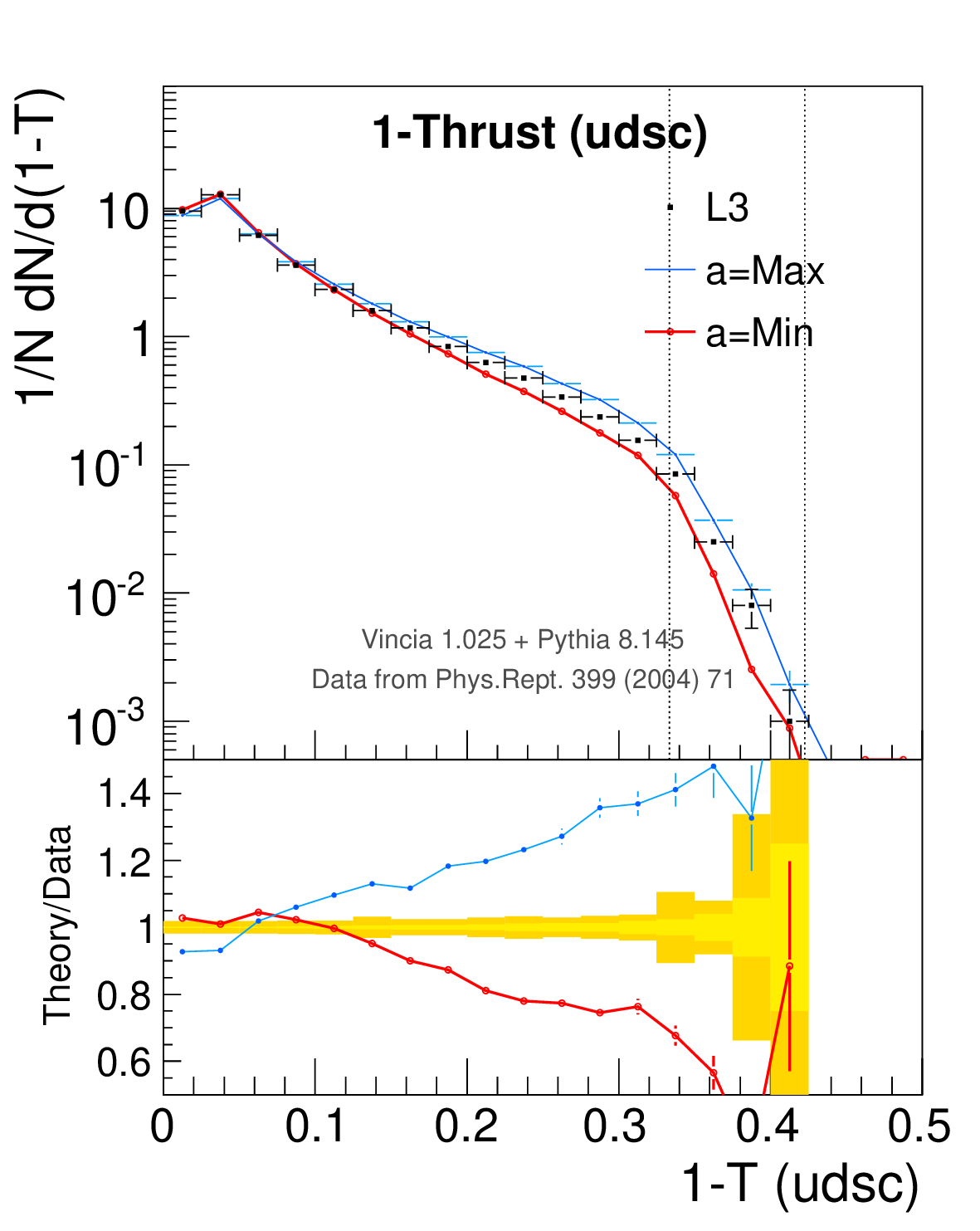}}
\vskip-1mm
\capt{
Comparison of {\it unmatched}
results from VINCIA's automatic uncertainty variations in the Thrust
observable around the default parameter set (left) with those from running the
generator for each variation separately (right), for variation of the
antenna-function finite terms. The L3 data taken from ref.~\cite{Achard:2004sv} 
  is shown for
comparison. The yellow band in the lower plots represents the
experimental uncertainties on the thrust measurement.
\label{fig:uBands2}}
\end{figure}

To test whether the uncertainty bands produced in this way really
reproduce what the shower model would have generated with different
settings, we show a few distributions in Figs.~\ref{fig:uBands1} and
\ref{fig:uBands2}, 
with default \Vc\ (thin blue line) 
plus an uncertainty variation (light blue band) on the left-hand side,
and \Vc\ run with the actual settings corresponding to that variation
on the right, for variations of the renormalization scale
(Fig.~\ref{fig:uBands1}) 
and of the antenna function finite terms (Fig.\ref{fig:uBands2}). 
In order to
maximize the result of the variations, all matching is switched off,
and hence the uncertainty bands are rather larger than would be the
case for default \Vc\ settings. The L3 data (black points)
\cite{Achard:2004sv} are included mostly to 
provide a constant reference across the plots; we postpone 
the discussion of the data comparison to \sect{sec:LEP}, where we discuss LEP
observables. The top panels of each the plots shows MC compared
to data, with both normalized to unity. The bottom panels show the
ratio MC/data, with the uncertainties on the data shown as yellow
shaded bands, the inner (lighter) one corresponding to the statistical
component only and the outer (darker) shade corresponding to statistical
plus systematic errors (added linearly, to be conservative). 

Comparing Figs.~\ref{fig:uBands1} and
\ref{fig:uBands2}, one observes that the two different variations lead
to qualitatively different shapes on the uncertainty predictions. The
renormalization scale uncertainty, Fig.~\ref{fig:uBands1}, 
produces an uncertainty band of
relatively constant size over the whole range of Thrust, whereas the
finite terms, Fig.~\ref{fig:uBands2},
only contribute to the uncertainty for large values of
$1-T$, as expected. Comparing left to right in both figures, 
we conclude that both the
features and the magnitude of the full uncertainty bands on the right are well
reproduced by the weight variations on the left.  

\paragraph{Available Variations:} 
So far, five types of 
automatic variations have been included in the \Vc\ code, starting
from version 1.025, via a simple on/off switch. These uncertainty
variations are:
\begin{itemize}
\item \Vc's default settings. This is obviously not a true uncertainty
  variation, but is provided as a useful comparison
  reference when the user has changed one or more parameters. 
\item MAX and MIN variations of the renormalization scale. The default
  variation is by a factor of 2 around $p_\perp$. 
\item MAX and MIN variations of the antenna function finite terms, as
  described in the online documentation of the code\footnote{Available at
      \texttt{http://projects.hepforge.org/vincia/}.}. 
\item Two variations in the ordering variable, one being closer to
  strong ordering in $p_\perp$ and the other to ordering in the
  $m_D$ variable. 
\item MAX and MIN variations of the subleading color corrections. The
  specific nature of the variation depends on whether subleading
  corrections are switched on in the shower or not. If not, the MAX
  variation uses $C_A$ for all gluon emission antennae and the MIN one
  $\hat{C}_F$. If switched on, the correction described in Section
  \ref{sec:nlc} is applied, but the correction itself is then modified
  by $\pm 50\%$ for the MAX and MIN variations here. 
\end{itemize}
These variations are provided as alternative weight sets for the
generated events, which are available through methods described in the
program's online manual. 
Limited user control over the
variations is also  included, such as the ability to change the
factor of variation of the renormalization scale.

When combining several variations to compute the total uncertainty, we
advise to take just the largest bin-by-bin 
deviations (in either direction) as representing the
uncertainty. 
We believe this is better than adding the individual terms together
either linearly or quadratically, since the latter would have to be
supplemented by a treatment of unknown correlations.
 With the maximal-deviation approach, we
are free to add as many uncertainty variations as we like, without
the number of variations by itself leading to an inflation of the error. 

We should also note that, in the \Vc\ code, matching coefficients etc.\ are
calculated for each  uncertainty variation separately. The size of
each band is therefore properly reduced, as expected, 
when switching on corrections that impact that particular source of
uncertainty.  

Finally, we note that, though the speed of the calculation is
typically not significantly affected by adding uncertainty variations, 
the code does run slightly faster without them. We therefore advise to keep them
switched off whenever they are not going to be used.

\section{Hadronization}
\label{sec:hadronization}

Since the \Vc\ code is a plug-in to \Py, it is (almost)
trivial to use \Py's string hadronization model with \Vc, as long as
one takes into account a few basic points:
\begin{itemize}
\item The matching of parton showers to hadronization models.
\item The ``tuning'' of the resulting shower+hadronization framework. 
\end{itemize}

Concerning the matching of parton showers to hadronization models, the
main issue to keep in mind is that this matching is performed at a
specific scale, the hadronization scale, which is implemented as a
lower cutoff in the perturbative shower evolution, usually at a scale
of order 1 GeV. Since no perturbative evolution is carried out below
that scale, the job of the hadronization model is then to give as good
a representation as possible of \emph{all} the physics that takes
place at lower scales. Since the hadronization model is inherently
non-perturbative, this means that the cutoff cannot be taken too high,
or else it would become apparent that the hadronization model does not
include a good description of the perturbative parts. Vice versa,
the cutoff cannot be taken too low, or else it would become apparent
that the perturbative modeling does not include a good description of
the non-perturbative parts. This is how one ends up with scales of
order 1 GeV as the matching point. 

In principle, if the cutoff is
varied around that point, both the perturbative shower and the
non-perturbative modeling parameters should obey evolution equations
that tell how each should scale, so that the end result would be
approximatively independent of the cutoff. 
These evolution equations
are nothing but an inclusive version of the shower evolution
equations, which the shower obviously respects by definition. 

But there is so far no formalism for
the non-perturbative modeling that allows us to take parameters
``tuned'' with one value of the cutoff and translate them for use with
another value for the cutoff. Hence each setting of the parameters of
the hadronization model are only valid for the exact cutoff value that
they were tuned with. If one uses a lower cutoff, then there would be
double counting between the shower, which now extends to lower scales,
and the hadronization model, whose tuning attempted to absorb those
same corrections as well as possible. Conversely, if one used a higher
cutoff, there would be a kind of ``dead zone'' unreachable by evolution, 
and where also the hadronization model
tuning did not attempt to absorb the corrections. 

\begin{figure}
\begin{center}
\scalebox{\figscale}{
\begin{tabular}{lclcl}
\textsc{Pythia 8} & & \textsc{Vincia} \\
\toprule
\includegraphics*[scale=0.5]{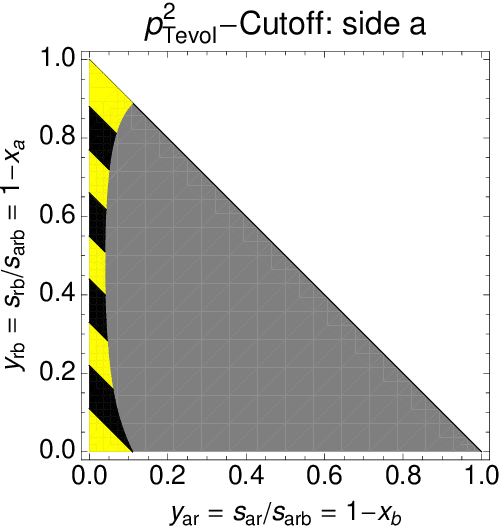} & & 
\includegraphics*[scale=0.5]{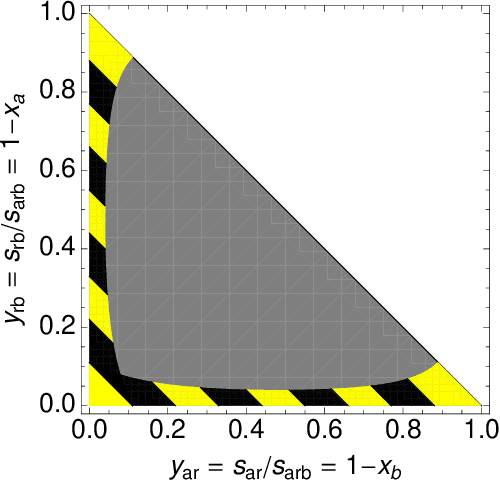}\\
\\[-4.6cm]
\hspace*{2.3cm}\includegraphics*[scale=0.2]{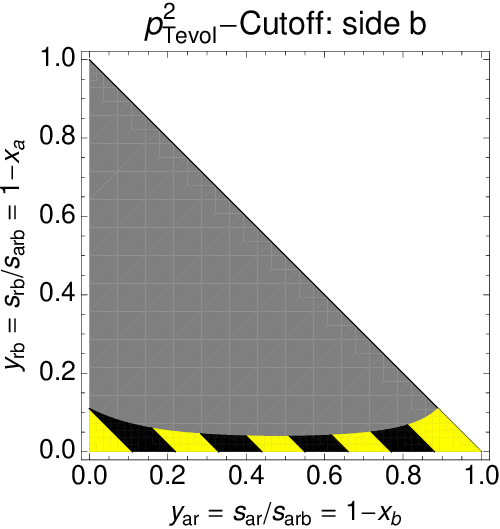} \\[2.1cm]
\bottomrule
\end{tabular}}
\capt{{\sl Left:} the hadronization cutoff in \pT{\mrm{evol}} as it is imposed 
  in \Py, with
  the main plot showing one side of the evolving dipole and the inset
  the other. {\sl Right:} in \Vc, the whole dipole-antenna evolves as one entity,
  and therefore the cutoff must be placed likewise. The
  remaining difference only contains non-singular contributions from 
  branchings in \Py\ that accidentally throw the radiated parton
  close to the recoiling one.\label{fig:had}}
\end{center}
\end{figure}

To use \Py's hadronization model directly with \Vc, we must therefore 
take the infrared cutoff to be at the same scale as the one used for the
\Py\ tuning, and since phase space is not one-dimensional, we also
need to make sure
it is in a variable which is as close to the one used by \Py\ as
possible. Contours corresponding to constant values of the 
\Py~8 evolution variable were already illustrated in the discussion of
evolution variables, fig.~\ref{fig:oldmc}, and is reproduced in the
left-hand pane of fig.~\ref{fig:had} with an explicit cut showing the
hadronization scale. 

In the dipole-antenna framework,
it is not possible to work with exactly the same variable; since we do
not keep the two sides of the dipole-antenna, $a$ and $b$, separate, it is
not possible to use a different form for the cutoff 
for the two sides, which is effectively what is done in \Py. The closest we
can come is to apply the cutoff if the smallest of the two scales,
$\pT{\mrm{evol,a}}$ and $\pT{\mrm{evol},b}$, is
below the chosen cutoff scale, illustrated on the right pane of
fig.~\ref{fig:had}. This will veto some branchings in \Vc\ that would
have been allowed in \Py, but since the \Py\ radiation functions are
not singular in those regions, and since the kinematics of the
corresponding phase-space points are near-collinear, we do not expect
this slight difference to have significant practical consequences. 

In addition to the possibility of using \Vc's own variables, $2\pT{}$ or
$m_D$, as cutoffs, the option to use 
this emulation of the \Py\ cutoff has therefore also 
been implemented in \Vc\ and should allow, to a first approximation, to use
the \Py\ hadronization model with any of the \Vc\ evolution settings, 
without retuning the
non-perturbative parameters, as long as one accepts that the resulting
answer will only be good up to perturbative uncertainties.  That is,
as long as the full \Vc\ uncertainty is estimated, the data should
still be compatible with the resulting uncertainty bands. 

We note that, in order to obtain a \emph{central} \Vc\ tune, 
one would still have to perform a dedicated tuning of the \Py\ 8
hadronization parameters using the particular \Vc\ shower model for
which the tuning is desired. This would also be necessary at
the point when \Vc's formal level of precision is higher than that of
\Py, in which case using hadronization settings tuned with \Py's
showers might actually result in incompatibility with the data, beyond
the allowed perturbative uncertainty bands. 

A first step towards getting a dedicated \Vc\ tuning
of the hadronization parameters was taken in April 2010 
when one of the authors acted as host for a 1-week ``industry
internship'' at CERN. 
Through the use of a specially developed runtime display
for  \Vc, and given some basic explanations about 
the effect of the different hadronization
parameters on the LEP distributions, M.~Jeppsson, a Danish
middle-school student, 
succeeded in making a tune of \Vc\ to LEP data, including Thrust, the
 $C$ and $D$ parameters, jet rates, identified particle production
 rates, and the inclusive fractional momentum distribution,
\begin{eqn}
x_{\mrm{particle}}=\frac{2|p_{\mrm{particle}}|}{\sqrt{s}}~.
\label{eq:x_particle}
\end{eqn}
 The final parameters he settled on
 appeared well motivated and physical, and now constitute the
 default in \Vc. They are given in appendix \ref{app:tune} for
 reference. 
 The runtime display, which is based on ROOT, 
 has subsequently been made publicly available as part of the
 \Vc\ package. To our knowledge, this study represents the first time 
 ``citizen science'' has been used for event generator tuning. 

\section{Comparison to LEP Data}
\label{sec:LEP}

In the following, we have used version 1.025 of the \textsc{Vincia}
plug-in and version 8.145 of the \textsc{Pythia}~8 generator, using 
default settings unless otherwise specified. Note that for \Vc, the 
default settings include a matching to tree-level matrix elements
through third order in QCD (via its \Mg\ interface), 
while \Py\ only formally includes a first-order matching. 

To keep questions of mass effects separate (the implementation of
which will be reported on in a separate paper \cite{GehrmannDeRidder:2011dm}), 
we shall here mainly compare to a useful data set presented by the L3
collaboration \cite{Achard:2004sv}, in which the contributions from light 
flavors
(defined as $u$, $d$, $s$, $c$) has been separated from that of events
containing $b$ quarks. 

Unfortunately, however, the L3 light-flavor data set does not contain 
jet observables. We therefore include comparisons also to ALEPH and
DELPHI jet observables that include all flavors, using a preliminary
implementation of mass effects in \Vc~\cite{GehrmannDeRidder:2011dm}. Since the
largest correction specific to $b$ quarks is simply the $B$ meson
decay, for which we rely on \Py's string hadronization and hadron
decay model, we believe these comparisons are still meaningful, even
if we must postpone a full discussion of them to the follow-up study
in ref.~\cite{GehrmannDeRidder:2011dm}. 

\begin{figure}[t!]
\begin{center}
\vskip-3mm\hspace*{-4mm}\scalebox{\figscale}{\includegraphics[scale=0.28]{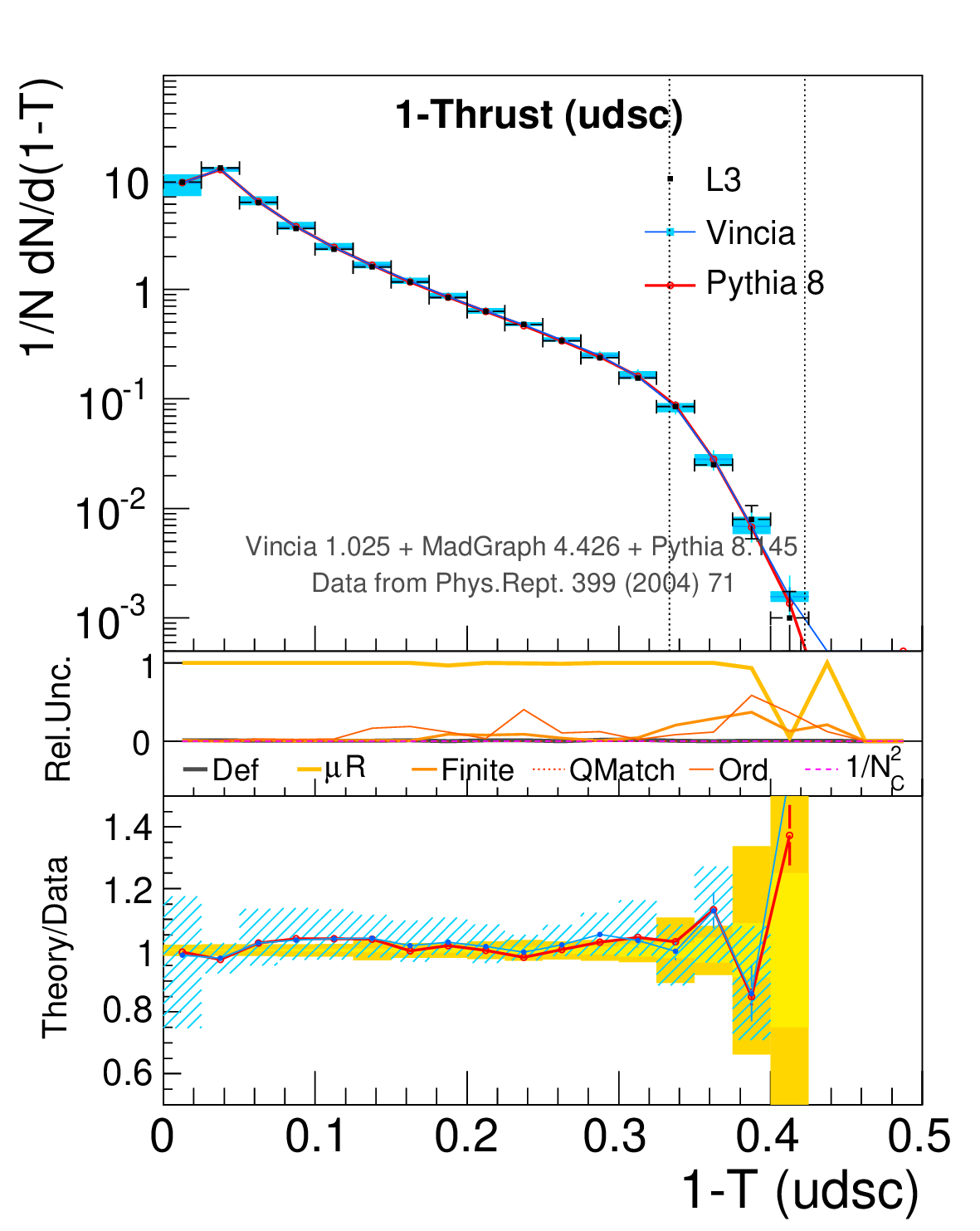}}%
\hspace*{-3mm}\scalebox{\figscale}{\includegraphics[scale=0.28]{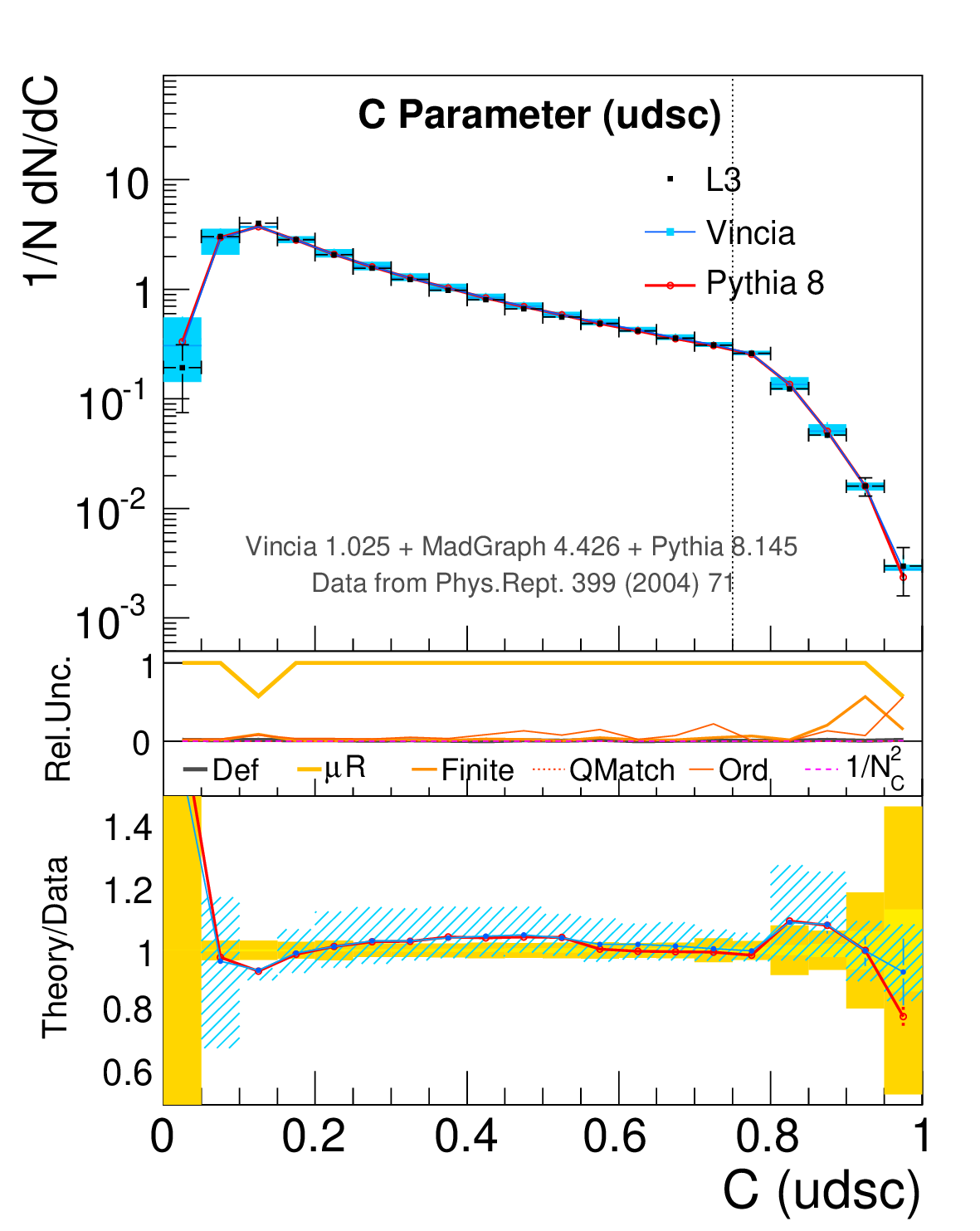}}%
\hspace*{-3mm}\scalebox{\figscale}{\includegraphics[scale=0.28]{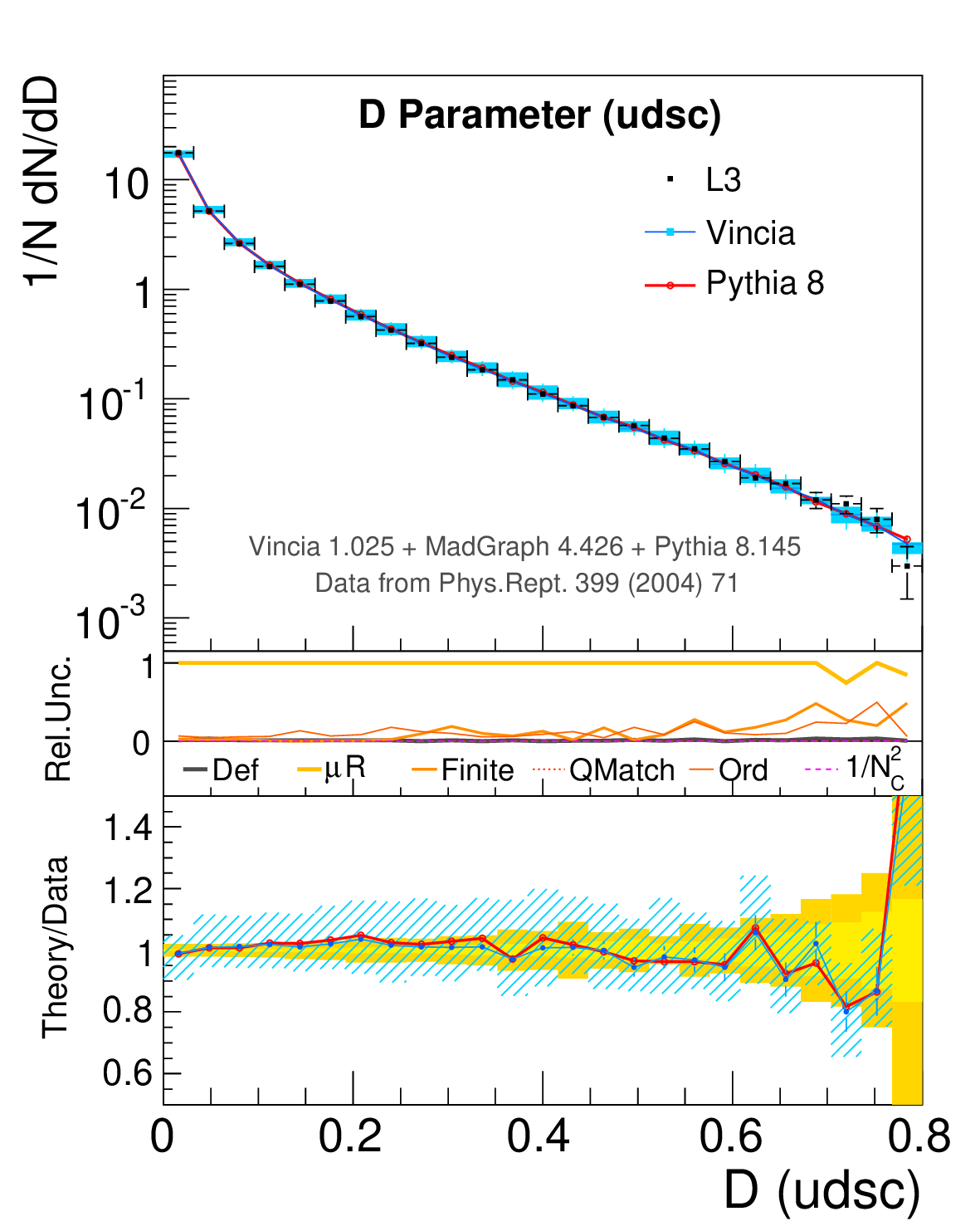}}%
\hspace*{-3mm}%
\capt{Comparison to the L3 light-flavor data set
  \cite{Achard:2004sv} (black points) at the $Z$
  pole for the $1-T$ (left), $C$ (middle), and $D$ (right) 
  event shape variables. 
  \Vc{} is shown in thin blue lines, with shaded light-blue bands
  representing the perturbative uncertainty estimate. The middle pane
  on each plot illustrates the relative composition of the
  \Vc\ uncertainty band. For comparison,
  the \Py 8 result is shown with a thick red line with open circles. 
The yellow bands in the bottom panels represent the
 experimental uncertainties on the measurement.
\label{fig:lep1}
}
\end{center}
\end{figure}

In Fig.~\ref{fig:lep1}, we compare default \Vc\ and \Py\ to the L3
light-flavor data for the Thrust (left) and the $C$ (middle) and
$D$ (right) event shape parameters 
\cite{Achard:2004sv}. Dashed vertical lines
indicate the boundaries between the 3- and 4-jet regions for the
Thrust and $C$ parameter (the right-most dashed line on the Thrust
plot indicates the boundary with the 5-jet region). 
The $D$ parameter measures the deviation
from planar events and is a 4-jet observable over its entire range. 
Despite substantial differences in the shower modeling,
matching level, and hadronization tune parameters, the two models give
almost identical results. Further, since \Py\ is already giving a
very good description of this data, there is little for the additional
matching in \Vc\ to improve on here. 

Still on Fig.~\ref{fig:lep1}, \Vc's uncertainty bands give about a
$\pm10\%$ uncertainty over most of the observable ranges, with larger
uncertainties near the edges of the distributions. The middle panels of the
plots show the relative composition of the uncertainty estimates, and
inform us that the renormalization scale variation is the dominant
source of uncertainty for all the observables, with other sources only
becoming competitive towards the 
right-hand extremes of the plots. This is an explicit consequence of
the tree-level matching in \Vc\ (by default imposed through third
order), which significantly reduces the allowed range of
finite-term, ordering-variable, and subleading-color uncertainty. 
The renormalization-scale uncertainty, however, is unaffected by
tree-level matching. Although the scale-dependence-reducing correction
described in Section \ref{sec:one-loop} is acting to reduce this dependence, the
residual uncertainty from scale variation is still larger than that
from any of the other sources. 

\begin{figure}[t!]
\begin{center}
\vskip-3mm\scalebox{\figscale}{\includegraphics[scale=0.28]{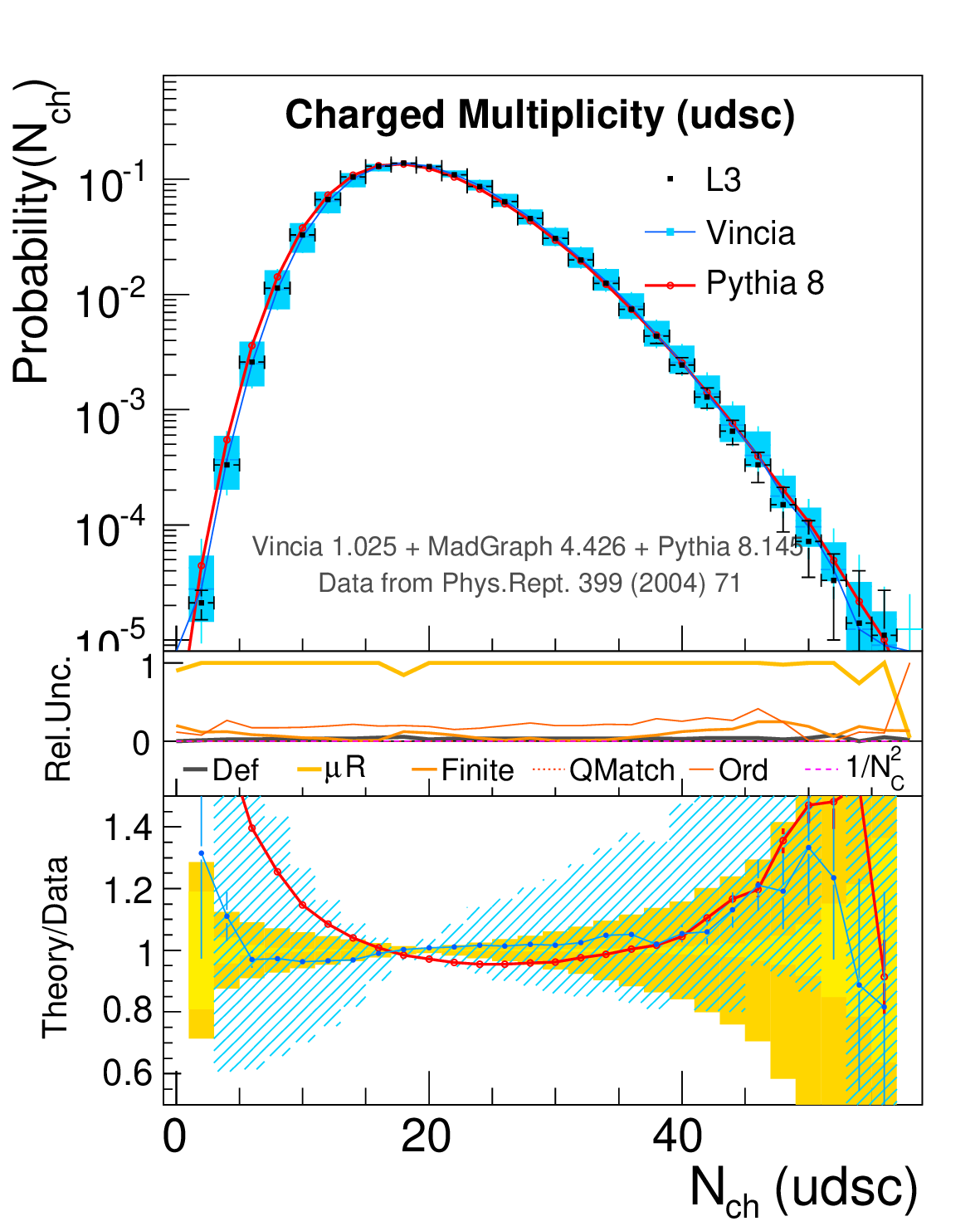}
\includegraphics[scale=0.28]{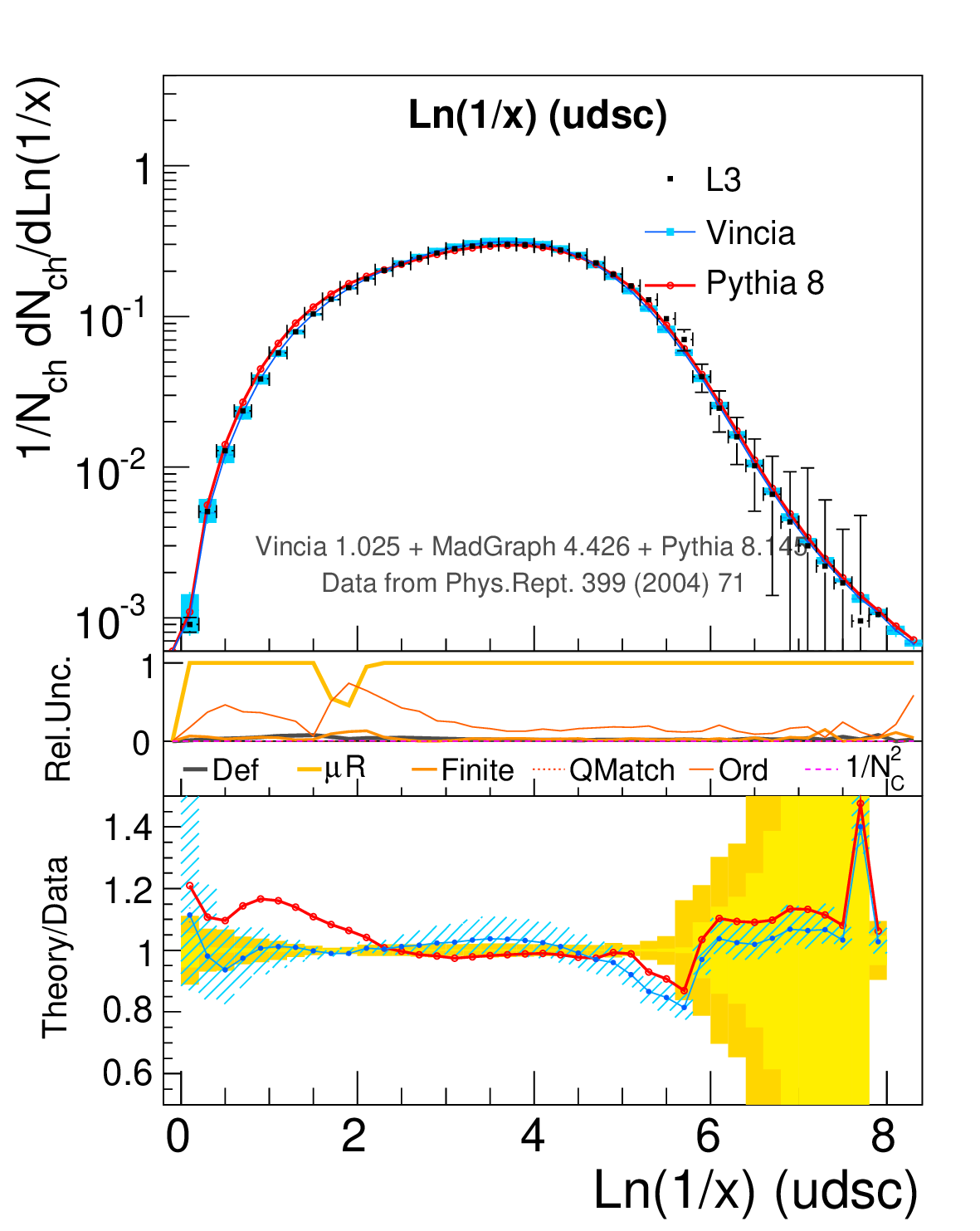}}
\capt{Comparison to the L3 light-flavor data set
  \cite{Achard:2004sv} (black points) at the $Z$
  pole for the charged track multiplicity (left) and
  fractional momentum (right) spectra. 
  \Vc{} is shown in thin blue lines, with shaded light-blue bands
  representing the perturbative uncertainty estimate. The middle pane
  on each plot illustrates the relative composition of the
  \Vc\ uncertainty band. For comparison,
  the \Py 8 result is shown with a thick red line with open circles. 
The yellow bands in the bottom panels represent the
 experimental uncertainties on the measurement.
\label{fig:lep2}
}
\end{center}
\end{figure}

In Fig.~\ref{fig:lep2}, we compare to two infrared-sensitive
observables also measured by L3 with light-flavor tagging, the
charged track multiplicity (left) and the fractional momentum
distribution (right), with the latter given by
\eq{eq:x_particle}. (Note that ref.~\cite{Achard:2004sv} uses the
notation $\xi = - \ln x$.) 
We conclude that \Py~8 was probably tuned on slightly different
observables, and hence the agreement obtained with \Vc\ is here
improved both by giving a slightly narrower multiplicity distribution,
with fewer low-multiplicity events and a slightly softer fragmentation
spectrum, with fewer particles carrying $x$ fractions very close to
unity. One also notes that \Vc's estimated uncertainty on the
individual bins of the charged-track multiplicity distribution is much
larger than the estimated uncertainty on the fragmentation
spectrum. Recall, however, that \Vc\ is only able to vary the
perturbative parameters --- variations of the string fragmentation
parameters would have to be included here to gain a better 
understanding of the full uncertainties. All we can say at this level
is that the charged-multiplicity distribution appears to suffer from a larger
perturbative uncertainty than the fragmentation spectrum.

\begin{figure}[t!]
\begin{center}
\vskip-3mm
\scalebox{\figscale}{\includegraphics[scale=0.62]{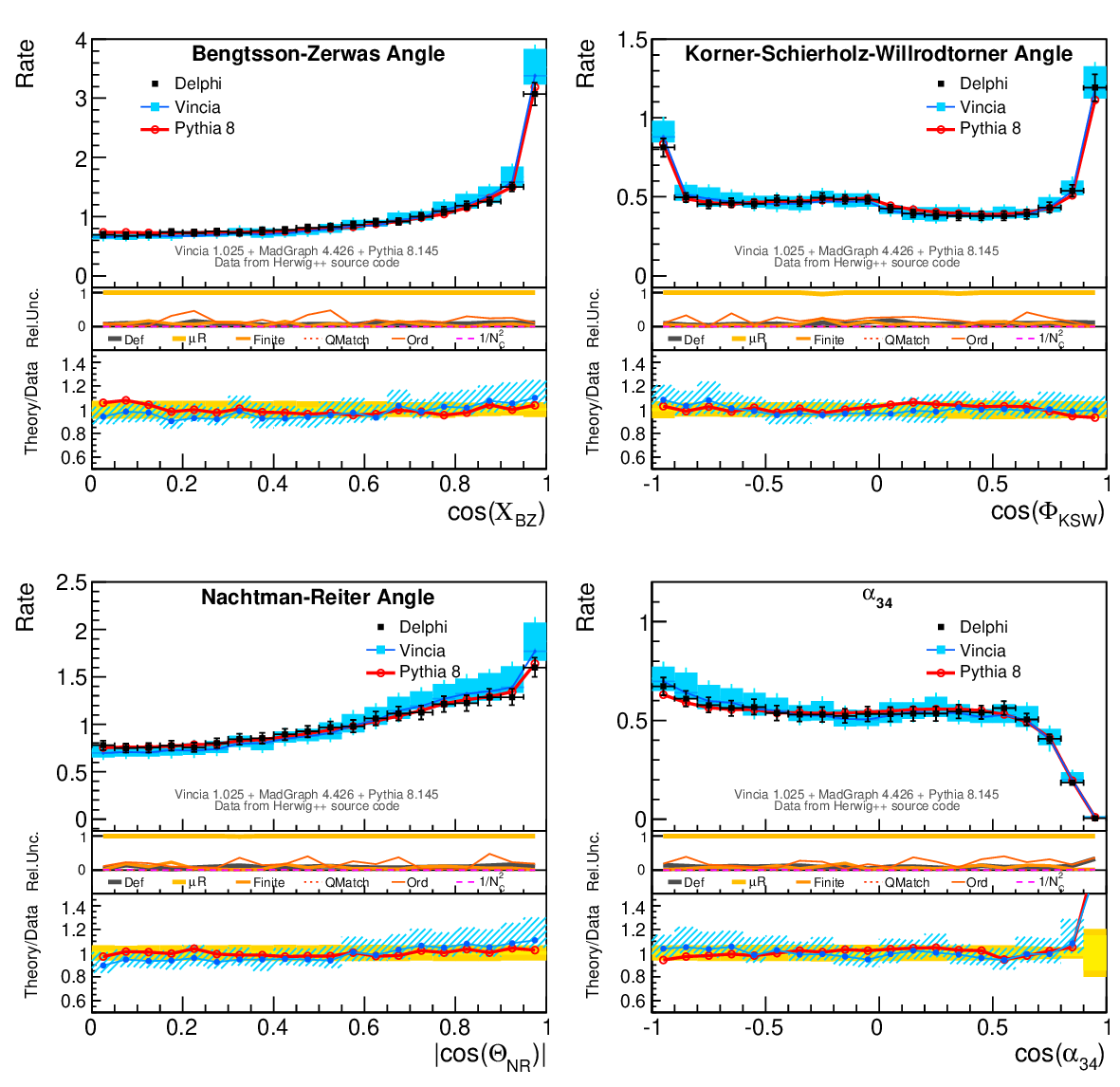}}
\capt{Comparison to DELPHI 4-jet angle measurements (black points) at the $Z$
  pole. \Vc{} is shown in thin blue lines, with shaded light-blue bands
  representing the perturbative uncertainty estimate. The middle pane
  on each plot illustrates the relative composition of the
  \Vc\ uncertainty band. For comparison,
  the \Py 8 result is shown with a thick red line with open circles. 
The yellow bands in the bottom panels represent the
 experimental uncertainties on the measurement.
\label{fig:lep3}
}
\end{center}
\end{figure}

A further set of variables that is interesting in the context of
differential multi-jet production are the so-called four-jet angles,
which were also measured at LEP. Not having found a public
data repository containing this particular data, however, we 
instead resorted to extracting the data point values from the \Hw++
source code \cite{Bahr:2008pv},  where it is encoded
for validation and tuning purposes. A comparison between this data and
default \Vc\ and \Py\ is shown in Fig.~\ref{fig:lep3}. Again, it is
clear that \Py\ itself is already doing a very good job. Since \Py\ is
not matched to 4-jet matrix elements and also does not contain
explicit spin correlations in the shower, this may at first be
surprising. However, \Py\ does correlate the production and decay planes
of gluons in the shower, and thereby includes the leading effect of
gluon polarization. The \Vc\ shower, on the other hand, contains no
polarization effects a priori. In \Vc's case, the effective
correlations of the four-jet angles are instead coming from matching
to the 4-parton matrix elements, and both codes are able to 
describe the 4-jet angles within a roughly 5\% margin, which is
comparable to the experimental precision.

\begin{figure}[t!p]
\begin{center}
\vskip-5mm\hspace*{-5mm}\scalebox{\figscale}{\includegraphics[scale=0.82]{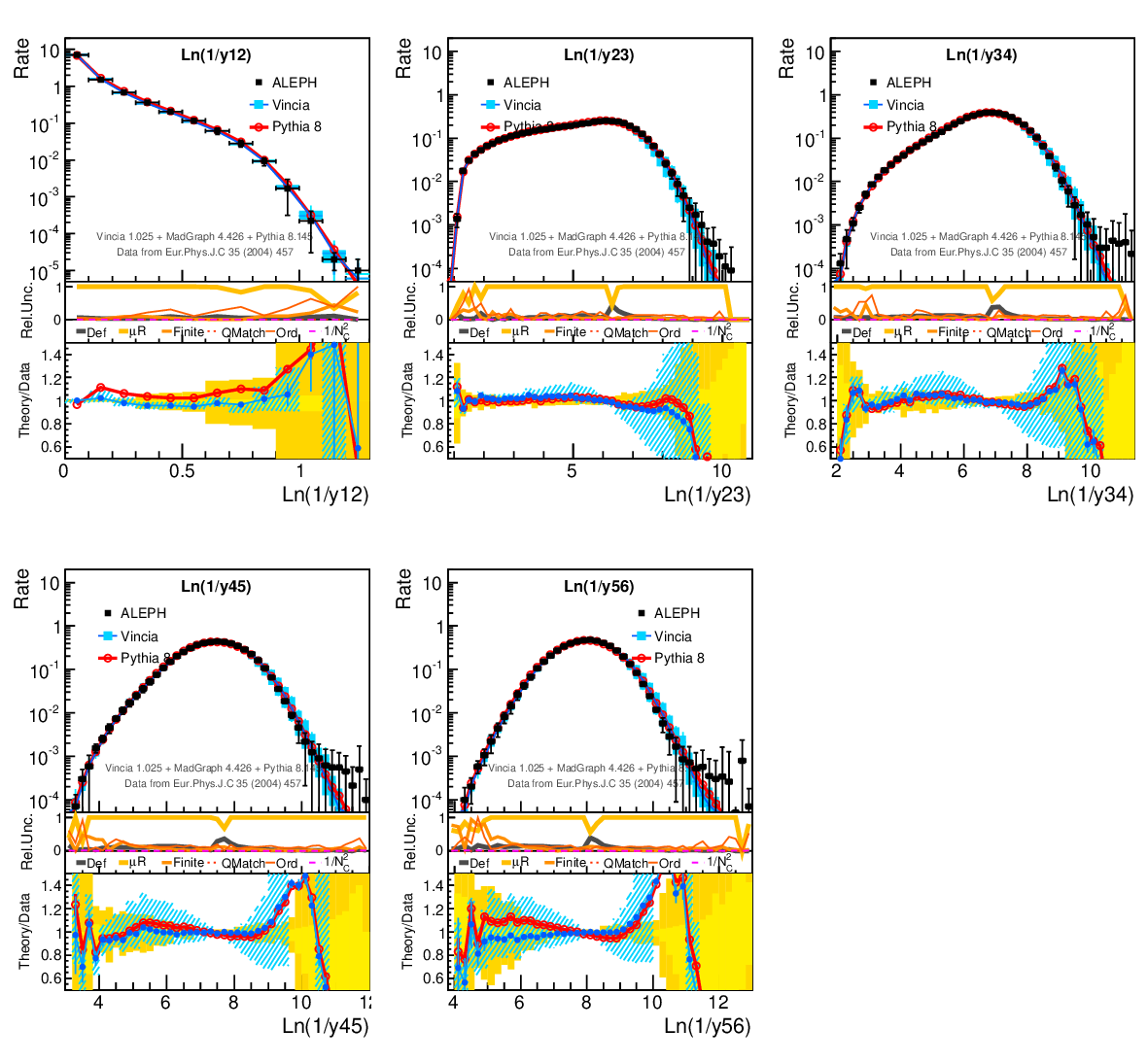}}\hspace*{-5mm}\vskip-3mm
\capt{Comparison to ALEPH jet resolution measurements
  \cite{Heister:2003aj} (black points) at the $Z$
  pole. \Vc{} is shown in thin blue lines, with shaded light-blue bands
  representing the perturbative uncertainty estimate. The middle pane
  on each plot illustrates the relative composition of the
  \Vc\ uncertainty band. For comparison,
  the \Py 8 result is shown with a thick red line with open circles. 
The yellow bands in the bottom panels represent the
 experimental uncertainties on the measurement.
\label{fig:lep4}
}
\end{center}
\end{figure}

Finally, in Fig.~\ref{fig:lep4}, we compare to the jet resolutions
measured by the ALEPH experiment \cite{Heister:2003aj}. Firstly, note
that pure \Py\ is basically able to describe all the distributions, within the
experimental accuracy, despite its being matched only to $Z\to 3$
partons. On the one hand, this is good, since it 
implies that the \Py\ 8 shower is delivering a quite good
approximation to QCD also beyond the matched 
orders. On the other hand, it also means that we are not really able
to quantify any significant improvement by matching using this data
alone. This may partly be due to the data having quite large uncertainties for
hard multi-jet configurations; at least $\pm 20\%$ for jet resolutions
$\ln(y_{45}) > -4$, which corresponds to \pT\ scales of order 10
GeV. Thus, this data does not really have enough power to make
precision tests in the hard multi-jet region. This relatively low
discriminating power of the hard multi-jet data will likely be an
even greater problem for testing any future one-loop matched schemes. 

We see four possible avenues to improve on this situation, requiring
various levels of additional experimental effort. One, combine the hard multi-jet
data from the four LEP experiments to increase the overall discriminating
power. Two, re-process the data focussing on observables more
specifically tailored to project out regions not dominated by leading
logs, e.g., by measuring $y_{nn+1}$ while imposing specific
constraints on the resolutions of the other $(n-1)$ harder jets in the
event. Three, increased statistics from a future facility such as
GigaZ. Four, measurements of hard jet rates and resolutions outside
the $e^+e^-$ environment, i.e., at hadron colliders, again possibly
placing more exclusive restrictions on the multi-jet structure of the
events. Especially the latter looks promising now in the dawn of the
LHC era, but of course comes at the price of introducing additional
uncertainties due to the colored initial states. LEP is therefore
likely to retain its position as our main jet fragmentation laboratory
for the foreseeable future, and with the official closing of the LEP
experiments this year, we wish to encourage those in a
position to do so to keep the LEP data `alive and well' for future
analysis studies that are likely to involve tests of models far more
sophisticated than those that were available ten or twenty years
ago. This makes it necessary to look at data much more differentially
and/or exclusively than is possible with, e.g., with the observables 
that were included in our comparisons here.

\section{Conclusions}
\label{sec:conclusions}

We have taken the next step in developing the formalism, started in
ref.~\cite{Giele:2007di}, for generic parton showers based on the
dipole-antenna formalism. Evolution equations for a wide class of 
 evolution variables, kinematics maps, and
radiation functions, have been presented, including all the necessary
steps to construct an explicit stochastic Markov-chain Monte Carlo code.

The basic ideas behind this shower model are similar to those behind
the existing \Ar\ program \cite{Lonnblad:1992tz}, to whose properties 
we make some comparisons. Aside from the more generic formalism, we also 
propose some systematic improvements, including suppressed 
unordered branchings
to cover the hard region of phase space and systematic
``next-to-leading-color'' (NLC) corrections. We compare explicitly to
matrix elements at both leading and subleading color for $Z\to4$, $5$,
and $6$ partons, to check the validity of our approximations.

We have also presented a new method for matching parton showers to
tree-level matrix elements at the multi-jet level, formulated in a
language appropriate to our shower framework. At lowest order, it
is similar to an older scheme by Sj\"ostrand and collaborators 
\cite{Bengtsson:1986et,Bengtsson:1986hr}, which has
recently been reformulated in a more generic NLO context called \Pw\ 
\cite{Frixione:2007vw}. Though our scheme is therefore similar to, and
compatible with, these existing methods, we here extend the method to
tree-level matching involving more than one emission. As such, the
method is at the same formal level of precision as the
\textsc{Menlops} approach \cite{Hamilton:2010wh}, 
but with the difference that the multi-parton
matrix element corrections are here exponentiated, which
should both improve the logarithmic accuracy of our Sudakov factors at
the same time as making the approach much less sensitive to 
so-called matching scales (scales below which the matrix-element
corrections are switched  off). We do still advocate imposing 
\emph{some} matching scale at high multiplicities, mostly to avoid
spending lots of time computing matrix elements for very soft
emissions that the unmatched shower is describing correctly
anyway. For the results reported on in this paper, a matching scale of
5 GeV (above the hadronization scale but well below typical jet
$p_T$s) was therefore imposed starting from third order in QCD 
(corresponding to $Z\to 5$ partons). 

By default, the generated events
 form one continuous sample with no specific separation between jet
multiplicities and with all events having unit weight.

The matched shower algorithm is implemented in the \Vc\ plug-in to the
\Py~8  Monte Carlo generator. Apart from its dependence on \Py~8, the
plug-in is self-contained with its own documentation and 
user-definable parameters. It also includes facilities for linking to
the \textsc{FastJet} \cite{Cacciari:2005hq} and \textsc{Root}
packages. In the latter case, a run-time interface has been developed
that allows to display \textsc{Root} histograms in real time during
the generator run, which can be useful both to give an immediate
sanity check that histograms are being filled correctly, and also 
to visualize the gradual improvement in MC statistics over the run. 
The generated events have similar physical properties as 
those generated by standard \Py~8 and can be passed through the
latter's string model for hadronization, and, e.g., to 
HEPMC~\cite{Dobbs:2001ck} for further processing, e.g., by analysis 
tools like \textsc{Rivet} \cite{Buckley:2010ar} or detector simulation
packages.

Finally, we have presented a new efficient and automatable 
method for the evaluation of uncertainty bands by parton shower
generators. The method draws on the 
unitarity property of shower calculations to compute several sets of
weights for a single generated event sample. This sample can
then be subjected to cuts, hadronized, passed through detector
simulations, etc., and the uncertainty variations can be obtained by
filling histograms with each of the different weight sets separately
at any stage during the processing. We have implemented this method in
\Vc, to provide an option for automatic evaluation of
renormalization-scale, finite-term, ordering, and 
subleading-color uncertainties.

\subsection*{Acknowledgments}
We thank L.~Dixon, A.~Gehrmann-de-Ridder,
L.~L\"onnblad, and M.~Ritzmann, for many useful comments on the
manuscript.
We also gratefully acknowledge the contributions of M.~Jeppsson to the
optimization of \Py's hadronization parameters for use with \Vc, by
performing a first hadron-level tune to LEP data. We thank the \Hw++
collaboration for help with extracting the DELPHI 4-jet
angle data from their source code. 
This work was supported in part by the Marie Curie FP6 research training
network ``MCnet'' (contract number MRTN-CT-2006-035606), by the
U.S. Department of Energy under contract No. DE-AC02-07CH11359,
and by the European
Research Council under Advanced Investigator Grant ERC--AdG--228301.

\clearpage
\appendix
\section{Comparisons to 2nd Order QCD Matrix Elements}
\label{app:comparison}

In the following, we present a series of figures comparing 
dipole-antenna showers --- ordered in various different evolution
variables --- to matrix elements, on the $R_4$
ratio defined in eq.~(\ref{eq:R4E}). As in \sect{sec:evolution}, 
these plots were
made on 20M random points in a flat phase-space scan, using
\textsc{Rambo} \cite{Kleiss:1985gy}. We use the default \Vc\ antenna functions and
kinematics maps in all cases and only vary the evolution variable.

On each plot, the quantity on the $x$ axis is a scale characterizing 
the first emission ($2\to3$), while the quantity on the $y$ axis is 
a scale characterizing the second ($3\to4$). Further, the quantity on the
$y$ axis has been normalized to the one on the $x$ axis, so that it 
really represents the ordering of the second emission, relative to the
first. Thus, phase-space points with hard initial $2\to3$ branchings will lie
to the right in the figures, while soft (strongly ordered) initial emissions lie to
the left. On the $y$ axis, points where the second scale is much
smaller than the first (i.e., strongly ordered relative to the first)
will be towards the bottom of the plot, while unordered points will be
towards the top of the plot. 

To help the eye, we have added a horizontal red line at $\ln(y)=0$ on
the plots, dividing the phase space for the 2nd emission into an
ordered part (below the line) and an unordered one. 
Similarly, the emphasized black box highlights the
region where $x$ is more than an order of magnitude below $M_Z$, and $y$
more than an order of magnitude below $x$, i.e., the 
doubly-ordered region. Finally, the dotted lines show contours of
constant $\ln(y/M_Z^2)$, or equivalently constant $y$. 

For each evolution variable, we show 4 plots, all with logarithmic $x$
and $y$ scales: 
\begin{itemize} 
\item
{\sl Top Left:}
  Average PS/ME ratio $\left<R_4\right>$ vs.\ the \pT{} scales of the
  two emissions. On the $y$ axis, the smallest of the two possible
  $\pT{}$ scales in the 4-parton configuration. On the $x$ axis, the
  \pT{} scale of the corresponding 3-parton configuration. 
\item {\sl Top Right:} The RMS width of the left-hand ratio, counting
  dead zones as having a factor 10 difference (otherwise the log of
  the ratio would be undefined). This
  helps illustrate whether a good average agreement on the left-hand
  side is just an accident of a wide distribution centered around
  unity, or whether the distribution itself is really narrow.
\item {\sl Bottom Left:} Same as top left, but with the $y$ axis
  showing the invariant mass of the two gluons. This projection is
  interesting since it allows us to isolate a particular
  double-collinear limit, as follows. If $m^2_{gg} \sim m_Z^2$, then
  the two gluons are well separated and carry all the energy
  of the original partons. This energy can only have been transferred
  to them by two successive extreme collinear branchings, one where the quark
  gives all its energy to the first gluon, and a second where the
  antiquark gives all its energy to the other gluon. To check if we
  get this non-trivial limit right, the upper edge of this distribution
  is therefore especially interesting. 
\item{\sl Bottom Right:} Same as top left, but
  with the ordering measure being invariant mass instead of
  $\pT{}$. This distribution is complementary to the top left one,
  showing the same ratio in a slightly different projection, 
  versus a measure of invariant masses, again in order to check
  whether any agreement observed in the above variables persists when
  doing a different cut through phase space. 
\end{itemize}

Note that, since we use the default \Vc\ antenna functions,
which reproduce the $Z\to 3$ matrix elements, we expect a good
agreement even when the first emission is not strongly ordered. It is
therefore mainly the ordering of the second emission with respect to
the first that is interesting. 

\begin{figure}[p]
\begin{center}
{\LARGE{\bf Phase-Space Ordering}}\\
{\large
\[
\displaystyle
Q_E^2 = \mrm{max}(s_{ij},s_{jk})
\]}
\scalebox{1.0}{\plotset{PSdivA4Avg-vc01kin4Sum-def-ev0-kin1-had1q1tun0-ord-p5-m1-T3-q3-grey.eps}
{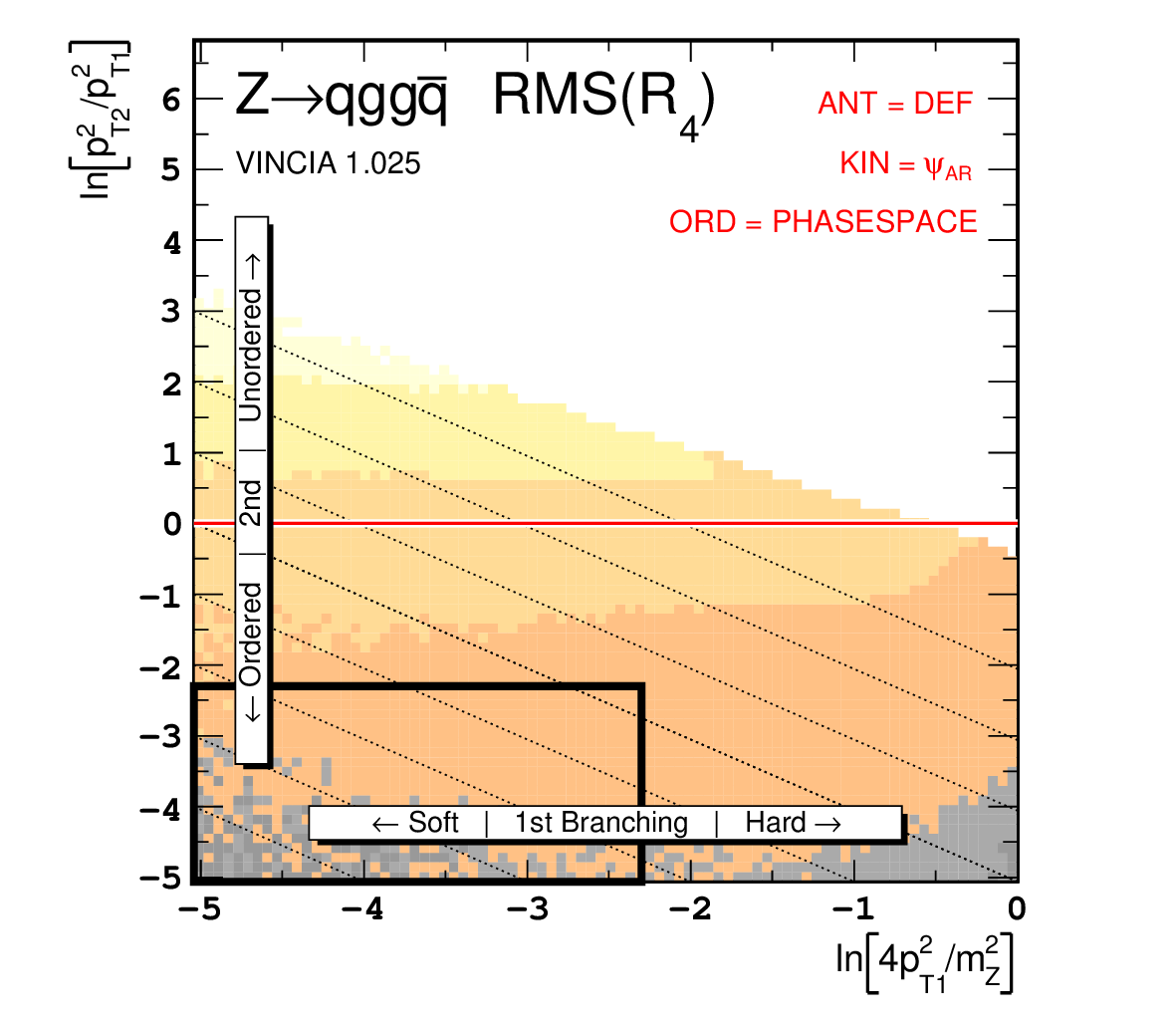}}
\scalebox{1.0}{
\plotset{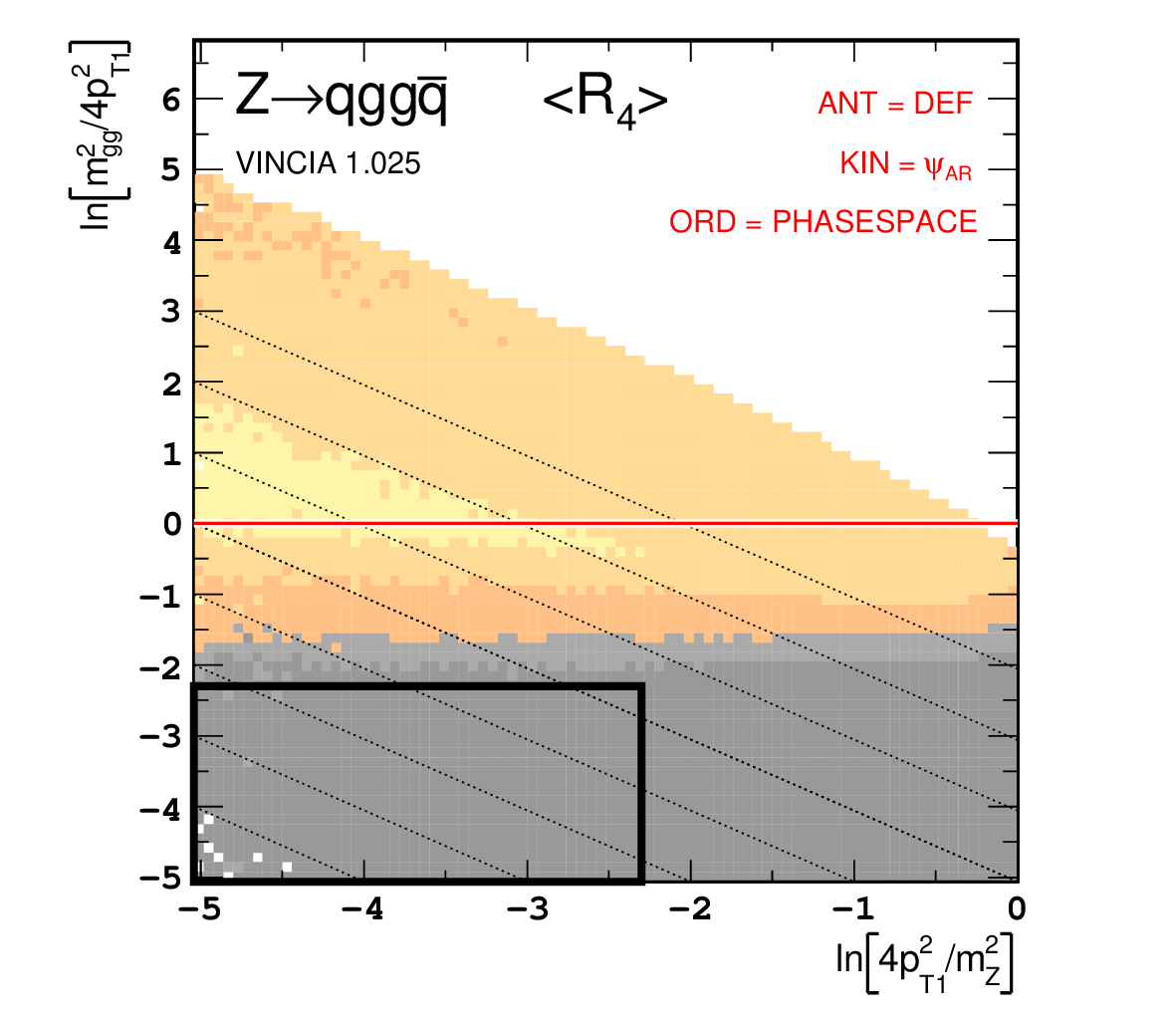}
{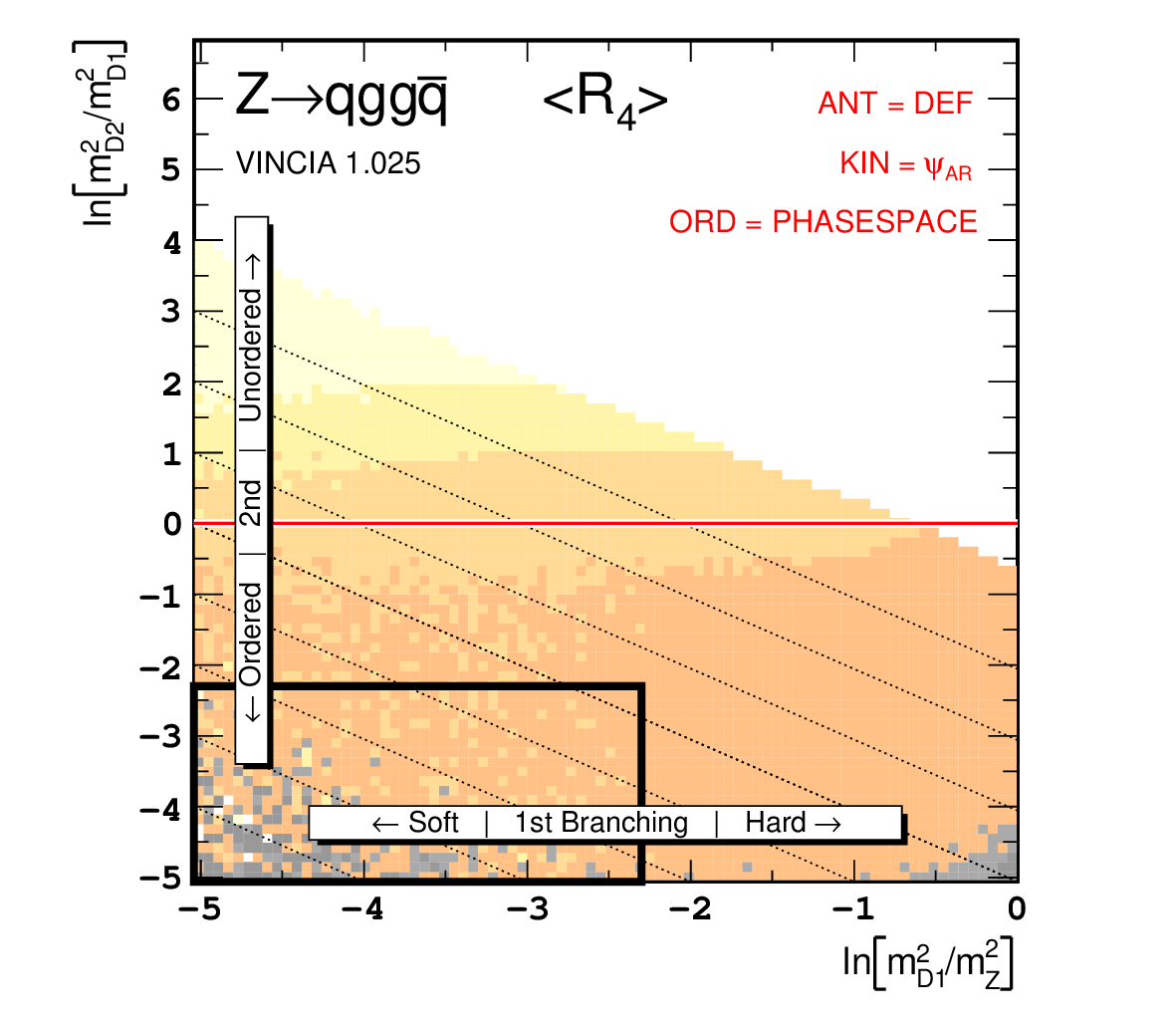}}
\capt{Phase-space-ordered antenna approximation compared to 2nd
  order QCD matrix elements. Note: this roughly corresponds to a
  mass-ordered parton shower without coherence. Although the
  double-soft limit is eventually reached, there is a large
  overcounting over most of phase space, reflecting a lack of
  coherence. Also, the double counting extends into the 
double-collinear region at the top of the lower left-hand plot. This
ordering, therefore, does not lead to the correct multiple-collinear singular
limit.} 
\end{center}
\end{figure}

\begin{figure}[p]
\begin{center}
{\LARGE{\bf Transverse-Momentum Ordering (ARIADNE)}}\\
{\large
\[
\displaystyle
\pT{}^2 = \frac{s_{ij}s_{jk}}{s}
\]}
\scalebox{1.0}{\plotset{PSdivA4Avg-vc01kin4Sum-def-ev1-kin1-had1q1tun0-ord-p5-m1-T3-q3-grey.eps}
{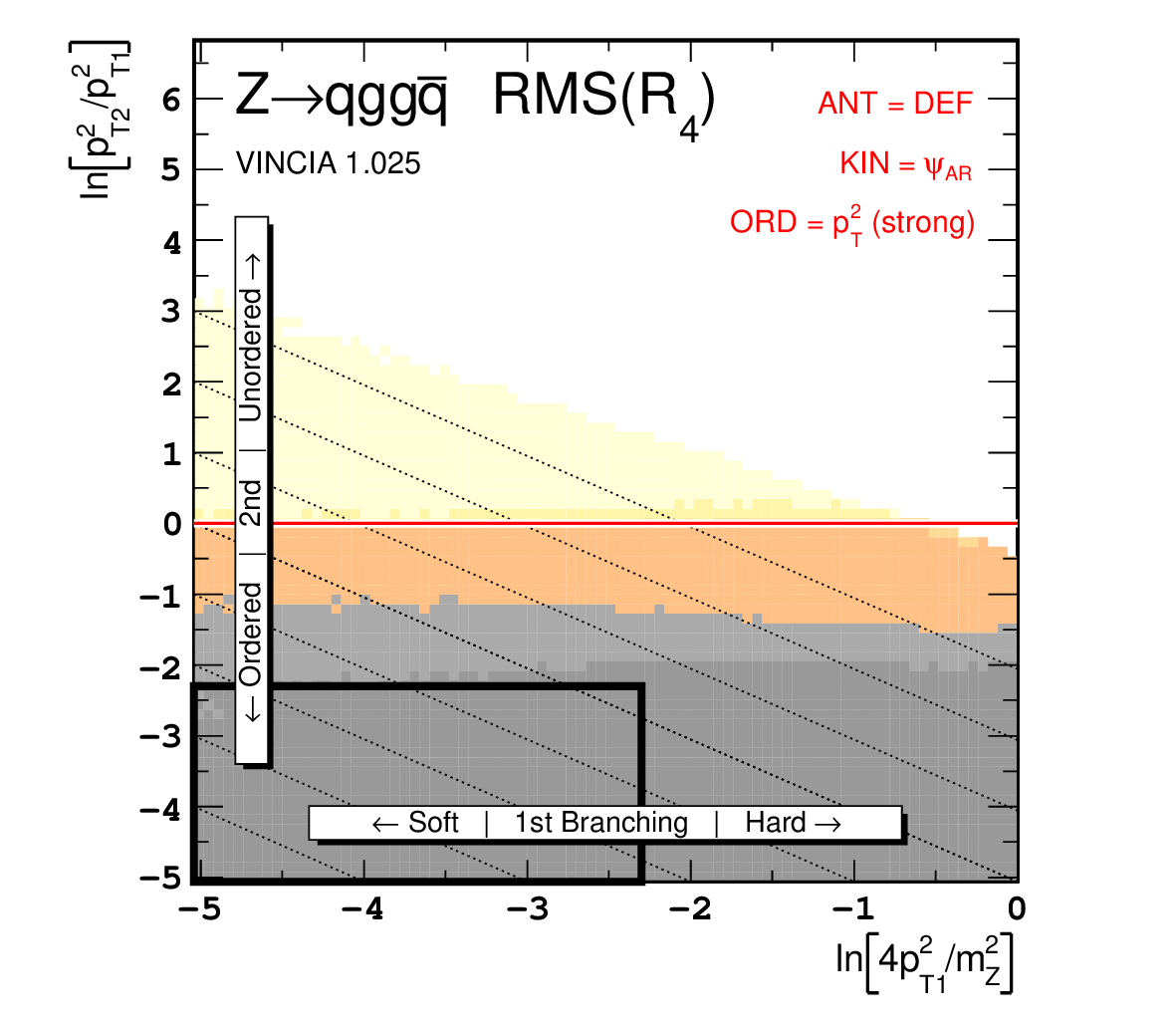}}
\scalebox{1.0}{
\plotset{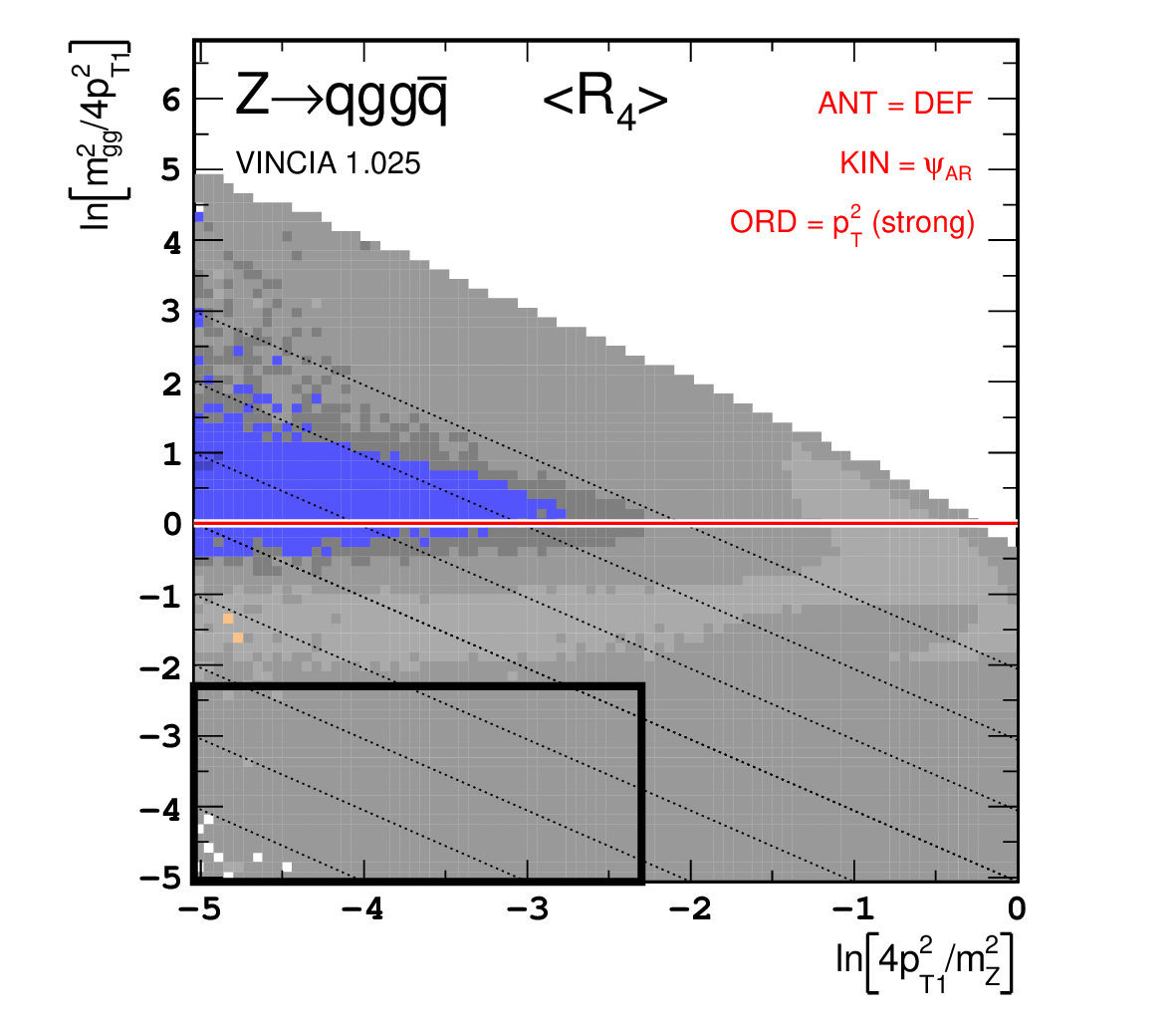}
{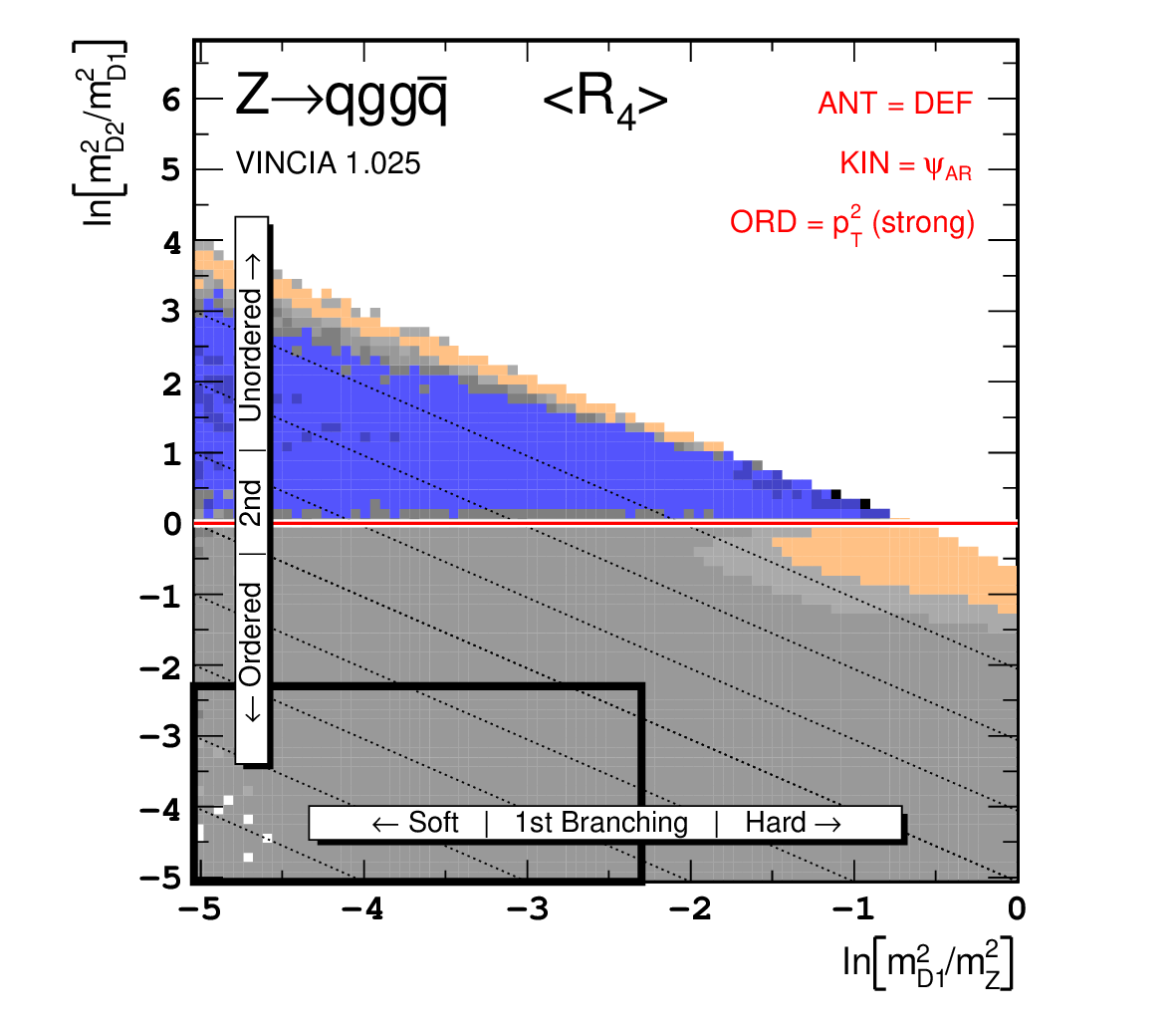}}
\capt{Transverse-Momentum-Ordered antenna approximation compared to
  2nd order QCD matrix elements, using the \Ar\ definition of \pT{}, 
which is also the default evolution variable in \Vc. Most of the
double-counting evident for phase-space ordering has been removed, and
the shower approximation now also gives the correct answer in the
double-collinear region at the top of the lower left-hand plot. The
price is the introduction of a dead zone, visible at the top of the
upper left-hand plot. The size of the dead zone in the flat
phase-space scan amounts to about 2\% of all sampled points.}
\end{center}
\end{figure}

\begin{figure}[p]
\begin{center}
{\LARGE{\bf Transverse-Momentum Ordering (PYTHIA)}}\\
{\large
\[
 \displaystyle 4\pT{\mrm{evol},I}^2  =   
    \displaystyle 4\frac{s_{ij}(s - s_{jk})(s_{ij} + s_{jk})}{(s +
      s_{ij})^2} 
\]}
\scalebox{1.0}{\plotset{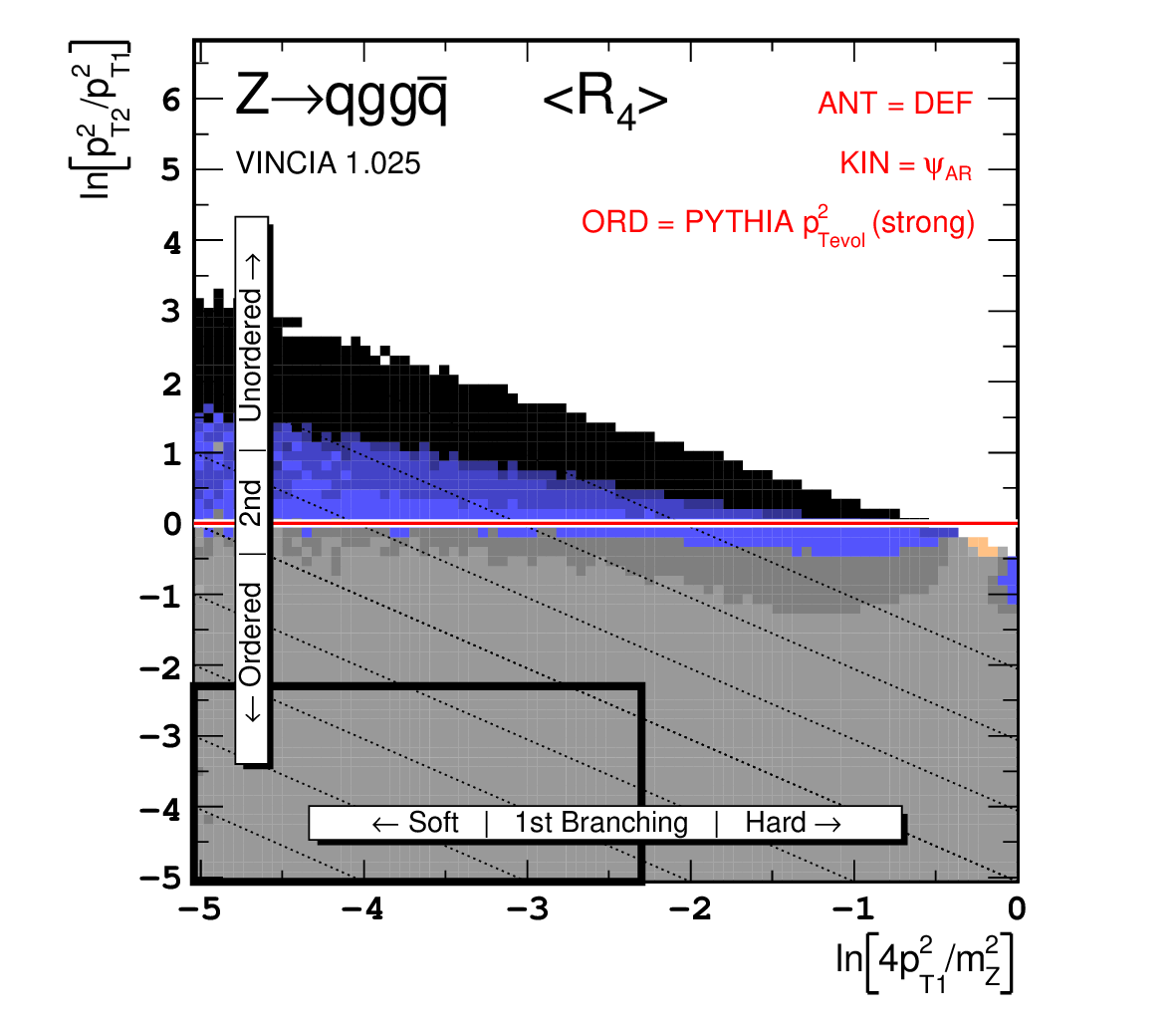}
{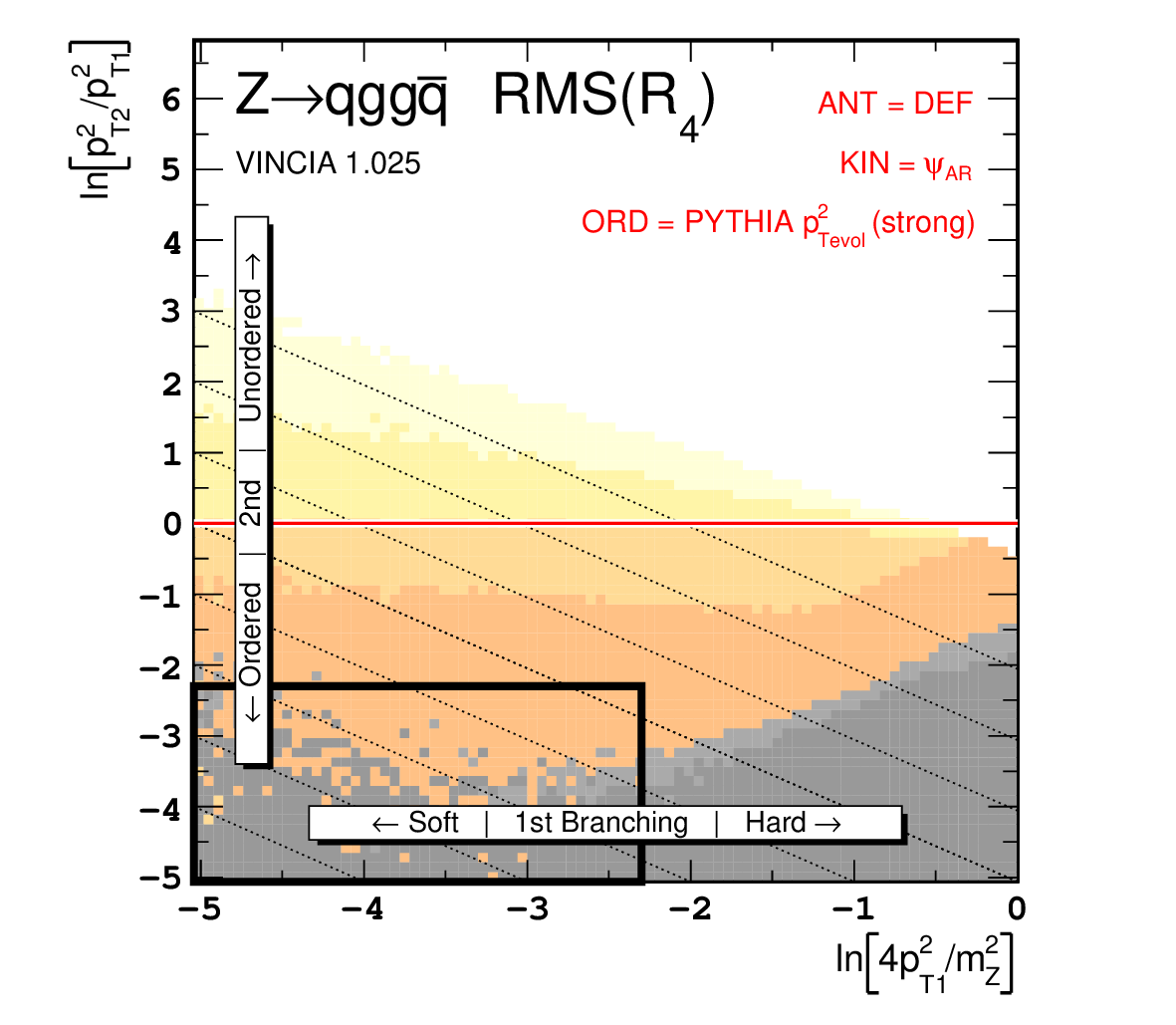}}
\scalebox{1.0}{
\plotset{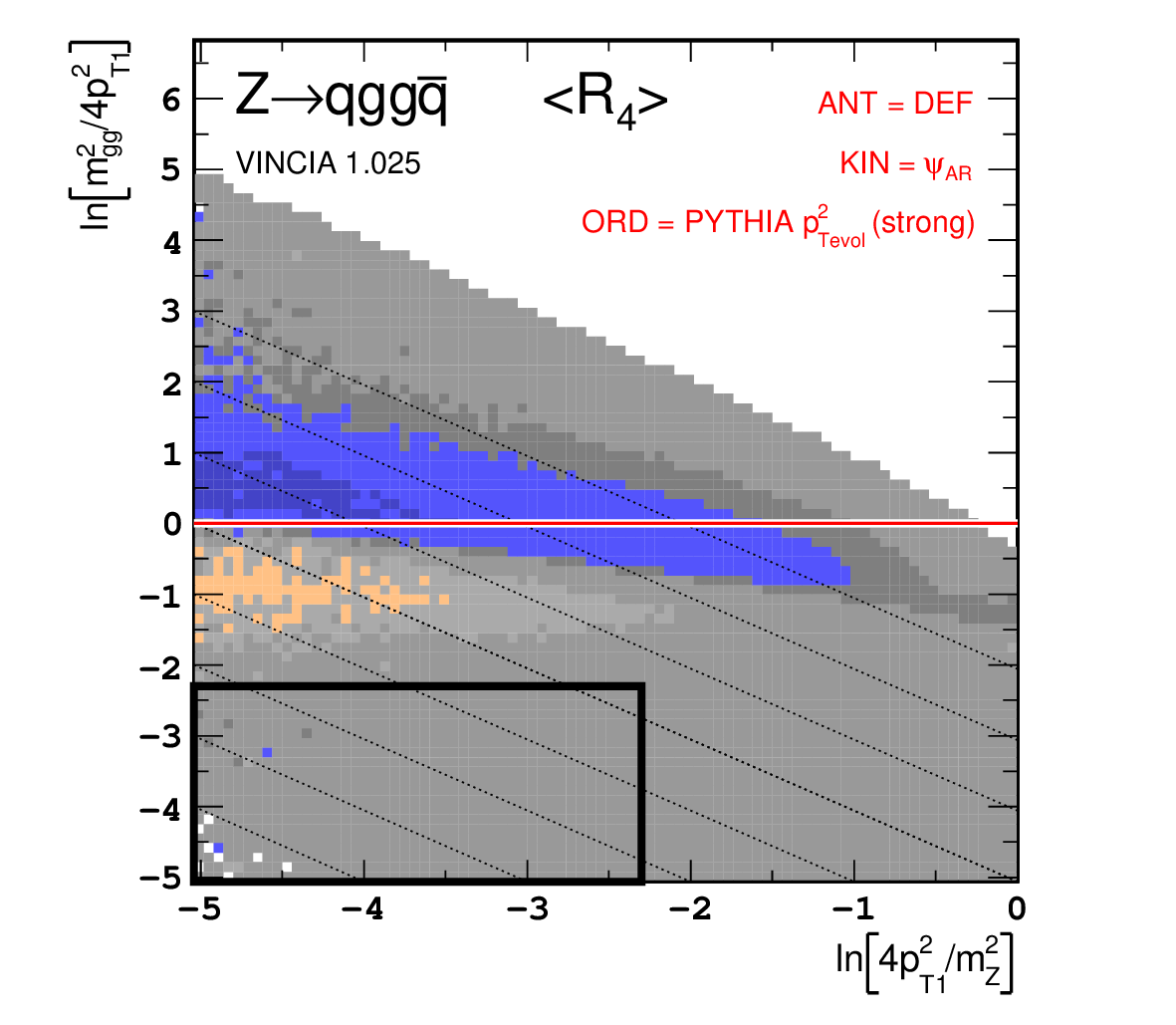}
{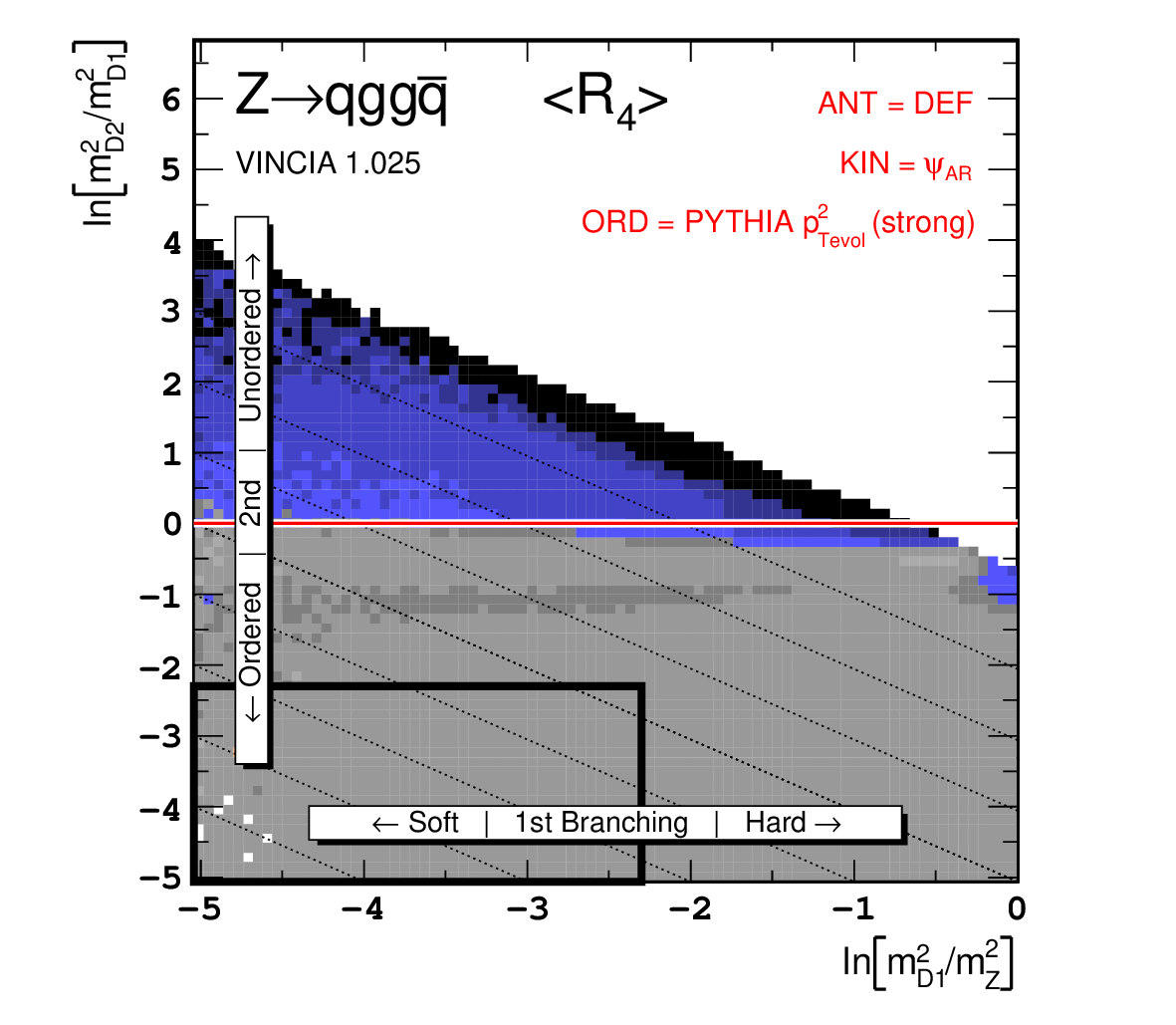}}
\capt{Transverse-Momentum-Ordered antenna approximation compared to
  2nd order QCD matrix elements, using \Py's definition of Transverse
  Momentum, \pT{\mrm{evol}}. Note: the antenna functions
  and kinematics maps are still the default \Vc\ ones, hence these results
  do not directly correspond to what would be obtained with the \Py\ 
  program. The \Py\ \pT{}-definition is a bit more
  restrictive and increases the size of the dead zone to $5\%$ of the
  sampled points. This in turn leads to a lower average ratio in some
  regions. The fact that the RMS distribution indicates a large spread
  of weights even on the edges of the 
  strongly-ordered region (black box) reflects the
  migration of a few dead-zone points to this region, due to the 
  mismatch between the ordering variable and the \pT{} definition on
  the axes.}
\end{center}
\end{figure}

\begin{figure}[p]
\begin{center}
{\LARGE{\bf Mass-Ordering (VINCIA)}}\\
{\large
\[
\displaystyle
m_D^2 = 2\mrm{min}(s_{ij},s_{jk})
\]}
\scalebox{1.0}{\plotset{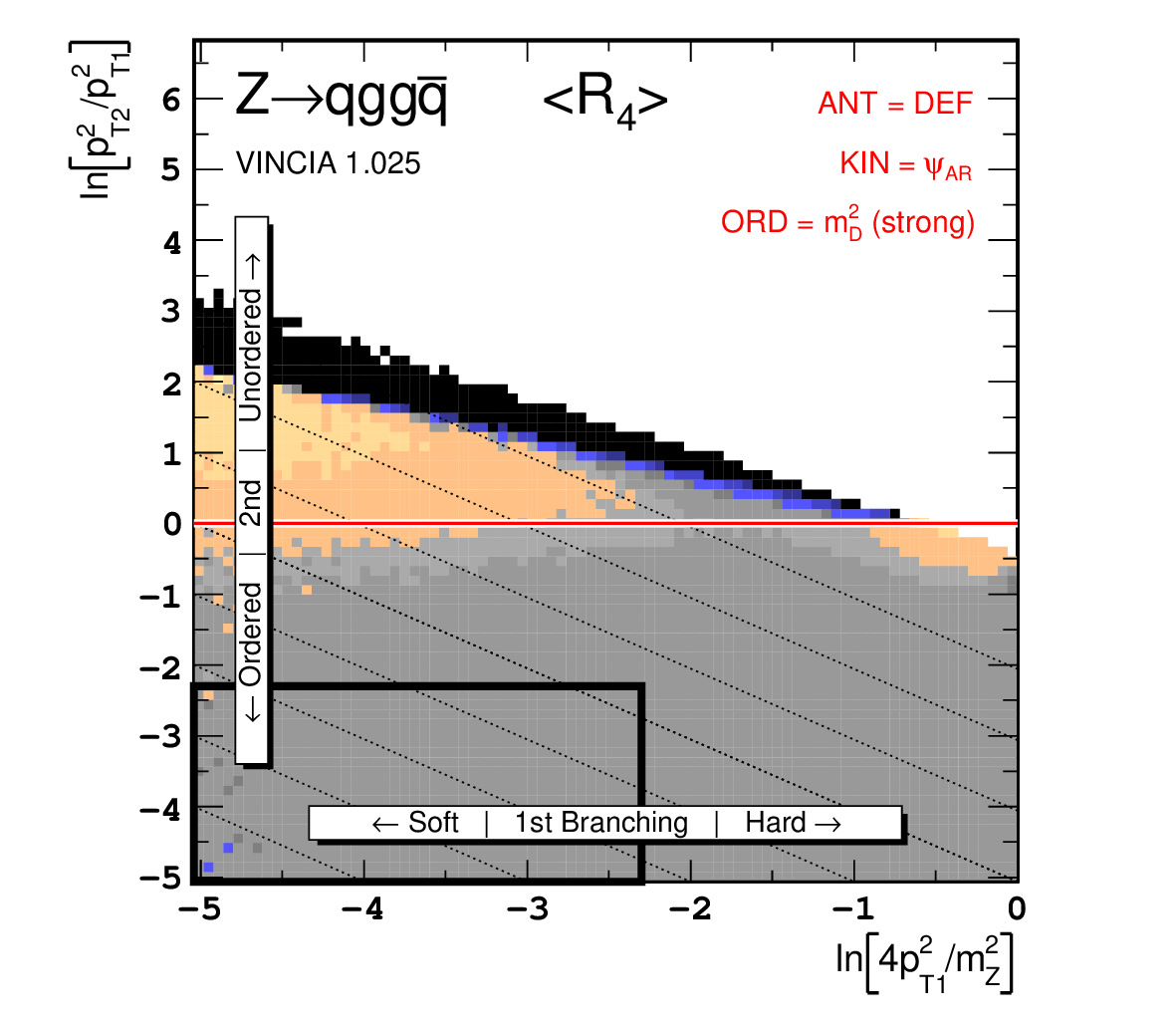}
{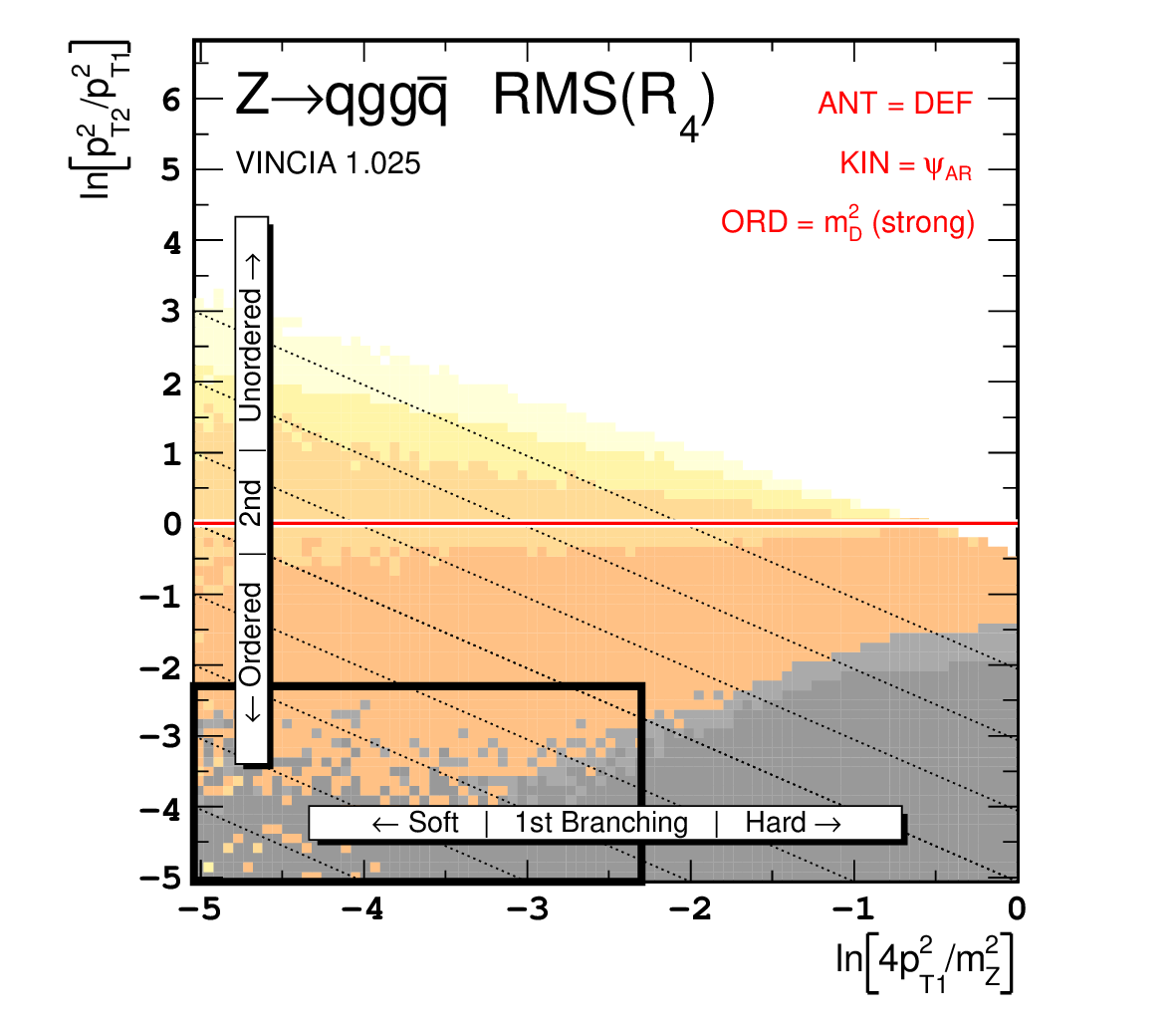}}
\scalebox{1.0}{
\plotset{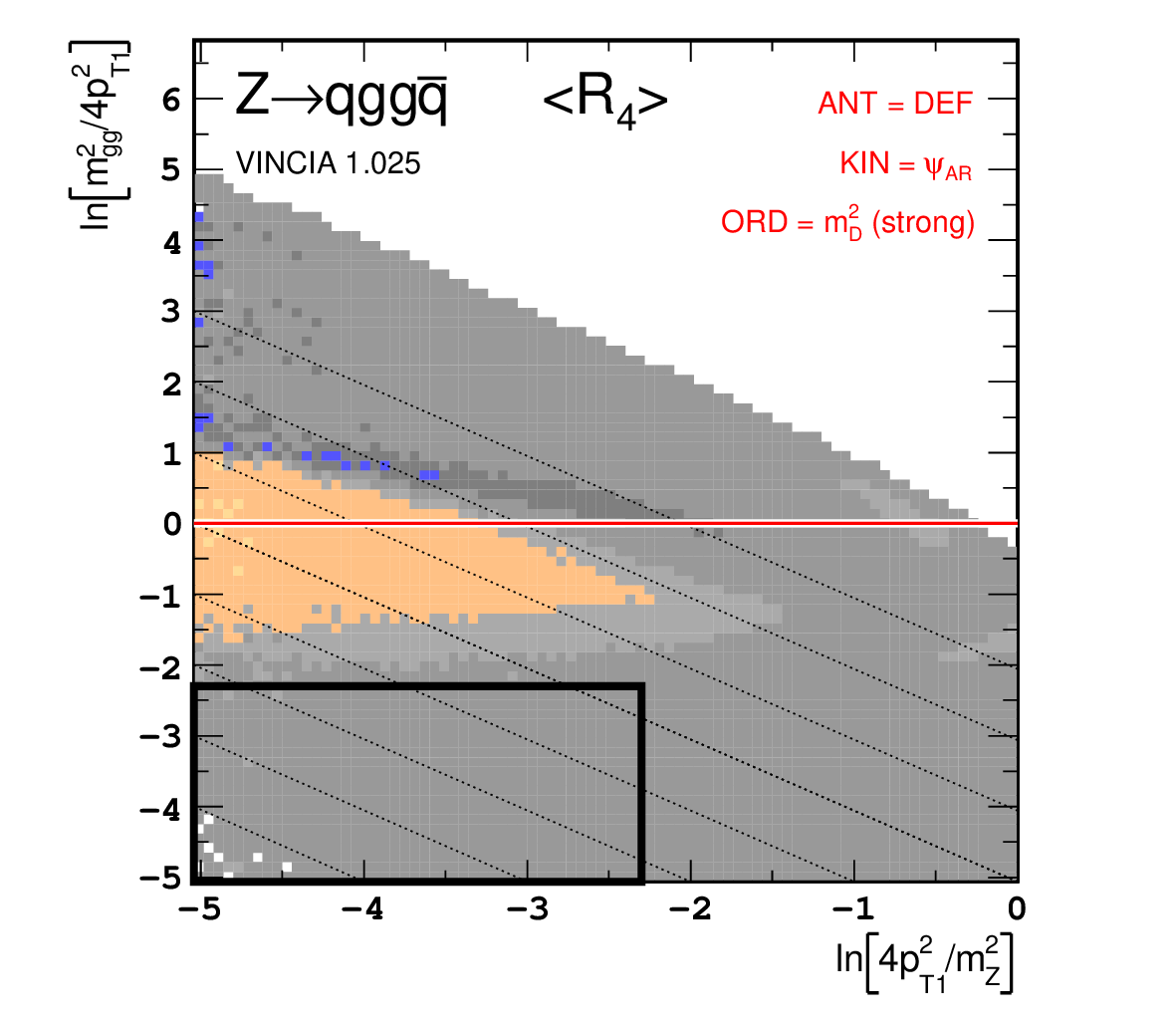}
{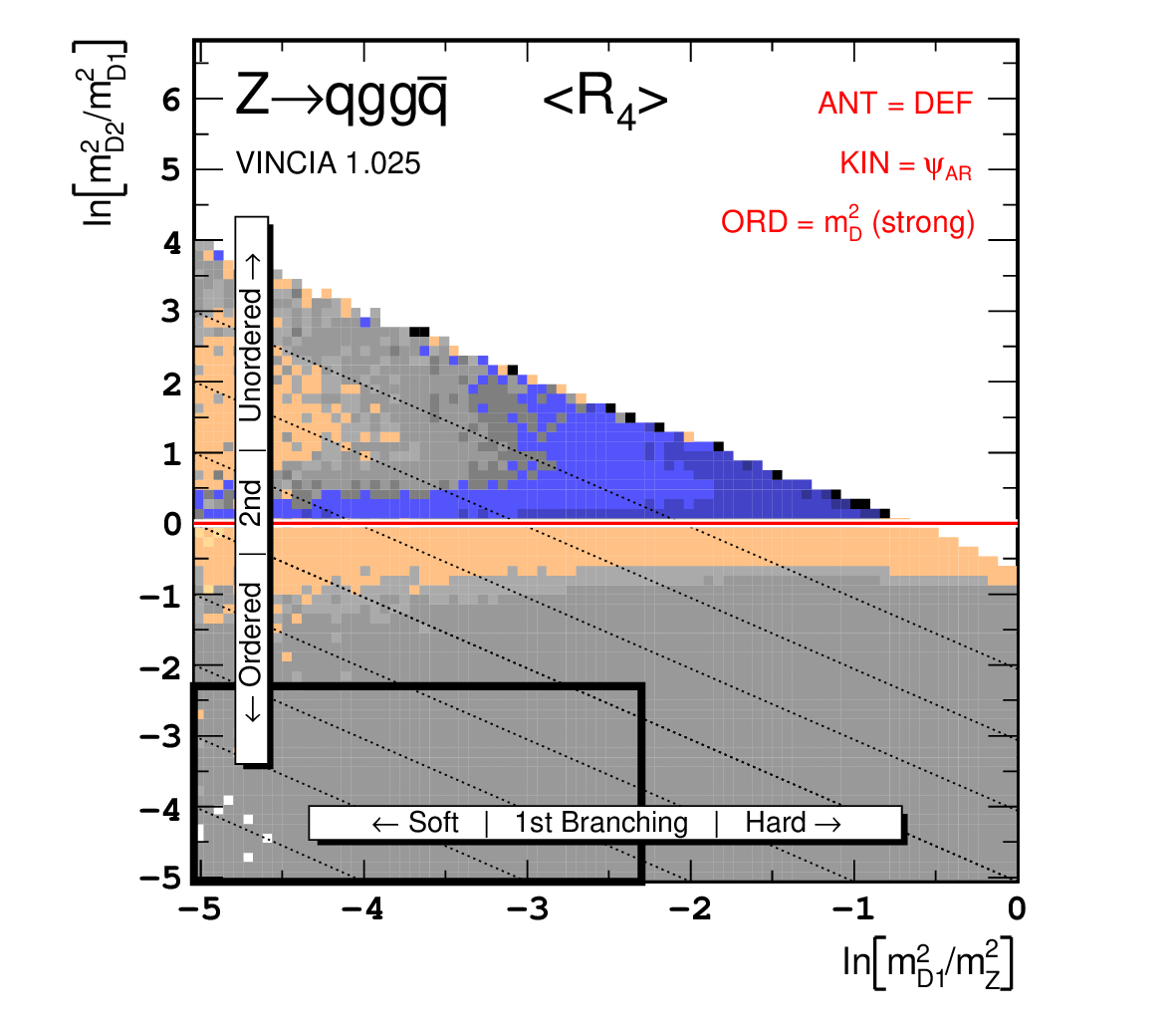}}
\capt{Mass-Ordered antenna approximation compared to
  2nd order QCD matrix elements, using \Vc's definition of 
Daughter Dipole Mass, $m_D$. This ordering is a bit less restrictive
than \pT{}, hence the size of the dead zone shrinks to about 1\% of
the sampled points. The tradeoff is that some overcounting remains
over some parts of phase space. As for the \Py\ \pT{}-ordering, the 
RMS distribution indicates a quite wide distribution extending 
inside the  doubly-ordered box, which we interpret as being 
due to the mapping of some dead-zone points to this region.}
\end{center}
\end{figure}

\begin{figure}[p]
\begin{center}
{\LARGE{\bf Energy-Ordering (DM)}}\\
{\large
\[
\displaystyle
4E^{*2}_j = \frac{(s_{ij}+s_{jk})^2}{s}
\]}
\scalebox{1.0}{\plotset{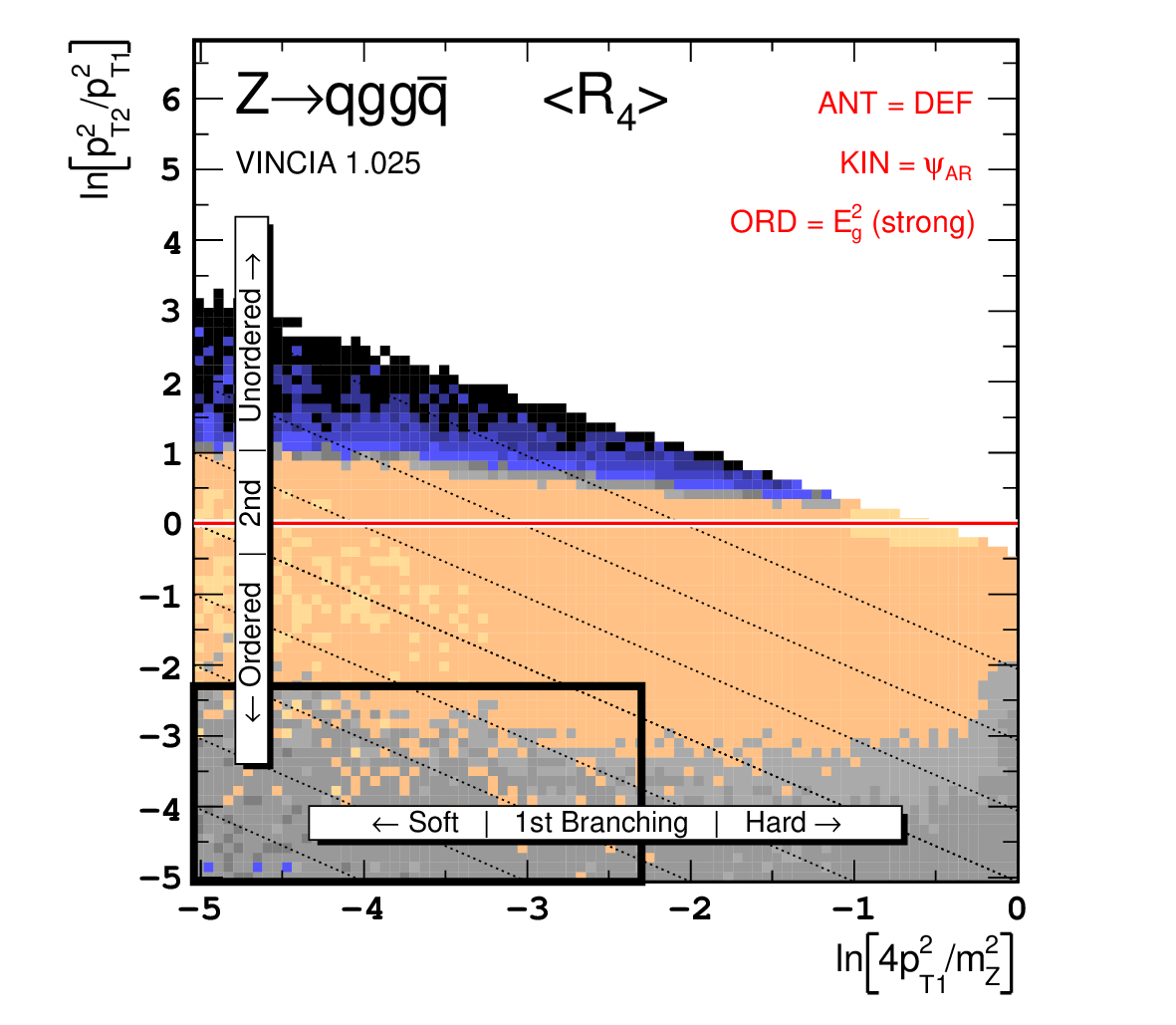}
{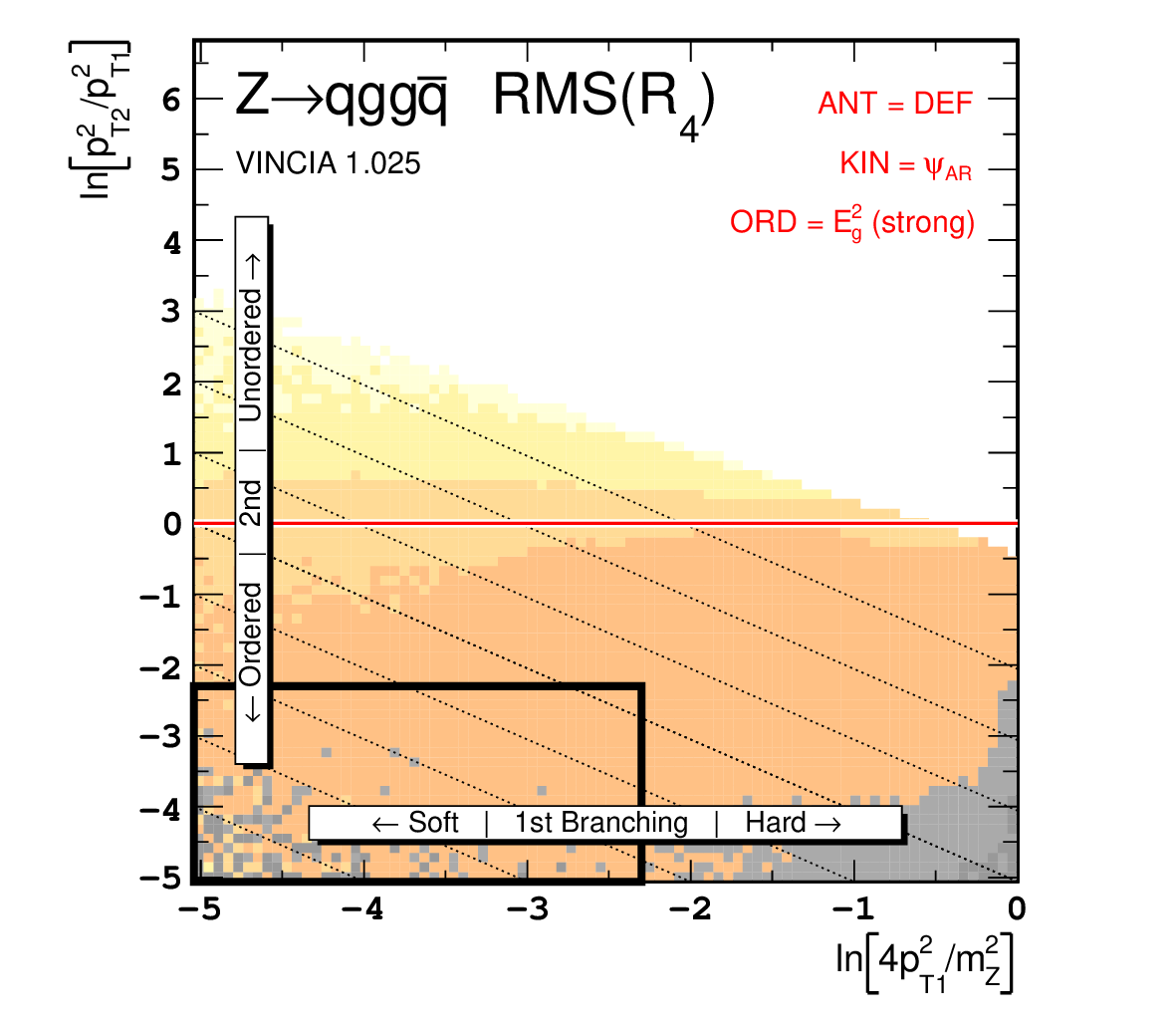}}
\scalebox{1.0}{
\plotset{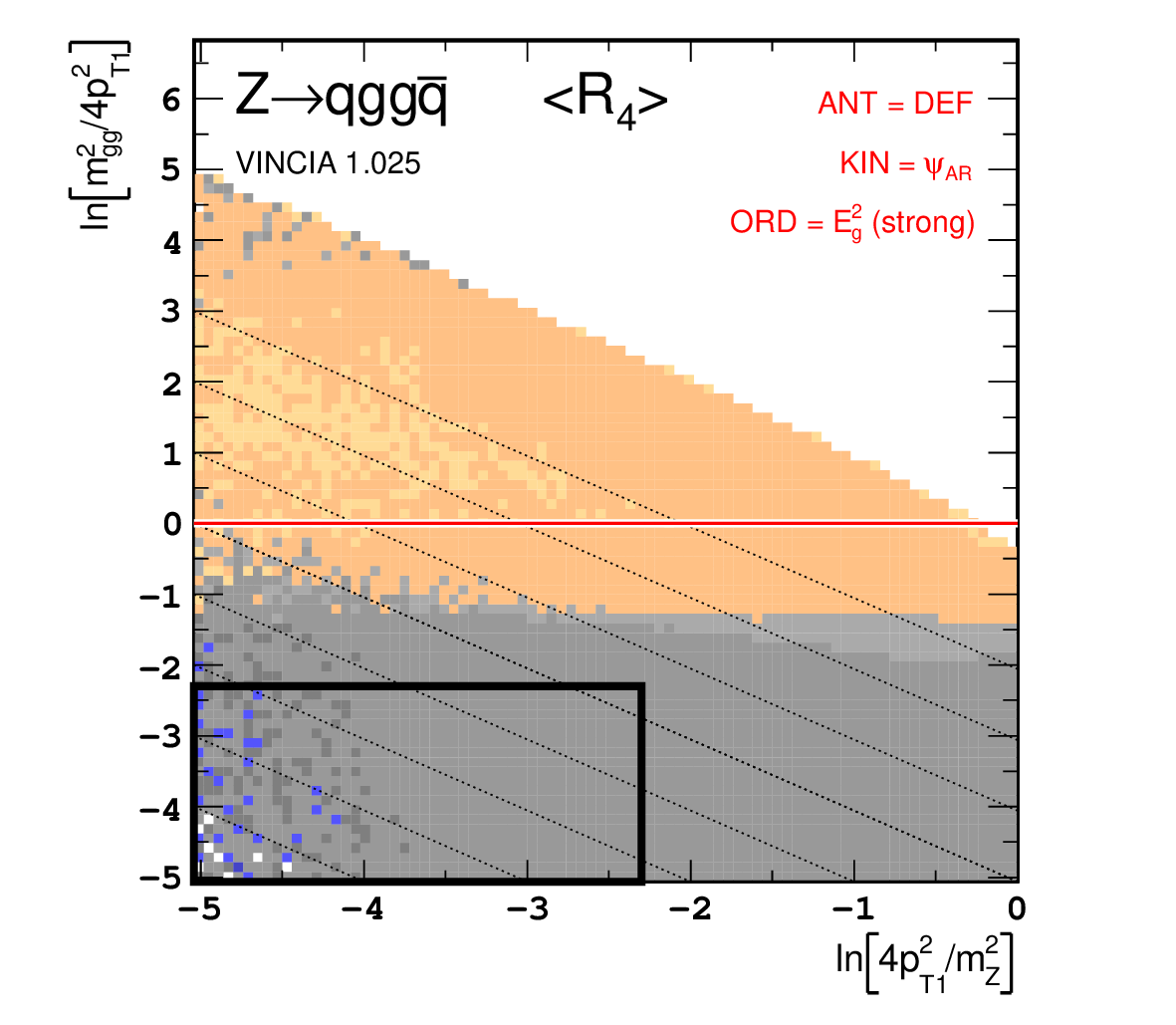}
{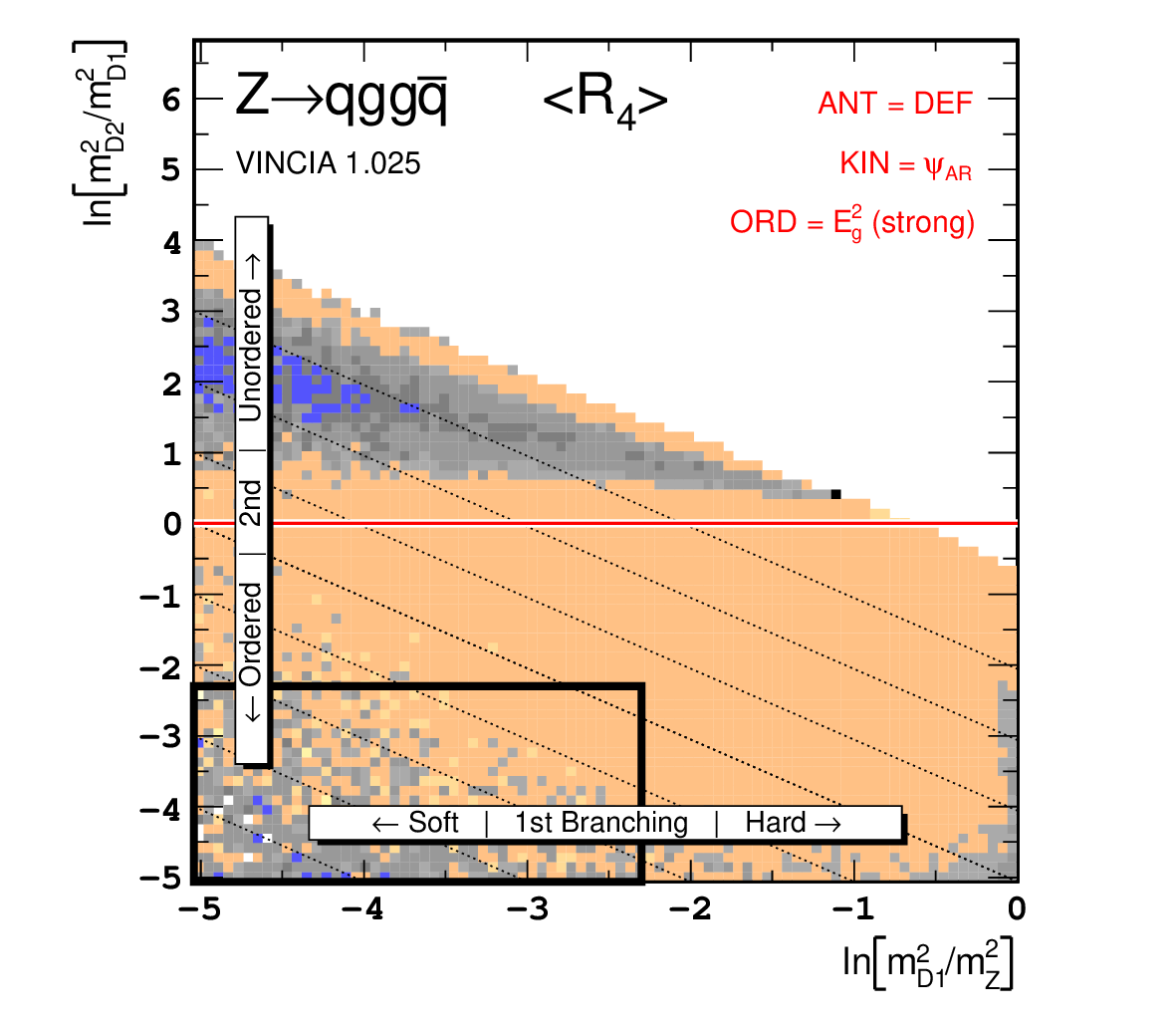}}
\capt{Energy-Ordered antenna approximation compared to
  2nd order QCD matrix elements, using a definition of energy 
a la Dokshitzer-Marchesini (DM). Although a small dead zone in the unordered
region still exists (0.6\% of the sampled points), 
there remains a very large overcounting over
significant parts of phase space, including the double-collinear
region mentioned before, at the top of the lower left-hand plot. We
conclude that this variable does not lead to the correct
multiple-collinear singular limit.}
\end{center}
\end{figure}

\begin{figure}[p]
\begin{center}
{\LARGE{\bf Modified Energy-Ordering (VINCIA)}}\\
{\large
\[
E_{T1}^{*2} = \frac{\sqrt{8s_{ij}s_{jk}(s_{ij}^2+s_{jk}^2)}}{s}
\]}
\scalebox{1.0}{\plotset{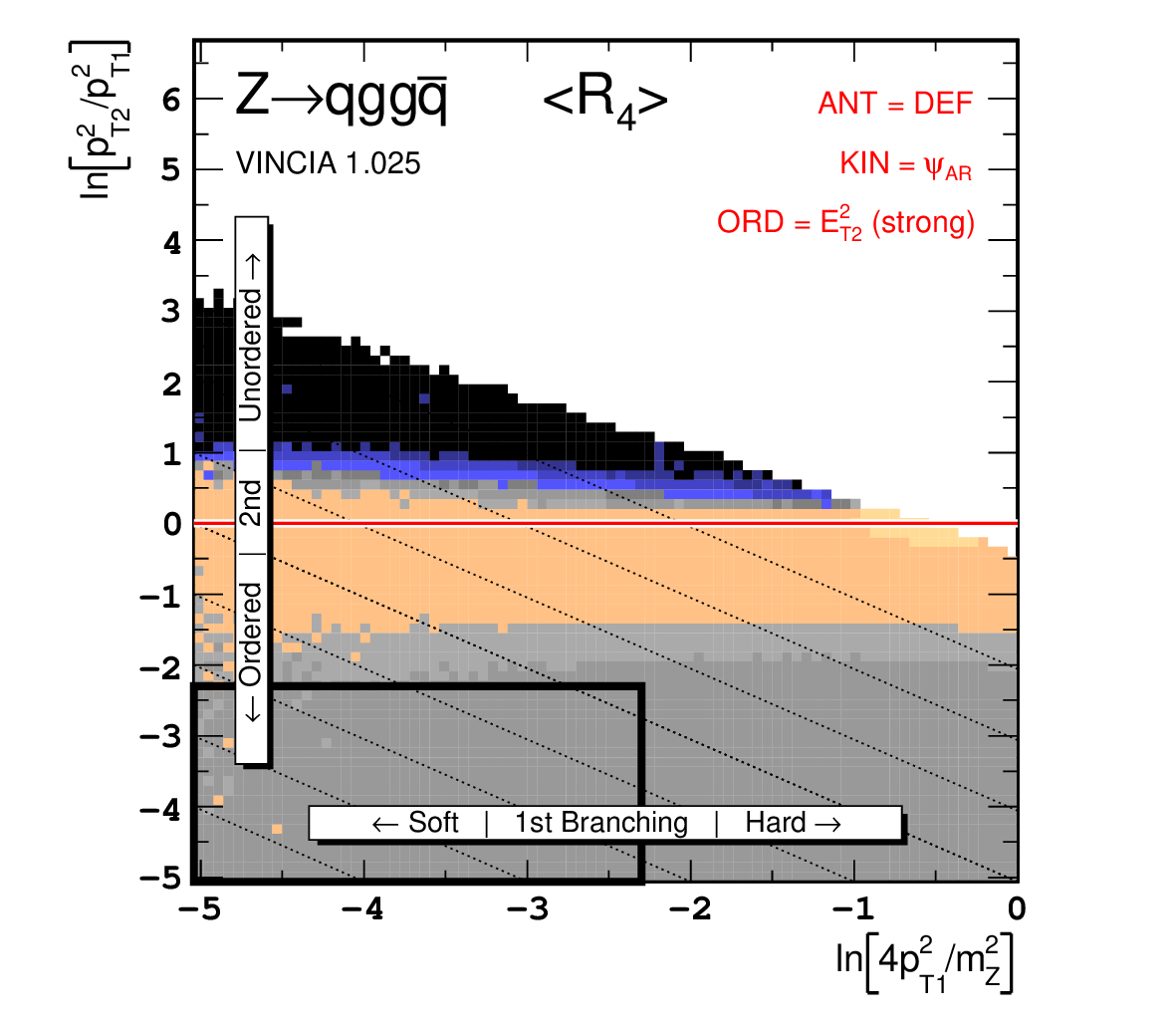}
{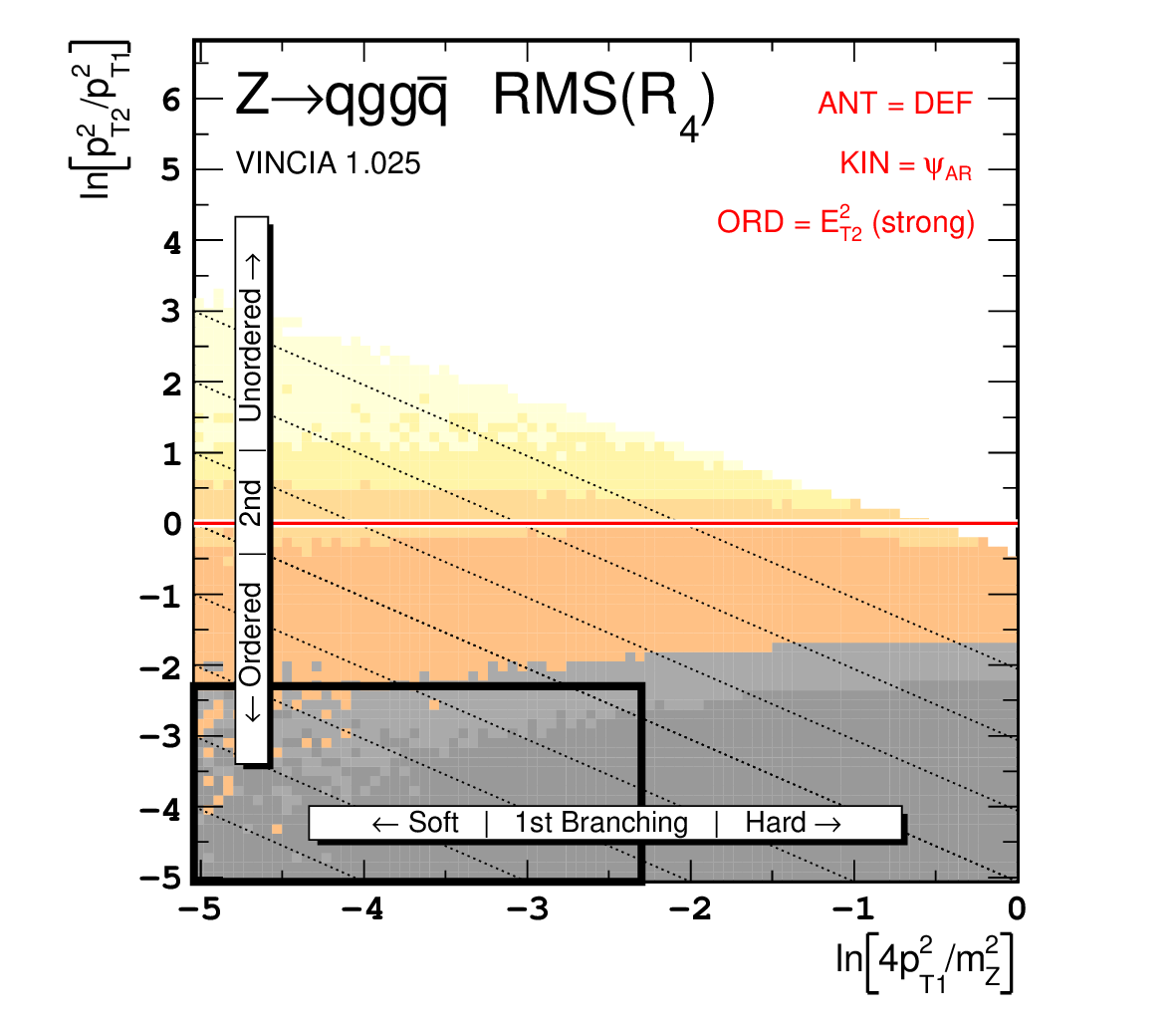}}
\scalebox{1.0}{
\plotset{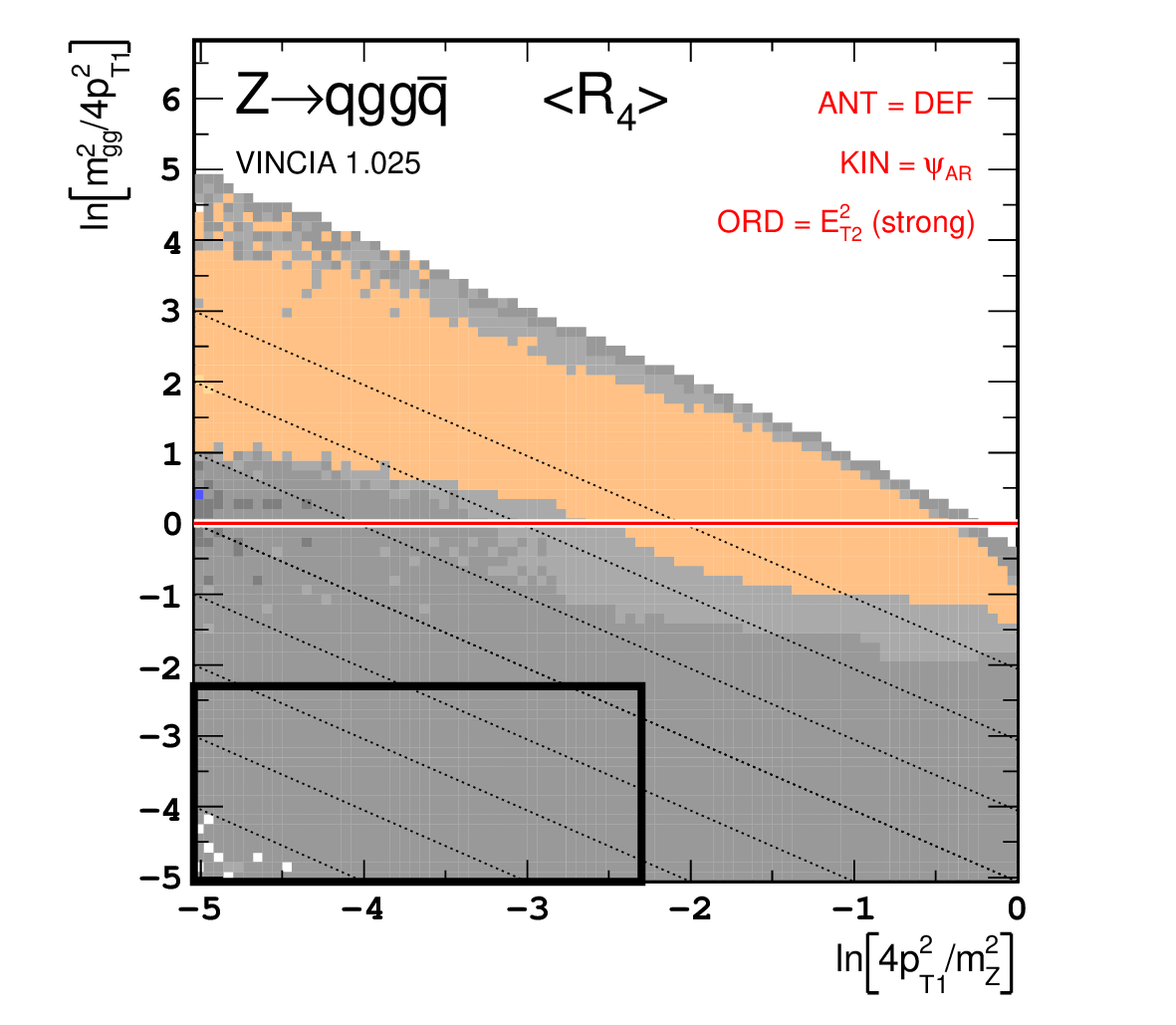}
{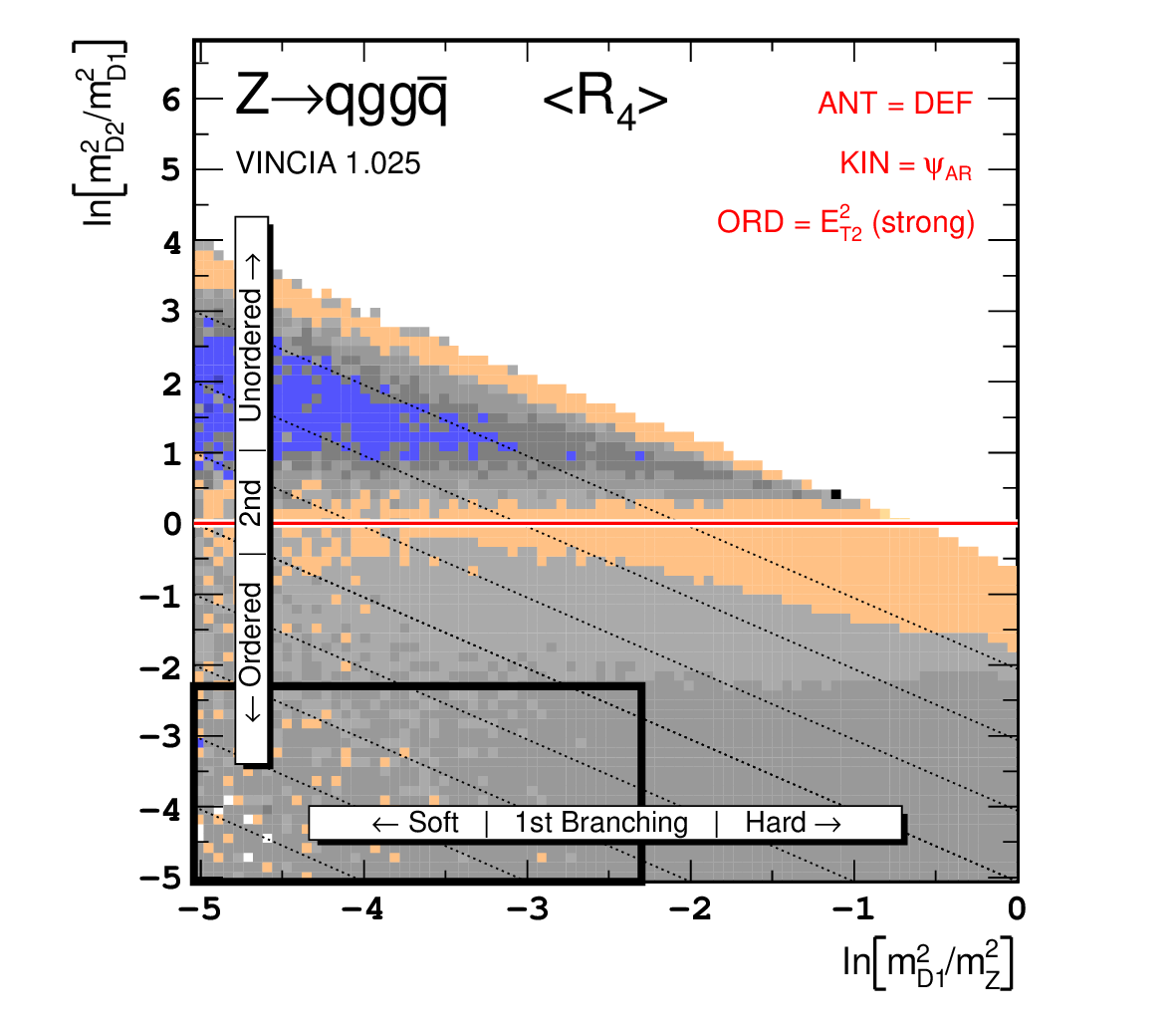}}
\capt{Energy-Ordered antenna approximation compared to
  2nd order QCD matrix elements, using a modified definition of energy 
that makes it explicitly sensitive to collinear emissions. The dead
zone is still quite small, ca.\ 0.7\% of the sampled points, but 
the approximation is improved with respect to energy ordering over
most of phase space, including the double-collinear region at the top
of the lower left-hand plot. 
}
\end{center}
\end{figure}

\begin{figure}[p]
\begin{center}
{\LARGE{\bf Angular-Ordering (HERWIG++)}}\\
{\large\[
 \displaystyle q_{\theta,I}^2  =   
    \displaystyle 
      4s\left(\frac{s s_{ij}}{(s - s_{jk})(s_{ij} + s_{jk})}\right)^2
      = 4s \left(\frac{1-x_k}{x_ix_j}\right)^2~
\]}
\scalebox{1.0}{\plotset{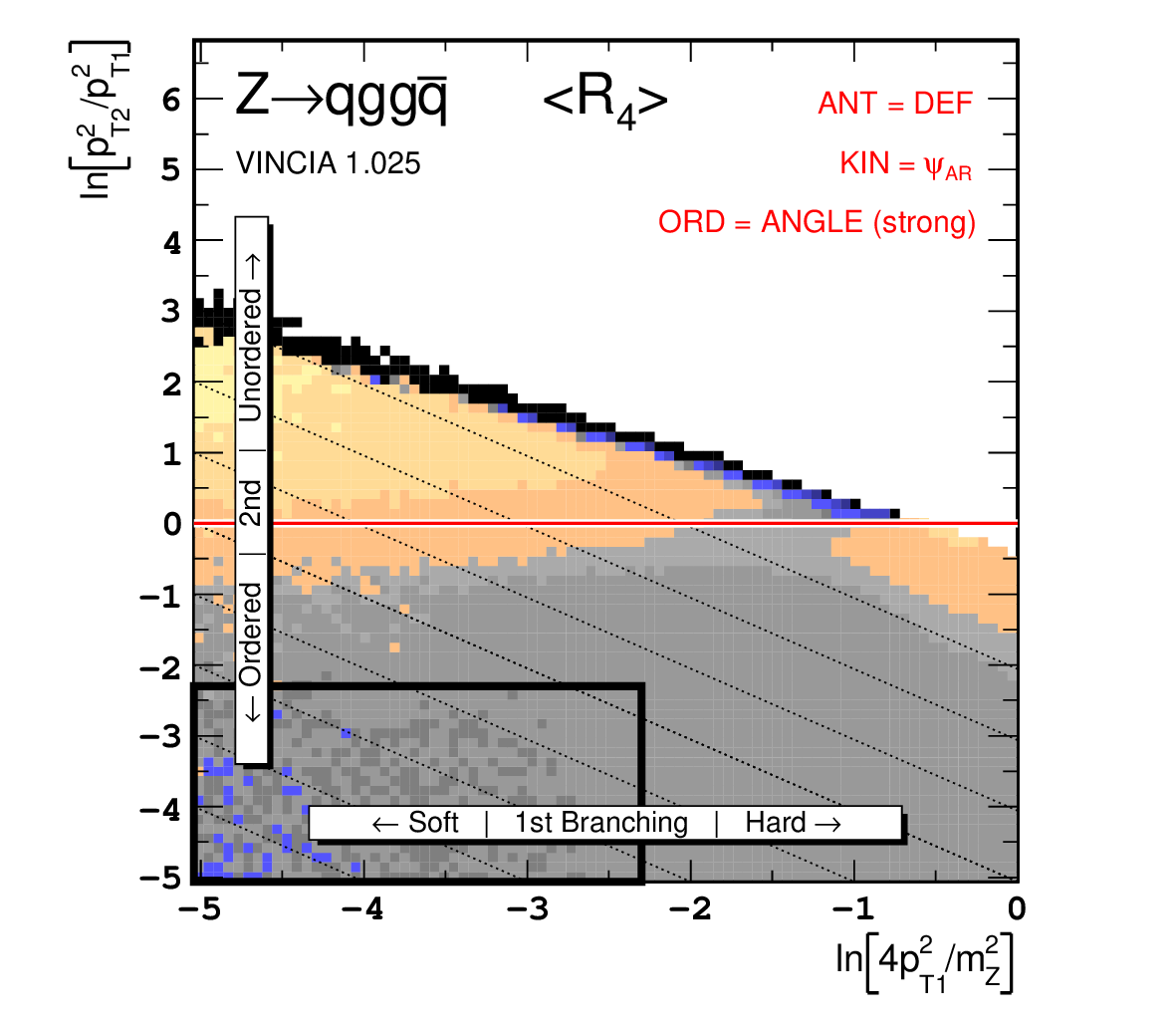}
{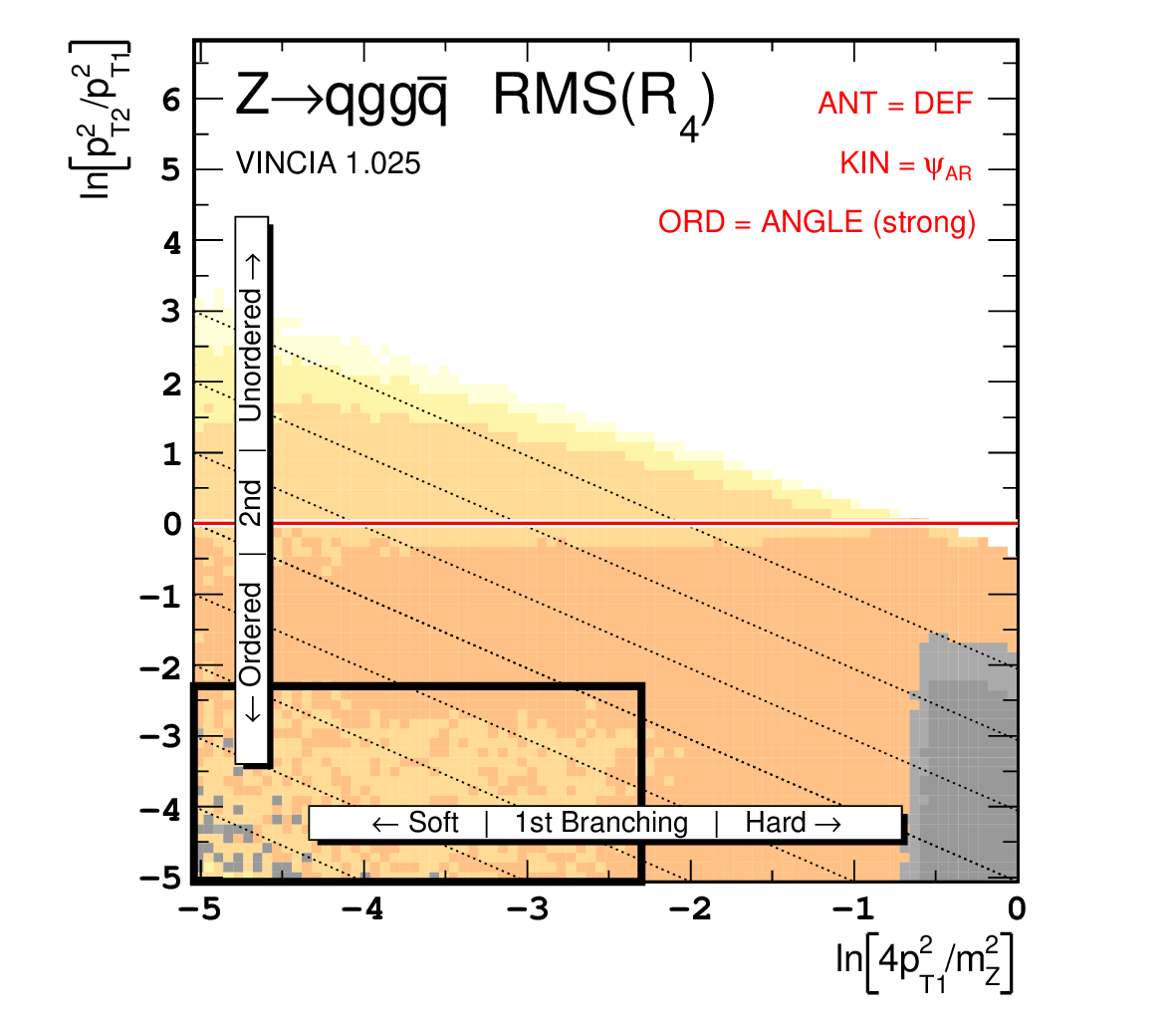}}
\scalebox{1.0}{
\plotset{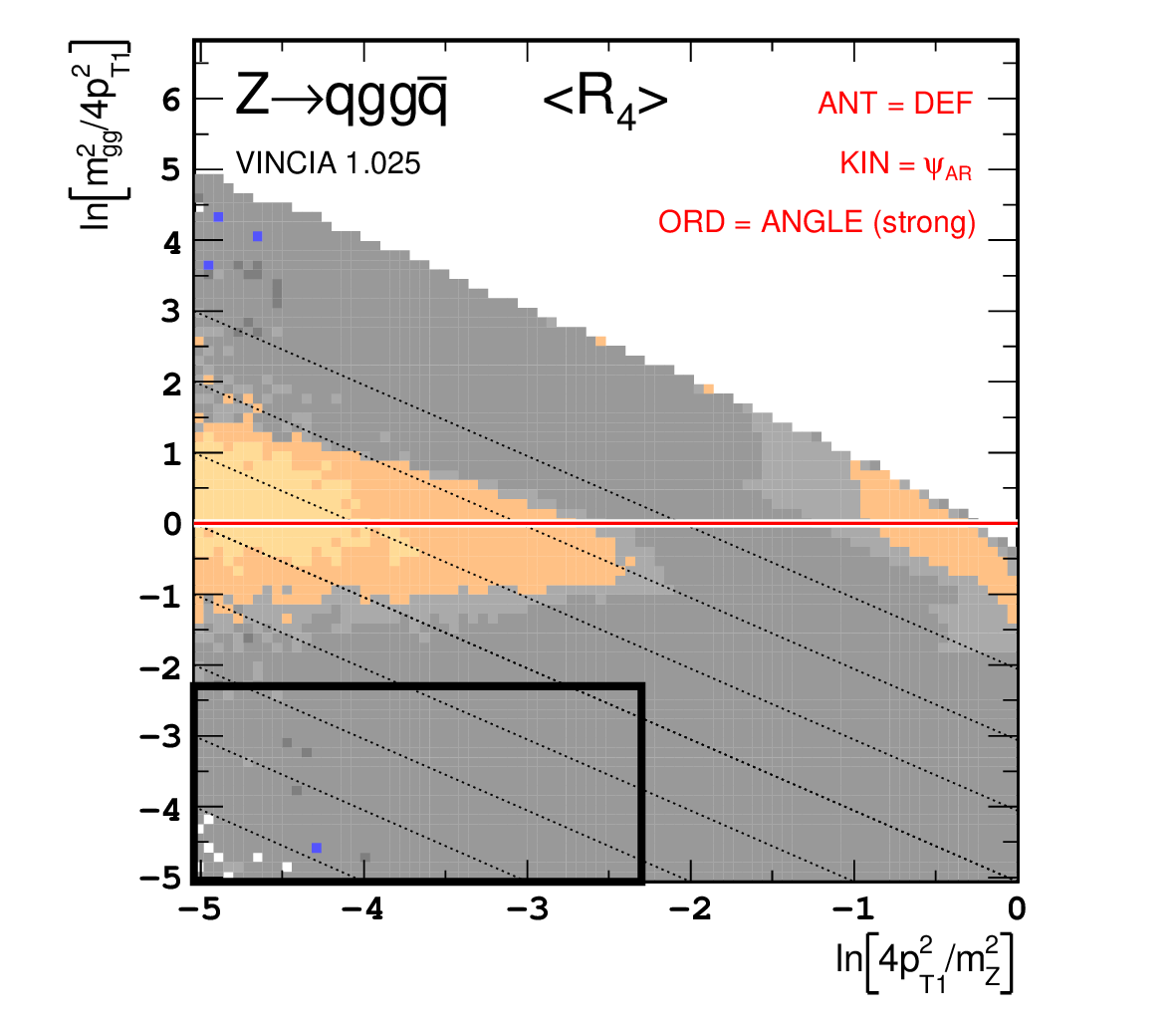}
{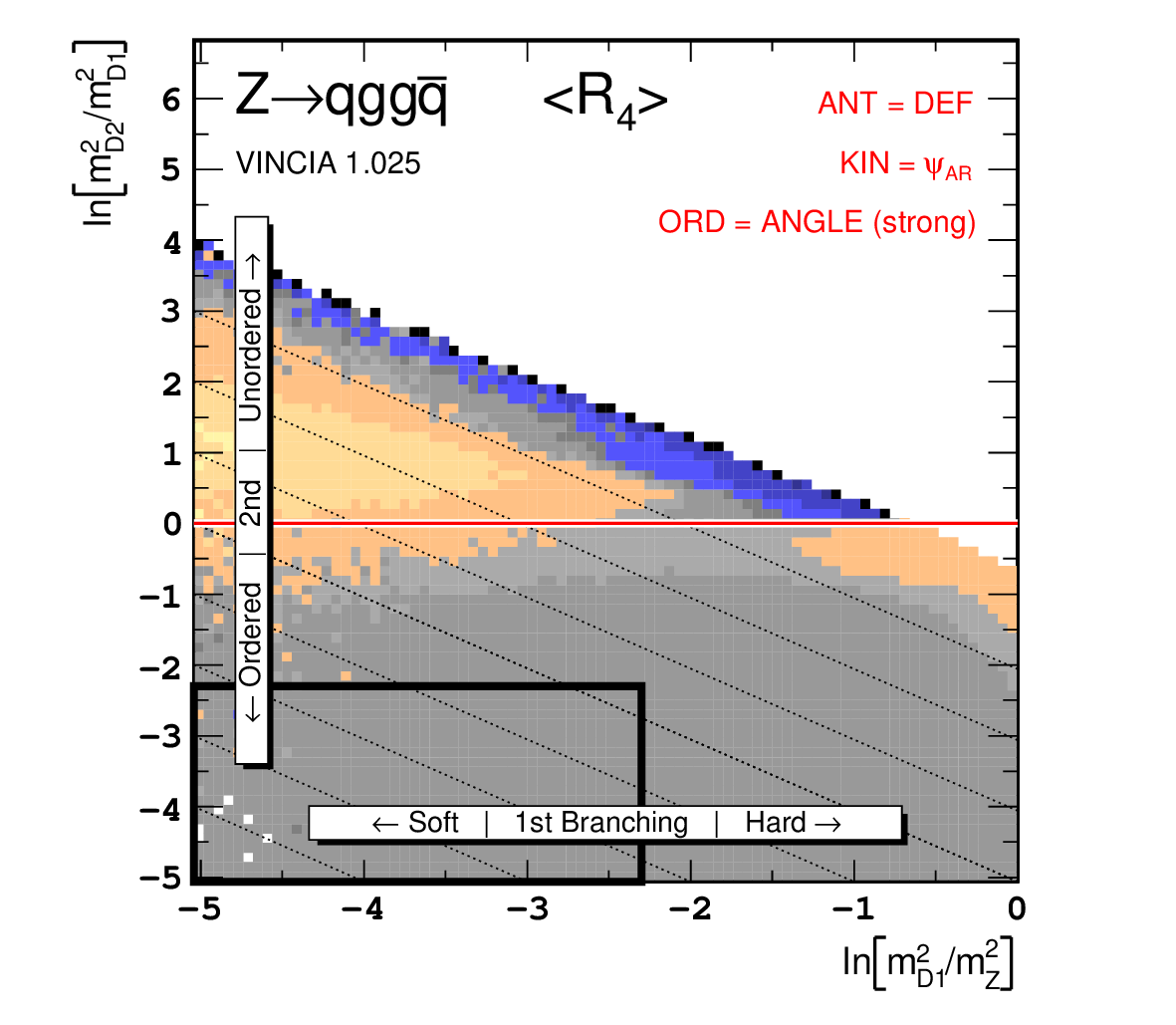}}
\capt{Angular-Ordered antenna approximation compared to 2nd order
  QCD matrix elements, using the \Hw++ definition of angles. 
  Note: the antenna functions
  and kinematics maps are still the default \Vc\ ones, hence these results
  do not directly correspond to what would be obtained with the \Hw++
  program. The dead zone in this variable amounts to 1\% of the sampled points. 
  Although the double-collinear region appears to be well
  approximated (see top of lower left-hand plot), 
  there appears to be a tendency toward untercounting of the 
  double-soft region (box in top left plot) . The RMS spread is also very
  large over substantial regions of phase space (top right plot),
  extending well beyond the region affected by the dead zone.}
\end{center}
\end{figure}

\begin{figure}[p]
\begin{center}
{\LARGE{\bf $\mathbf{V}$-Ordering (VINCIA)}}\\
{\large
\[
V^2 = s\left(\sqrt{\frac{s_{ij}+s_{jk}}{s}}
- \sqrt{\frac{|s_{ij}-s_{jk}|}{s}}\right)
\]}
\scalebox{1.0}{\plotset{PSdivA4Avg-vc01kin4Sum-def-ev5-kin1-had1q1tun0-ord-p5-m1-T3-q3-grey.eps}
{PSdivA4Rms-vc01kin4Sum-def-ev5-kin1-had1q1tun0-ord-p5-m1-T3-q3-grey.eps}}
\scalebox{1.0}{
\plotset{PSdivA4AvgM2Min-vc01kin4Sum-def-ev5-kin1-had1q1tun0-ord-p5-m1-T3-q3-grey.eps}
{PSdivA4AvgMD-vc01kin4Sum-def-ev5-kin1-had1q1tun0-ord-p5-m1-T3-q3-grey.eps}}
\capt{Angular-Ordered antenna approximation compared to
  2nd order QCD matrix elements, using \Vc's modified angular
  ordering, $V_S$, which is explicitly sensitive to soft and collinear
  emissions. The overall picture is quite similar to that for the
  \Hw++ definition of angular-ordering. 
}
\end{center}
\end{figure}

\begin{figure}[p]
\begin{center}
{\LARGE{\bf Smooth Transverse-Momentum Ordering (VINCIA)}}\\
{\large
\[
\pT{}^2 = \frac{s_{ij}s_{jk}}{s}
 \displaystyle 
\]}
\scalebox{1.0}{\plotset{PSdivA4Avg-vc01kin4Sum-def-ev1-kin1-had1q1tun0-uqe-p5-m1-T3-q3-grey.eps}
{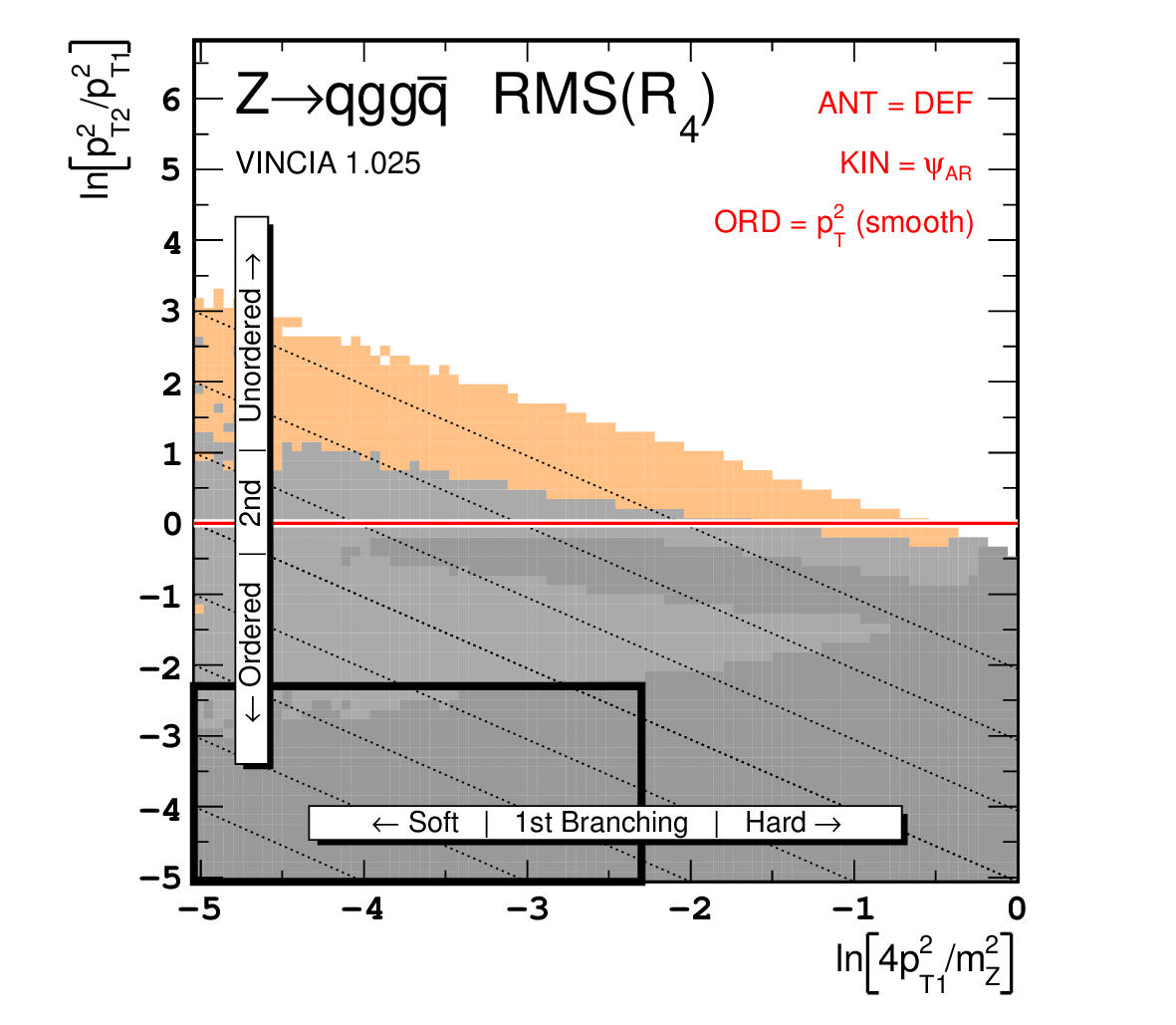}}
\scalebox{1.0}{
\plotset{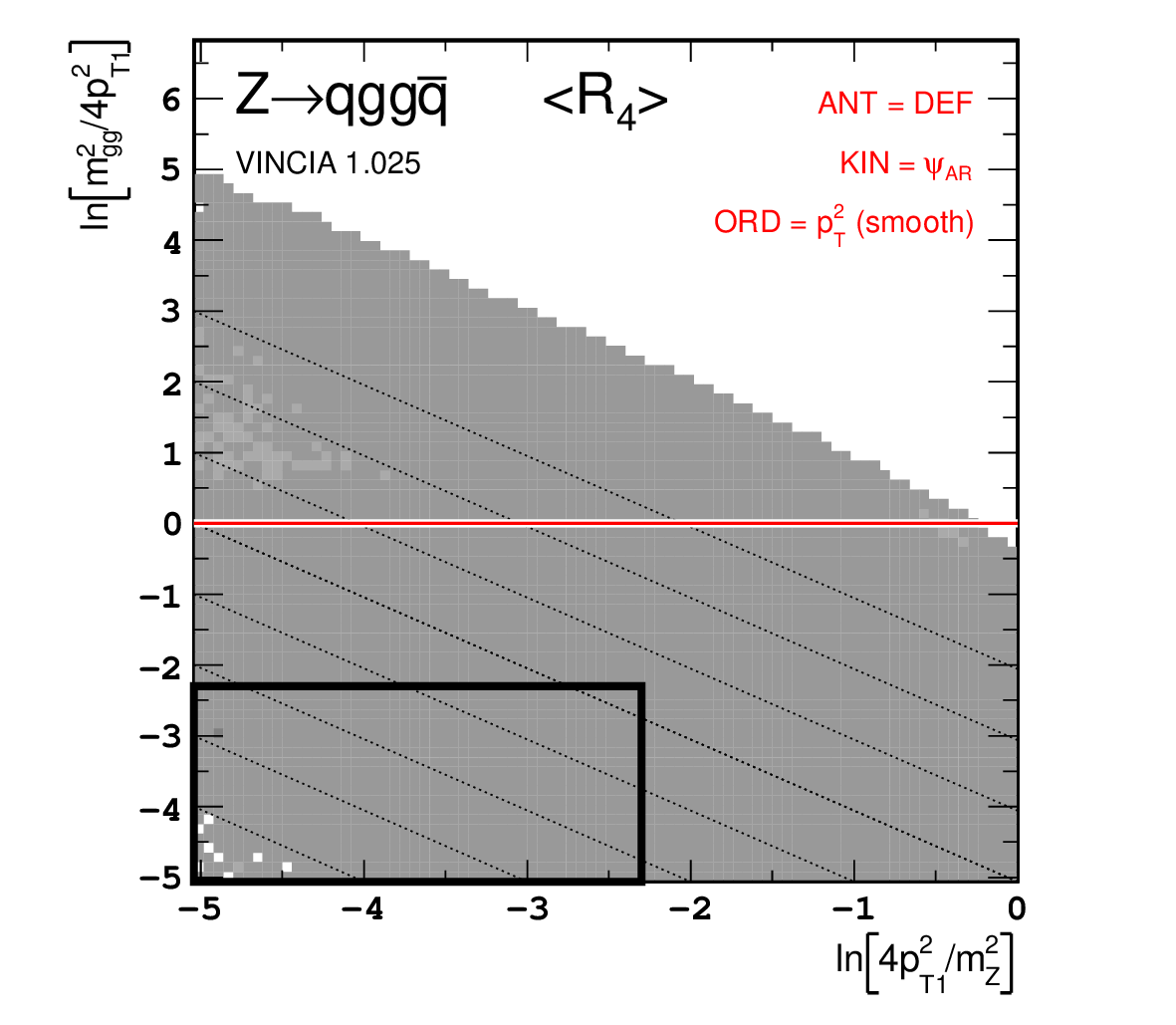}
{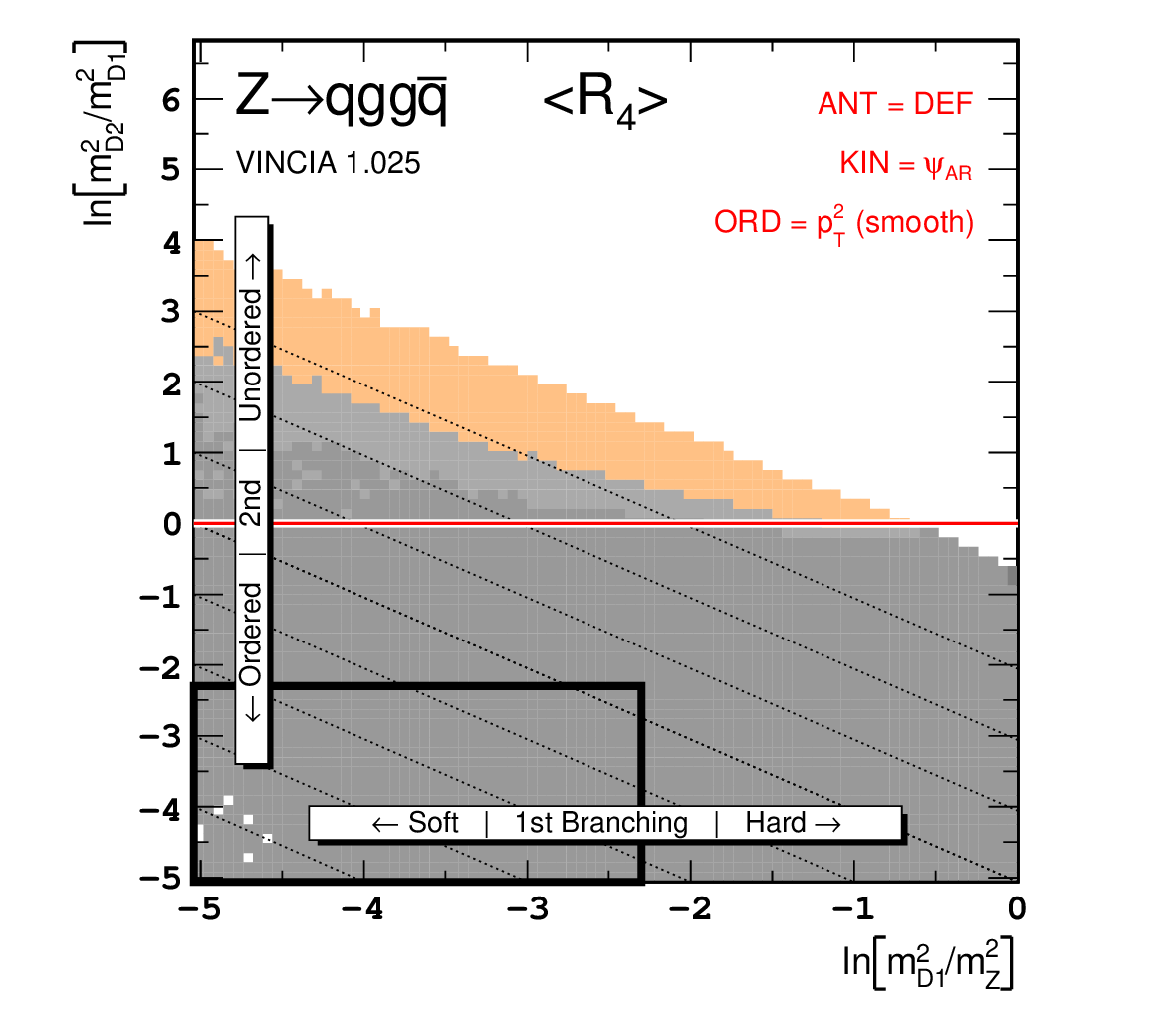}}
\capt{Transverse-momentum-ordered antenna approximation compared to 2nd order
  QCD matrix elements, using \Ar's definition of \pT{} and 
  \textsc{Vincia}'s smooth suppression factor 
  instead of the usual strong ordering condition. This corresponds to
  the default in \Vc\ without matching. (Note: by default, matching
  to $Z\to 4$ is on in \Vc, over all of phase space, 
  and hence these ratios are all equal unity).}   
\end{center}
\end{figure}

\begin{figure}[p]
\begin{center}
{\LARGE{\bf Smooth Mass-Ordering (VINCIA)}}\\
{\large
\[
 \displaystyle 
m_D^2 = 2\mrm{min}(s_{ij},s_{jk})
\]}
\scalebox{1.0}{\plotset{PSdivA4Avg-vc01kin4Sum-def-ev2-kin1-had1q1tun0-uqe-p5-m1-T3-q3-grey.eps}
{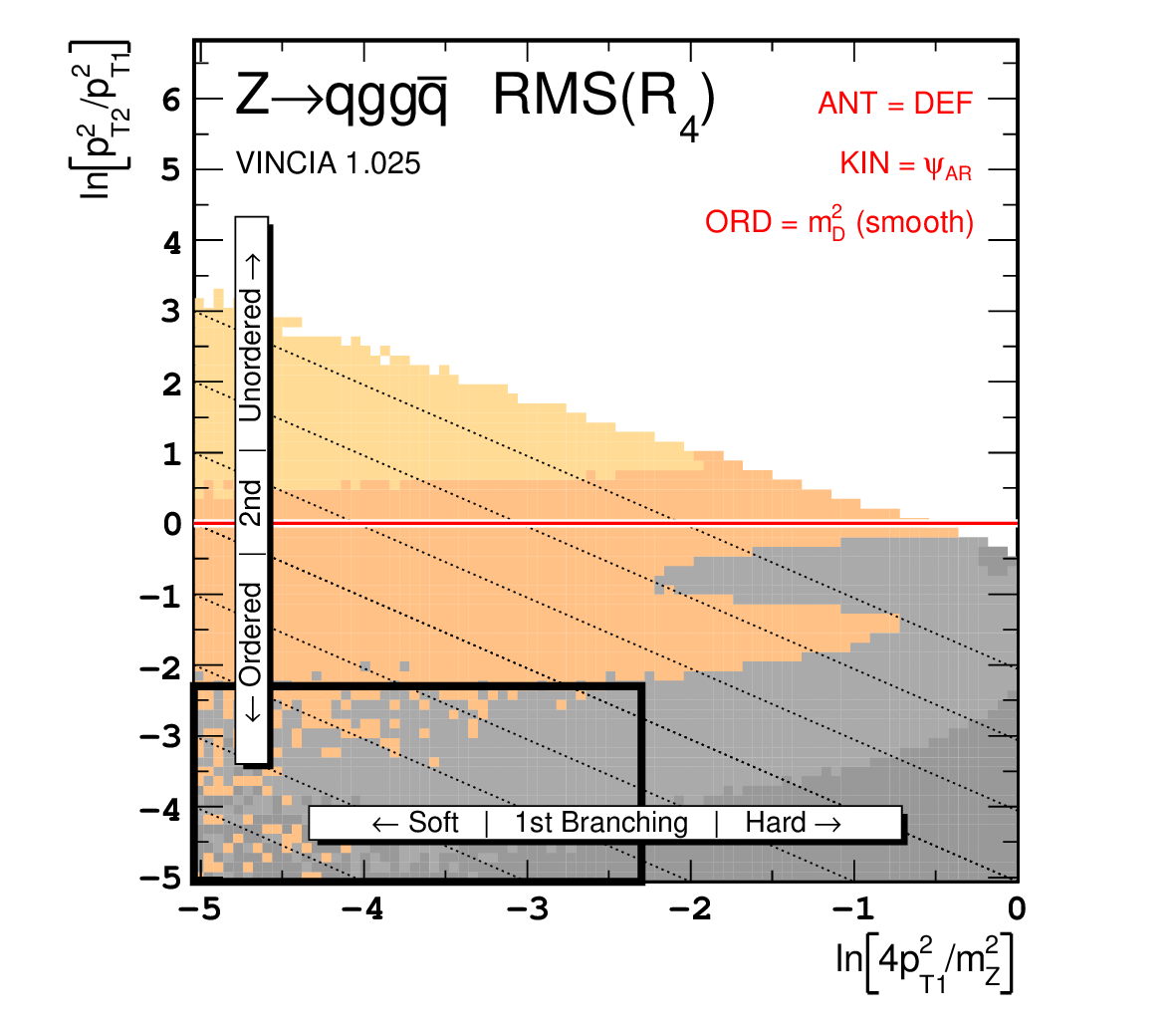}}
\scalebox{1.0}{
\plotset{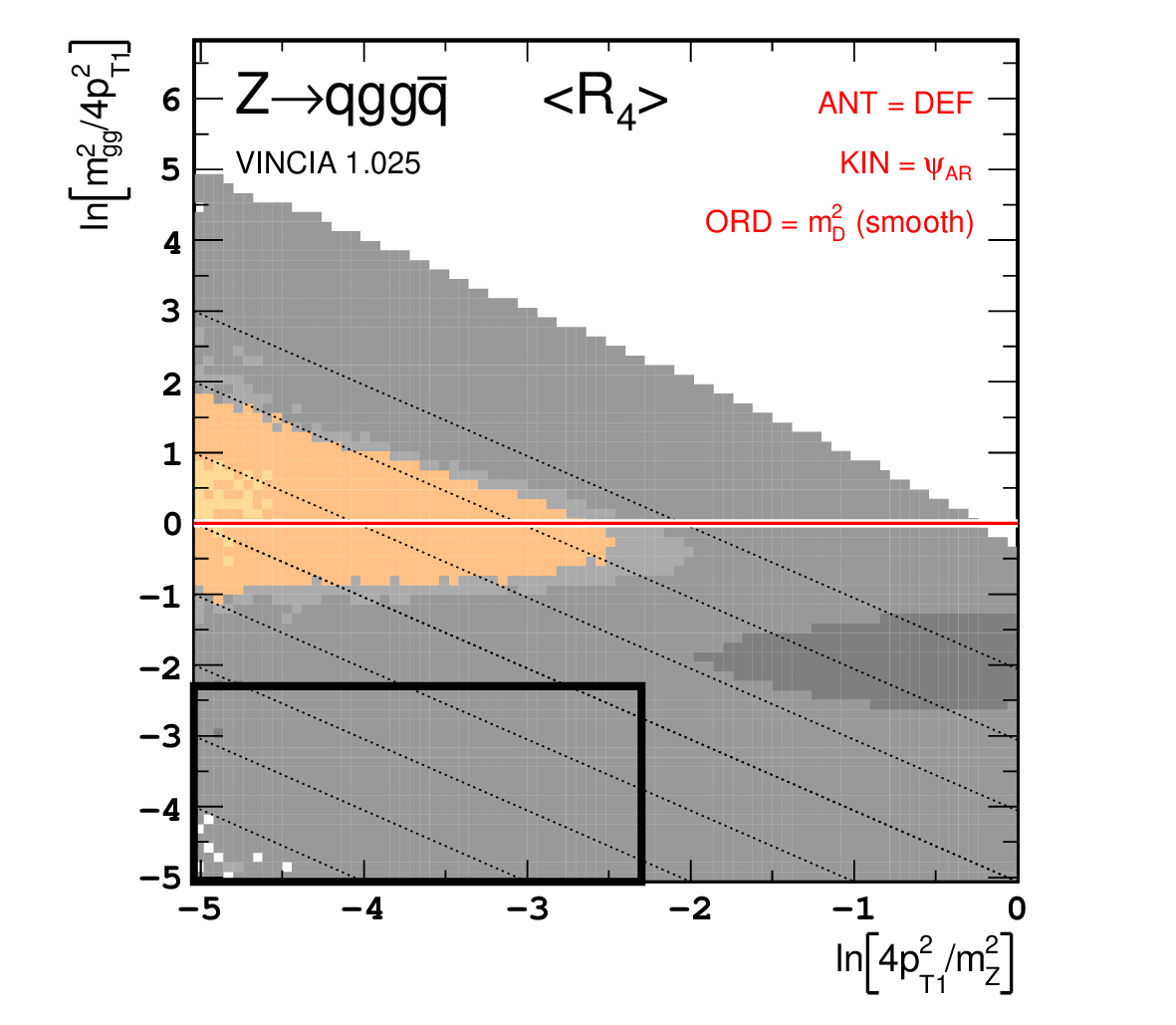}
{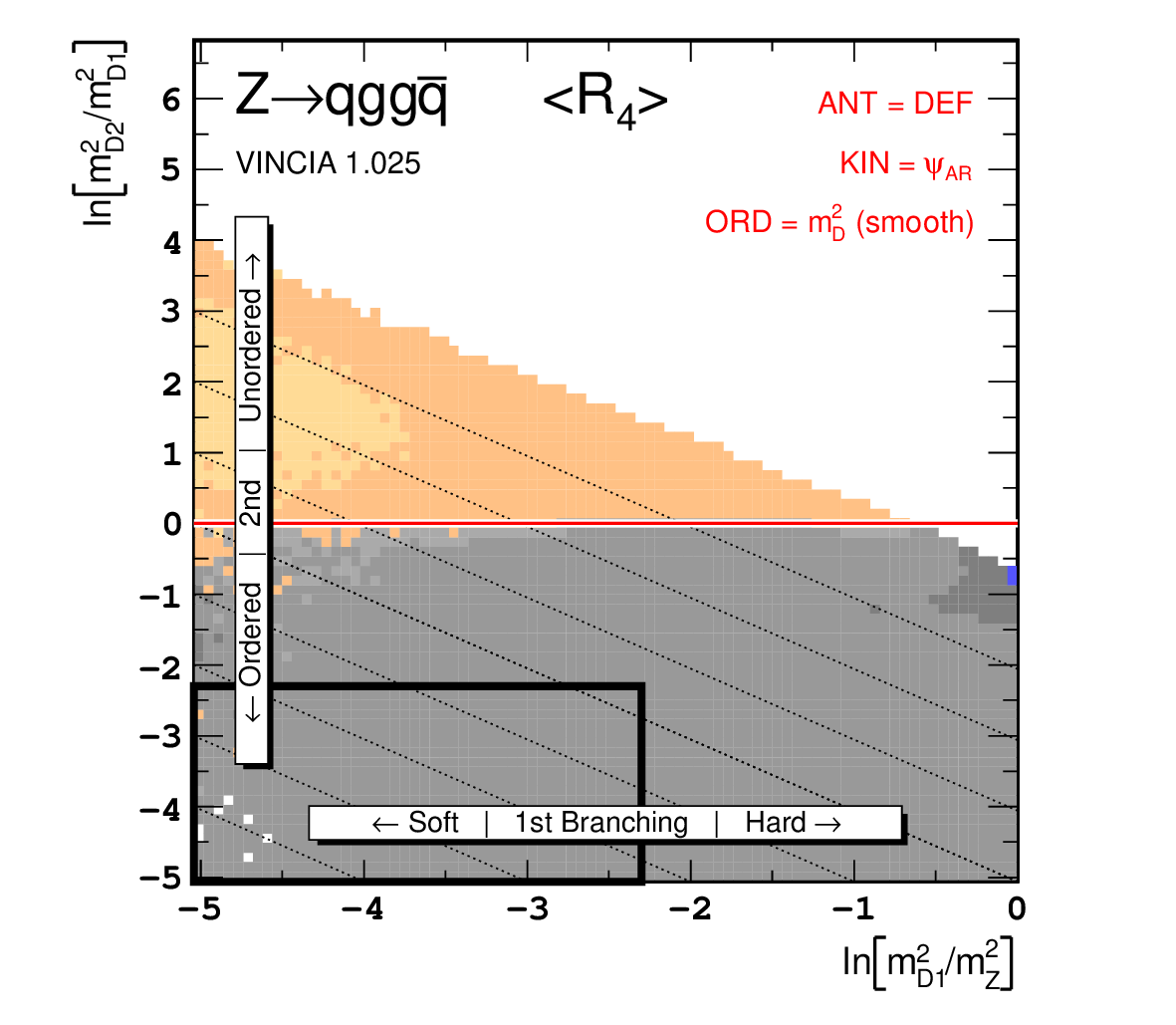}}
\capt{Mass-ordered antenna approximation compared to 2nd order
  QCD matrix elements, using \Vc's definition of $M_D$ applied as a 
  smooth suppression factor instead of the usual strong ordering
  condition. Although the dead zone has been removed without
  introducing a catastrophic over-counting, the results are less
  impressive than for \pT{}-ordering. Note: the default in \Vc\ is therefore to
  apply the smooth suppression factor in \pT{} regardless of the
  actual choice of evolution variable.}
\end{center}
\end{figure}

\section{Jeppsson Tune Parameters \label{app:tune}}

The following table gives an overview of the parameters used for
\Vc\ and \Py\ for the results obtained in this paper. The default
settings in \Py~8.145 were obtained by a fit to a large amount of LEP
data, using the \textsc{Rivet}+\textsc{Professor} framework
\cite{Buckley:2009bj,Buckley:2010ar}. The \Vc~1.025 parameters were
tuned manually, as reported on in \sect{sec:hadronization}.
The latter included the total charged particle multiplicity
and $\xi=-\log(x)$ 
   distributions, mean multiplicities of light mesons and baryons, event shapes 
   ($1-T$, Major, Minor, $C$, $D$, Oblateness), and jet resolution
   scales ($y_{23}$, $y_{34}$, $y_{45}$, $y_{56}$) extracted from the
   measurements contained in the 
   \textsc{HepData} \cite{Lafferty:1995jt} repository.
\begin{center}
\begin{tabular}{lcc}
\toprule
Parameter & \Py\ 8.145 Default & \Vc\ 1.025 \\
\cmidrule{1-3}
\multicolumn{3}{c}{Shower Parameters}\\
\cmidrule{1-3}
Evolution Variable & \pT{\mrm{evol}} & \pT{} \\
Renormalization Scale & \pT{\mrm{evol}} & \pT{} \\
$\alpha_s(M_Z)$ & 0.1383 & 0.139 \\
$\alpha_s$ loop order & 1 & 1\\
Shower Cutoff Variable & \pT{\mrm{evol}} & 2\pT{} \\
Shower Cutoff Scale in GeV & 0.5 & 1.0 \\
\cmidrule{1-3}
\multicolumn{3}{c}{String Breakup Parameters}\\
\cmidrule{1-3}
\texttt{StringZ:aLund} & 0.30 & 0.28 \\
\texttt{StringZ:bLund} & 0.80 & 0.55 \\
\texttt{StringPT:sigma} & 0.304 & 0.275 \\
\texttt{StringPT:enhancedFraction} & 0.01 & 0.01 \\
\texttt{StringPT:enhancedWidth} & 2.0 & 2.0 \\
\cmidrule{1-3}
\multicolumn{3}{c}{Flavor Parameters}\\
\cmidrule{1-3}
\texttt{StringFlav:probStoUD} & 0.19 & 0.21\\
\texttt{StringFlav:mesonUDvector} & 0.62 & 0.40\\
\texttt{StringFlav:mesonSvector} & 0.725 & 0.6\\
\texttt{StringFlav:probQQtoQ} & 0.09 & 0.079\\
\texttt{StringFlav:probSQtoQQ} & 1.0 & 1.0\\
\texttt{StringFlav:probQQ1toQQ0} & 0.027 & 0.035\\
\texttt{StringFlav:decupletSup} & 1.0 & 1.0\\
\texttt{StringFlav:etaSup} & 0.63 & 0.60\\
\texttt{StringFlav:etaPrimeSup} & 0.12 & 0.075\\
\bottomrule
\end{tabular}
\end{center}
\mciteSetSublistMode{n}
\bibliography{formalism}

\end{document}